\pdfoutput=1
\documentclass[12pt]{article}

\usepackage[top=80pt,bottom=85pt,left=60pt,right=60pt]{geometry}
\usepackage{amssymb,amsmath,xypic}
\usepackage{slashed}
\usepackage{graphicx,color,xcolor}
\usepackage{float}
\usepackage{sectsty}
\usepackage{cite}
\usepackage{amsmath,amsfonts,amsthm,amssymb,dsfont}
\usepackage{graphicx,wrapfig,lipsum}
\usepackage{esvect}
\usepackage[section]{placeins}
\usepackage{float}
\usepackage{multicol}
\usepackage{xcolor}
\usepackage{braket}
\usepackage{comment}
\usepackage{amsmath}
\usepackage{graphicx,bm,times}
\usepackage{enumitem}
\usepackage{amssymb}
\usepackage{float}
\usepackage{multicol}
\usepackage{braket}
\usepackage{esvect}
\usepackage{caption}
\usepackage{color}
\usepackage{slashed}
\usepackage{amsmath}
\usepackage{subcaption}
\usepackage{tikz}
\usetikzlibrary{3d}
\usepackage{rotating}
\usetikzlibrary{positioning}
\usetikzlibrary{arrows}
\newcommand{\midarrow}{\tikz \draw[-triangle 90] (0,0) -- +(.1,0);}

\usetikzlibrary{shapes.geometric}
\usetikzlibrary{arrows.meta}
\usetikzlibrary{decorations.pathreplacing,calligraphy}

\tikzset{
    process1/.style={rectangle, minimum width=1cm, minimum height=1cm, text width=1cm,text centered, draw=black},
    process2/.style={rectangle, minimum width=6cm, minimum height=1cm, text width=5cm,text centered, draw=black},
    decision/.style={diamond, text centered, draw=black, aspect=2, inner xsep=-2mm},
    stop/.style={rectangle, rounded corners, minimum width=3cm, minimum height=1cm,text centered, draw=black},
    arr/.style={thick,-stealth}
    }

\usepackage[debug,pageanchor=false]{hyperref}
\definecolor{link}{rgb}{.8,.15,.1}
\hypersetup{colorlinks=true,linkcolor=link,citecolor=link,urlcolor=link,linktocpage}

 \allowdisplaybreaks

\makeatletter
\@addtoreset{equation}{section}
\makeatother

\newcommand{\Vol}{\text{Vol}}
\newcommand{\beq}{\begin{equation}}
\newcommand{\eeq}{\end{equation}}
\newcommand{\bea}{\begin{eqnarray}}
\newcommand{\eea}{\end{eqnarray}}
\newcommand{\nn}{\nonumber}
\usepackage{rotating}

\begin{document}

\begin{titlepage}

\begin{center}

\vskip .5in %.3in 
\noindent

% A Web of $\mathcal{N}=1$ Solutions
{\Large \bf{Marginally deformed AdS$_5$/CFT$_4$ Backgrounds in Type IIB%: A web of $\mathcal{N}=1$ Solutions
}}

		\bigskip\medskip
Paul Merrikin\footnote{p.r.g.merrikin.2043506@swansea.ac.uk},\\

\bigskip\medskip
{\small  Department of Physics, Swansea University, Swansea SA2 8PP, United Kingdom }\vskip 3mm

		\vskip 1.5cm 
		\vskip .9cm %.6cm
		{\bf Abstract }

		\vskip .1in
	
	\noindent 
Multi-parameter families of $\mathcal{N}=0$ Type IIA and Type IIB AdS$_5$ solutions are presented, promoting to $\mathcal{N}=1$ in some special cases. The G-Structure description of each $\mathcal{N}=1$ solution is given, requiring an Abelian T-Duality of the G-Structure conditions and Pure Spinors. Investigations at the boundaries are performed for a two-parameter family of Type IIA and a three-parameter family of Type IIB solutions, finding the presence of orbifold singularities in some backgrounds. All parameters drop out of the Holographic Central Charge calculation, pointing to marginal deformations in the dual CFT description. 
\end{center}
\vskip .1in

\noindent

\noindent

\vfill
\eject

\end{titlepage}

\tableofcontents

\section{Introduction}
Since it's inception by Maldacena in 1997 \cite{Maldacena:1997re}, the AdS/CFT correspondence has proved very fruitful in the study of Supergravity backgrounds (with an AdS$_{d+1}$ factor) and their dual CFTs (of $d$ spatial dimensions). In this paper, we focus our attention on AdS$_5$ backgrounds of both Type IIA and Type IIB. 

The work presented in this paper is complementary to the recent work of \cite{us}, where a new two-parameter family of Type IIA solutions were derived via dimensional reduction of the Gaiotto-Maldacena backgrounds \cite{Gaiotto:2009gz}, following an $SL(3,\mathds{R})$ transformation amongst its three $U(1)$ directions.  In \cite{us}, it was shown that these solutions give rise to $\mathcal{N}=(2,1,0)$ backgrounds, depending on the choice of the parameters. Central to this analysis was the method of G-Structures, which we employ again in this work. In the cases of the $SU(2)\times U(1)$ preserving $\mathcal{N}=0$, and $U(1)\times U(1)$ preserving $\mathcal{N}=1$ solutions, it was found that these additional parameters change the quantization of D brane charge from integer to rational. Following careful study of the boundaries, this was nicely interpreted as branes back-reacted onto a Spindle and it's higher dimensional analogue, in the two cases respectively. The Holographic central charge demonstrated that these transformation parameters drop out neatly, suggesting that such deformations are in fact marginal in the dual CFT. Some stability analysis of the backgrounds was then presented.

\begin{figure}[h!]
   \centering
 \begin{subfigure}[b]{0.4\textwidth}
        \centering
       \begin{tikzpicture}[scale=0.8]
\draw[-stealth, line width=0.53mm] (-3.3,0)--(3.3,0) node[right ]{$\xi$};
\draw[-stealth, line width=0.53mm] (0,-3.3)--(0,3.3) node[above ]{$\zeta$};
%\draw[black] (7,0) node[above] {$\Longrightarrow$};
%\draw[black] (7,-0.6) node[above] {ATD};

%\draw[black] (0,-4.5) node[above] {Type IIA};
   \draw[red ] (2.7,0.65) node[above] {\textbf{D4}};
     \draw[red ] (1,0) node[above] {$\mathbf{S^2}$};
   \draw[->,line width=0.2mm,red](2.3,0.7)--(1.8,0.2);
  %  \draw[green!60!black  ] (-1.8,-2.5) node[above] {\textbf{NS5,~D6}};
       \draw[black ] (-2.2,-1.45) node[above] {\textbf{NS5 }};
    \draw[black ] (-1.8,-2.15) node[above] {\textbf{D6}$\in \mathds{Q}$};
     \draw[green!50!black ] (-1.2,-3) node[above] {\textbf{D6 }$\mathbf{\in\mathds{Z}}$};
%\draw[ blue,line width=0.93mm] (-6,-4.5)--(-5.5,-4.5);
%\draw (-5.5,-4.5) node[right]{$ \mathcal{N}=1$ U(1)$_R$ Preserving};
%\draw[ red,line width=0.93mm] (2,-4.5)--(2.5,-4.5);
%\draw  (2.5,-4.5) node[right]{$ \mathcal{N}=0$ SU(2) Preserving};
%\fill[violet!60!magenta] (-0.3,-4.5) circle(.1);
%\draw(-0.3,-4.5) node[right]{$~\mathcal{N}=2$};
%\draw[violet!60!magenta] (0,-0.5) node[left]{$\mathcal{N}=2$};
\clip (-3,-3) rectangle (3,3);
\begin{scope}[cm={0.5,-0.5,  50,50,  (0,0)}]  
%\draw[green!70!black,%loosely 
%dashed] (-6,-6) grid (6,6);
\draw[fill=green,opacity=0.07] (-6,-6) -- (-6,6) -- (6,6) -- (6,-6) -- cycle;
\end{scope}

\draw[blue,line width=0.35mm](-3,3)--(3,-3);
\draw[line width=0.5mm,red](-3,0)--(3,0);
\draw[line width=0.5mm,green!60!black ](0,-3)--(0,3);
\fill[violet!60!magenta] (0,0) circle(.2);
  \node[violet!60!magenta ,rotate=0] at (-1.1,-0.5) {$\mathcal{N}=2$};
\draw[green!60!black ] (1.45,1.45) node[above] {$\mathcal{N}=0$};
 \node[blue ,rotate=-45] at (1.9,-1.4) {$\zeta=-\xi$};
  \node[blue ,rotate=-45] at (1.25,-1.85) {$\mathcal{N}=1$};
\end{tikzpicture}
        \caption{Type IIA}
        \label{fig: TypeIIA}
    \end{subfigure}~~~~~~~~~
     \begin{subfigure}[b]{0.05\textwidth}
        \centering
       \begin{tikzpicture}[scale=0.8]
       \draw[black] (0,0) node[above] {$\Longrightarrow$};
	\draw[black] (0,-0.6) node[above] {ATD};
	\draw[black] (0,-3) node[above] {$~$};
       \end{tikzpicture}
       \caption*{}
           \end{subfigure}%
    ~~~~~~~~~~~~
       \begin{subfigure}[b]{0.4\textwidth}
        \centering
 \begin{tikzpicture}[scale=0.5,y={(1cm,0.5cm)},x={(-1cm,0.5cm)}, z={(0cm,1cm)}]
%\draw[black] (0,-9,0) node[above] {$~$};
%\draw[black] (0,0,-7) node[above] {Type IIB};
%\draw[fill=red,opacity=0.3] (-3,0,0) -- (3,0,0) -- (3,3,0) --
%(3,-3,0) -- cycle;
\draw[-stealth, line width=0.53mm] (0,0,-3.3)--(0,0,3.3) node[right ]{$\mathbf{\gamma}$};
\draw[green!60!black ] (2,2,2) node[above] {$\mathcal{N}=0$};
%\draw[green!60!black ] (2,-2.9,-2.5) node[above,rotate=-25] {\textbf{NS5, D7}};
\draw[black ] (2.4,-2.9,-2) node[above,rotate=-25] {\textbf{NS5}};
\draw[black ] (2,-2.9,-3) node[above,rotate=-25] {\textbf{D7} $\in \mathds{Q}$};
\draw[green!60!black ] (-1.8,0,-3) node[above,rotate=-25] {\textbf{D7} $\in \mathds{Z}$};
\draw[blue ] (-2.7,2,-2) node[above] {$\mathcal{N}=1$};
\begin{scope}[canvas is yx plane at z=0]
\draw[-stealth, line width=0.53mm] (-3.3,0)--(3.3,0) node[right ]{$\xi$};
\draw[-stealth, line width=0.53mm] (0,-3.3)--(0,3.3) node[above ]{$\zeta$};
\clip (-3,-3) rectangle (3,3);
\begin{scope}[cm={0.5,-0.5,  50,50,  (0,0)}]  
%\draw[green!70!black,%loosely 
%dashed] (-6,-6) grid (6,6);
\end{scope}
%\draw[blue,line width=0.35mm](-3,3)--(3,-3);
\draw[line width=0.5mm,red](-3,0)--(3,0);
%\fill[violet!60!magenta] (0,0) circle(.2);
%\draw[green!60!black ] (1.45,1.45) node[above] {$\mathcal{N}=0$};
 %\node[blue ,rotate=-45] at (1.25,-1.75) {$\zeta=-\xi$};
\end{scope}
\begin{scope}[canvas is yx plane at z=-1]
\draw[blue,line width=0.4mm](-3,3)--(3,-3);
 \node[blue ,rotate=0] at (3,-1.5) {$\mathbf{\zeta=-\xi}$};
  \node[blue ,rotate=0] at (-1.75,-0.15) {$\gamma=-1$};
\end{scope}

\draw[fill=red,opacity=0.2] (0,-3,-3) -- (0,3,-3) -- (0,3,3) -- (0,-3,3) -- cycle;
\draw[fill=green,opacity=0.075] (-3,0,-3) -- (3,0,-3) -- (3,0,3) -- (-3,0,3) -- cycle;
%\draw[fill=green,opacity=0.1] (-3,-3,-3) -- (-3,3,-3) -- (-3,3,3) -- (-3,-3,3)  --  cycle;

\draw[fill=green,opacity=0.05] (-3,-3,-3) -- (3,-3,-3) -- (3,-3,3) -- (-3,-3,3) -- cycle;
\draw[fill=green,opacity=0.05] (-3,-3,-3) -- (-3,-3,3) -- (-3,3,3) -- (-3,3,-3) -- cycle;
\draw[fill=green,opacity=0.05] (-3,-3,3) -- (-3,3,3) -- (3,3,3) -- (3,-3,3) -- cycle;
\begin{scope}[canvas is yz plane at x=1.5]
\coordinate (Origin) at (0,0);
\shade[ball color=red] (Origin) circle (0.05cm);
 % \node[blue ,rotate=30] at (-1,0) {D5};
\end{scope}
\begin{scope}[canvas is yz plane at x=0]
  \node[red ,rotate=20] at (2,2) {\textbf{D5}};
   \node[red ,rotate=0] at (1.5,0.5) {$\mathbf{S^2}$};
\end{scope}
\end{tikzpicture}
        \caption{Type IIB}
        \label{fig: TypeIIB}
    \end{subfigure}
 \caption{In (a), for arbitrary values of $(\xi,\zeta)$ represented in green, all SUSY is broken. In this generic situation, there are NS5 and D6 branes present, with rational D6 charge. Along the $\xi=0$ axis, given in dark green, this rational D6 charge becomes integer. The point $(\xi,\zeta)=(0,0)$, in purple, recovers the infinite family of $\mathcal{N}=2$ solutions (with an $SU(2)_R\times U(1)_r$ R-symmetry). When $\zeta=-\xi$, along the blue line, $\mathcal{N}=1$ SUSY is preserved (with a $U(1)_R$ R-symmetry). Along the $\zeta=0$ axis (given in red), the SUSY is still broken, however one recovers the $SU(2)$ isometry ($S^2$). Only along this axis are D4 branes present, recovering an integer charge via a gauge transformation!\\
   In (b), following an ATD to IIB, an additional parameter, $\gamma$, is picked up. For arbitrary values of $(\xi,\zeta,\gamma)$, given in green, the SUSY is still broken. Now there are NS5 and D7 branes, with rational D7 charge (now becoming integer along the $\xi=0$ plane). There are no $\mathcal{N}=2$ solutions present here. The $\mathcal{N}=1$ SUSY solutions, along the blue line, require the additional condition $\gamma=-1$ and so is now offset from the $\gamma=0$ plane. The $SU(2)$ isometry is preserved only along the $(\zeta,\gamma)=(0,0)$ axis, in red, with SUSY still broken. The D5 branes exist only along the $\zeta=0$ plane in red (so do not require a preserved $S^2$), and have an integer charge recovered by a gauge transformation. Only where the red plane and blue line intersect do we have $\mathcal{N}=1$ solutions with D5 branes, namely when $\gamma=-1,~(\xi,\zeta)=(0,0)$.}

    \label{fig:MainFigure}
\end{figure}
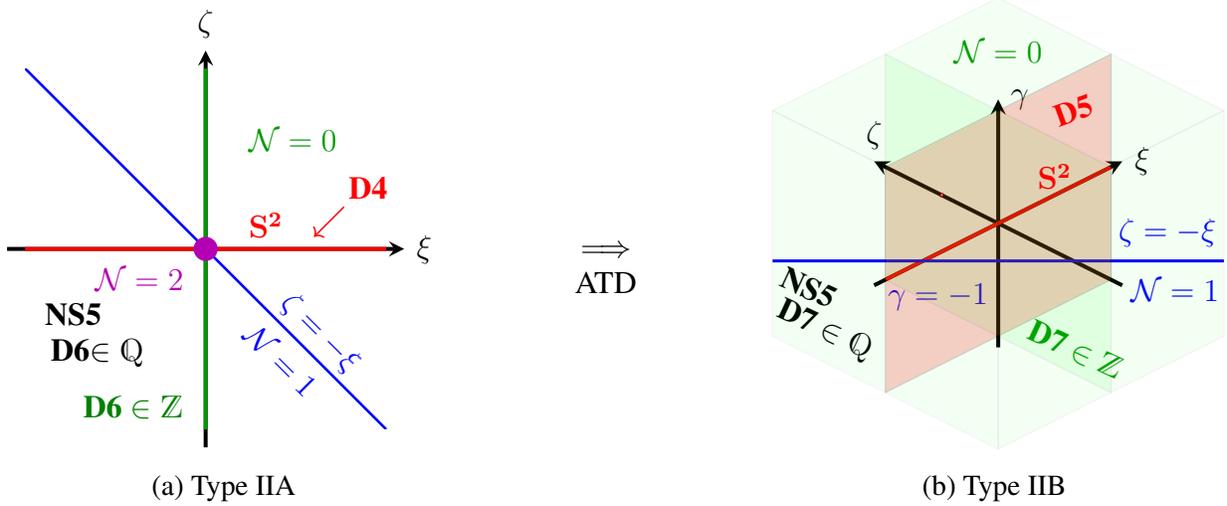

In this work, we begin with the boundary analysis given in \cite{us}, but for the full two-parameter family of solutions. We will show that both NS5 and D6 branes are present in all backgrounds, but D4 branes only appear for a preserved $S^2$. In addition, by introducing a new term to the large gauge transformation of $B_2$, we are able to eliminate the effect of the Spindle on the D4 charge, restoring an integer quantization. This is of course not possible for the D6 branes. However, as we will show in this paper, the D6 charge has the same rational form for all of the solutions - becoming integer when one of the parameters, $\xi$, is fixed to zero. See Figure \ref{fig: TypeIIA} for a plot of the two parameters, summarising these solutions nicely.

We then move on to investigate Type IIA backgrounds which are derived via dimensional reductions along the other two $U(1)$ directions with respect to the case of \cite{us}, and see that this gives rise to a few new and unique solutions. Amongst these is a one-parameter family of $\mathcal{N}=0$ solutions which enhances to a new and unique zero-parameter $\mathcal{N}=1$ solution, when the parameter is fixed to zero. The G-Structure description of this background is also given. 

The main focus of this work then follows, investigating the Abelian T-Duality (ATD) of the two-parameter family of Type IIA solutions given in \cite{us}. This derives a three-parameter family of Type IIB solutions, picking up the additional non-trivial parameter in the T-Duality.  This three-parameter solution contains within it a new one-parameter family of $\mathcal{N}=1$ Type IIB backgrounds. Throughout this analysis, the supersymmetry is kept track of by G-Structures. The transformation generating new backgrounds involved performing an ATD of the G-Structure forms and conditions. This derived new expressions for the Pure Spinors in Type IIB, playing a central role in the analysis of the $\mathcal{N}=1$ solutions (presented in detail in Appendix \ref{sec:GStructureCalcs}).  It is worth noting here that this infinite family of solutions have zero five-form flux, which is an interesting result given that very few supersymmetric solutions are known with this property. The first two examples of such backgrounds were found in \cite{Macpherson:2014eza}, evading the prior classification of AdS$_5$ solutions considered in \cite{Gauntlett:2005ww} (which had a non-vanishing five-form flux). This led to the work of \cite{Couzens:2016iot}, where the classification was completed.%\footnote{I thank Christopher Couzens for this observation.}.}

We go on to investigate the boundaries of the three parameter family, and see the presence of certain orbifold singularities. However, as a consequence of the ATD from type IIA, they do not appear to be as straightforward as the Spindles found in \cite{us}. It would be interesting to study this topology further in the future. In analogy with the IIA analysis, we find the presence of NS5 and D7 branes in all backgrounds. In the case of the D5 branes, they only exist in the T-Dual solutions of the $S^2$ preserved IIA backgrounds. However, if the $S^2$ is broken under the ATD itself, the D5 branes still remain. Once again, by including an additional term in the gauge transformation of $B_2$, one can recover integer quantization of D5 charge. See Figure \ref{fig: TypeIIB} for a plot of the three parameters, summarising the above discussion.

Finally, some additional Type IIB backgrounds are presented corresponding to ATDs of different dimensional reductions. Amongst these we find a new and unique zero-parameter family of $\mathcal{N}=1$ solutions (corresponding to the ATD of the unique $\mathcal{N}=1$ Type IIA mentioned above). We again note that this background is another supersymmetric example with zero five-form flux. The G-Structure forms are again included. Some topics are left for future study, including the stability analysis conducted in \cite{us}. By deriving the Holographic central charge, we note that the deformations presented throughout this paper are again marginal in the dual CFT, but do not recover the $\mathcal{N}=2$ solution in Supergravity. 

The content of this paper is as follows:
\begin{itemize}
\item In Section \ref{sec:Review} we review the $\mathcal{N}=2$ Gaiotto-Maldacena solutions in M-Theory. We include a brief summary of some relevant G-Structure theory, setting up a discussion of the G-Structures specific to the Gaiotto-Maldacena solutions, it's IIA reductions and subsequent IIB ATD. The new material here is the 11D G-Structure forms corresponding to a simple dimensional reduction along the other two $U(1)$ directions to \cite{us} (namely, $\chi$ and $\phi$), and the IIB G-Structure Pure Spinors and conditions derived in Appendix \ref{sec:GStructureCalcs}. We conclude the section with a brief review of the $SL(3,\mathds{R})$ transformation being performed in M-Theory, along with the two-parameter family presented in \cite{us}. We then perform the boundary analysis for the full two-parameter family, finding the presence of NS5 and D6 branes in all cases, with the D4 branes only appearing for a preserved $S^2$.
\item In Section \ref{sec:furtherIIA} we present some more IIA results, derived via dimensional reductions along the alternative $U(1)$ directions. Most notable is the zero-parameter $\mathcal{N}=1$ background. These results are largely just presented, further analysis is left to future study. However, the G-Structure forms are included for the $\mathcal{N}=1$ solution. The full calculations are given in Appendix \ref{sec:DimRed}. 
\item In Section \ref{sec:IIB}, we present the three-parameter family of Type IIB solutions. We then study the boundary of the $(\sigma,\eta)$ space, following the analysis which gave rise to the Spindle interpretation given in \cite{us}. We again find the presence of orbifold singularities, the topology of which would be nice to study further. We find analogous behaviour with the IIA backgrounds, with the existence of NS5 and D7 branes in all solutions. The D5 branes are only present for descendants of the $S^2$ preserved IIA solutions, but do not require a preserved $S^2$ themselves. We then present the one-parameter family of $\mathcal{N}=1$ solutions and zone in on a one-parameter family of $S^2$ preserved $\mathcal{N}=0$ backgrounds. In these examples, the charge quantization is broken without the presence of an orbifold singularity in the metric, a possible physical interpretation of this is suggested in Appendix \ref{sec:RotatingBranes} - corresponding to rotated D-branes. Following this, some additional IIB solutions are given, including a zero-parameter $\mathcal{N}=1$ background. The G-structure forms and conditions are derived and presented at length in Appendix \ref{sec:GStructureCalcs}. Once again, further analysis is needed for these backgrounds, this is left for future work. The full ATD calculations are given at length in Appendix \ref{sec:ATD}, and along the way, the TST solution of \cite{Nunez:2019gbg} is re-derived (and argued to be $\mathcal{N}=0$ in general, promoting to $\mathcal{N}=1$ when $\gamma=-1$). Finally in Appendix \ref{sec:TSTGM}, a one-parameter M-Theory deformation of the Gaiotto-Maldacena solution is given, and believed to break the supersymmetry completely. This is derived by reducing to IIA, performing a TST transformation and uplifting.  It proved highly useful to perform an ATD and TST once for a general form, allowing one to easily derive the ATD solutions of many backgrounds at once. These forms are derived and presented in Appendix \ref{sec:ATD}.
\item Finally, Section \ref{sec:conclusions} gives some conclusions and an outline of future analysis required. This work is complimented by lengthy Appendices which will hopefully be of use.
\end{itemize}

%\newpage

%%%%%%%%%%%%%%%%%%%%%%
%%%%%%%%%%%%%%%%%%%%%%
%%%%%%%%%%%%%%%%%%%%%%

   \section{Review of Type IIA reductions of Gaiotto-Maldacena }\label{sec:Review}
   
   We begin with a brief review of the $\mathcal{N}=2$ AdS$_5$ class of solutions found by Gaiotto-Maldacena (GM)  in $d=11$ \cite{Gaiotto:2009gz}, along with it's G-Structure description. The method of G-Structures allows one to easily keep track of the supersymmetry under reduction, a feature which will be further exploited throughout this paper. We then proceed to review the $SL(3,\mathds{R})$ transformations and the corresponding two parameter family of dimensional reductions discussed in \cite{us}, finishing the section by investigating the boundary.
   \subsection{Gaiotto-Maldacena Background}
         If one begins with the more general AdS$_5$ $\mathcal{N}=2$ solutions of Lin-Lunin-Maldacena (LLM) \cite{Lin:2004nb}, with bosonic isometry group $SO(4,2)\times SU(2)_R\times U(1)_r$\footnote{Where $SU(2)_R\times U(1)_r$ is the $\mathcal{N}=2$ R-Symmetry.} and magnetic four-form, $G_4=dC_3$, one can derive the following Electrostatic form of the Gaiotto-Maldacena Background via the B$\ddot{\text{a}}$cklund transformation\footnote{Noting that the GM U(1) directions $(\chi,\beta)$ and the LLM U(1) directions $(\bar{\chi},\bar{\beta})$ are related as follows
 \beq
 (\chi,\beta)\rightarrow (\bar{\chi}+ \bar{\beta},~ -\bar{\beta}).
 \eeq   
 This is discussed in \cite{Bah:2019jts,Bah:2022yjf,Couzens:2022yjl} and shown in detail in \cite{us}.}. See \cite{us} for a more in depth review. The resulting metric and four form are then defined as follows
 
 \begin{equation}\label{eqn:GM}
\hspace{-1cm}
\begin{gathered}
ds_{11}^2=f_1\Bigg[4ds^2(\text{AdS}_5) +f_2ds^2(\text{S}^2)+f_3d\chi^2+f_4\big(d\sigma^2 +d\eta^2\big)+f_5\Big(d\beta +f_6d\chi\Big)^2\Bigg],\\
C_3=\Big(f_7d\chi +f_8 d\beta \Big)\,\wedge\text{Vol}(\text{S}^2),\\
%ds^2(AdS_5) = e^{2\rho}dx_{1,3}^2+d\rho^2,~~~~~~~~~~~~~ds^2(S^2)=d\theta^2+\sin^2\theta d\bar{\phi}^2,~~~~~~~~~~~~~~~d\Omega_2=\sin\theta d\theta \wedge d\bar{\phi},
        \\f_1=\kappa^{\frac{2}{3}}\bigg(\frac{\dot{V}\tilde{\Delta}}{2V''}\bigg)^{\frac{1}{3}},~~~~f_2 = \frac{2V''\dot{V}}{\tilde{\Delta}},~~~~~~~~f_3=\frac{4\sigma^2}{\Lambda},~~~~~~~~f_4 = \frac{2V''}{\dot{V}},~~~~~~~~f_5=\frac{2\Lambda V''}{\dot{V}\tilde{\Delta}},
      \\f_6=\frac{2\dot{V}\dot{V}'}{V''\Lambda},~~~~f_7=-\frac{4\kappa \dot{V}^2V''}{\tilde{\Delta}},~~~~~~~f_8=2\kappa\bigg(\frac{\dot{V}\dot{V}'}{\tilde{\Delta}}-\eta\bigg),
        \\ \tilde{\Delta} =\Lambda(V'')^2+(\dot{V}')^2,~~~~~~~~~\Lambda=\frac{2\dot{V}-\ddot{V}}{V''},
\end{gathered}
\end{equation}
where the warp factors $f_i=f_i(\eta,\sigma)$ depend on a constant $\kappa$ and the function $V(\eta,\sigma)$, with 
        \begin{equation}
\dot{V}\equiv \sigma \partial_\sigma V,~~~~~~~~ V' \equiv \partial_\eta V.
    \end{equation}
    %The Supergravity equations are satisfied by imposing 
    The above class of solutions are defined by
    the following cylindrically symmetric Laplace equation for $V(\eta,\sigma)$
    \begin{equation}\label{eqn:laplace}
\frac{1}{\sigma}\partial_\sigma (\sigma \partial_\sigma V)+\partial_\eta^2\equiv\ddot{V}+\sigma^2V''=0,
\end{equation}
  along with the Boundary Conditions \cite{Nunez:2019gbg}  (where $0<\sigma<\infty$ and  $0<\eta<P$, with $P$ finite)
\begin{equation}\label{eqn:BCs}
\dot{V}\Big|_{\eta=0,P} =0,~~~~~~~~~~~\dot{V}\Big|_{\sigma=0} =\mathcal{R}(\eta).
\end{equation}
One now has an Electrostatic description of the background \eqref{eqn:GM}, where a line of charge is extended along $\eta$. The charge density is defined by the ``Rank Function'' of the dual quiver, $\mathcal{R}(\eta)$, with Fourier decomposition 
\begin{equation}\label{eqn:rank}
\mathcal{R}(\eta) = \sum_{n=1}^\infty \mathcal{R}_n \sin\bigg(\frac{n\pi}{P}\eta\bigg),
\end{equation}
which for quantized fluxes, must be a continuous, linear-by-pieces and convex function. Hence, in general, it takes the form
\begin{equation}\label{eqn:Rkoriginal}
\mathcal{R}(\eta) =
    \begin{cases}
     ~~~~~~~~~~~~~~~ N_1\eta & \eta\in[0,1]\\
      N_k+(N_{k+1}-N_k)(\eta-k) & \eta\in[k,k+1]\\
   ~~~~~~~~~   N_{P-1}(P-\eta)  & \eta\in[P-1,P]
    \end{cases}      
\end{equation}
where $\mathcal{R}(0)=\mathcal{R}(P)=0$. One can find an appropriate solution to \eqref{eqn:laplace} using the separation of variables, 

\begin{equation}\label{eqn:potential}
V(\sigma,\eta) =-\sum_{n=1}^\infty \mathcal{R}_n \sin\bigg(\frac{n\pi}{P}\eta\bigg) K_0\bigg(\frac{n\pi}{P}\sigma\bigg),~~~~~~~~~~~\mathcal{R}_n=\frac{1}{P}\int_{-P}^P \mathcal{R}(\eta)\sin\bigg(\frac{n\pi}{P}\eta\bigg)d\eta ,
\end{equation}
 which indeed satisfies the boundary conditions \eqref{eqn:BCs}. Here, $K_0$ is a modified Bessel function of the second kind. 
 
 It proves useful when investigating the boundary to introduce an alternative form for $\dot{V}$, following the arguments of \cite{us}, and shown to be equivalent in \cite{Reid-Edwards:2010vpm}
\begin{align}
\dot{V}&=\frac{\pi}{P}\sum_{n=1}^{\infty}{\cal R}_n\sigma\sin\bigg(\frac{n\pi}{P}\eta\bigg) K_1\bigg(\frac{n\pi}{P}\sigma\bigg),\label{eq:alternative}\\[2mm]
&=\frac{1}{2}\sum_{m=-\infty}^{\infty}\sum_{k=1}^Pb_k\left(\sqrt{\sigma^2+(\eta-2m P+k)^2}-\sqrt{\sigma^2+(\eta-2m P-k)^2}\right),\nn
\end{align}
where $b_k=2N_k-N_{k+1}-N_{k-1}$ (see \cite{us} for a more in depth discussion).\\\\
 We now summarise some key material on the method of G-Structures.
 \subsection{G-Structure description and review}
 In order to easily keep track of the supersymmetry when dimensionally reducing $\mathcal{N}=2$ Gaiotto-Maldacena to Type IIA, it proves highly useful to use the method of G-Structures. We now provide a brief review, with a focus on the GM description used throughout this paper. This will set up the next section, where we discuss the G-Structure formalism for the Gaiotto-Maldacena background, it's IIA reduction and subsequent IIB T-Dual solution.\\
 
 %\subsubsection*{G-Structure Review}
 
% \textcolor{red}{
% \begin{itemize}
 %\item Basic principle - spinor vs forms
% \item $SU(d/2)$ in $d$ even dimensions - SU(2), SU(3) in 6 and 7 d
% \item $SU(3)\times SU(3)$
% \item bilinears
% \end{itemize}
% }
 
Supersymmetric solutions of M-Theory or Type II Supergravity must satisfy the necessary supersymmetry conditions, usually given in terms of a spinor and metric, as well as the Bianchi identities\footnote{Where we follow the convention that $d_H$ is defined with a minus, as follows
 \begin{equation*}
 d_H\equiv d-H\wedge\, .
 \end{equation*}}
 \beq\label{eqn:SUSYconditions}
 \begin{gathered}
 \partial_\epsilon \psi_M=0,~~~~~~~~~~~~~~~~~~~~~\partial_\epsilon \lambda=0,\\
 d_HF=0,~~~~~~~~~~~dH_3=0,
 \end{gathered}
 \eeq
 where $\psi_m$ are the Gravitinos, $\lambda$ the Dilatini and $\epsilon^a$ are two 11D Majorana-Weyl spinors (with $a=(1,2)$). %\textcolor{red}{Include the exact equations here?    Mention the M Theory Bianchi above...  Bilinear vs Pure spinor???}

The technique of G-structures allows one to recast the above supersymmetry equations (involving a spinor and metric) in terms of non-spinorial and geometrical objects (forms). For our purposes, we will focus on the supersymmetry conditions for Gaiotto-Maldacena and it's daughter Supergravity solutions, namely Mink$_4\times M_7$ and Mink$_4\times M_6$ backgrounds. The interested reader is directed to \cite{Grana:2005sn,Grana:2005jc,Gauntlett:2004zh,Kaste:2003zd,Gauntlett:2003cy,TomTextbook,Grana:2006kf,TomsNotes} for more detailed and insightful discussions on the topic.

%\subsubsection*{Supergravity}
%\begin{itemize}
%\item \textbf{Mink$_4\times $M$^7$ M-Theory}\\
%\subsubsection*{Mink$_4\times $M$^7$ M-Theory}
In this discussion, we are interested in the internal %(compact)
 space, writing the metric as a Warp product. In the Mink$_4\times M_7$ M-Theory case, we have
\beq\label{eqn:MTheorymetric}
ds_{11}^2=e^{2\hat{A}}ds^2(\text{Mink}_4)+ds^2(M_7),
\eeq
where $\hat{A}$ is a function of the internal space, $M_7$, coordinates. We decompose the spinors on the %non-compact/compact 
external/internal parts of the space, as $\epsilon = \psi_+\otimes \theta_++\psi_-\otimes \theta_-$ (following the notation of \cite{Kaste:2003zd})
%\beq
%\eeq
where $\theta$ are 7 dimensional spinors, with $\theta_+=\theta_-^*$.  %\textcolor{red}{add more details and references etc, redefine theta as eta so consistent?? }
%\item \textbf{Mink$_4\times $M$^6$ Supergravity}\\
%Before returning to this discussion, we first review the Supergravity case. 
%\subsubsection*{Mink$_4\times $M$^6$ Supergravity}

In the case of Mink$_4\times M_6$ Supergravity, we have
\beq\label{eqn:M6metric}
ds_{10}^2=e^{2A}ds^2(\text{Mink}_4)+ds^2(M_6),
\eeq
where $A$ is a function of the $M_6$ coordinates. In addition, the fields depend only on the internal space coordinates, with the Poly-form
\beq\label{eqn:relstuff}
\begin{gathered}
F=g+e^{4A}\text{Vol}(\text{Mink}_4)\wedge *_6\lambda(g),\\
\text{Type IIA: }~~~~~~~~~F_{tot} = \sum_{j=0}^5F_{2j},~~~~~~~~~~~~~\lambda(g)=g_0-g_2+g_4-g_6,\\
\text{Type IIB: }~~~~~~~~~F_{tot} = \sum_{j=0}^4F_{2j+1},~~~~~~~~~\lambda(g)=g_1-g_3+g_5.~~~~~~~~~~
\end{gathered}
\eeq
%with the Bianchi identities becoming
%\beq
%dH_3=0,~~~~~~~~~~~~d_Hg=0,~~~~~~~~~~~~~~~~d_H(e^{4A}*_6g)=0
%\eeq
%\textcolor{red}{check the last equation here..... should there be a lambda and also its a plus in the tomaselo notes not a minus?!}\\
Decomposing the spinor yields\footnote{In IIA, $\epsilon^1$ has + chirality and $\epsilon^2$ has - chirality. In IIB, they are both + chirality.} (adopting the notation given in \cite{Grana:2005sn})
%\beq
%\epsilon =\zeta \otimes \eta
%\eeq
\beq
\begin{gathered}
\text{Type IIA: }~~~~~~~~~\epsilon^1=\zeta_+ \otimes \eta_+^1+\zeta_-\otimes \eta_-^1,~~~~~~~~~~~~~~~~~~~~~
\epsilon^2=\zeta_+ \otimes \eta_-^2+\zeta_-\otimes \eta_+^2,\\
\text{Type IIB: }~~~~~~~~~\epsilon^i=\zeta_+ \otimes \eta_+^i+\zeta_-\otimes \eta_-^i,
\end{gathered}
\eeq
% \textcolor{red}{add more details here.......}
where $\eta$ lives on $M_6$. Plugging these decompositions back into the SUSY conditions given in \eqref{eqn:SUSYconditions}, allows one to factor out $\zeta$ - leading to six fairly ugly equations in terms of $\eta^{1,2}$. One can however re-write such conditions in a more elegant manner in the language of generalized complex geometry and differential forms. %The spinor $\eta$ naturally defines a G-Structure as it's stabilizer group (the group of rotations which keep $\eta$ invariant). In addition, a G-Structure is typically equivalent to a set of invariant tensors, allowing one to swap the spinor with non-spinorial objects (forms) - this is the upshot of the method of G-Structures. 
\\%\textcolor{red}{define a G-Structure somewhere? maybe at the start of the section....}
%\textcolor{red}{Mention torsion stuff}\\
%\textcolor{red}{In the absence of flux}, 
The supersymmetry conditions can be split into 2 parts (see for example \cite{Grana:2005sn} for a nice discussion)
\begin{itemize}
\item \textbf{Algebraic part}: %The existence of a non-vanishing, globally well defined spinor, $\eta$. 
This is a topological requirement on the manifold\footnote{That $T\otimes T^*$ must have an $SU(3)\times SU(3)$ structure.}, and implies the existence of two nowhere vanishing, globally defined Clifford$(6,6)$ pure spinors, $\Psi_+,\Psi_-$
\beq
\Psi_+\equiv \eta_+^1 \otimes \eta_+^{2\dagger},~~~~~~~~~~~~~~~\Psi_-\equiv \eta_+^1 \otimes \eta_-^{2\dagger}.
\eeq
These are sums of even or odd forms, respectively.
\item \textbf{Differential part}: %The spinor, $\eta$, is covariantly constant, defining the Killing Spinor equation. The amount of supersymmetry depends on the number of spinors satisfying this equation, with the R-Symmetry of the theory (on the non-compact space) rotating the Killing Spinors into each other. 
The preservation of supersymmetry then imposes differential conditions on the metric, saying that the two pure spinors should satisfy %(see \cite{Grana:2005sn})
\beq\label{eqn:diffcons}
e^{-2A+\Phi}d_H(e^{2A-\Phi}\Psi_1)=0,~~~~~~~~~~~~~~~~~~e^{-2A+\Phi}d_H(e^{2A-\Phi}\Psi_2)=dA\wedge \bar{\Psi}_2+F,
\eeq
with an additional normalization requirement. The more supersymmetry, the more structure is required on the manifold. These conditions contain the same information as the supersymmetry variations from \eqref{eqn:SUSYconditions}%. Once again, to find a vacuum, they 
, and must be supplemented with the Bianchi identities and equations of motion for the fluxes.\\
 Writing this more schematically as in \cite{Barranco:2013fza}, we have
 
 \beq
 \begin{gathered}
 d_{H_3}\Psi_1=0,~~~~~~~~~~d_{H_3}\Psi_2=F_{RR},%~~~~~~~~~~~~~d_{H_3}\equiv d+H_3\wedge ,
 \end{gathered}
 \eeq
 with 
 \beq
 \text{Type IIA:}~~~~~~\Psi_1=\Psi_+,~\Psi_2=\Psi_-,~~~~~~~~~~~~~~~~~~~~  \text{Type IIB:}~~~~~~\Psi_1=\Psi_-,~\Psi_2=\Psi_+,
 \eeq
given that $F_{RR}$ has an even form in IIA and an odd form in IIB. Clearly the pure spinors must transform in essentially the same manner under T-Duality as the Ramond fields, this will become important to us later. 
\end{itemize}

\subsubsection*{Example Structures}

We now summarise some key (and relevant) examples. See \cite{Grana:2005sn}\cite{TomTextbook}\cite{Grana:2006kf} for further details.\\\\
Having two spinors, $\eta^{1,2}$, describes an $SU(3)\times SU(3)$ structure. When $\eta^1$ and $\eta^2$ are parallel, they describe an $SU(3)$ structure - now defined by a single nowhere vanishing spinor $\eta$ (which is covariantly constant in the case of a Calabi-Yau 3-fold). When they are nowhere parallel, they define what is called a static $SU(2)$ structure.

\begin{itemize}
\item \textbf{$SU(3)$ Structure}\\
An $SU(3)$ structure on an $M_6$ can be defined in three ways: via a metric and spinor $(g,\eta)$; a metric and a complex, decomposable and non-degenerate 3-form $(g,\Omega)$; or a real 2-form and the complex 3-form $(J,\Omega)$. Hence there is a bijection between the descriptions, notably $(g,\eta)\leftrightarrow (J,\Omega)$\footnote{Using the Fierz identities, one gets the following mapping
\begin{equation*}
\Omega_{mnp}=-\eta_-^\dagger \gamma_{mnp}\eta_+,~~~~~~~\eta_-=(\eta_+)^c,~~~~~~~~~~~~J_{mn}=-i\,\eta_+^\dagger \gamma_{mn}\eta_+.
\end{equation*}
%\textcolor{red}{Be clearer on this and with the middle equation....}
}. The following conditions are necessary
\beq\label{eqn:JOmegacondition}
J\wedge \Omega=0,~~~~~~~~~~~~J\wedge J\wedge J=\frac{3}{4}i\,\Omega \wedge \bar{\Omega},
\eeq
with the positive definite metric defined as $g=-J I_\Omega$.\footnote{Where $I_\Omega$ is an Almost Complex Structure - a tensor $I^m_n$ with $I^2=-1$. This allows one to embed the $GL(d/2,\mathds{C})\in GL(d,\mathds{R})$. The stabilizer group is given as follows
\begin{equation*}
\text{Stab}_{GL(d,\mathds{R})}(I)=GL(d/2,\mathds{C}),~~~~~~~~~~~~\text{Stab}(I)=O(d)\,\cap \,GL(d/2,\mathds{C}) \cong U(d/2)
\end{equation*}
so adding a metric on $M_6$ means the stabilizer group of $I$ is $U(3)$. One then defines $J\equiv gI$, or alternatively, $g=-JI$. To move to an $SU(3)$ one requires a nowhere vanishing holomorphic 3-form $\Omega$, with $\Omega\rightarrow \text{det}(U)\Omega$ and $det(U)=1$. See \cite{TomTextbook} for further details.}

One can then build the normalized pure spinors, as follows
\beq\label{eqn:SU(3)spinors}
\Psi_+=\eta_+\otimes \eta_+^\dagger =\frac{1}{8}e^{-iJ},~~~~~~~~~~~~~~~~~~~~\Psi_-=\eta_+\otimes \eta_-^\dagger = -\frac{i}{8}\Omega.
\eeq

To generalize this to an $SU(3)$ structure on an $M_7$, one must add a real one-form, $K$, where
\beq\label{eqn:ds7defn}
ds_7^2=ds_6^2+K^2,
\eeq
so an $SU(3)$ structure on an $M_7$ is described by $(J,\Omega,K)$. This is the case for Gaiotto-Maldacena, we will return to this case in the next section.

\item \textbf{$SU(d)$ in $d=$ even Structure }\\
More generally, this is extended to $SU(d/2)$ in $d=$ even dimensions. Here the structure is specified by a real 2-form $J$ and a complex $d/2$ form, $\Omega$ 
\beq\label{eqn:JOmegadefs}
\Omega=E^1\wedge ...\wedge E^{\frac{d}{2}},~~~~~~~~~~~~~~~~J=\frac{i}{2}\sum_{a=1}^{d/2}E^a \wedge \bar{E}^a,
\eeq
where $E^a$ is a holomorphic vielbein $E^a=e^a+i\,e^{a+d/2}$, and the metric is defined by
\beq\label{eqn:gdefn}
g=\sum_{a=1}^{d/2}E^a\bar{E}^a.
\eeq
As a noteworthy example, an $SU(2)$ structure on $M_4$ is defined by a real 2-form, $j$, and complex 2-form, $\omega$. The following conditions must then be met
\beq\label{eqn:jomegaconditions}
j\wedge \omega=\omega\wedge \omega=0,~~~~~~~~~~~~\omega\wedge \bar{\omega}=2j \wedge j.
\eeq

To generalize to an $SU(2)$ structure on an $M_6$, one must now add a complex one-form, $z$, with
\beq
ds_6^2=ds_4^2+z\bar{z},~~~~~~~~~~~~~~~~~~~~~z\equiv u+i\,v.
\eeq
The pure spinors are then constructed as follows
\beq \label{eqn:Psi}
\Psi_+=\frac{1}{8} e^{\frac{1}{2}z\wedge \overline{z}}\wedge \omega,~~~~~\Psi_-=\frac{i}{8}  e^{-i j}\wedge z.
\eeq
This description will become relevant when dimensionally reducing Gaiotto-Maldacena, once again, we return to this in the next section.

\item \textbf{$SU(3)\times SU(3)$ Structure}\\
In general, having two spinors, $\eta^{1,2}$, describing an $SU(3)\times SU(3)$ structure, we have
\beq\label{eqn:SU(3)SU(3)}
\Psi_+=\eta_+^1 \otimes \eta_+^{2\dagger} = \frac{1}{8}  e^{\frac{1}{2}z\wedge \bar{z}} \wedge (\bar{c}e^{-ij}-i\omega) ,~~~~~~~~~~~~~~~~~\Psi_-=\eta_+^1\otimes \eta_-^{2\dagger}=-\frac{1}{8}(e^{-ij}+ic\,\omega)\wedge z.
\eeq

Hence, for static $SU(2)$, one derives \eqref{eqn:Psi}. %with $c=0$ \textcolor{red}{Make the 2 equations consistent with each other!!!...}. 
For $SU(3)$, one gets \eqref{eqn:SU(3)spinors} with
\beq
J=j+\frac{i}{2}z\wedge \bar{z},~~~~~~~~~~\Omega=\omega\wedge z.
\eeq
%\textcolor{red}{The spinor $\eta$ naturally defines a G-Structure as it's stabilizer group (the group of rotations which keep $\eta$ invariant). In addition, a G-Structure is typically equivalent to a set of invariant tensors, allowing one to swap the spinor with non-spinorial objects (forms) - this is the upshot of the method of G-Structures. }

\end{itemize}

%In the following, $\Psi_\pm$ are normalised, as described in \textcolor{red}{0505212}

 %       \begin{table}[h!]
  %   \begin{center}
%\begin{tabular}{c || c |c c c  }
%$$ &$\Psi_+$&$\Psi_-$  \\
%\hline
%$ SU(3)\times SU(3)$ & $(\bar{c} e^{-ij}-i\,\omega)\wedge e^{-i v \wedge u}$ & $- (e^{-ij}+i\,c\,\omega)\wedge (v+iu)$\\
%\hline
%$SU(2)$&$\omega \wedge e^{-iv \wedge u}$&$e^{-ij}\wedge (v+iu)$ \\
%\hline
%$SU(3)$&$e^{-iJ}$&$\Omega$ 
%\end{tabular}
%\end{center}
%\caption{...}
%\label{table:Gstructures}
%\end{table}

%\end{itemize}

 %\textcolor{red}{NEED TO BE CAREFUL WITH THE DEFINITION OF $d_H$ THROUGHOUT}
 
 % \paragraph{SU(d/2) in d even and the stuff that follows}

%  \paragraph{SU(3) times SU(3) and the two spinor stuff} 
 
 %\paragraph{Generalized complex geometry form}

  \subsection{Gaiotto-Maldacena G-Structures}
  We will now discuss the G-Structure formalism for the Gaiotto-Maldacena background, which describes an SU(3) structure on an $M_7$. The $\mathcal{N}=1$ G-Structure conditions for the Mink$_4$ solutions of the form given in \eqref{eqn:MTheorymetric} were derived in \cite{Kaste:2003zd}. They are defined by a real 2-form, $J$, a holomorphic 3-form, $\Omega$, (giving an SU(3) structure on $d=6$), with an additional (orthogonal and unit normed) real one-form, $K$ (to move to $d=7$). The G-Structure conditions then read
\beq\label{eqn:Geqns}
\begin{aligned}
&d(e^{2\hat{A}} K)=0,~~~~~~~~~~~~~~~~~~
d(e^{4\hat{A}} J)=e^{4\hat{A}} *_7G_4,\\
&d(e^{3\hat{A}} \Omega)=0,~~~~~~~~~~~~~~~~~~
d(e^{2\hat{A} }J\wedge J )=-2 e^{2\hat{A} }G_4\wedge K,
\end{aligned}
\eeq
  where $e^{\hat{A}}$ and $G_4$ live on $M_7$, with the condition given in \eqref{eqn:JOmegacondition}.
%  \beq
%J\wedge \Omega=0,~~~~J\wedge J\wedge J=\frac{3}{4}i\Omega \wedge \overline{\Omega}
%\eeq
One must then impose the Bianchi identities separately. These conditions allow $(J,\Omega)$ to be written in the manner outlined in \eqref{eqn:JOmegadefs}, written here explicitly for clarity
     \beq\label{eqn:OmegaJ}
\Omega= E^1\wedge E^2\wedge E^3  ,~~~~~~~~~~~~~~~ J=\frac{i}{2}(E^1\wedge \overline{E}^1+E^2\wedge \overline{E}^2+E^3\wedge \overline{E}^3),
\eeq
where the complex vielbein, $E^a$, is orthogonal to $K$ (with $a=1,2,3$). Hence, for the GM background, using %the \textcolor{red}{Poincare patch}
\begin{equation}
ds^2(AdS_5)=e^{2\rho}ds^2(\text{Mink}_4) +d\rho^2,
\end{equation}
we have (recalling \eqref{eqn:MTheorymetric}, \eqref{eqn:ds7defn} and \eqref{eqn:gdefn})

\begin{equation}\label{eqn:metriccondition}
\begin{aligned}
ds^2 &=e^{2\hat{A}}ds^2(\text{Mink}_4) +ds_7^2,~~~~~~~~~~~~~~~e^{2\hat{A}}=4f_1e^{2\rho},\\
ds_7^2 &= \sum_{a=1}^{3}E^a \bar{E}^{\bar{a}}+K^2= f_1\bigg[4d\rho^2 +f_2 ds^2(S^2) +f_3 d\chi^2 + f_4(d\sigma^2 + d\eta^2) +f_5(d\beta + f_6 d\chi)^2\bigg].
\end{aligned}
\end{equation}
For the $\beta$ reduction frame, one derives the following results (given in \cite{us})\footnote{Where the reduction frame can be read off directly from $E_3$ using the reduction formula
\begin{equation*} 
e^{-\frac{2}{3}\Phi_{IIA}}ds_{IIA}^2 = ds^2 -  e^{\frac{4}{3}\Phi_{IIA}}( d\psi+C_1)^2.
\end{equation*}
In the case of \eqref{eqn:GstructureForms}, $\psi=\beta$,  $C_1=f_6 d\chi$ and $e^{\frac{4}{3}\Phi_{IIA}}=f_1f_5$.}
%\textcolor{red}{CHANGE THE E2 OVER HERE SO IT MATCHES ....
   \begin{align}\label{eqn:GstructureForms}
&~~~~~~~~~~~~~~~~~~~~~~~~~~~~~~~~~~~~~~~~~~   \textbf{$\beta$ reduction frame}\nn\\
&K= \frac{\kappa\, e^{-2\rho}}{f_1}d(\cos\theta e^{2\rho}\dot{V}),~~~~~~~~~~~~~~~~~~~~~~~~~~
E_1 =  -\sqrt{f_1f_3}\bigg(\frac{1}{\sigma} d\sigma+ d\rho   +i d\chi\bigg),\nn\\[2mm]
&E_2= e^{i\phi} \bigg[ \frac{\kappa}{f_1} e^{-2\rho}d(\sin \theta e^{2\rho}\dot{V})+i\, \sqrt{f_1f_2}\sin \theta d\phi\bigg],\\[2mm]
%E_2= \frac{\kappa \,e^{i\phi} }{f_1}\bigg[e^{-2\rho}d(\sin \theta e^{2\rho}\dot{V})+i\,\sin \theta \,\dot{V}d\phi\bigg],\\%~~~~~~~~~~~~~~~
&E_3 =-e^{i\chi}\sqrt{f_1f_5} \,\Bigg[-\frac{1}{4}f_3 \frac{\dot{V}'}{\sigma} d\sigma -V''d\eta +f_6 d\rho  + i \Big(d\beta +f_6d\chi\Big)\Bigg],\nn
\end{align}
where $K$ and $E_2$ are independent of $(\beta,\chi)$, so remain constant under a rotation to the $\chi$ reduction frame. The $(E_1,E_3)$ vielbeins in this frame then read (now labelled with a tilde for clarity)
  \begin{align}\label{eqn:GM-G}
&~~~~~~~~~~~~~~~~~~~~~~~~~~~~~~~~~~~~~~~~~~   \textbf{$\chi$ reduction frame}\nn\\
%K= \frac{\kappa\, e^{-2\rho}}{f_1}d(\cos\theta e^{2\rho}\dot{V}),~~~~~~~~~~~~~~~~~~~~~~~~~~
&\tilde{E}_1 = \sqrt{\frac{f_1f_3}{f_6^2 +\frac{f_3}{f_5}}}  \bigg( d(V') -i\,d\beta\bigg),\\[2mm]
%E_2= e^{i\phi} \bigg[ \frac{\kappa}{f_1} e^{-2\rho}d(\sin \theta e^{2\rho}\dot{V})+i\, \sqrt{f_1f_2}\sin \theta d\phi\bigg],\\
&\tilde{E}_3 =-e^{i\chi} \sqrt{f_1 f_5\Big(f_6^2+\frac{f_3}{f_5}\Big)} \,\Bigg[ d\rho - \frac{V''}{(\dot{V}')^2-\ddot{V}V''} d(\dot{V})+ i\,\bigg(d\chi +  \frac{f_6}{f_6^2 +\frac{f_3}{f_5}}  d\beta\bigg)\Bigg],\nn
\end{align}
where, from the conditions in \eqref{eqn:OmegaJ} and \eqref{eqn:metriccondition}, one can check that the following relations indeed hold
\beq
\begin{aligned}
E_1\wedge \bar{E_1}+E_3\wedge \bar{E_3}&=\tilde{E}_1\wedge \bar{\tilde{E}}_1+\tilde{E}_3\wedge \bar{\tilde{E}}_3,\\
E_1\wedge E_3&=\tilde{E}_1\wedge \tilde{E}_3,\\
E_1\bar{E_1}+E_3\bar{E_3}&=\tilde{E}_1\bar{\tilde{E}}_1+\tilde{E}_3\bar{\tilde{E}}_3.
\end{aligned}
\eeq

For a final rotation to the $\phi$ reduction frame, $K$ is independent of $\phi$, so once again remains untouched. The $E_2$ given in \eqref{eqn:GstructureForms} does however depend on $\phi$, and has the appropriate form for a $\phi$ reduction $E_3$ (with $C_1=0$). Hence, performing a trivial relabelling of the vielbeins, this $E_2$ now becomes the $E_3$ for the $\phi$-reduction frame. The $E_1$ and $E_2$ are then simply either $(E_1,E_3)$ from \eqref{eqn:GstructureForms} or $(\tilde{E}_1,\tilde{E}_3)$ from \eqref{eqn:GM-G}. See Appendix \ref{sec:GMGstructures} for further details.\\

It is worth noting, using \eqref{eqn:OmegaJ}, \eqref{eqn:GstructureForms} and \eqref{eqn:GM-G}, we can read off the $U(1)_r$ component of the $SU(2)_R\times U(1)_r$ R-Symmetry directly from the overall phase of $\Omega$, namely $\chi+\phi$. Hence, as it stands, the only KK reduction which will preserve any supersymmetry is a $\beta$ reduction. This in fact leads to the Type IIA solution given in \cite{Reid-Edwards:2010vpm} (see also \cite{Nunez:2019gbg}), and is the only background which preserves the full $\mathcal{N}=2$ supersymmetry. However, as discussed in \cite{us}, by performing an $SL(3,\mathds{R})$ transformation prior to reduction, further possibilities emerge (including the additional $\mathcal{N}=1$ Type IIA and Type IIB solutions which we uncover in this paper).\\\\
Before reviewing this $SL(3,\mathds{R})$ transformation, we must first discuss the G-Structure analysis of the Type IIA and Type IIB daughter backgrounds.

  \subsubsection*{IIA reduction}
  
  Now we turn to the G-Structure description of the daughter Type IIA backgrounds. By performing a dimensional reduction along a $U(1)$ which lies strictly outside $K$, the resulting theory has an SU(2) structure on an $M_6$. In this case, the G-structure analysis will be defined by a real two-form, $j$, a holomorphic two-form, $\omega$, (defining an SU(2) structure on $d=4$), with an additional complex one-form, $z$ (to move to $d=6$). Hence, the G-structure analysis of the supersymmetric background outlined here will be defined by $(j,\omega,z)$\footnote{If the dimensional reduction was performed along $K$, the background would define an $SU(3)$ on $M_6$, described by $(J,\Omega)$ from \eqref{eqn:SU(3)spinors} - namely, a real 2-form and holomorphic 3-form, respectively. }.\\\\
  These forms can easily be calculated from the results of the previous section using the standard reduction formula\footnote{
One could of course use the generalised reduction formula (for reductions along $n \psi$, with $n \in \mathds{Z}$), 
\begin{equation*}
e^{-\frac{2}{3}\Phi_{IIA}}ds_{IIA}^2 = ds^2 - \frac{1}{n^2}e^{\frac{4}{3}\Phi_{IIA}}(n \,d\psi+n\, C_1)^2,
\end{equation*}
where one can easily see
\begin{equation*}
C_1\rightarrow n C_1, ~~~~~~~~~~~~~e^{\frac{4}{3}\Phi}\rightarrow \frac{1}{n^2}e^{\frac{4}{3}\Phi},~~~~~~~~~~~~B_2\rightarrow \frac{1}{n}B_2,
\end{equation*}
however, this serves no obvious purpose. We thus fix $n=1$.
}
\begin{equation}\label{eqn:DimRed}
 ds^2 = e^{-\frac{2}{3}\Phi }ds_{IIA}^2+ e^{\frac{4}{3}\Phi }(d\psi+C_1)^2 ,
\end{equation}
and the following relations given in \cite{Macpherson:2015tka}
\begin{equation}\label{eqn:11Dto10Dforms}
\begin{gathered}
\Omega  = \omega \wedge \Big(e^{-\Phi}u +i(d\psi+C_1)\Big)%= (\hat{E}_1 \wedge \hat{E}_2)\wedge \hat{E}_3, \\
~~~~~~~~~~~~~~~~J = j\,e^{-\frac{2}{3}\Phi} +e^{\frac{1}{3}\Phi}u \wedge (d\psi+C_1),~~~~~~~~~~~~~~~~K= v\,  e^{-\frac{1}{3}\Phi},\\
%\text{where }~~~~~~~~~~~~~~~~~~~~~~~~~~~~~~~~~~~~~~~~~~~~~~~~~~~~~~~~~~~~~~~~~~~~~~~~~~~~~~~~~~~~~~~~~~~~~~~~~~~~~~~~~~~~~~~~~~~~~~~~~~~~~~~~~~~~~~~~~~~~~~~~~~~~~~~~~~~
z=u+i\,v.
\end{gathered}
\end{equation}

  The G-structure conditions for type II Mink$_4$ solutions of the form given in \eqref{eqn:M6metric} and \eqref{eqn:relstuff} \footnote{With $\lambda(\alpha_n)=(-1)^{\frac{n}{2}(n-1)}\alpha_n$
ensuring the condition $F_n=(-1)^{\frac{n}{2}(n-1)}*_{10}F_{10-n}$ still holds.}
%\beq\label{eqn:relstuff}
%ds^2= e^{2A}ds^2(\text{Mink}_4)+ ds^2(\text{M}_6),~~~~F=g +e^{4A}\text{Vol}_4 \wedge *_6\lambda(g),
%\eeq
were derived in \cite{Grana:2005sn}, with $(A,\Phi,H_3,g$) living on $M_6$. We note the following relation with the GM Warp factor
\beq
e^A=e^{\hat{A}+\frac{1}{3}\Phi},
\eeq
with $\Phi$ the dilaton.
The differential conditions on the metric, equivalent to \eqref{eqn:diffcons} from before, are expressed in terms of the SU(2) Pure Spinors $\Psi_\pm$ given in \eqref{eqn:Psi}
and read \footnote{Where the Pure spinors can be re-written here in a more convenient form
  
    \begin{equation*}
    \begin{aligned}
    \Psi_{+}&=\frac{1}{8}\Big(1+\frac{1}{2}z\wedge \bar{z}\Big)\wedge \omega,\\
 \Psi_{-}&= \frac{i}{8}z \wedge \Big(1-i\,j-\frac{1}{2}j\wedge j\Big)=\frac{1}{8}\bigg[u\wedge j-v\wedge\Big(1-\frac{1}{2}j\wedge j\Big)\bigg]+\frac{i}{8}\,\bigg[u\wedge \Big(1-\frac{1}{2}j\wedge j\Big)+v\wedge j\bigg],
    \end{aligned}
    \end{equation*}
    with the real and imaginary parts of $ \Psi_{-}$ made clear.
    }
\begin{samepage}
 \begin{subequations}
\begin{align}
d_{H_3}(e^{3A-\Phi}\Psi_+)&=0,\\
d_{H_3}(e^{2A-\Phi}\text{Re}\Psi_-)&=0,\\
d_{H_3}(e^{4A-\Phi}\text{Im}\Psi_-)&=\frac{e^{4A}}{8}*_6\lambda(g),\label{eqn:Calibrationform}
\end{align}
 \end{subequations}
 \end{samepage}
with $g$ the total internal RR flux, and with the conditions given in \eqref{eqn:jomegaconditions},
%\beq
%j\wedge \omega=0,~~~~~~~~~~~~~~~~~~~~j\wedge j=\frac{1}{2}\omega\wedge \bar{\omega},
%\eeq
where $(j,\omega)$ are orthogonal to $z$. These conditions are equivalent to a set of equations written explicitly in terms of $(j,\omega,z)$, as outlined in appendix D of \cite{Macpherson:2015tka}.\\

%Finally, we note the following Type IIA relations.
%\beq
%\begin{gathered}
%\text{Type IIA: }~~~~~~~~~F_{tot} = \sum_{j=0}^5F_{2j},~~~~~~~~~~~~~\lambda(g)=g_0-g_2+g_4-g_6,
%\end{gathered}
%\eeq
%\textcolor{red}{where by supersymmetry, $g_6=0$}. 
From \eqref{eqn:relstuff} we have $e^{4A}\text{Vol}_4 \wedge *_6\lambda(g)= F_6+F_8+F_{10}$, which allows one to re-write \eqref{eqn:Calibrationform} as
\beq\label{eqn:newform}
  \text{Vol}_4\wedge d_{H_3}(e^{4A-\Phi}\text{Im}\Psi_-)=    \frac{1}{8}( F_6+F_8+F_{10})
\eeq
We will utilise this form in the derivation of the Type IIB G-Structure conditions, which we now summarise. See Appendix \ref{sec:IIBGstructures} for the full calculations.

      \subsubsection*{IIB Abelian T-Dual}
      
      Finally, we investigate the G-Structure description of the Abelian T-Dual solution. As mentioned at the beginning of this section, and made more explicit in \eqref{eqn:newform}, the pure spinors will transform under T-Duality in the same manner as the Ramond fields. In the G-Structure conditions, the roles of $\Psi_\pm$ switch when moving from Type IIA to Type IIB. This is because Ramond fields are even in Type IIA and odd in Type IIB.
% \beq
%\begin{aligned}
%&\text{Type IIB: }~~~~~~~~~F_{tot} = \sum_{j=0}^4F_{2j+1},~~~~~~~~~\lambda(g)=g_1-g_3+g_5.
%\end{aligned}
%\eeq

      Using the rules given in \cite{Kelekci:2014ima} (see also \cite{Legramandi:2020des}), first make the following decomposition in Type IIA
  \begin{equation}
  ds_{10}^2=ds_{9,\mathcal{A}}^2+e^{2C}(dy+A_1)^2,~~~~~~~B=B_2+B_1 \wedge dy,~~~~~~~~~F=F_{\perp}+F_{||}\wedge E^y,
  \end{equation}
  with $E^y =e^{C}(dy+A_1)$. Then the Type IIB T-dual theory is defined as follows
  \begin{equation} \label{eqn:Tdualeq}
  \begin{gathered}
  ds_{9,\mathcal{B}}^2=ds_{9,\mathcal{A}}^2,~~~~~\Phi^\mathcal{B}=\Phi^\mathcal{A}-C^\mathcal{A},~~~~~~~C^\mathcal{B}=-C^\mathcal{A},\\
  B_2^\mathcal{B}=B_2^\mathcal{A}+A_1^\mathcal{A} \wedge B_1^\mathcal{A},~~~~~~~~A_1^\mathcal{B}=-B_1^\mathcal{A},~~~~~~~~B_1^\mathcal{B}=-A_1^\mathcal{A},\\
  F_{\perp}^\mathcal{B}=e^{C^\mathcal{A}}F_{||}^\mathcal{A},~~~~~~~F_{||}^\mathcal{B}=e^{C^\mathcal{A}}F_{\perp}^\mathcal{A}.
  \end{gathered}
  \end{equation}
 These rules then justify the following decompositions
 \beq
\begin{gathered}
 \Psi_\pm^\mathcal{A}=\Psi_{\pm_{\perp}}^\mathcal{A} + \Psi_{\pm_{||}}^\mathcal{A}\wedge E^y_\mathcal{A},\\
\omega^\mathcal{A}=\omega^\mathcal{A}_{\perp}+\omega^\mathcal{A}_{||}\wedge E^y_\mathcal{A},~~~~~~~~~~~~~~~~~
j^\mathcal{A}=j^\mathcal{A}_{\perp}+j^\mathcal{A}_{||}\wedge E^y_\mathcal{A},~~~~~~~~~~~~~
z^\mathcal{A}=z^\mathcal{A}_{\perp}+z^\mathcal{A}_{||}\wedge E^y_\mathcal{A},\\
z^\mathcal{A}_{\perp}=u^\mathcal{A}_{\perp}+i\,v^\mathcal{A}_{\perp},~~~~~~~~~~~~~~~~~~z^\mathcal{A}_{||}=u^\mathcal{A}_{||}+i\,v^\mathcal{A}_{||}.
\end{gathered}
\eeq  
%which, after switching the roles of $\Psi_\pm$, gives
 %\beq
 % \Psi_\mp^\mathcal{B}=e^{C^\mathcal{A}}\Psi_{\pm_{||}}^\mathcal{A} + \Psi_{\pm_{\perp}}^\mathcal{A}\wedge (dy-B_1^\mathcal{A}).
 %\eeq
 
One can then use the rules in \eqref{eqn:Tdualeq} to transform the G-Structure conditions and Pure Spinors. After switching the roles of $\Psi_\pm$, we get
\begin{equation}
e^{-\Phi_\mathcal{B}}  \Psi_\mp^\mathcal{B}=e^{-\Phi_\mathcal{A}}\Big[e^{C^\mathcal{A}}\Psi_{\pm_{||}}^\mathcal{A} + \Psi_{\pm_{\perp}}^\mathcal{A}\wedge (dy-B_1^\mathcal{A})\Big],
\end{equation}
where the factors of $\Phi_\mathcal{A}$ and $\Phi_\mathcal{B}$ were introduced to derive the following forms for the IIB G-Structure conditions (solely in terms of IIB quantities)
\begin{equation}
\begin{aligned}
d_{H^\mathcal{B}_3}(e^{3A-\Phi_\mathcal{B}}\Psi^\mathcal{B}_-)&=0,\\
d_{H^\mathcal{B}_3}(e^{2A-\Phi_\mathcal{B}}\text{Re}\Psi^\mathcal{B}_+)&=0,\\
d_{H^\mathcal{B}_3}(e^{4A-\Phi_\mathcal{B}}\text{Im}\Psi^\mathcal{B}_+)&=\frac{e^{4A}}{8}*_6\lambda(g). 
\end{aligned}
\end{equation}
Working through the calculation, one gets the following forms

\beq
\begin{aligned}
  \Psi_-^\mathcal{B} &=\frac{1}{8}e^{\Phi_\mathcal{B}-\Phi_\mathcal{A}}\bigg[e^{\frac{1}{2}z^\mathcal{A}_{\perp}\wedge \overline{z}^\mathcal{A}_{\perp}}\wedge\Big( e^{C^\mathcal{A}}\omega^\mathcal{A}_{||}+   \omega^\mathcal{A}_{\perp}\wedge (dy-B_1^\mathcal{A})\Big)+e^{C^\mathcal{A}}\frac{1}{2}( z^\mathcal{A}_{\perp}\wedge \overline{z}^\mathcal{A}_{||} -z^\mathcal{A}_{||}\wedge \overline{z}^\mathcal{A}_{\perp})\wedge \omega^\mathcal{A}_{\perp}\bigg], \\
   \Psi_+^\mathcal{B}&=\frac{i }{8} e^{\Phi_\mathcal{B}-\Phi_\mathcal{A}}e^{-ij^\mathcal{A}_{\perp}}\wedge \bigg[ (e^{C^\mathcal{A}}z^\mathcal{A}_{||}+ z^\mathcal{A}_{\perp}\wedge (dy-B_1^\mathcal{A}))+i  e^{C^\mathcal{A}} j^\mathcal{A}_{||}\wedge z^\mathcal{A}_{\perp}\bigg],
\end{aligned}
\eeq
which for our purposes, in which $z^\mathcal{A}_{||}=\bar{z}^\mathcal{A}_{||}=0$, reduces to\footnote{Comparing these results with \eqref{eqn:Psi} (and \eqref{eqn:SU(3)SU(3)} more generally), from the form of $\Psi_+^\mathcal{B}$ this would suggest the following relations for an $SU(2)$ structure (noting $\frac{1}{2}z^\mathcal{B}\wedge \bar{z}^\mathcal{B}=iv^\mathcal{B}\wedge u^\mathcal{B}$) 
\beq
v^\mathcal{B}\wedge u^\mathcal{B} = -j_\perp^\mathcal{A},~~~~~~~\omega^\mathcal{B}=i\,  e^{\Phi_\mathcal{B}-\Phi_\mathcal{A}}\, z^\mathcal{A}_{\perp}\wedge \Big( (dy-B_1^\mathcal{A})-i    e^{C^\mathcal{A}} j^\mathcal{A}_{||}\Big).
\eeq
The $\Psi_-^\mathcal{B}$ comparison is less clear but the dimensions of the Polyform match as they should.}
\beq
\begin{aligned}
  \Psi_-^\mathcal{B} &=\frac{1}{8}e^{\Phi_\mathcal{B}-\Phi_\mathcal{A}} e^{\frac{1}{2}z^\mathcal{A}_{\perp}\wedge \overline{z}^\mathcal{A}_{\perp}}\wedge\Big( e^{C^\mathcal{A}}\omega^\mathcal{A}_{||}+   \omega^\mathcal{A}_{\perp}\wedge (dy-B_1^\mathcal{A})\Big),  \\
   \Psi_+^\mathcal{B}&=\frac{i }{8} e^{\Phi_\mathcal{B}-\Phi_\mathcal{A}}e^{-ij^\mathcal{A}_{\perp}}\wedge    z^\mathcal{A}_{\perp}\wedge \Big( (dy-B_1^\mathcal{A})-i    e^{C^\mathcal{A}} j^\mathcal{A}_{||}\Big).
\end{aligned}
\eeq

%\beq
%\begin{aligned}
%  \Psi_-^\mathcal{B}  &=\frac{1}{8} e^{\Phi_\mathcal{B}-\Phi_\mathcal{A}} e^{\frac{1}{2}z^\mathcal{A}_{\perp}\wedge \overline{z}^\mathcal{A}_{\perp}}\wedge\Big(  \tilde{ \omega}^\mathcal{B}+e^{C^\mathcal{A}}\frac{1}{2}( z^\mathcal{A}_{\perp}\wedge \overline{z}^\mathcal{A}_{||} -z^\mathcal{A}_{||}\wedge \overline{z}^\mathcal{A}_{\perp})\wedge \omega^\mathcal{A}_{\perp}\Big)  \\
%    \Psi_+^\mathcal{B}&=\frac{i }{8} e^{\Phi_\mathcal{B}-\Phi_\mathcal{A}}e^{-ij^\mathcal{A}_{\perp}}\wedge \bigg[ (e^{C^\mathcal{A}}z^\mathcal{A}_{||}+ z^\mathcal{A}_{\perp}\wedge (dy-B_1^\mathcal{A}))+i  e^{C^\mathcal{A}} j^\mathcal{A}_{||}\wedge z^\mathcal{A}_{\perp}\bigg]
%\end{aligned}
%\eeq
%where
%\beq
%\tilde{ \omega}^\mathcal{B}= e^{C^\mathcal{A}}\omega^\mathcal{A}_{||}+   \omega^\mathcal{A}_{\perp}\wedge (dy-B_1^\mathcal{A})
%\eeq
 %is now an odd polyform (indicated here by a tilde). The real and imaginary components of $ \Psi_+^\mathcal{B}$ can then be read off independently as follows.....

See Appendix \ref{sec:IIBGstructures} for more details. 

%\textcolor{red}{Analyse the G-Structures here - $SU(3)\times SU(3)$? .....}

\subsection{$SL(3,\mathds{R})$ Transformations}
Now that we have reviewed the Gaiotto-Maldacena background and the necessary G-Structure analysis, we should introduce the $SL(3,\mathds{R})$ transformations which generate the various solutions presented in this paper (along with the results of \cite{us}).

Given that there are three $U(1)$ directions in the GM background $(\beta,\chi,\phi)$, one can make the following $SL(3,\mathds{R})$ transformation
   \begin{equation}\label{eqn:S2breakingdefns}
 \begin{gathered}
d\beta =a\, d\chi+b\, d\beta +c\, d\phi,~~~~~~~~~~~~~~~~~d\chi = p\, d\chi +q\, d\beta + m \,d\phi,~~~~~~~~~~~~~~~~~~d\phi = s\, d\chi +v\,d\beta +u\,d\phi,\\
\begin{vmatrix}
~p&q&m~\\
~a&b&c~\\
~s&v&u~\\
\end{vmatrix}
=p(bu-vc)-q(au-sc)+m(av-sb)=1.
\end{gathered}
\end{equation}

The utility of this transformation becomes clear under dimensional reductions to Type IIA\footnote{These nine parameters must reduce to three free parameters (corresponding to the three $U(1)$ directions being mixed). For our purposes, we fix $(p,b,u)=1$ by trivially absorbing them into the definitions of $(\chi, \beta,\phi)$, respectively. This avoids re-defining the three $U(1)$ directions amongst themselves and immediately eliminates three of the nine parameters. However, as we note where appropriate, there can be some utility in choosing different values for these parameters.}, with the $\beta$ reduction case already considered in \cite{us}. In that case, one can derive $\mathcal{N}=(2,1,0)$ Type IIA backgrounds%, a brief review of which will be presented in the following subsection
. However, as we will investigate in this paper, the $\chi$ and $\phi$ reduction cases lead only to $\mathcal{N}=1$ and $\mathcal{N}=0$ solutions. %As already discussed, we will utilise the method of G-Structures to keep track of the supersymmetry under reduction% (following the approach outlined in \cite{us})
Using the forms given in \eqref{eqn:GstructureForms}, one can see that the $U(1)$ component of the $SU(2)_R\times U(1)_r$ R-symmetry of the Gaiotto-Maldacena background now becomes
\beq\label{eqn:U(1)}
\begin{aligned}
U(1)_r&=\chi+\phi \\&\rightarrow (p+s)\chi +(q+v)\beta + (m+u)\phi.
\end{aligned}
\eeq
This immediately provides more possibilities to preserve supersymmetry under dimensional reduction, i.e. in the case of a $\beta$ reduction, one can now preserve this $U(1)_r$ component by fixing $v=-q$. Throughout this paper we use the reduction formula given in \eqref{eqn:DimRed}. We now review the $\beta$ reduction case.
 
\subsection*{The $\beta$ reduction case}

As outlined in \cite{us}, by performing the following 11D transformations (with $q=\xi,~v=\zeta$)
\beq\label{eqn:exacttransformation}
d\beta \rightarrow   d\beta  ,~~~~~~~~~~~~~~~~~d\chi \rightarrow d\chi +\xi\, d\beta  ,~~~~~~~~~~~~~~~~~~d\phi \rightarrow d\phi + \zeta\,d\beta,
\eeq
before dimensionally reducing along $\beta$, one gets the following two parameter family of solutions (re-written below in a form which will be useful for investigating the boundary)\footnote{the $\Delta$ here should not be confused with $\tilde{\Delta}$ from \eqref{eqn:GM}.}

 %      \begin{equation}\label{eqn:generalresult1}
  %  \begin{gathered}
    %    ds_{10,st}^2=e^{\frac{2}{3}\Phi}f_1\bigg[4ds^2(\text{AdS}_5)+f_2d\theta^2+f_4(d\sigma^2+d\eta^2)\bigg]+f_1^2 e^{-\frac{2}{3}\Phi}  ds^2_2,\\
   % ds^2_2 = f_3f_5 d\chi^2 + \sin^2\theta f_2\bigg[f_3 (\xi d\phi-\zeta d\chi )^2 + f_5\Big( -\zeta f_6 d\chi+(\xi f_6+1)d\phi\Big)^2\bigg]
    %    \\e^{\frac{4}{3}\Phi}= f_1 \bigg[  f_5 (1 +\xi f_6)^2 +\xi^2 f_3 + \zeta^2 f_2 \sin^2\theta\bigg] ,\\
    %    B_2  = \sin\theta \,\bigg[\zeta  f_7 d\chi  - (f_8+\xi f_7 ) d\phi \bigg] \wedge d\theta,\\
    %   C_1= f_1  e^{-\frac{4}{3}\Phi} \bigg[\bigg(f_5 f_6(1 + \xi f_6) +\,\xi f_3 \bigg)d\chi +  \zeta \sin^2\theta f_2  d\phi \bigg],~~~~~~~~~~~~~~~ C_3=f_7  d\chi \wedge \Vol(S^2),
   % \end{gathered}
   % \end{equation} 
    
           \begin{align}\label{eqn:generalresult1}
     &   ds_{10,st}^2=f_1^{\frac{3}{2}}f_5^{\frac{1}{2}}\sqrt{\Xi}\bigg[4ds^2(\text{AdS}_5)+f_4(d\sigma^2+d\eta^2)+ ds^2(M_3)\bigg],\nn\\[2mm]
   &     ds^2(M_3)=f_2 \bigg(d\theta^2+ \frac{\Delta}{\Xi}\sin^2\theta D\phi^2\bigg) +\frac{f_3}{\Delta}d\chi^2 =f_2 \bigg(d\theta^2+ \frac{1}{\Pi}\sin^2\theta d\phi^2\bigg) + \frac{\Pi}{\Xi} f_3 D\chi^2,\nn\\[2mm]
%    ds^2_2 = f_3f_5 d\chi^2 + \sin^2\theta f_2\bigg[f_3 (\xi d\phi-\zeta d\chi )^2 + f_5\Big( -\zeta f_6 d\chi+(\xi f_6+1)d\phi\Big)^2\bigg]
    &    B_2  = \sin\theta \,\bigg[\zeta  f_7 d\chi  - (f_8+\xi f_7 ) d\phi \bigg] \wedge d\theta,~~~~~~~~~~~~~~~~        e^{\frac{4}{3}\Phi}= f_1 f_5 \Xi ,\nn\\[2mm]
  &     C_1=\frac{1}{\Xi}\bigg(\Big(f_6(1 + \xi f_6) +\,\xi \frac{f_3}{f_5} \Big)d\chi +  \zeta  \frac{f_2}{f_5}  \sin^2\theta d\phi \bigg),~~~~~~~~~~~~~~~ C_3=f_7  d\chi \wedge \Vol(S^2),\nn\\[2mm]
&  \Xi=\Delta+\zeta^2\frac{f_2}{f_5}\sin^2\theta,~~~\Delta=(1+ \xi f_6)^2+\xi^2\frac{f_3}{f_5},~~~\Pi=1+\zeta^2f_2\frac{f_3+f_5 f_6^2}{f_3 f_5}\sin^2\theta,\nn\\[2mm]
&D\phi =d\phi -\frac{\zeta}{\Delta}\bigg(f_6(1+\xi f_6)+\xi \frac{f_3}{f_5}\bigg) d\chi,~~~~~~~D\chi=d\chi -\frac{\zeta}{\Pi}\frac{f_2}{f_3}\bigg(f_6(1+\xi f_6) +\xi \frac{f_3}{f_5}\bigg)\sin^2\theta\,d\phi,
    \end{align} 
  leading to the backgrounds given in Table 1 of \cite{us}. %\footnote{With $(\xi,\zeta)$ free parameters, with the third free parameter set to zero without loss of generality. See Appendix \ref{sec:DimRed}.}.
%        \begin{table}[h!]
 %    \begin{center}
%\begin{tabular}{c | c c c c  }
%$\beta$- Reduction&$\mathcal{N}$&$U(1)_r$&$SU(2)_R$  \\
%\hline
%$\xi=\zeta=0$&$ 2$ &$\checkmark$&$\checkmark$ \\
%$\xi=-\zeta\neq 0$&$ 1$ &$\checkmark$&$\times$  \\
%$\xi\neq 0,~\zeta=0$&$ 0$ &$\times$&$\checkmark$  \\
%$\xi=0,~\zeta\neq0$&$ 0$ &$\times$ &$\times$  \\
%$\xi\neq0,~\zeta\neq 0$&$ 0$ &$\times$&$\times$ 
%\end{tabular}
%\end{center}
%\caption{$\beta$ Reduction }
%\label{table:1}
%\end{table}
The metric has the nice property that all dependence on the transformation parameters drops out neatly when calculating the Holographic Central charge. Hence, one derives the following (parameter independent) result
  \beq\label{eqn:HCC}
c_{hol}=\frac{\kappa^3}{\pi^4}\sum_{k=1}^\infty P \mathcal{R}_k^2.
\eeq
This means that the two parameters correspond to marginal deformations of the $d=4$ $\mathcal{N}=2$ long linear quiver dual CFTs, see \cite{us} for a detailed calculation.

%giving
%\beq
%\hat{F}_4=%-d\bigg[\frac{4\kappa \big(\dot{V}-\dot{V}' (\eta-k)\big)(1+\xi \dot{V}')\sin\theta}{2(1+\xi \dot{V}')^2+\zeta^2 \dot{V}V'' \sin^2\theta}\bigg] \wedge d\theta \wedge d\phi \wedge d\chi
%\eeq
In \cite{us} the $SU(2)\times U(1)$ preserved $\mathcal{N}=0$ background (with $\zeta=0$) and the $\mathcal{N}=1$ case (with $\zeta=-\xi$) were investigated in further detail. Studying the boundary of the $(\sigma,\eta)$ coordinates, one can show that these two backgrounds contain branes which are backreacted on a Spindle and it's higher dimensional analogue, respectively. In the next subsection we will review this analysis by following the same procedure for the general two-parameter family.

%As we will investigate later in this paper, this $\mathcal{N}=1$ Type IIA solution just described leads to an $\mathcal{N}=1$ Type IIB solutions under appropriate Abelian T-Dualities. %See \cite{us} for the G-Structure description (or Appendix \ref{sec:oneparamfam} with $\gamma=0$).
   %The $\mathcal{N}=1$ solutions in question is derived by fixing $\zeta=-\xi$ in \eqref{eqn:generalresult1}, with $\xi \in \mathds{Z}\text{\textbackslash} \{0\}$ (recovering the $\mathcal{N}=2$ solution of \cite{Nunez:2019gbg} for $\xi=0$). 

\subsubsection*{Investigations at the Boundary}
We now follow the procedure of \cite{us} for the more general two-parameter solution, using the limits of the Warp factors which are quoted for convenience in Appendix \ref{sec:fs}. This analysis will follow closely that of the $\mathcal{N}=1$ case, with the parameter $\zeta$ left general.

First, for general values of $(\eta,\sigma,\theta)$, we see that $(\Xi,\Pi)$ are non-zero and finite. The deformed $S^2$ given by $(\theta,\phi)$ has $\Pi\rightarrow 1$ at the poles, which given the expression for $M_3$ in \eqref{eqn:generalresult1}, means it still behaves topologically as an $S^2$. This follows the argument given in \cite{us}, where $D\chi\rightarrow d\chi$ up to leading order at the poles. We now turn to the behaviour at the boundaries. In the following analysis, to be consistent with \cite{us} we will use the following large gauge transformation
 \beq\label{eqn:LGT}
  B_2\rightarrow B_2+2\kappa k\text{Vol}(S^2),
  \eeq
  but we will note a potential alternative which leads to quantized D4 charges in the $\zeta=0$ case.
\begin{itemize}
\item At the $\sigma\rightarrow \infty$ boundary, all parameters drop out, so the behaviour remains the same as the cases studied in \cite{us}. We summarise the presentation for completeness. One finds
 %  \subsubsection*{At $\sigma\rightarrow \infty$} 
%   \textcolor{red}{MOVE DEFNS TO AN APPENDIX? - PERHAPS BETTER TO PUT THE LIMITS IN APPENDIX AND BE MORE BRIEF IF NOTHING NEW COMPARED TO N=1 CASE...}\\

%re-deriving the $\mathcal{N}=2$ case, with all parameters dropping out
    \begin{align}
 &      ds^2=\kappa  \Bigg[4\sigma \Big(ds^2(\text{AdS}_5)+d\chi^2\Big) +\frac{2P}{\pi}\bigg(d\Big(\frac{\pi}{P}\sigma\Big)^2+d\Big(\frac{\pi}{P}\eta\Big)^2 +  \sin^2\bigg(\frac{\pi\eta}{P}\bigg) ds^2(S^2)\bigg)  \Bigg],\nn\\[2mm]
  &e^{-\Phi}= \frac{\mathcal{R}_1\pi^2}{2P^{\frac{3}{2}}\sqrt{\kappa}}e^{-\frac{\pi}{P}\sigma}\Big(\frac{\pi}{P}\sigma\Big)^{-\frac{1}{2}} ,~~~~~~~~~~H_3= - \frac{4\kappa P}{\pi} \sin^2\Big(\frac{\pi}{P}\eta\Big)  d\Big(\frac{\pi}{P}\eta\Big)\wedge \text{Vol}(S^2),
  \end{align}  
where we see that $(\eta,S^2)$ span a unit-radius 3-sphere. %\textcolor{red}{To sub-leading order in $\sigma$, we have $4\kappa\sigma \left(ds^2(\text{AdS}_5)+d\chi^2\right)\to ds^2(\text{Mink}_6)$. Introducing a new coordinate $ \tilde{r}= e^{-\frac{\pi}{P}\sigma}(\frac{\pi}{P}\sigma)^{-\frac{1}{2}}$, we have
%\beq
%ds^2=ds^2(\text{Mink}_6)+\frac{2P \kappa}{\pi \tilde{r}^2}\bigg(d\tilde{r}^2+\tilde{r}^2ds^2(\text{S}^3)\bigg),~~~H_3=-\frac{4\kappa P}{\pi}\text{vol}(\text{S}^3),~~~e^{-\Phi}=\frac{{\cal R}_1\pi^2}{2 P^{\frac{3}{2}}\sqrt{\kappa}}\tilde{r},
%\eeq}
   Hence, at this boundary the $\mathcal{N}=2$ case is recovered, giving a stack of $P$ NS5 branes (after fixing $2\kappa=\pi$)\footnote{The Page Charges are calculated using the following relations
 \begin{align}
&Q_{Dp/NS5}=\frac{1}{(2\pi)^{7-p}}\int_{\Sigma_{8-p}}\hat{F}_{8-p}, ~~~~~~~~~~\hat{F}^{\text{Page}}=F^{\text{Max}}\wedge e^{-B_2} = d(C\wedge e^{-B_2} ),~~~~~~~~~F_{p+2}^{\text{Max}}=dC_{p+1}-H_3\wedge C_{p-1}.\nn
%&\hat{F}_{8-p}=F_{8-p}-B_2 \wedge F_{6-p}+\frac{1}{2}B_2^2\wedge F_{4-p}-\frac{1}{6}B_2^3\wedge F_{2-p}+\frac{1}{24}B_2^4\wedge F_2,\nn
\end{align}}
\beq
Q_{NS5}=-\frac{1}{(2\pi)^2}\int_{S^3}H_3=P.
\eeq
%Hence, there is a stack of $P$ NS5 branes at this boundary.

\item At $\eta=0$ with $\sigma\neq 0$, using \eqref{eqn:feq}, one finds using the second form for $M_3$ that $(\eta,\theta,\phi)$ vanish as $\mathds{R}^3$ in polar coordinates, namely
\beq
f_4 d\eta^2+f_2 \bigg(d\theta^2+ \frac{1}{\Pi}\sin^2\theta d\phi^2\bigg)  = \frac{2|\dot{f}|}{\sigma\,f}\Big( d\eta^2+\eta^2 ds^2(S^2) \Big),
\eeq
this once again matches the specific cases studied in \cite{us}.
%In the next section we investigate some results derived via dimensional reductions along  $\chi$ and $\phi$, and present among others, a unique $\mathcal{N}=1$ background. \\\\

\item At $\sigma=0,~\eta\in (k,k+1)$ we have that along the $\sigma=0$ boundary, $\ddot{V}=0$ to leading order. Using \eqref{eqn:fsat0} we find
\begin{align}
  &  \Xi=l_k^2+ \frac{1}{2}\zeta^2 \mathcal{R} V'' \sin^2\theta,~~~\Delta\rightarrow l_k^2,~~~D\phi =d\phi -\frac{\zeta (N_{k+1}-N_k)}{l_k} d\chi,~~~~l_k=1+\xi  (N_{k+1}-N_k),
    \end{align} 
   giving an $\mathds{R}^2/\mathds{Z}_{l_k}$ orbifold singularity in $(\sigma,\chi)$, as follows
   \beq\label{eqn:orbifold1}
   f_4 d\sigma^2+\frac{f_3 }{\Delta}  d\chi^2  \rightarrow   \frac{2V''}{\mathcal{R}}\bigg( d\sigma^2 +\frac{\sigma^2}{l_k^2}d\chi^2 \bigg).% + \frac{2\mathcal{R}V''}{2\mathcal{R}V''+(\mathcal{R}')^2}d\theta^2 + \frac{2\mathcal{R}V''+(\mathcal{R}')^2}{4\kappa^2\mathcal{R}^2 \hat{l}_k^2 \sin^2\theta}D\chi^2
     \eeq
     %\textcolor{red}{understand the $D\phi$ here- why not important?}
     
    \item At $\sigma=0,~\eta=0$, to approach this boundary one makes the following coordinate change  \\$(\eta=r \cos\alpha,~\sigma=r \sin\alpha)$, expanding about $r=0$. Using \eqref{eqQdef} we find
  \begin{align}
 & \Xi\rightarrow \Delta \rightarrow l_0^2 ,~~~~~~l_0=1+\xi N_1,~~~~~~~~~D\phi=d\phi-\frac{\zeta N_1}{l_0}d\chi,\\[2mm]
%&D\phi =d\phi -\frac{\zeta}{\Delta}\bigg(f_6(1+\xi f_6)+\xi \frac{f_3}{f_5}\bigg) d\chi,~~~~~~~D\chi=d\chi -\frac{\zeta}{\Pi}\frac{f_2}{f_3}\bigg(f_6(1+\xi f_6) +\xi \frac{f_3}{f_5}\bigg)\sin^2\theta\,d\phi.
 & f_4(d\sigma^2+d\eta^2) +ds^2(M_3) \rightarrow \frac{2Q}{N_1}\bigg(dr^2+r^2d\alpha^2+r^2\cos^2\alpha \bigg(d\theta^2 +\sin^2\theta \,D\phi^2 \bigg) + r^2\sin^2\alpha \frac{d\chi^2}{l_0^2}\bigg),\nn
  \end{align}
  where we say that the internal space vanishes as $\mathds{R}^5/\mathds{Z}_{l_0}$.
 
\item At $\sigma=0,~\eta=k$, using the coordinate change $(\eta=k-r \cos\alpha,~\sigma=r \sin\alpha)$ and \eqref{eqn:fsfork}, we see that the term $f_2/f_5$ dominates, unless we are at a pole of the deformed $S^2$. We first assume we are away from a pole, finding
\begin{align}
&\Xi\rightarrow \frac{b_k\zeta^2 N_k\sin^2\theta}{4r},~~~~~~~~~\Delta\rightarrow  \Delta_k,~~~~~~~~~~~   
\Delta_k\equiv  \frac{1}{4}\xi^2 b_k^2 \sin^2\alpha +\Big(1+ \xi g(\alpha)\Big)^2,\nn\\[2mm]
&\Delta_k(\alpha=0)=l_{k-1}^2,~~~~~~\Delta_k(\alpha=\pi)=l_{k}^2.
\end{align}
Using $r=z^2$ we get a very similar result to the $\mathcal{N}=1$ case studied in \cite{us}, 
\begin{align}\label{eqn:B3eq}
& ds^2= \zeta \kappa \sin\theta \Big[N_k\Big(4ds^2(AdS_5) +d\theta^2\Big) +4b_k\Big(dz^2 +z^2 ds^2(\mathds{B}_3)\Big)\Big],~~~~~~e^{4\Phi}= \kappa^2 \zeta^6 N_k^2\sin^6\theta, \nn\\[2mm]
&ds^2(\mathds{B}_3)= \frac{1}{4}\Big(d\alpha^2 + \frac{\sin^2\alpha}{\Delta_k}d\chi^2\Big)+\frac{\Delta_k}{\zeta^2 b_k^2}(d\phi+\mathcal{A}_k)^2,~~~~\mathcal{A}_k=-\frac{\zeta}{\Delta_k}\Big[g(\alpha)\Big(1+\xi g(\alpha)\Big)+\frac{\xi}{4}b_k^2\sin^2\alpha\Big]d\chi, \nn\\[2mm]
&B_2= -2\kappa N_k\,\sin\theta \Big(\zeta  d\chi - \xi  d\phi\Big)\wedge d\theta,~~~~~~~C_1=\frac{1}{\zeta}d\phi,~~~~~C_3=-2\kappa N_k\sin\theta  d\theta\wedge d\phi\wedge d\chi.
\end{align}
Now
\begin{align}
&ds^2(\mathds{B}_3)\bigg|_{\alpha\sim 0} =  \frac{1}{4}\Big(d\alpha^2 + \frac{\alpha^2}{l_{k-1}^2}d\chi^2\Big)+\frac{l_{k-1}^2}{\zeta^2 b_k^2}\bigg(d\phi-\frac{\zeta (N_k-N_{k-1})}{l_{k-1}}  d\chi\bigg)^2,\nn\\[2mm]
&ds^2(\mathds{B}_3)\bigg|_{\alpha\sim \pi} =  \frac{1}{4}\Big(d\alpha^2 + \frac{(\pi-\alpha)^2}{l_{k}^2}d\chi^2\Big)+\frac{l_{k}^2}{\zeta^2 b_k^2}\bigg(d\phi-\frac{\zeta (N_{k+1}-N_{k})}{l_{k}}  d\chi\bigg)^2,\nn\\[2mm]
&\frac{1}{2\pi}\int_{\mathds{W}\mathds{C}\mathds{P}_{[l_{k-1},l_k]}}d\mathcal{A}=\frac{\zeta b_k}{l_{k-1}l_k},~~~~~~~~~~~~~~~~~b_k=2N_k-N_{k+1}-N_{k-1}.
\end{align}
So we see that the general case doesn't differ too much from the $\mathcal{N}=1$ case in \cite{us}. Hence, we interpret $\mathds{B}_3$ as $U(1)$ fibration over $\mathds{W}\mathds{C}\mathds{P}^1_{[l_{k-1},l_k]}$. 

Notice here that $C_3-C_1\wedge B_2=0$, meaning that there is no D4 branes present (for $\zeta\neq0$), as $\hat{F}_4=d(C_3\wedge e^{-B_2})$. However, in the $S^2$ preserved case presented in \cite{us} (with $\zeta=0$), D4 branes are in fact recovered.

\begin{itemize}
\item \underline{$\zeta=0$}: We now make an aside and quote the $\zeta=0$ case given in \cite{us} (keeping the parameter $r$). We first present the $B_2$ with no gauge transformation
\begin{align}
&\frac{ds^2}{2\kappa\sqrt{N_k}}= \sqrt{\Delta_k}\bigg[\frac{1}{\sqrt{\frac{b_k}{r}}}\bigg(4ds^2(\text{AdS}_5)+ ds^2(\text{S}^2)\bigg)+ \frac{\sqrt{\frac{b_k}{r}}}{N_k}\bigg(dr^2+ r^2 \left(d\alpha^2+\frac{\sin^2\alpha}{\Delta_k} d\chi^2\right)\bigg)\bigg],\nn\\[2mm]
&e^{-\Phi}=\left(\frac{N_kb_k^3}{2^6\kappa^2r^3 }\right)^{\frac{1}{4}}\Delta_k^{-\frac{3}{4}},~~~~~
C_1=  \frac{\frac{1}{4}\xi k^2b_k^2\sin^2\alpha+ g(\alpha)(1+\xi g(\alpha))}{\Delta_k}d\chi,\label{eq:neq0D6s}\\[2mm]
&C_3=-2\kappa N_k d\chi\wedge \text{Vol}(\text{S}^2),~~~~~~~~~B_2=-2\kappa (k+\xi N_k) \text{Vol}(S^2).
\end{align}
Now, using the large gauge transformation of \cite{us}, we have (with $2\kappa=\pi$)
\begin{align}
&  B_2\rightarrow B_2+2\kappa k\text{Vol}(S^2),~~~~~~~~B_2=-2\kappa \xi N_k\text{Vol}(\text{S}^2),\nn\\[2mm]
&Q_{D4}=-\frac{1}{(2\pi)^2}\int_{S^2\times \mathds{W}\mathds{C}\mathds{P}^1_{[l_{k-1},l_k]}}\hat{F}_4= \frac{N_k}{l_k}-\frac{N_{k-1}}{l_{k-1}}.
\end{align}
However, if we add an additional term to the gauge transformation, noting $(\xi,N_k)\in \mathds{Z}$,
\begin{align}
&  B_2\rightarrow B_2+2\kappa (k+ \xi N_k)\text{Vol}(S^2),~~~~~~~~B_2=0,\nn\\[2mm]
&Q_{D4}=-\frac{1}{(2\pi)^2}\int_{S^2\times \mathds{W}\mathds{C}\mathds{P}^1_{[l_{k-1},l_k]}}\hat{F}_4= N_k - N_{k-1} ,
\end{align}
we recover the $\mathcal{N}=2$ quantization. This is a potential oddity given that we are integrating over the $(\alpha,\chi)$ Spindle. It would be interesting to investigate this further in future. %\footnote{\textcolor{red}{CHECK FOR IIB too: One can in fact recover a D5 charge for the $\zeta\neq0$ case  by adding an additional $2\kappa \zeta N_k \sin\theta d\chi\wedge d\theta$ term to the $B_2$ gauge transformation, however, although this make sense locally, it may not be well defined globally.}}
\end{itemize}

We now return to our general analysis and study this limit when approaching the pole $(\sigma=0,$ $\eta=k,\sin\theta=0)$. Following \cite{us}, we use
\beq
\eta=k-\rho \cos\alpha \sin^2\mu,~~~~~\sigma=\rho \sin\alpha \sin^2\mu,~~~~~~\sin\theta = 2\sqrt{\frac{b_k\rho}{N_k}}\cos\mu,
\eeq
%(where $r\rightarrow \rho \sin^2\mu$)
noting that $r= \rho \sin^2\mu$, and expanding about $\rho=0$. Now,
 \beq
 \sin^2\mu \,\Xi \rightarrow \tilde{\Xi}_k =\Delta_k \sin^2\mu + \zeta^2 b_k^2 \cos^2\mu,~~~~~~~~~~~ B_2\rightarrow B_2+\frac{4\kappa \,k\,b_k \cos^2\mu}{N_k}  d\rho\wedge d\phi,
 \eeq
 giving
 \begin{align}\label{eqn:B4eq}
 & \frac{ds^2}{ 2\kappa\sqrt{N_k} } =\sqrt{\tilde{\Xi}_k} \Bigg[\frac{4}{\sqrt{\frac{b_k}{\rho}}}ds^2(AdS_5)+ \frac{\sqrt{\frac{b_k}{\rho}}}{N_k}\bigg(d\rho^2+4\rho^2 ds^2(\mathds{B}_4)\bigg)\Bigg],~~~~~~~~~~e^{-\Phi}=\bigg(\frac{b_k^3N_k}{2^6\kappa^2\tilde{\Xi}_k^3\rho^3}\bigg)^{\frac{1}{4}},\nn\\[2mm]
 &ds^2(\mathds{B}_4) = d\mu^2 +\frac{1}{4}\sin^2\mu \bigg(d\alpha^2 + \frac{\sin^2\alpha}{\Delta_k}d\chi^2\bigg) +\frac{\sin^2\mu \cos^2\mu\,\Delta_k}{\tilde{\Xi}_k}(d\phi+\mathcal{A}_k)^2,\nn\\[2mm]
 &B_2=4\kappa b_k \cos^2\mu \Big(\xi d\phi -\zeta d\chi\Big)\wedge d\rho,~~~~~~~~~~~~C_3=-4\kappa\, b_k \cos^2\mu \,d\rho\wedge d\phi\wedge d\chi,\nn\\[2mm]
 &C_1=\frac{1}{\tilde{\Xi}_k}\bigg[b_k^2 \zeta \cos^2\mu\, d\phi + \bigg(g(\alpha)\Big(1+\xi g(\alpha)\Big)+\frac{1}{4}\xi b_k^2 \sin^2\alpha\bigg) \sin^2\mu\,d\chi\bigg].
 \end{align}
\end{itemize}

We now find that we simply re-derive the $\mathds{B}_4$ of \cite{us} with all $\zeta$ dependence given in $\Delta_k,~\tilde{\Xi}_k$ and $\mathcal{A}_k$. Hence, following the arguments given there, we say that $\mathds{B}_4$ is topologically a $\mathds{C}\mathds{P}^2$, but with additional orbifold singularities. Calculating the Euler Characteristic, using
\beq
\chi_E=\frac{1}{32\pi^2}\int_{\mathds{B}_4}(R_{abcd}R^{abcd}-4R_{ab}R^{ab}+R^2)\text{Vol}(\mathds{B}_4),
\eeq
from the Chern-Gauss-Bonnet theorem, leads to
\beq
\chi_E=\frac{\zeta l_{k-1}^2+\xi l_{k-1}l_k-\zeta l_k^2}{\zeta l_{k-1}(l_{k-1}-l_k)l_k }=3-\Big(1-\frac{1}{l_{k-1}}\Big)-\Big(1-\frac{\xi}{\zeta}\frac{1}{l_{k-1}-l_k}\Big)-\Big(1-\frac{1}{l_{k}}\Big).
\eeq
where $\zeta (l_{k-1}-l_k)/\xi=\zeta \,b_k\in \mathds{Z}$. For a round $\mathds{C}\mathds{P}^2$, $\chi_E=3$ - hence, $\mathds{B}_4$ is the weighted projective space $\mathds{W}\mathds{C}\mathds{P}^2_{[l_{k-1},l_k,\frac{\zeta}{\xi}(l_{k-1}-l_k)]}$.
Now calculating the charge at $\mu=\frac{\pi}{2}$, we notice that all $\zeta$ dependence drops out of the calculation, re-deriving the D6 charge given in \cite{us}
\beq
Q_{D6}=-\frac{1}{2\pi}\int_{\mathds{W}\mathds{C}\mathds{P}^1_{[l_{k-1},l_k]}}F_2 =-\frac{1}{2\pi}\int_{\chi=0}^{\chi=2\pi}C_1\Big|_{\alpha=0}^{\alpha=\pi} =\frac{2N_k-N_{k+1}-N_{k-1}}{l_kl_{k-1}}.
\eeq
This rational charge is a consequence of the Spindle, see for example \cite{Ferrero:2021etw}. We now see that for all values of $(\xi,\zeta)$ we have NS5 and D6 branes, with the respective charges given above. The case of the D4 branes is special, they are only present when $\zeta=0$ - that is, when there is a preserved $S^2$. Picturing the $(\xi,\zeta)$ plot given in \cite{us}, the D4 branes are only present along the $\zeta=0$ axis - see Figure \ref{fig: TypeIIA} for an updated version of this plot, and a pictorial summary of this discussion. Strangely, the rational charge quantization of the D4 branes, which derives from integrating over the spindle, can be eliminated by an additional term in the gauge transformation of $B_2$.

%Note that once again, $C_3-C_1\wedge B_2=0$, so there are no D4 branes present for $\zeta\neq0$.

%\subsubsection{$\mathcal{N}=1$ Type IIA - Background}
%The $\mathcal{N}=1$ Type IIA background given in \cite{us}, derived by performing the following 11D transformation
%\beq
%\begin{gathered}
%\chi \rightarrow \chi+ \xi\, \beta~~~~~~~~~~\beta\rightarrow \beta,~~~~~~~\phi\rightarrow\phi -\xi \,\beta, \\
%U(1)_r\rightarrow \chi+\phi.
%\end{gathered}
%\eeq
%and a dimensional reduction along $\beta$, is given by fixing $\zeta=-\xi$ in \eqref{eqn:generalresult1}.

%\begin{align}
%ds^2&= f_1^{\frac{3}{2}} f_5^{\frac{1}{2}}\sqrt{\Xi}\bigg[4ds^2(\text{AdS}_5)+f_4(d\sigma^2+d\eta^2)+ds^2(\text{M}_3)\bigg],~~~~e^{\frac{4}{3}\Phi}=  f_1 f_5\Xi\label{eqn:N=1main}\\[2mm]
%H_3 &= df_8\wedge \text{vol}(\text{S}^2)+ \xi \sin\theta df_7\wedge d\theta\wedge (d\phi+d\chi),\nn\\[2mm]
%C_1&=  \frac{\left(f_6+\xi\left(f_6^2+\frac{f_3}{f_5}\right)d\chi-\xi \frac{f_2}{f_5}\sin^2\theta d\phi\right)}{\Xi},~~~~ C_3=f_7 d\chi\wedge\text{vol}(\text{S}^2),\nn 
%\end{align}
%where
%\beq
%\Xi=\Delta+\xi^2\frac{f_2}{f_5}\sin^2\theta,~~~\Delta=(1+ \xi f_6)^2+\xi^2\frac{f_3}{f_5},~~~\Pi=1+\xi^2f_2\frac{f_3+f_5 f_6^2}{f_3 f_5}\sin^2\theta.
%\eeq
%with M$_3$ given by
%\begin{align}
%ds^2(\text{M}_3)&=f_2\left(d\theta^2+\frac{\Delta}{\Xi}\sin^2\theta D\phi^2\right)+\frac{f_3}{\Delta} d\chi^2=f_2\left(d\theta^2+\frac{1}{\Pi}\sin^2\theta d\phi^2\right)+\frac{\Pi}{\Xi}f_3D\chi^2,\nn\\[2mm]
%D\phi&= d\phi+\frac{\Delta-1-\xi f_6}{\Delta}d\chi,~~~D\chi=d\chi+\frac{\Pi-1+\frac{f_2 f_6}{f_3}\sin^2\theta}{\Pi}d\phi.\label{eqn:M3}
%\end{align}

  \section{More Type IIA solutions}\label{sec:furtherIIA}
%The primary focus of this section will be to present the results of dimensional reductions of Gaiotto-Maldacena along the $\chi$ and $\phi$ directions in turn, accompanying the $\beta$ reduction case considered in \cite{us}. We will then include some further analysis of the $\beta$ reduction case to finish off the section. The reader is referred to Appendix \ref{sec:DimRed} for a more in-depth derivation of the backgrounds presented in this section.

%We begin this section by investigating the one-parameter family of solutions which was not analysed in \cite{us}, namely the $\mathcal{N}=0$ solution with a broken $SU(2)\times U(1)$ isometry group (with $\xi=0,\zeta \neq0$ in Table \ref{table:1}). We then proceed to investigate some noteworthy solutions derived from dimensionally reducing Gaiotto-Maldacena along the $\chi$ and $\phi$ directions. The reader is referred to Appendix \ref{sec:DimRed} for a more in-depth derivation and discussion of the backgrounds presented in this section.

%\subsection{$\beta$ Reduction: $SU(2)\times U(1)$ breaking $\mathcal{N}=0$ deformation}

%\subsubsection*{$SU(2)\times U(1)$ breaking $\mathcal{N}=0$ deformation}

%\textcolor{red}{To add (could maybe not include this at all... only if I have something worth saying}

In this section we present some noteworthy solutions derived from dimensionally reducing Gaiotto-Maldacena along the $\chi$ and $\phi$ directions. The reader is referred to Appendix \ref{sec:DimRed} for a more in-depth derivation and discussion of the backgrounds presented in this section. 

The form of the metrics presented here match \eqref{eqn:generalresult1}, with all parameter dependence in the dilaton and $ds_2^2$. By the same calculation, the Holographic Central charge remains the same, given in \eqref{eqn:HCC}. These solutions then correspond to marginal deformations in the dual CFT. However, in the following cases, fixing all parameters to zero does not recover the $\mathcal{N}=2$ solution in Supergravity.

\subsection{$\chi$ Reduction}
One can derive a two-parameter family of solutions by dimensionally reducing along $\chi$. This solution can be mapped to \eqref{eqn:generalresult1} by the transformations given in \eqref{eqn:mapping1} and \eqref{eqn:mapping1-1}. Under this mapping, one requires $\xi\rightarrow 1/\xi$. Hence, by fixing $\xi=0$ we derive a unique background, which reads

     \begin{align}\label{eqn:IIAnew1} 
   &     ds_{10,st}^2=e^{\frac{2}{3}\Phi}f_1\bigg[4ds^2(\text{AdS}_5)+f_2d\theta^2+f_4(d\sigma^2+d\eta^2)\bigg]+ f_1^2 e^{-\frac{2}{3}\Phi}  ds^2_2,\nn\\[2mm]
   &        ds^2_2 = \big(f_3f_5 +\zeta^2  \sin^2\theta f_2 f_5\big) d\beta^2 +\sin^2\theta f_2\big(f_3+f_5 f_6^2\big)d\phi^2 -2\zeta \big( \sin^2\theta f_2  f_5f_6 \big)d\beta d\phi,\nn\\[2mm]
    &    e^{\frac{4}{3}\Phi}= f_1 \Big(f_3+f_5 f_6^2 + \zeta^2 \sin^2\theta f_2\Big) ,~~~~~~~~~~~~~
       B_2 = \sin\theta \big(  \zeta f_8 \,d\beta - f_7 d\phi\big) \wedge \, d\theta,\nn\\[2mm]
    &   C_1=  f_1 e^{-\frac{4}{3}\Phi} \bigg(f_5 f_6 d\beta  +\zeta \sin^2\theta f_2 d\phi\bigg)  ,~~~~~~~~
       C_3=  f_8  \sin\theta \,d\beta  \wedge \, d\theta \wedge d\phi.
    \end{align} 
This is a one-parameter family of $\mathcal{N}=0$ solutions, promoting to $\mathcal{N}=1$ for $\zeta=-1$ (see Appendix \ref{sec:chireduction} for further details).

\subsubsection{Unique $\mathcal{N}=1$ IIA Solution}
One now derives the following (unique) $\mathcal{N}=1$ solution (corresponding to the $\xi=0$ case of the one-parameter family of $\mathcal{N}=1$ solutions given in \eqref{eqn:chiN=1})\footnote{Here $(p,b,u)=1,(m,q,c)=0,v\equiv \gamma=0,a\equiv \xi=0,s\equiv\zeta=-1$.}

      \begin{align}\label{eqn:N=1IIA2}
&    ds_{10,st}^2=e^{\frac{2}{3}\Phi}f_1\bigg[4ds^2(\text{AdS}_5)+f_2d\theta^2+f_4(d\sigma^2+d\eta^2)\bigg]+ f_1^2 e^{-\frac{2}{3}\Phi}  ds^2_2,\nn\\[2mm]
  &    ds^2_2 = \big(f_3f_5 +  \sin^2\theta f_2 f_5\big) d\beta^2 +\sin^2\theta f_2\big(f_3+f_5f_6^2\big)d\phi^2 +2  \sin^2\theta f_2  f_5f_6 d\beta d\phi,\nn\\[2mm]
&    e^{\frac{4}{3}\Phi}= f_1f_5 \Big(f_6^2 + \frac{f_3}{f_5}+ \frac{f_2}{f_5}\sin^2\theta \Big),~~~~~~~~~~~~~%\Sigma=\Big(f_6^2 + \frac{f_3}{f_5}+ \frac{f_2}{f_5}\sin^2\theta \Big)\\
    B_2 =- \sin\theta \big(   f_8 d\beta + f_7d\phi\big) \wedge \, d\theta,\nn\\[2mm]
 &    C_1= f_1 e^{-\frac{4}{3}\Phi} \bigg(f_5 f_6 d\beta  -  f_2 \sin^2\theta d\phi\bigg)  ,~~~~~~~~
     C_3=  f_8  \sin\theta \,d\beta  \wedge \, d\theta \wedge d\phi.
  \end{align}

%         \begin{equation}\label{eqn:N=1IIA2}
 %    \hspace{-2cm}
 %   \begin{gathered}
  %      ds_{10,st}^2=f_1^{\frac{3}{2}}f_5^{\frac{1}{2}}\sqrt{\Sigma}\bigg[4ds^2(\text{AdS}_5)+f_2d\theta^2+f_4(d\sigma^2+d\eta^2)+ \frac{1}{ f_5  \Sigma} ds^2_2\bigg],\\
  %        ds^2_2 = \big(f_3f_5 +  \sin^2\theta f_2 f_5\big) d\beta^2 +\sin^2\theta f_2\big(f_3+f_5f_6^2\big)d\phi^2 +2  \sin^2\theta f_2  f_5f_6 d\beta d\phi,\\
   %     e^{\frac{4}{3}\Phi}= f_1f_5 \Sigma,~~~~~~~~~~~~~\Sigma=\Big(f_6^2 + \frac{f_3}{f_5}+ \frac{f_2}{f_5}\sin^2\theta \Big)\\
   %    B_2 =- \sin\theta \big(   f_8 d\beta + f_7d\phi\big) \wedge \, d\theta,~~~~~~~~
  %     C_1= f_1 e^{-\frac{4}{3}\Phi} \bigg(f_5 f_6 d\beta  -  f_2 \sin^2\theta d\phi\bigg)  ,\\
  %     C_3=  f_8  \sin\theta \,d\beta  \wedge \, d\theta \wedge d\phi,
  %  \end{gathered}
 %   \end{equation} 
    
    See Appendix \ref{sec:UniqIIAGstructures} for the G-Structure description.
 
\subsection{$\phi$ Reduction}

As one would expect, dimensionally reducing along $\phi$ again leads to a two-parameter family of solutions (given in \eqref{eqn:phieqmatch}). Once again, this solution can be mapped to \eqref{eqn:generalresult1} using the transformations outlined in \eqref{eqn:mappingphi1} and \eqref{eqn:mappingphi2}. In this case, the mapping involves transformations with both $1/\xi$ and $1/\zeta$. We therefore derive the following backgrounds.

When $\zeta=0$, one gets the following one-parameter family of $\mathcal{N}=0$ solutions
   \begin{align}\label{eqn:newIIA2}
 &       ds_{10,st}^2=e^{\frac{2}{3}\Phi}f_1\bigg[4ds^2(\text{AdS}_5)+f_2d\theta^2+f_4(d\sigma^2+d\eta^2)\bigg]+f_1^2 e^{-\frac{2}{3}\Phi}  ds^2_2,\nn\\[2mm]
 &           ds^2_2 = \Big( \xi^2 f_3f_5  +\sin^2\theta f_2( f_3   +f_5 f_6^2)\Big) d\chi^2 +   f_2 f_5 \sin^2\theta d\beta^2 +2 f_2 f_5 f_6   \sin^2\theta d\chi d\beta,\nn\\[2mm]
 &       e^{\frac{4}{3}\Phi}=  f_1 \Big(\xi^2 f_5 + \sin^2\theta f_2\Big) ,~~~~~~~~~~~~~ C_1= \xi  f_1 f_5  e^{-\frac{4}{3}\Phi} \Big(  f_6  d\chi+ d\beta\Big),\nn\\[2mm]
  &     B_2= \sin\theta\Big(   f_7 d\chi+  f_8 d\beta\Big)\wedge d\theta,~~~~~~~~~~~   C_3 =0.
    \end{align}

When $\xi=0$, one now derives %(\textcolor{red}{less sure this one is unique - need to double check...})
   \begin{align}\label{eqn:newIIA3}
  &      ds_{10,st}^2=e^{\frac{2}{3}\Phi}f_1\bigg[4ds^2(\text{AdS}_5)+f_2d\theta^2+f_4(d\sigma^2+d\eta^2)\bigg]+f_1^2 e^{-\frac{2}{3}\Phi}  ds^2_2,\nn\\[2mm]
   &         ds^2_2 =  f_2( f_3   +f_5 f_6^2) \sin^2\theta\, d\chi^2 +\Big(\zeta^2 f_3f_5  +\sin^2\theta f_2 f_5\Big) d\beta^2 +2 f_2  f_5 f_6   \sin^2\theta d\chi d\beta,\nn\\[2mm]
 &       e^{\frac{4}{3}\Phi}=  f_1 \Big[\zeta^2  (f_5 f_6^2 +  f_3)+\sin^2\theta f_2\Big] ,~~~~~~~~~~~~~~~        C_1= \zeta  f_1 e^{-\frac{4}{3}\Phi} \Big[  \Big(  f_3 + f_5f_6^2 \Big)  d\chi+ f_5 f_6 d\beta\Big],\nn\\[2mm]
  &     B_2= \sin\theta\Big(   f_7 d\chi+  f_8 d\beta\Big)\wedge d\theta,~~~~~~~~~~~   C_3 =0.
    \end{align} 
   This is a one-parameter family of $\mathcal{N}=0$ solutions which enhances to the $\mathcal{N}=1$ background given in \eqref{eqn:N=1IIA2} when $\zeta=-1$ (following appropriate gauge transformations - see \eqref{eqn:N1trans} with $\xi=0$).

  \section{Type IIB T-Duals}\label{sec:IIB}
  We now present IIB solutions derived from Abelian T-Dualising the IIA reductions. In all backgrounds presented throughout this paper, the transformation parameters drop out exactly when calculating the Holographic Central Charge (see Appendix \ref{sec:HCC}), matching the IIA result in  \eqref{eqn:HCC}. This is not particularly surprising given the arguments presented in Section 4.3 of \cite{Macpherson:2014eza}. Hence, the transformation parameters once again correspond to marginal deformations in a dual CFT.
  We now present a three-parameter family of solutions.
  \subsection{Three Parameter Family}
From the $U(1)_r$ component given in \eqref{eqn:U(1)}, fixing $(p,b,u)=1,~(a,c,m)=0,~q\equiv \xi,v\equiv \zeta,s\equiv \gamma$, we have
\beq\label{eqn:newU(1)orig}
\begin{aligned}
U(1)_r&= (1+\gamma)\chi +(\xi+\zeta)\beta + \phi.
\end{aligned}
\eeq
Performing a dimensional reduction along $\beta$, followed by an Abelian T-Duality along $\chi$, one gets the following three-parameter family of Type IIB solutions\footnote{See also the 3-parameter families given in \eqref{eqn:ATD2}, \eqref{eqn:ATD3} and \eqref{eqn:ATD6}, which can all be mapped to \eqref{eqn:ATD1} as outlined there.}

    \begin{align}\label{eqn:ATD1}
 &      ds_{10,B}^2= f_1^{\frac{3}{2}}f_5^{\frac{1}{2}}\sqrt{\Xi}\Bigg[4ds^2(\text{AdS}_5)+f_4(d\sigma^2+d\eta^2) +ds^2(M_3)\Bigg],\nn\\[2mm]
&   ds^2(M_3)=f_2\bigg(d\theta^2  + \frac{1}{\Pi} \sin^2\theta\, d\phi^2\bigg)+\frac{1}{f_1^3f_3f_5\Pi }\,D\chi^2,~~~~~~~~~~~~~~
%\Xi_1 = f_3f_5+f_2\big[f_3(\gamma\xi-\zeta)^2 +f_5\big(\gamma +(\gamma\xi-\zeta)f_6\big)^2\big]\sin^2\theta,\\
  e^{2\Phi_\mathcal{B}}=\frac{f_5}{f_3}\frac{\Xi^2}{\Pi } ,\nn\\[2mm]%~~~~~~~~~~~~~~ e^{\frac{4}{3}\Phi_\mathcal{A}}= f_1f_5\Xi,\\
&  C_0=%  f_1 e^{-\frac{4}{3}\Phi_\mathcal{A}}\Big(f_5 f_6 +\xi f_3 + \xi f_5f_6^2 +\gamma \zeta \sin^2\theta f_2\Big),
  \frac{1}{\Xi}\bigg(f_6(1+\xi f_6) +\xi \frac{f_3}{f_5}+\gamma \zeta \frac{f_2}{f_5}\sin^2\theta\bigg),~~~~~~~~~~~~D\chi=d\chi -\big((\gamma \xi -\zeta )f_7 +\gamma f_8 \big)\sin\theta  d\theta, \\[2mm]
&B_2=-\frac{1}{\Pi} \bigg( f_8+\xi f_7 -\zeta \frac{f_2}{f_3} \Big((\gamma\xi-\zeta)\frac{f_3}{f_5}f_8 -(f_7-f_6f_8)\big(\gamma +(\gamma\xi-\zeta)f_6\big)\Big)\sin^2\theta \bigg) \sin\theta\,d\phi \wedge d\theta \nn\\[2mm]
&~~~~~~~~-  \frac{1}{\Pi} \frac{f_2}{f_3}\Big(\xi (\gamma\xi-\zeta)\frac{f_3}{f_5} +(1+\xi f_6)\big(\gamma+(\gamma\xi-\zeta)f_6\big)\Big)\sin^2\theta  d\phi\wedge d\chi ,\nn\\[2mm]
&  C_2=\frac{1}{\Pi}\bigg( f_7-\gamma \frac{f_2}{f_3}\Big((\gamma\xi-\zeta)\frac{f_3}{f_5}f_8+(f_6f_8-f_7)\big(\gamma+(\gamma\xi-\zeta)f_6\big)\Big)\sin^2\theta \bigg)\sin\theta  d\theta \wedge d\phi\nn\\[2mm]
&~~~~~~~  - \frac{1}{\Pi}  \frac{f_2}{f_3}\Big( (\gamma\xi-\zeta)\frac{f_3}{f_5} +f_6\big(\gamma+(\gamma\xi-\zeta)f_6\big)\Big)\sin^2\theta  d\phi\wedge d\chi,\nn\\[2mm]
&\Pi= 1+\frac{f_2}{f_3}\bigg(\frac{f_3}{f_5}(\gamma\xi-\zeta)^2 +\big(\gamma +(\gamma\xi-\zeta)f_6\big)^2\bigg)\sin^2\theta,~~~~~~~~~~~~~\Xi=(1+\xi f_6)^2 +\xi^2\frac{f_3}{f_5}+\zeta^2 \frac{f_2}{f_5}\sin^2\theta.
  \end{align}
   We can preserve the $U(1)_r$ component given in \eqref{eqn:newU(1)orig} by first fixing $\zeta=-\xi$ (as in the IIA case), followed by fixing $\gamma=-1$ under the T-Duality. These conditions together mean $\gamma\xi-\zeta=0$. Due to the transformation given in \eqref{eqn:S2breakingdefns}, such conditions break the $SU(2)_R$ component of the R-Symmetry - leading to an $\mathcal{N}=1$ background. All other solutions are of course $\mathcal{N}=0$ backgrounds.
One can recover the $S^2$ by enforcing $\gamma=\zeta=0$, giving a one-parameter family of $S^2$ preserved solutions (corresponding to the ATD of the $S^2$ preserved $\mathcal{N}=0$ solution studied in \cite{us}). Unlike in the IIA case however, this is $\mathcal{N}=0$ for all values of $\xi$ (including $\xi=0$). This is still a noteworthy example, so we present it explicitly in a later subsection. We also present the $\mathcal{N}=1$ solution explicitly. One could split \eqref{eqn:ATD1} into two categories of 2-parameter families, each promoting to the $\mathcal{N}=1$ solution in a different manner\footnote{It may prove useful to relabel $\gamma\equiv \hat{\gamma}-1$ such that the $\mathcal{N}=1$ solution is recovered for $\hat{\gamma}=0$, however given that $\gamma=0$ can lead to a recovered $S^2$, we leave as is. }
  \begin{itemize}
  \item Case 1: with $\zeta=-\xi$ (or $\zeta=\gamma\xi$) and $\gamma\in\mathds{Z}$ (promoting to $\mathcal{N}=1$ for $\gamma=-1$), 
   \item Case 2: with $\gamma=-1$ and $(\zeta,\xi)\in \mathds{Z}$ (promoting to $\mathcal{N}=1$ for $\zeta=-\xi$).
  \end{itemize}
  We will however focus on a more general discussion of the full three parameter family, investigating the boundary by the same procedure as the IIA solutions.
%We will however focus directly on the three parameter family, attempting to investigate it's behaviour at the boundaries of the $(\sigma,\eta)$ space - copying the analysis conducted in \cite{us}.\\

\subsection*{Investigations at the Boundary}
We first note, for general values of $(\eta,\sigma,\theta)$, $(\Xi,\Pi)$ are non-zero and finite. The deformed $S^2$ given by $(\theta,\phi)$ has $\Pi\rightarrow 1$ at the poles, which given the expression for $M_3$ in \eqref{eqn:ATD1}, means it still behaves topologically as an $S^2$. This follows the argument given in \cite{us}. We now turn to the behaviour at the boundaries, keeping all three parameters non-zero. We will now use the large gauge transformation which includes the additional term introduced in the IIA discussion
 \beq\label{eqn:LGT2}
  B_2\rightarrow B_2+2\kappa( k+\xi N_k)\text{Vol}(S^2).
  \eeq
  
\begin{itemize}
\item At $\sigma\rightarrow \infty$, we use utilise \eqref{eqn:Vlimit} and \eqref{eqn:finf} to find
    \begin{equation}
  \begin{gathered}
       ds_{10,B}^2=\kappa  \Bigg[4\sigma ds^2(\text{AdS}_5) +\frac{2P}{\pi}\bigg(d\Big(\frac{\pi}{P}\sigma\Big)^2+d\Big(\frac{\pi}{P}\eta\Big)^2+  \sin^2\bigg(\frac{\pi\eta}{P}\bigg) (d\theta^2+ \sin^2\theta\, d\phi^2)\bigg) \\
     ~~~~~~~~~~~~~~~~~~~~~~~~~~~~+\frac{1}{4\kappa^2\sigma } \,\bigg(d\chi +\gamma \frac{\kappa P}{\pi} \Big(\frac{2\pi}{P}\eta-\sin\Big(\frac{2\pi}{P}\eta\Big) \Big)\sin\theta  d\theta\bigg)^2 \Bigg],
       \\
  e^{2\Phi_\mathcal{B}}=\frac{P^2}{\pi^3\mathcal{R}_1^2 }e^{\frac{2\pi \sigma}{P}} ,%~~~~~~~~~~~~~~~~~~~~~~~~~~~~ 
%  C_0=   \sqrt{\frac{2}{P\sigma}} \pi\mathcal{R}_1 e^{-\frac{\pi \sigma}{P}} \cos\Big(\frac{\pi \eta}{P}\Big),\\
~~~~~~~~~~%B_2= - \kappa\Big(2\eta-\frac{P}{\pi}\sin\Big(\frac{2\pi}{P}\eta\Big) \Big) \sin\theta\,d \theta\wedge d\phi  
H_3= - \frac{4\kappa P}{\pi} \sin^2\Big(\frac{\pi}{P}\eta\Big)\sin\theta\,  d\Big(\frac{\pi}{P}\eta\Big)\wedge d \theta\wedge d\phi  .
  \end{gathered}
  \end{equation}  
  It is immediately clear that $\gamma$ plays a special role here, the only parameter of the three which remains in this limit - when $\gamma=0$, the $S^2$ is recovered. So too is the $S^3$ spanned by $(\eta,S^2)$. This makes sense given that the $\mathcal{N}=1$ solution (with a broken $SU(2)$) is derived when $\gamma=-1$.   The background tends to a stack of NS5 branes. The $S^1$ shrinks in this limit, in contrast to the IIA cases where it grows alongside the AdS$_5$ - this makes intuitive sense following the T-Duality. Again we find $P$ NS5 branes, as follows (with $2\kappa=\pi$)
  \beq
Q_{NS5}=-\frac{1}{(2\pi)^2}\int_{(\eta,\theta,\phi)}H_3=P.
\eeq
  \item At $\eta=0$ with $\sigma\neq 0$, using the Warp factor limits in \eqref{eqn:feq}, gives 
%        \begin{align}
%  &     ds^2=  \sqrt{\frac{2f+|\dot{f}|}{|\dot{f}|}}\sqrt{\Xi} \Bigg[ 4\kappa \sigma \,ds^2(\text{AdS}_5)+\frac{1}{4\kappa\sigma }\bigg(d\chi +2\kappa \eta \Big(2(\gamma \xi-\zeta ) \frac{|\dot{f}|}{\sigma^2}+\gamma \Big(1 -\frac{1}{f}\Big) \Big)\sin\theta d\theta\bigg)^2 \nn\\[2mm]
%&  ~~~~~~~~~~~~~~~~~~     +\kappa  \frac{2|\dot{f}|}{\sigma f}\bigg(d\sigma^2+d\eta^2 +\eta^2 (d\theta^2+\sin^2\theta d\phi^2)\bigg)\Bigg],~~~~~~~~~~~  \Xi=\bigg(1+\frac{2\xi f^2}{2f+|\dot{f}|}\bigg)^2 +8\xi^2 \frac{|\dot{f}|}{f^3}
%
%  \end{align}  
        \begin{align}
  &   f_4(d\sigma^2+d\eta^2)+ds^2(M_3)=\frac{1}{\sigma^2}\bigg[   \frac{2|\dot{f}|}{ f}\bigg(d\sigma^2+d\eta^2 +\eta^2 (d\theta^2+\sin^2\theta d\phi^2)\bigg)+\frac{1}{4\kappa^2}D\chi^2\bigg], \nn\\[2mm]
&D\chi=d\chi +2\kappa\, \eta \Big[2(\gamma \xi-\zeta ) \frac{|\dot{f}|}{\sigma^2}+\gamma \Big(1 -\frac{1}{f}\Big) \Big]\sin\theta d\theta.
  \end{align}  
  %\textcolor{red}{UNDERTSAND f8 HERE....IS THE TERM IN D$\chi$ TRIVIAL, EITHER TOPOLOGICALLY OR BECAUSE ETA=0 THERE?}
    In this limit we notice that the $\mathds{R}^3$ of $(\eta,S^2)$ is recovered only when all three parameters are zero.%$\gamma\xi-\zeta=0$. This condition is of course trivially satisfied for $(\gamma,\xi,\zeta)=0$, but more interestingly, under the $\mathcal{N}=1$ condition - where $\gamma=-1$ and $\zeta=-\xi$.
    
\item At $\sigma=0,~\eta\in (k,k+1)$, recall that along the $\sigma=0$ boundary, $\ddot{V}=0$ to leading order. We now see
     \beq
     \begin{gathered}
     \Xi \rightarrow l_k^2 +\frac{1}{2}\zeta^2 \mathcal{R}V'' \sin^2\theta,~~~~~~~~~~~~~\frac{\Pi}{f_2}\rightarrow  \frac{\mathcal{R}}{2\sigma^2 V''}\hat{l}_k^2 \sin^2\theta,\\ %\hat{l}_k^2 \frac{1}{2}  \mathcal{R}V'' \sin^2\theta,\\
     l_k\equiv 1+\xi (N_{k+1}-N_k),~~~~~~~~~~~~~~\hat{l}_k \equiv \gamma+(\gamma \xi-\zeta )(N_{k+1}-N_k) =\gamma \, l_k-\zeta (N_{k+1}-N_k),
     \end{gathered}
     \eeq
     where
   \beq
   f_4(d\sigma^2+d\eta^2) +\frac{f_2 }{\Pi}\sin^2\theta\, d\phi^2  \rightarrow   \frac{2V''}{\mathcal{R}}\bigg(d\eta^2+\Big(d\sigma^2 +\frac{\sigma^2}{\hat{l}_k^2}d\phi^2\Big)\bigg).% + \frac{2\mathcal{R}V''}{2\mathcal{R}V''+(\mathcal{R}')^2}d\theta^2 + \frac{2\mathcal{R}V''+(\mathcal{R}')^2}{4\kappa^2\mathcal{R}^2 \hat{l}_k^2 \sin^2\theta}D\chi^2
     \eeq
     So there is a $\mathds{R}^2/\mathds{Z}_{\hat{l}_k}$ orbifold singularity in $(\sigma,\phi)$, with $\hat{l}_k$. Recall, in the IIA case, this orbifold singularity was over $(\sigma,\chi)$, with $l_k$ - see \eqref{eqn:orbifold1}. So in moving from IIA to IIB, $\chi$ has been replaced by $\phi$, and the orbifold singularity made a little more general ($l_k\rightarrow \hat{l}_k$). %, depending on the parameters chosen. 
     It is worth noting that the case of $(\gamma,\zeta)=0$ must be treated separately because $\hat{l}_k=0$ (this is the $S^2$ preserving condition). Now we turn to the remaining internal metric component
%   \textcolor{red}{ Let's rewrite the following component as follows
   %  \beq
    % \begin{gathered}
    % f_2 d\theta^2  +\frac{1}{f_1^3f_5^2\Pi }\,D\chi^2 =  \frac{f_2}{f_1^3f_5^2f_2\Pi+\Theta^2 \sin^2\theta}d\chi^2 + \frac{f_1^3f_5^2f_2\Pi+\Theta^2 \sin^2\theta}{f_1^3f_5^2\Pi}D\theta^2,\\
   %  D\chi = d\chi -\Theta \sin\theta  d\theta,~~~~~~~~~~~D\theta = d\theta - \frac{\Theta }{f_1^3f_5^2f_2\Pi+\Theta^2 \sin^2\theta}\sin\theta d\chi,\\
    % \Theta \equiv (\gamma \xi -\zeta )f_7 +\gamma f_8.
   %  \end{gathered}
   %  \eeq}
     \beq
          \begin{gathered}
       f_2 d\theta^2  +\frac{1}{f_1^3f_3f_5\Pi}\,D\chi^2 \rightarrow \frac{2\mathcal{R}V''}{2\mathcal{R}V''+(\mathcal{R}')^2}d\theta^2 + \frac{2\mathcal{R}V''+(\mathcal{R}')^2}{4\kappa^2\mathcal{R}^2}\frac{1}{ \hat{l}_k^2 \sin^2\theta}D\chi^2,\\
       %   D\chi\rightarrow d\chi +\frac{2\kappa }{2\mathcal{R}V''+(\mathcal{R}')^2}\bigg(2(\gamma\xi-\zeta)\mathcal{R}^2 V'' +\gamma \Big(\eta\big(2\mathcal{R}V''+(\mathcal{R}')^2\big)-\mathcal{R}\mathcal{R}'\Big)\bigg)\sin\theta d\theta\\
                D\chi\rightarrow d\chi +2\kappa\bigg( \gamma\, \eta +\frac{\mathcal{R} }{2\mathcal{R}V''+(\mathcal{R}')^2} \Big(2(\gamma\xi-\zeta)\mathcal{R} V'' -\gamma \mathcal{R}' \Big)\bigg)\sin\theta d\theta    .
               \end{gathered}
     \eeq
    Notice that for the $\mathcal{N}=1$ condition, $\gamma\xi-\zeta=0$, we have $\hat{l}_k=\gamma=-1$.

  \item At $\sigma=0,~\eta=0$, we again adopt the coordinate change \eqref{eqQdef}, where
    \beq
    \Xi\rightarrow l_0^2,~~~~~~~~~~ \Pi \rightarrow 1+ \hat{l}_0^2 \,\text{cot}^2\alpha\sin^2\theta,~~~~~~~~~~~~l_0=1+\xi N_1,~~~~~~~~~~~\hat{l}_0 = \gamma\,l_0-\zeta  N_1 . %\frac{N_1}{2r^2Q \sin^2\alpha }\hat{l}_0^2\sin^2\theta
    \eeq
 %Once again, we would need to treat the $S^2$ preserving case separately.  
 We note in general
  \begin{align}
 & f_4(d\sigma^2+d\eta^2) +ds^2(M_3) \rightarrow \frac{2Q}{N_1}\bigg(dr^2+r^2d\alpha^2+r^2\cos^2\alpha \bigg(d\theta^2 +\frac{\sin^2\theta \,d\phi^2}{1+ \hat{l}_0^2\, \text{cot}^2\alpha\sin^2\theta}\bigg)\bigg) \nn\\[2mm]
& ~~~~~~ +\frac{D\chi^2}{4\kappa^2  r^2 \sin^2\alpha ( 1+\hat{l}_0^2\, \text{cot}^2\alpha \sin^2\theta)},~~~~~~~~~D\chi=d\chi+4\kappa (\gamma\xi-\zeta)Qr^3 \cos^3\alpha \sin\theta d\theta.
  \end{align}
%  \begin{itemize}
%\item \underline{$\gamma=-1,\zeta=-\xi$}:  In the $\mathcal{N}=1$ situation, $D\chi\rightarrow d\chi$ with $\hat{l}_0=-1$.
%%    \beq
%%  \begin{gathered}
%%  f_4(d\sigma^2+d\eta^2) +ds^2(M_3)\rightarrow\frac{2Q}{N_1}\bigg(dr^2+r^2d\alpha^2+r^2\cos^2\alpha \bigg(d\theta^2 +\frac{\sin^2\theta \,d\phi^2}{1+   \text{cot}^2\alpha\sin^2\theta}\bigg)\bigg) \\
% % +\frac{N_1^2}{4\kappa^2  r^2 \sin^2\alpha ( 1+ \text{cot}^2\alpha \sin^2\theta)}d\chi^2
 %% %\\\rightarrow %\frac{2Q}{N_1}\bigg(dr^2+r^2d\alpha^2+r^2\cos^2\alpha d\theta^2 + r^2\sin^2\alpha d\phi^2 
%% % +\frac{N_1^3 d\chi^2}{8\kappa^2 Q r^2 (\sin^2\alpha +\cos^2\alpha \,\sin^2\theta)} \bigg)
% % \end{gathered}
 % %\eeq
 % \item \underline{$(\gamma,\zeta)=0$}: In this case $\hat{l}_0=0$, giving
 %   \beq
 % \begin{gathered}
 % f_4(d\sigma^2+d\eta^2) +ds^2(M_3) \rightarrow \frac{2Q}{N_1}\bigg(dr^2+r^2d\alpha^2+r^2\cos^2\alpha ds^2(S^2)\bigg) 
%  +\frac{N_1^2}{4\kappa^2  r^2 \sin^2\alpha}d\chi ^2.
 % \end{gathered}
 % \eeq
%  \end{itemize}
So we see, as one would expect following a T-Duality, the internal space no longer vanishes as $\mathds{R}^5/\mathds{Z}_{l_0}$. The orbifold singularity here is also less clean.
Note that for the $\mathcal{N}=1$ case, $D\chi\rightarrow d\chi$ and again, $\hat{l}_0=-1$.

\item At $\sigma=0,~\eta=k$,  we use \eqref{eqn:fsfork} as in the IIA case, recalling that in this limit, the term $f_2/f_5$ term dominates away from the pole. We first assume then that we are indeed away from the pole, finding %We need to split this into two sections. We start with the case $\zeta\neq0$ before looking at the $\zeta=0$ situation.
\begin{align}
&\Xi\rightarrow \frac{b_k\zeta^2 N_k\sin^2\theta}{4r},~~~~~~~\Pi\rightarrow \frac{N_k\sin^2\theta}{b_k r\sin^2\alpha}\Pi_k(\alpha),~~~~~~~~\Pi_k\equiv (\gamma\xi-\zeta)^2 \frac{b_k^2}{4}\sin^2\alpha +\Big(\gamma+(\gamma\xi-\zeta) g(\alpha)\Big)^2,\nn\\[2mm]
&\Pi_k(\alpha=0) = \hat{l}_{k-1}^2,~~~~~~~\Pi_k(\alpha=\pi) =\hat{l}_k^2.
\end{align}
We make the same coordinate transformation as the IIA case, $r=z^2$, giving    
    \begin{align}\label{eqn:zeq}
 &      ds_{10,B}^2=\zeta \kappa N_k\sin\theta \Big(4ds^2(\text{AdS}_5)+ds^2(\mathds{B}_5)\Big),\nn\\[2mm]
%      ds^2(B_5) = \frac{b_k}{N_k r}(dr^2+r^2d\alpha^2) +%ds^2(M_3) \\
 %  ds^2(M_3)= 
 %d\theta^2  +  \frac{b_k  }{N_k\Pi_k}\bigg(  r\sin^2\alpha \,  d\phi^2+\frac{1}{4\kappa^2 r\sin^2\theta} (d\chi +\mathcal{A}_k )^2 \bigg),\\
&       ds^2(\mathds{B}_5) =d\theta^2  + \frac{4b_k}{N_k }\bigg( dz^2+\frac{z^2}{4} \Big(d\alpha^2 +   \frac{\sin^2\alpha  }{\Pi_k}\,  d\phi^2 \Big)+ \frac{1}{z^2\sin^2\theta}\frac{ (d\chi +\mathcal{A}_k )^2 }{16 \,\kappa^2 \Pi_k}  \bigg),\nn\\[2mm]
&  e^{2\Phi_\mathcal{B}}=\frac{\zeta^4 N_k b_k\sin^2\theta  }{ 4z^2 \Pi_k},~~~~~~~~~~~~~~\mathcal{A}_k =2\kappa \big((\gamma \xi -\zeta )N_k+\gamma k\big)\sin\theta  d\theta ,~~~~~~~~~~C_0=  \frac{ \gamma }{\zeta}\nn\\[2mm]
%&B_2=\frac{1}{\Pi_k}    \bigg(\Big(\frac{1}{4}\xi (\gamma\xi-\zeta)b_k^2 \sin^2\alpha +\big(1+\xi g(\alpha)\big)\big(\gamma+(\gamma\xi-\zeta)g(\alpha) \big)\Big) d\chi\wedge d\phi      \nn\\[2mm]
%&~~~~~~~~~~~~~~~~~~+  2\zeta \kappa   \Big( (\gamma\xi-\zeta)\frac{k}{4}b_k^2 \sin^2\alpha  - (N_k-k\,g(\alpha))\big(\gamma +(\gamma\xi-\zeta)g(\alpha)\big)\Big)  \sin\theta\,d\theta\wedge d\phi\bigg) , \nn\\[2mm]%prior to large gauge trans..
&B_2=\frac{1}{\Pi_k} \bigg(\Big(1+\xi g(\alpha)\Big)\Big(\gamma+(\gamma\xi-\zeta)g(\alpha)\Big) +\frac{1}{4}\xi b_k^2 (\gamma \xi-\zeta)\sin^2\alpha\bigg) \nn\\[2mm]
&~~~~~~~~~~~~~~~~~~~~~~~~\times \bigg(2\kappa \Big(\gamma\,k +(\gamma \xi-\zeta)N_k\Big)\sin\theta d\theta \wedge d\phi -d\phi\wedge d\chi\bigg), \nn\\[2mm]
&   C_2=\frac{1}{\Pi_k } \bigg(\Big( (\gamma\xi-\zeta)\frac{1}{4}b_k^2\sin^2\alpha +g(\alpha)\big(\gamma+(\gamma\xi-\zeta)g(\alpha)\big)\Big)  d\chi\wedge d\phi   ~~~~~~~~~~~~~~~~~~~~~~~~  \\[2mm]
&~~~~~~~~~~~~~~~~~~+2\kappa\gamma \Big((\gamma\xi-\zeta)\, \frac{ k}{4}\,  b_k^2\sin^2\alpha+( k\,g(\alpha)-N_k)\big(\gamma+(\gamma\xi-\zeta)g(\alpha)\big)\Big) \sin\theta  d\theta \wedge d\phi\bigg).\nn
  \end{align} 
We note that $\mathds{B}_5$ corresponds to the ATD of the $(\theta,~z,~\mathds{B}_3)$ components of \eqref{eqn:B3eq} (using the appropriate gauge transformation for $B_2$).  Now, we find
  \begin{align}
 &    ds^2(\mathds{B}_5) \Big|_{\alpha\sim0}= d\theta^2  + \frac{4b_k}{N_k }\bigg( dz^2+\frac{z^2}{4} \Big(d\alpha^2 +   \frac{\alpha^2  }{\hat{l}_{k-1}^2}\,  d\phi^2 \Big)+ \frac{1}{z^2\sin^2\theta}\frac{ (d\chi +\mathcal{A}_k )^2 }{16 \,\kappa^2 \hat{l}_{k-1}^2}  \bigg),\\[2mm]
  &    ds^2(\mathds{B}_5) \Big|_{\alpha\sim\pi}= d\theta^2  + \frac{4b_k}{N_k }\bigg( dz^2+\frac{z^2}{4} \Big(d\alpha^2 +   \frac{(\pi-\alpha)^2  }{\hat{l}_{k}^2}\,  d\phi^2 \Big)+ \frac{1}{z^2\sin^2\theta}\frac{ (d\chi +\mathcal{A}_k )^2 }{16 \,\kappa^2 \hat{l}_{k}^2}  \bigg).\nn
  \end{align}
  %and noting \textcolor{red}{the periods of $\chi$ and $\theta$ differ by a factor of 2 so the connection term is not topologically trivial but $d\mathcal{A}_k=0$}.
  Note the presence of a $\sin\theta$ out the front of the whole metric! We observe that $(\alpha,\phi)$ form a $\mathds{W}\mathds{C}\mathds{P}^1_{[\hat{l}_{k-1},\hat{l}_k]}$, with Euler characteristic\footnote{%The $+$ signs inside the two brackets are technically $\pm$ contributions, coming from taking square-roots. However, by fixing both signs to be $+$, 
  Where we have taken $\sqrt{\hat{l}_k^2}\equiv |\hat{l}_k|$, such that in the $\mathcal{N}=1$ case (where $\hat{l}_k=\hat{l}_{k-1}=-1$), we find $\chi_E=2$ - which recovers the necessary $S^2$ (or $\mathds{C}\mathds{P}^1$). Without taking the absolute value here, one would instead (incorrectly) find either $\mathcal{\chi}_E=0$ or $\mathcal{\chi}_E=-2$, describing a genus 1 or genus 2-torus, respectively. This subtlety arises because it is only $\hat{l}_k^2$ which shows up in the metric, not $\hat{l}_k$ itself. Alternatively, one could define $\hat{l}_k \equiv |\gamma \, l_k-\zeta (N_{k+1}-N_k)|$ from the outset. }
\beq
\chi_E=\frac{1}{2\pi}\int_{\mathds{W}\mathds{C}\mathds{P}^1_{[\hat{l}_{k-1},\hat{l}_k]}}R \,\text{Vol}_2 = 2-\bigg(1- \frac{1}{|\hat{l}_{k-1}|}\bigg)-\bigg(1-\frac{1}{|\hat{l}_{k}|}\bigg).
\eeq
%Where the two $\pm$ contributions come from taking square-roots. Taking both signs to be $+$ would mean, in the $\mathcal{N}=1$ case (where $\hat{l}_k=\hat{l}_{k-1}=-1$), we find $\chi_E=2$ - which recovers the $S^2$ (or $\mathds{C}\mathds{P}^1$). Note that $\mathcal{\chi}_E=-2$ describes a genus 2-Torus.
 \\ Now we see that $C_2-C_0 B_2=\frac{1}{\zeta}d\phi\wedge d\chi$, meaning that there are no D5 branes present for $\zeta\neq0$, noting $\hat{F}_3=d(C_2\wedge e^{-B_2})$. Before we approach the pole, let's make an aside and look at the $\zeta=0$ case in this limit.
  \begin{itemize}
  \item \underline{$\zeta=0$}: When $\gamma\neq0$ we have $\Pi_k\rightarrow \gamma^2\Delta_k$, hence
  \begin{align}
  & \Xi\rightarrow \Delta_k,~~~~~~~~\Pi\rightarrow \gamma^2 \frac{N_k\sin^2\theta}{b_k r\sin^2\alpha}\Delta_k,~~~~~~~~\Delta_k\equiv  \frac{1}{4}\xi^2 b_k^2 \sin^2\alpha +\Big(1+ \xi g(\alpha)\Big)^2,
  \end{align}
  we then see (with the same $\mathcal{A}_k$ as above and skipping the gauge transformation in $B_2$ for the moment)
  \begin{align}
  &ds^2= 2\kappa \sqrt{\frac{N_k}{b_k}}\sqrt{\Delta_k}\, z \Big[4ds^2(AdS_5)+ ds^2(\mathds{B}_5)\Big], ~~~~~~~e^{2\Phi_{\mathcal{B}}} =\frac{4z^2 \Delta_k}{\gamma^2b_k N_k \sin^2\theta}, \nn\\[2mm]
  &ds^2(\mathds{B}_5) = d\theta^2 +\frac{4b_k}{N_k}\bigg(dz^2+\frac{1}{4}z^2 \Big(d\alpha^2 + \frac{\sin^2\alpha}{\gamma^2 \Delta_k}d\phi^2\Big) +\frac{1}{ z^2}\frac{(d\chi+\mathcal{A}_k)^2}{16\gamma^2 \kappa^2\Delta_k \sin^2\theta} \bigg),\nn\\[2mm]
  &%B_2=-\frac{1}{\gamma}d\phi\wedge d\chi +2\kappa (k+\xi N_k)\sin\theta d\theta \wedge d\phi,
 B_2=-\frac{1}{\gamma}d\phi\wedge d\chi , ~~~~~~~C_0=\frac{1}{\Delta_k}\Big(g(\alpha)\Big(1+\xi g(\alpha)\Big) +\frac{1}{4}b_k^2 \xi \sin^2\alpha\Big),\nn\\[2mm]
  &C_2=\frac{1}{\Delta_k} \bigg[2\kappa \bigg(\Big(1+\xi g(\alpha)\Big)\Big(k g(\alpha) -N_k\Big)+\frac{1}{4}b_k^2 \xi k \sin^2\alpha\bigg) \sin\theta d\theta \wedge d\phi \nn\\[2mm]
  &~~~~~~~~~~~~~~~~~~~-\frac{1}{\gamma} \bigg(g(\alpha)\Big(1+\xi g(\alpha)\Big) +\frac{1}{4} b_k^2 \xi \sin^2\alpha \bigg) d\phi\wedge d\chi \bigg].
  \end{align}
  Note, similar to the IIA case, the $S^2$ has been replaced by a Spindle here. 
  Now we again impose the two gauge transformations separately. Using \eqref{eqn:LGT}, we have % beg \eqref{eqn:LGT} instead of \eqref{eqn:LGT2}, we instead have 
  \begin{align}\label{eqn:D5charge}
  &  B_2\rightarrow B_2+2\kappa k\text{Vol}(S^2), ~~~~~~~~~~~~~~~B_2=-\frac{1}{\gamma}d\phi\wedge d\chi +2\kappa k \sin\theta d\theta \wedge d\phi,\nn \\[2mm]
   &Q_{D5}=-\frac{1}{(2\pi)^2} \int_{(\alpha,\theta,\phi)} \hat{F}_3 =- \frac{1}{(2\pi)^2} \int_{\phi=0}^{\phi=2\pi}\int_{\theta=0}^{\theta=\pi} \Big[C_3-C_1\wedge B_2\Big]_{\alpha=0}^{\alpha=\pi} =\frac{ N_k}{l_k}-\frac{N_{k-1}}{l_{k-1}}.
  \end{align}
  Alternatively, as in the IIA case, we can include an additional term in the gauge transformation
  \begin{align}
  & B_2\rightarrow B_2+2\kappa (k+\xi N_k)\text{Vol}(S^2), ~~~~~~~~~~~~~~~B_2=-\frac{1}{\gamma}d\phi\wedge d\chi +2\kappa (k+\xi N_k) \sin\theta d\theta \wedge d\phi,\nn \\[2mm]
& Q_{D5} =-\frac{1}{(2\pi)^2} \int_{(\alpha,\theta,\phi)} \hat{F}_3 = N_k-N_{k-1}.
  \end{align}

 So this additional term in the gauge transformation recovers the quantization of D5 charge,  eliminating the effect of the orbifold singularity present. The same result is true in the $\gamma=0$ case (with $\zeta=0$). In the IIA background, D4 branes were only present when $\zeta=0$. Here we see the same behaviour with the D5 branes, but now the $S^2$ preserved condition is $(\gamma,\zeta)=0$. Therefore, in the IIB case, we can have D5 branes without the requirement of a preserved $S^2$ (as long as the $S^2$ is broken under the ATD itself).
  \end{itemize}
  
 We now return to the general case at this boundary. As in the IIA case, we approach the pole $(\sigma=0,\eta=k,\sin\theta=0)$ using
\beq
\eta=k-\rho \cos\alpha \sin^2\mu,~~~~~\sigma=\rho \sin\alpha \sin^2\mu,~~~~~~\sin\theta = 2\sqrt{\frac{b_k\rho}{N_k}}\cos\mu,
\eeq
%(where $r\rightarrow \rho \sin^2\mu$)
recalling that $r= \rho \sin^2\mu$, and expanding about $\rho=0$. Now
 \begin{align}
&\sin^2\alpha\sin^2\mu\,\Pi\rightarrow \sin^2\alpha \sin^2\mu+4\cos^2\mu\,\Pi_k,~~~~~~~~~~ \sin^2\mu \,\Xi \rightarrow \tilde{\Xi}_k =\Delta_k \sin^2\mu + \zeta^2 b_k^2 \cos^2\mu,\nn\\[2mm]
&\Delta_k\equiv  \frac{1}{4}\xi^2 b_k^2 \sin^2\alpha +\Big(1+ \xi g(\alpha)\Big)^2,%\nn\\[2mm]%~~~~~~
%\Delta_k(\alpha=0)=l_{k-1}^2,~~~~~~\Delta_k(\alpha=\pi)=l_{k}^2,
~~~~~~~~~~~ B_2\rightarrow B_2+\frac{4\kappa b_k \cos^2\mu}{N_k}(k+\xi N_k)d\rho\wedge d\phi,
 \end{align}
 giving
 \begin{align}\label{eqn:B5eq}
 &\frac{ds^2}{2\kappa \sqrt{N_k}}=\sqrt{\tilde{\Xi}_k}\Bigg[\frac{4}{\sqrt{\frac{b_k}{\rho}}}ds^2(AdS_5)+\frac{\sqrt{\frac{b_k}{\rho}}}{N_k}ds^2(\mathds{B}_5)\Bigg],~~~~~~~e^{\Phi_{\mathcal{B}}} =\frac{2\,\tilde{\Xi}_k}{b_k\sin\mu \sqrt{\sin^2\alpha\sin^2\mu +4\cos^2\mu\,\Pi_k}},\nn\\[2mm]
& ds^2(\mathds{B}_5) = d\rho^2 +4\rho^2 \bigg[d\mu^2+\frac{1}{4}\sin^2\mu\bigg( \,d\alpha^2 +\frac{\sin^2\alpha}{\Pi_k+\frac{1}{4}\sin^2\alpha\tan^2\mu} d\phi^2\bigg)\bigg]\nn\\[2mm]
&~~~~~~~~~~~~~~~~~~+\frac{1}{\rho}\frac{N_k}{4\kappa^2b_k\sin^2\mu} \frac{(d\chi+\mathcal{A}_k)^2}{(\sin^2\alpha\sin^2\mu+4\,\cos^2\mu\,\Pi_k)},\nn\\[2mm]
&\mathcal{A}_k=\frac{4\kappa b_k \cos\mu}{N_k}\Big(\gamma k+(\gamma \xi-\zeta)N_k\Big)\Big(\cos\mu\,d\rho-2\rho \sin\mu\,d\mu\Big),\nn\\[2mm]
%&B_2=\frac{-4}{\sin^2\alpha\sin^2\mu +4\cos^2\mu\,\Pi_k} \bigg[\frac{b_k\kappa \cos^2\mu}{N_k}\bigg(k \sin^2\alpha\sin^2\mu-4k\zeta(\gamma \xi-\zeta)\cos^2\mu \Big(\frac{1}{4}b_k^2\sin^2\alpha+g(\alpha)^2\Big)\nn\\[2mm]
%&~~~~~~+N_k(\xi \sin^2\alpha\sin^2\mu +4\gamma\zeta \cos^2\mu) -4\zeta  \,g(\alpha) \,\cos^2\mu\Big(\gamma k +(\gamma\xi-\zeta)N_k\Big)\bigg)d\rho\wedge d\phi \nn\\[2mm]
%&~~~~~~~+\cos^2\mu \bigg(\Big(1+\xi g(\alpha)\Big)\Big(\gamma+ (\gamma \xi-\zeta)g(\alpha)\Big) +\frac{1}{4}b_k^2\xi (\gamma\xi-\zeta)\sin^2\alpha\bigg)d\phi\wedge d\chi\bigg]\nn\\[2mm]%pre-LGT
&B_2=\frac{4\cos^2\mu}{\sin^2\alpha\sin^2\mu +4\cos^2\mu\,\Pi_k} \bigg(\Big(1+\xi g(\alpha)\Big)\Big(\gamma+(\gamma\xi-\zeta)g(\alpha)\Big)+\frac{1}{4}\xi b_k^2 (\gamma\xi-\zeta)\sin^2\alpha\bigg)\nn\\[2mm]
&~~~~~~~~~~~~~~~\times \bigg(\frac{4\kappa b_k}{N_k} \cos^2\mu \Big(\gamma k +(\gamma \xi-\zeta )N_k\Big) d\rho\wedge d\phi -d\phi \wedge d\chi\bigg), \nn\\[2mm]
&C_0=\frac{1}{\tilde{\Xi}_k}\bigg(g(\alpha)\Big(1+\xi g(\alpha)\Big)\sin^2\mu +\frac{1}{4}b_k^2 \Big(4\gamma \zeta\,\cos^2\mu +\xi \sin^2\alpha\sin^2\mu\Big)\bigg)\nn\\[2mm]
&C_2=\frac{4}{\sin^2\alpha\sin^2\mu +4\cos^2\mu\,\Pi_k} \bigg[\frac{b_k\kappa \cos^2\mu}{N_k} \bigg(\gamma b_k^2k (\gamma\xi-\zeta) \cos^2\mu \sin^2\alpha +4\gamma k (\gamma \xi-\zeta)g(\alpha)^2\, \cos^2\mu \nn\\[2mm]
&~~~~~~~ -N_k (\sin^2\alpha\sin^2\mu +4\gamma^2 \cos^2\mu) +4\gamma \,g(\alpha)\,\cos^2\mu \Big(\gamma k -(\gamma\xi-\zeta)N_k\Big)\bigg)d\rho\wedge d\phi  \nn\\[2mm]
&~~~~~~~-\cos^2\mu \bigg(\frac{1}{4}b_k^2\sin^2\alpha (\gamma\xi-\zeta) +g(\alpha) \Big(\gamma+(\gamma\xi-\zeta)g(\alpha)\Big)\bigg)d\phi\wedge d\chi\bigg].
 \end{align}
 So again, we see that the $S^2$ has inherited orbifold singularities. On first sight, these appear more complicated than the Spindle found in IIA (in the general case at least). This is because $\mathds{B}_5$ in \eqref{eqn:B5eq} is the ATD of the $(\rho,~\mathds{B}_4)$ components of \eqref{eqn:B4eq} (using the appropriate gauge transformation in $B_2$). One can then conclude that $z$ and $\rho$, in \eqref{eqn:zeq} and \eqref{eqn:B5eq} respectively, aren't a part of the orbifolds themselves but the cones over them. %\textcolor{red}{This is the consequence of T-dualising the $(\rho,~\mathds{B}_4)$ components of \eqref{eqn:B4eq}, which derives the $\mathds{B}_5$ in \eqref{eqn:B5eq} (using the appropriate gauge transformation in $B_2$).} 
 It would be nice to understand this further topologically in the future.  %Note however that the Euler characteristic of an odd-dimensional compact manifold, such as $\mathds{B}_5$, is known to be zero.
 Now we calculate the charge of the D7 branes at $\mu=\frac{\pi}{2}$ on one of the components of the `Spindle like' space, $\alpha$, as follows
 \beq\label{eqn:D7charge}
Q_{D7}=- \int_\alpha F_1=C_0\Big|_{\alpha=0}^{\alpha=\pi}=\frac{2N_k-N_{k+1}-N_{k-1}}{l_kl_{k-1}}.
 \eeq
 This is the result for all values of $(\gamma,\zeta)$ - including the $\mathcal{N}=1$ background. See Figure \ref{fig: TypeIIB} for a pictorial representation of the solutions, summarising the above analysis. It is important to note that this broken quantization of charge is due to the $\tilde{\Xi}_k$ (and hence $\Delta_k$) in the denominator of $C_0$, and not $\Pi_k$ which is what gives rise to singularities in the internal manifold. It appears then that this breaking of quantization is simply a remnant of the IIA orbifold singularity. As a result of this, one can find situations where the background has the above rational charge without having orbifold singularities in it's metric. In other words, by T-dualising within the Spindle like orbifold in Type IIA, we can break the orbifold structure completely in Type IIB whilst still inheriting the rational charge. The $\mathcal{N}=1$ and $S^2$ preserved $\mathcal{N}=0$ cases are two such example, which we now write explicitly.
 
  \end{itemize}

  \subsubsection{$\mathcal{N}=1$ Type IIB - Background}
  Let's now present the one-parameter $\mathcal{N}=1$ family as promised, where $\gamma=-1,\zeta=-\xi$
     \begin{align}\label{eqn:IIBN=1-1}
&       ds_{10,st}^2=e^{\frac{2}{3}\Phi_\mathcal{A}}f_1\bigg[4ds^2(\text{AdS}_5)+f_2d\theta^2+f_4(d\sigma^2+d\eta^2)+ \frac{f_2 f_3}{f_3+f_2\sin^2\theta} \sin^2\theta d\phi^2+\frac{(d\chi +f_8\sin\theta  d\theta)^2}{f_1^3(f_3f_5  +\sin^2\theta f_2   f_5) }\bigg], \nn\\[2mm]
&       e^{\frac{4}{3}\Phi_\mathcal{A}}= f_1 \Big[  f_5(1+\xi f_6)^2+\xi^2f_3 + \xi^2 f_2 \sin^2\theta \Big],\nn\\[2mm]
& e^{2\Phi^\mathcal{B}}=\frac{  \big( f_5(1+\xi f_6)^2+\xi^2f_3 + \xi^2 \sin^2\theta f_2\big)^2}{ f_3f_5  +\sin^2\theta f_2   f_5},~~~~~~~~~~~~~~  C_0=\frac{  f_5 f_6(1+\xi f_6) +\xi f_3 +\xi f_2 \sin^2\theta }{  f_5(1+\xi f_6)^2+\xi^2f_3 + \xi^2  f_2 \sin^2\theta},\nn\\[2mm]
&B_2=\bigg(\frac{f_2f_8(1+\xi f_6)\sin^2\theta}{f_3+f_2\sin^2\theta} -(f_8+\xi f_7)\bigg)\sin\theta d\phi \wedge d\theta+\frac{f_2(1+\xi f_6)\sin^2\theta }{f_3+f_2\sin^2\theta}d\phi\wedge d\chi ,\nn\\[2mm]
&  C_2=\frac{f_3 f_7+f_2(f_7-f_6f_8)\sin^2\theta}{f_3+f_2\sin^2\theta}\sin\theta d\theta \wedge d\phi + \frac{f_2f_6\sin^2\theta}{f_3+f_2\sin^2\theta} d\phi\wedge d\chi.
  \end{align}
See Appendix \ref{sec:oneparamfam} for the G-Structure description. As already discussed, this background has NS5 branes and D7 branes (with charge given in \eqref{eqn:D7charge}). There are only D5 branes present when $\xi=0$ - namely, where the red plane and blue line in Figure \ref{fig: TypeIIB} intersect. As previously mentioned, there is no orbifold singularity present in this metric, but the D7 branes still inherit the broken quantization of charge from its type IIA ancestor. 
    \subsubsection{$S^2$ preserved  $\mathcal{N}=0$ Solutions}
   In this case, fixing $(\gamma,\zeta)=0$ preserves the $S^2$ whilst breaking the supersymmetry. Of course, to get the non-deformed solution, we fix $\xi=0$ in the following discussion. The background reads
      \begin{align}\label{eqn:N=0IIBcase1}
 &      ds_{10,B}^2=e^{\frac{2}{3}\Phi_A}f_1\Bigg[4ds^2(\text{AdS}_5)+f_2 ds^2(S^2)+f_4(d\sigma^2+d\eta^2) +\frac{1}{f_1^3f_3f_5}\,d\chi^2\Bigg],\nn\\[2mm]
 & e^{2\Phi_B}=\frac{1}{f_3f_5}\Big[f_5(1+\xi f_6)^2 +\xi^2f_3\Big]^2,~~~~~~~~~~~~~~~~~ e^{\frac{4}{3}\Phi_A}= f_1 \Big[f_5(1+\xi f_6)^2 +\xi^2f_3\Big],\nn\\[2mm]
&  C_0=  \frac{f_5 f_6(1+\xi f_6) +\xi f_3 }{ f_5(1+\xi f_6)^2 +\xi^2f_3},~~~~~~~~~~~B_2=(f_8+\xi f_7)\text{Vol}(S^2),~~~~~~~~~~~~~~~~~~~
  C_2=f_7 \text{Vol}(S^2).
  \end{align}

%  Using the particular solution given in equation \eqref{eqn:potential}, 
  %and the following limits 
%  \footnote{Noting
%\begin{equation}
%\begin{gathered}
%\lim_{x\rightarrow 0} [x^n K_n(x)]= \prod_{l=1}^{n-1} 2l,~~~~~n>1\in \mathds{Z},~~~~~~~~~~~~~~~~~~~~~~~~~~\lim_{x\rightarrow 0} x K_0(x)=  \lim_{x\rightarrow 0} x^2 K_0(x)= 0,\\
%\lim_{x\rightarrow 0} x K_1(x)=1,~~~~~~~~~~~~~~~~~~~~~~~~
%\lim_{x\rightarrow \infty}  K_j(x)\sim\sqrt{\frac{\pi}{2}}\frac{e^{-x}}{\sqrt{x}},~~~~j\in(0,1,2,....).
%\end{gathered}
%\end{equation}}

% \begin{equation}\label{eqn:sigma0limit}
%\begin{gathered}
%\underline{\textbf{Close to $\sigma=0$}}\\
%f_7 \sim -2\kappa \mathcal{R}(\eta),~~~~~~~~(f_8+2\kappa\Theta) = -2\kappa (\eta-\Theta),~~~~~~~~~~f_6 \sim  \mathcal{R}'(\eta),\\
%f_5 \sim \frac{1}{\text{log}\,\sigma} \rightarrow 0  ,~~~~~~~~~~~~~~f_3 \sim \sigma \rightarrow 0,~~~~~~~~~~~~~~\frac{f_3}{f_5} \sim \sigma \text{log}\,\sigma \sim 0,~~~~~~~~~~~~\frac{f_2}{f_5}\rightarrow \infty,\\
%\underline{\textbf{Close to $\sigma\rightarrow \infty$}}\\
%\frac{f_2}{f_5}\rightarrow 0,~~~~~~~~~~~~\frac{f_3}{f_5}\rightarrow 0,~~~~~~~~~~~~~f_6\rightarrow 0,~~~~~~~~~~~~~f_7\rightarrow 0,~~~~~~~~~(f_8+2\kappa\Theta) \Big|_k^{k+1}= -2\kappa (\eta-\Theta) \Big|_k^{k+1}
%\end{gathered}
%\end{equation}

We can see from the metric in this case that there are no orbifold singularities present here. However, from the boundary analysis just presented, we know that the charges have the same broken quantization as backgrounds containing spindles - in both the D7 branes and (depending on the gauge transformation chosen) in the D5 branes. Therefore, perhaps a different interpretation may be required, one possible proposal is the idea of rotating the D branes - see Appendix \ref{sec:RotatingBranes}.

%\textcolor{red}{Inserting the standard Rank Function form in equation \eqref{eqn:Rkoriginal} into the Page charges from equation \eqref{eqn:pagecharges} and evaluating in each interval $[k,k+1]$, one gets (with $Q_{NS5}^{[k,k+1]}=1$ unaffected by $\xi$)
%\beq\label{eqn:rotated}
%\begin{aligned}
%Q_{D5}^{[k,k+1]} &= \frac{ N_k}{1+\xi(N_{k+1}-N_k)},\\
%Q_{D7} &=   = \frac{2N_{k+1}-N_k -N_{k+2}}{\big( 1+\xi (N_{k+2}-N_{k+1}) \big)\big(1+\xi (N_{k+1}-N_{k}) \big)}
%\end{aligned}
%\eeq
%}

\subsection{More Solutions}
%\subsection*{Descendants frl other three-parameter families}
In a similar vein to the IIA discussion, the three-parameter family of solutions given in \eqref{eqn:ATD1} is not the only one which can be derived. We have in addition \eqref{eqn:ATD2}, \eqref{eqn:ATD3} and \eqref{eqn:ATD6} which all map to \eqref{eqn:ATD1} under the transformations given in \eqref{eqn:IIBtrans1}, \eqref{eqn:IIBtrans2} and \eqref{eqn:IIBtrans3}, respectively. As in the IIA section, one can still derive unique solutions from these backgrounds because of the nature of the required mappings. We refrain from including ATDs along $\phi$ here, as typically performing a T-Duality along the $U(1)$ of an $S^2$ can lead to singularities in the dual description. Perhaps the same is true in the Spindle case (see \eqref{eqn:newIIBcase1} and \eqref{eqn:chiredphiatd} for the solutions).
  
  \subsubsection*{ATD along $\beta$ of the $\chi$ reduction}
Performing an ATD along $\beta$ of the $\chi$ reduction, one derives \eqref{eqn:ATD3}, which requires $\xi\rightarrow 1/\xi$ to map to \eqref{eqn:ATD1}. Hence, by fixing $\xi=0$, we derive another unique solution
  \begin{align}\label{eqn:newIIBcase2}
   &    ds_{10,st}^2=e^{\frac{2}{3}\Phi_\mathcal{A}}f_1\bigg[4ds^2(\text{AdS}_5)+f_2d\theta^2+f_4(d\sigma^2+d\eta^2)+ \frac{1}{\hat{\Xi}}\bigg(f_2 f_3f_5 \sin^2\theta d\phi^2~~~~~~~~~~~~~~~~~~~~~~~~~~~~~~~~~~~~~~~~~~~~~~~~~~~~~~~~~~~~~~~~~~\nn\\[2mm]
&  ~~~~~~~~~~~~~~~~~~~~~~~~~~~~ ~~~~~~~~~~~~ ~~~~~~~~~~~~~~~~~~~~~~~~~~~~~~~  ~~~~~ ~~~~~~~~~~~~~      +\frac{1}{f_1^3 }\Big(d\beta +\big(\zeta f_8-\gamma f_7\big)\sin\theta  d\theta\Big)^2\bigg)\bigg] \nn\\[2mm]
&e^{2\Phi_\mathcal{B}}=\frac{ 1}{\hat{\Xi}}\big(  f_5 f_6^2+f_3 + \zeta^2f_2 \sin^2\theta\big)^2,~~~~~~~~~~~~~~    e^{\frac{4}{3}\Phi_\mathcal{A}}= f_1 \Big[  f_5 f_6^2+f_3 + \zeta^2f_2 \sin^2\theta \Big], \nn\\[2mm]
&B_2= \sin\theta \frac{1}{\hat{\Xi}} \bigg[f_2\Big(\gamma f_3 +f_5f_6(\gamma f_6-\zeta)\Big)\sin\theta d\beta   ~~~~  ~~~~~~~~~~~~~~~~~~~~~~~~~~~~~  ~~~~~~~~~~~~~~~~~~~~~~~~~~~~~~~~~~~~ \nn\\[2mm]
&  ~~~~~~~~~~~~~~~~~~~~~~~+ \Big(f_3f_5f_7+\zeta f_2\Big(\gamma f_3f_8+f_5(f_6f_8-f_7)(\gamma f_6-\zeta)\Big)\sin^2\theta\Big) d\theta\bigg] \wedge d\phi ,\nn\\[2mm]
&  C_2=  \sin\theta \frac{1}{\hat{\Xi}}\bigg[f_2f_5(\gamma f_6-\zeta)\sin\theta d\beta +\Big(f_3f_5f_8 +\gamma f_2 \Big(\gamma f_3f_8+f_5(f_6f_8-f_7)(\gamma f_6-\zeta)\Big)\sin^2\theta\Big)d\theta\bigg] \wedge d\phi ,\nn\\[2mm]
  &    C_0=f_1    e^{-\frac{4}{3}\Phi_\mathcal{A}}\Big( f_5 f_6 +\gamma \zeta f_2\sin^2\theta\Big) ,~~~~~~~~~~~~~   \hat{\Xi}=f_3f_5+f_2\big(\gamma^2f_3 +f_5(\gamma f_6-\zeta)^2\big)\sin^2\theta.
  \end{align}
This time, the background is a two-parameter family of $\mathcal{N}=0$ solutions which enhances to $\mathcal{N}=1$ for $\zeta=-1,~\gamma=0$. We now turn to this solution.

\subsubsection{Unique $\mathcal{N}=1$ IIB Solution}
We now derive a new and unique $\mathcal{N}=1$ background (corresponding to the $\xi=0$ case of \eqref{eqn:IIBN=1-2}\footnote{Notice that the transformations given in \eqref{eqn:N=1transforms} and \eqref{eqn:N=1transforms2} also demonstrate that the $\xi=0$ solution is unique. Here $(p,b,u)=1,~(q,c,m)=0,~a\equiv \xi=0,~s\equiv \zeta=-1,~v\equiv\gamma=0$.}).

     \begin{align}\label{eqn:uniqueN1IIB}
&       ds_{10,st}^2=e^{\frac{2}{3}\Phi_A}f_1\bigg[4ds^2(\text{AdS}_5)+f_2d\theta^2+f_4(d\sigma^2+d\eta^2)+ \frac{f_2 f_3}{f_3+f_2\sin^2\theta} \sin^2\theta d\phi^2+\frac{(d\beta -f_8\sin\theta  d\theta)^2}{f_1^3(f_3f_5  +\sin^2\theta f_2   f_5) }\bigg] \nn\\[2mm]
  &     e^{\frac{4}{3}\Phi_A}= f_1 \Big[  f_5 f_6^2+f_3 + f_2 \sin^2\theta \Big],~~~~~~~~~~~~~~e^{2\Phi^B}=\frac{ \big(  f_5f_6^2+f_3 +  \sin^2\theta f_2\big)^2}{ f_3f_5  +\sin^2\theta f_2   f_5},\nn\\[2mm]
&B_2=\frac{\sin\theta}{f_3+f_2\sin^2\theta} \bigg(f_2f_6\sin\theta d\beta + \Big(f_3f_7+f_2(f_7-f_6f_8)\sin^2\theta\Big) d\theta\bigg) \wedge d\phi ,\nn\\[2mm]
  &C_2=\frac{\sin\theta}{f_3+f_2\sin^2\theta}\Big(f_2\sin\theta d\beta +f_3f_8 d\theta\Big) \wedge d\phi   ~~~~~~~~~~~~~~  C_0=\frac{  f_5f_6 }{  f_5 f_6^2+f_3 +  \sin^2\theta f_2}.
  \end{align}
This solution corresponds to the SUSY preserving abelian T-Duality along $\beta$ of the unique IIA $\mathcal{N}=1$ solution given in \eqref{eqn:N=1IIA2}.

\subsubsection*{ATD along $\beta$ of the $\phi$ reduction}

Performing an ATD along $\beta$ of the $\phi$ reduction, one derives \eqref{eqn:ATD6}. To map this solution to \eqref{eqn:ATD1} we require transformations involving $1/\xi,~1/\zeta, 1/\gamma$ and $1/(\zeta-\gamma\xi)$ (see \eqref{eqn:IIBtrans3}). 

Fixing $\xi=0$, we get
     \begin{align}\label{eqn:newIIBcase3}
 &      ds_{10,st}^2=e^{\frac{2}{3}\Phi_\mathcal{A}}f_1\bigg[4ds^2(\text{AdS}_5)+f_2d\theta^2+f_4(d\sigma^2+d\eta^2)~~~~~~~~~~~~~~~~~~~~~~~~~~~~~~~~~~~~~~~~~~~~~~~~~~~~~~~~~  \nn\\[2mm]
 &    ~~~~~~  +\frac{1}{\hat{\Xi}}\bigg(  f_2 f_3 f_5 \sin^2\theta d\chi^2  +\frac{1}{f_1^3  }\Big(d\beta +(f_8+\gamma f_7)\sin\theta  d\theta\Big)^2\bigg)\bigg],  ~~~~~~~e^{\frac{4}{3}\Phi_\mathcal{A}}= f_1 \Big(  \zeta^2 (f_5f_6^2+ f_3) + f_2 \sin^2\theta \Big), \nn\\[2mm]
&e^{2\Phi_\mathcal{B}}=\frac{1}{\hat{\Xi}}\big(   \zeta^2 (f_5f_6^2+ f_3) + f_2 \sin^2\theta \big)^2 ,~~~~~~~~~~~~~~   \hat{\Xi}= \zeta^2f_3f_5  +f_2  \Big( \gamma^2 f_3+f_5(1+\gamma f_6)^2\Big)\sin^2\theta ,          \nn\\[2mm]
&B_2=\frac{1}{\hat{\Xi}} \bigg[ f_2(\gamma f_3+f_5f_6(1+\gamma f_6))\sin^2\theta d\beta~~~~~~~~~~~~~~~~~~~~~~~~~~~~~~~~~~~~~~~~~~~~~~~~~~~~~~~~~~~~~~~   \nn\\[2mm]
&~~~~~~~~~~~~~~~~~~~~~-\sin\theta\Big(\zeta^2 f_3f_5 f_7  +f_2 \Big(-\gamma f_3f_8 +f_5(f_7-f_6f_8)(1+\gamma f_6)\Big)\sin^2\theta\Big) d\theta\bigg] \wedge d\chi ,  \nn\\[2mm]
 & C_0= \zeta f_1   e^{-\frac{4}{3}\Phi_\mathcal{A}} \Big( \gamma f_3+f_5 f_6 (1+\gamma f_6)  \Big) ,~~~~~~~~~~~~~~~~~ C_2=-\frac{\zeta}{\hat{\Xi}}f_3f_5 \Big( d\beta +(f_8+\gamma f_7) \sin\theta d\theta\Big) \wedge d\chi .
  \end{align}
  Fixing $\zeta=0$, we find

       \begin{align}\label{eqn:newIIBcase4}
 &      ds_{10,st}^2=e^{\frac{2}{3}\Phi_\mathcal{A}}f_1\bigg[4ds^2(\text{AdS}_5)+f_2d\theta^2+f_4(d\sigma^2+d\eta^2)~~~~~~~~~~~~~~~~~~~~~~~~~~~~~~~~~~~~~~~~~~~~~~~~~~~~~~~~~   \nn\\[2mm]
  &   ~~~~~  +\frac{1}{\hat{\Xi}}\bigg(  f_2 f_3 f_5 \sin^2\theta d\chi^2  +\frac{1}{f_1^3  }\Big(d\beta +(f_8+\gamma f_7)\sin\theta  d\theta\Big)^2\bigg)\bigg],~~~~~~~~      e^{\frac{4}{3}\Phi_\mathcal{A}}= f_1 (\xi^2  f_5  + f_2 \sin^2\theta ),    \nn\\[2mm]
&e^{2\Phi_\mathcal{B}}=\frac{1}{\hat{\Xi}}\big(\xi^2   f_5   + f_2 \sin^2\theta \big)^2 ,~~~~~~~~~~  \hat{\Xi}=\gamma^2\xi^2f_3f_5  +f_2  \Big( \gamma^2 f_3+f_5(1+\gamma f_6)^2\Big)\sin^2\theta     \nn\\[2mm]
&B_2=\frac{1}{\hat{\Xi}} \bigg[\Big(\gamma\xi^2 f_3f_5 +f_2(\gamma f_3+f_5f_6(1+\gamma f_6))\sin^2\theta\Big) d\beta~~~~~~~~~~~~~~~~~~~~~~~~~~~~~~~~~~~~~~~~~~~~~~~~~~~~~~~~~~    \nn\\[2mm]
&~~~~~~+ \sin\theta\Big(\gamma\xi^2 f_3f_5 f_8 -f_2 \Big(-\gamma f_3f_8 +f_5(f_7-f_6f_8)(1+\gamma f_6)\Big)\sin^2\theta\Big) d\theta\bigg] \wedge d\chi ,   \nn\\[2mm]
 & C_0=\xi\, f_1    f_5   (1+\gamma f_6) \, e^{-\frac{4}{3}\Phi_\mathcal{A}}  , ~~~~~~~~~~~~~
   C_2=\frac{\gamma\xi}{\hat{\Xi}}f_3f_5 \Big( d\beta +(f_8+\gamma f_7) \sin\theta d\theta\Big) \wedge d\chi , 
  \end{align}
Fixing $\gamma=0$, we get
       \begin{align}\label{eqn:newIIBcase5}
 &      ds_{10,st}^2=e^{\frac{2}{3}\Phi_\mathcal{A}}f_1\bigg[4ds^2(\text{AdS}_5)+f_2d\theta^2+f_4(d\sigma^2+d\eta^2) +\frac{1}{\hat{\Xi}}\bigg(  f_2 f_3 f_5 \sin^2\theta d\chi^2  +\frac{1}{f_1^3  }\Big(d\beta + f_8 \sin\theta  d\theta\Big)^2\bigg)\bigg]    \nn\\[2mm]
&e^{2\Phi_\mathcal{B}}=\frac{1}{\hat{\Xi}}\big(   f_5( \zeta f_6+\xi)^2+\zeta^2 f_3 + f_2 \sin^2\theta \big)^2 ,~~~~~~~~~~        e^{\frac{4}{3}\Phi_\mathcal{A}}= f_1 \Big[  f_5( \zeta f_6+\xi)^2+\zeta^2 f_3 + f_2 \sin^2\theta \Big],   \nn\\[2mm]
&B_2=\frac{1}{\hat{\Xi}} \bigg[f_5 (-\xi \zeta f_3 +f_2 f_6 \sin^2\theta ) d\beta + \sin\theta\Big( -\zeta f_3f_5(\zeta f_7+\xi f_8) -f_2   f_5(f_7-f_6f_8)  \sin^2\theta\Big) d\theta\bigg] \wedge d\chi ,   \nn\\[2mm]
 & C_0= f_1   f_5  (\zeta f_6+\xi)  \, e^{-\frac{4}{3}\Phi_\mathcal{A}}  ,  ~~~~~~~~~~  C_2=-\frac{ \zeta}{\hat{\Xi}}f_3f_5 \Big( d\beta +f_8 \sin\theta d\theta\Big) \wedge d\chi ,~~~~~~~~~~\hat{\Xi}= \zeta^2f_3f_5  +f_2 f_5 \sin^2\theta .
  \end{align}
Then, by setting $\zeta=-1$ in this background, one re-derives the $\mathcal{N}=1$ solution in \eqref{eqn:IIBN=1-2} (following the transformations given in footnote \ref{footnote_ref1}). \\\\
Finally, fixing $\zeta=\gamma\,\xi$, we find
       \begin{align}\label{eqn:newIIBcase6}
 &      ds_{10,st}^2=e^{\frac{2}{3}\Phi_\mathcal{A}}f_1\bigg[4ds^2(\text{AdS}_5)+f_2d\theta^2+f_4(d\sigma^2+d\eta^2)~~~~~~~~~~~~~~~~~~~~~~~~~~~~~~~~~~~~~~~~~~~~~~~~~~~~~~~~~   \nn\\[2mm]
  &   ~~~~~~~~~~~~~~~~~~~~~~~~~~~~~~~~~~~~  +\frac{1}{\hat{\Xi}}\bigg(  f_2 f_3 f_5 \sin^2\theta d\chi^2  +\frac{1}{f_1^3  }\Big(d\beta +(f_8+\gamma f_7)\sin\theta  d\theta\Big)^2\bigg)\bigg]    \nn\\[2mm]
&e^{2\Phi_\mathcal{B}}=\frac{1}{\hat{\Xi}}\big( \xi^2  f_5( \gamma f_6+1)^2+\xi^2\gamma^2 f_3 + f_2 \sin^2\theta \big)^2 ,~~~~~~~~~~        e^{\frac{4}{3}\Phi_\mathcal{A}}= f_1 \Big[ \xi^2  f_5( \gamma f_6+1)^2+\xi^2\gamma^2 f_3 + f_2 \sin^2\theta \Big],   \nn\\[2mm]
&B_2=\frac{1}{\hat{\Xi}} \bigg[\Big( f_2(\gamma f_3+f_5f_6(1+\gamma f_6))\sin^2\theta\Big) d\beta + \sin\theta\Big( f_2 \Big(\gamma f_3f_8 -f_5(f_7-f_6f_8)(1+\gamma f_6)\Big)\sin^2\theta\Big) d\theta\bigg] \wedge d\chi ,   \nn\\[2mm]
 & C_0= \xi f_1   e^{-\frac{4}{3}\Phi_\mathcal{A}} \Big( \gamma^2 f_3+f_5(1+\gamma f_6)^2 \Big) , ~~~~~~~~~~~  C_2=0,~~~~~~~~~~~~   \hat{\Xi}= f_2  \Big( \gamma^2 f_3+f_5(1+\gamma f_6)^2\Big)\sin^2\theta  .
  \end{align}

  \subsubsection*{ATD along $\chi$ of the $\phi$ reduction}
  Performing an ATD along $\chi$ of the $\phi$ reduction leads to the following $\mathcal{N}=0$ family of solutions
       \begin{align}\label{eqn:ATD2param6}
 &      ds_{10,st}^2=e^{\frac{2}{3}\Phi_A}f_1\bigg[4ds^2(\text{AdS}_5)+f_2d\theta^2+f_4(d\sigma^2+d\eta^2)~~~~~~~~~~~~~~~~~~~~~~~~~~~~~~~~~~~~~~~~~~~  \nn\\[2mm]
  &  ~~~~~~~~~~~~~~~~~~~~~~~   + \frac{1}{\hat{\Xi}}\bigg( f_2 f_3 f_5\sin^2\theta d\beta^2 +\frac{1}{f_1^3  }(d\chi +f_7\sin\theta  d\theta)^2\bigg)\bigg]   \nn\\[2mm]
& e^{2\Phi^B}=\frac{1}{\hat{\Xi}}\big(  f_5(\zeta f_6+\xi)^2+\zeta^2f_3 + f_2 \sin^2\theta \big)^2 ,~~~~~~~~~      e^{\frac{4}{3}\Phi_A}= f_1 \Big[  f_5(\zeta f_6+\xi)^2+\zeta^2f_3 + f_2 \sin^2\theta \Big],  \nn\\[2mm]
&  C_0= f_1 e^{-\frac{4}{3}\Phi_A}\Big(f_5f_6 (\zeta f_6+\xi) + \zeta  f_3 \Big) ,~~~~~~~~~~~~~~~~~~~      \hat{\Xi} = \xi^2  f_3f_5+f_2(f_3+f_5f_6^2)\sin^2\theta,  \nn\\[2mm]
&B_2=-\frac{1}{\hat{\Xi}} \bigg[f_5(\zeta \xi f_3-f_2f_6\sin^2\theta) d\chi~~~~~~~~~~~~~~~~~~~~~~~~~~~~~~~~~~~~~~~~~~~~~~~~~~~~~~~~~~~~~~~~~~~~~~~~~~~~~~~~~~~~~~~~~~~~~~~~~   \nn\\[2mm]
&~~+ \Big[\xi f_3f_5 (\zeta f_7+\xi f_8)+f_2\Big(f_3f_8+f_5f_6(f_6f_8-f_7)\Big)\sin^2\theta\Big] \sin\theta d\theta\bigg] \wedge d\beta +\gamma d\chi\wedge d\beta,  \nn\\[2mm]
&  C_2=-\frac{\xi }{\hat{\Xi}}f_3f_5 \Big(d\chi +f_7 \sin\theta d\theta \Big)\wedge d\beta .
  \end{align}

%\textcolor{red}{Need to double check that all these are indeed unique - finish off}

    \section{Conclusions and future study}\label{sec:conclusions}
We now briefly summarize the new results presented throughout this paper.
\begin{itemize}
\item After reviewing the Gaiotto-Maldacena solutions in M-Theory and some relevant material on the method of G-Structures, we included the M-Theory G-Structure forms corresponding to a reduction along $\chi$ and $\phi$. We then presented the new IIB G-Structure conditions in terms of the Pure Spinors $\Psi_\pm$, derived in detail in Appendix \ref{sec:GStructureCalcs}.  
\item Investigations at the boundary were then conducted for the general two-parameter family of Type IIA solutions presented in \cite{us}, following the analysis of the $\mathcal{N}=1$ solution. We show that the space orthogonal to the branes is in general a higher dimensional analogue of the spindle. We find NS5 and D6 branes in all backgrounds, with the D4 branes only present for $S^2$ preserved solutions. Introducing an additional parameter into the gauge transformation of $B_2$, we find the integer quantization of D4 charge is recovered, negating the effect of the Spindle. This is not possible for the D6 branes, but the rational charge has the same form for all backgrounds.
\item Following an $SL(3,\mathds{R})$ transformation in M-Theory, by performing a dimensional reduction along $\chi$ and $\phi$ in turn, some new IIA solutions are presented (including a new $\mathcal{N}=1$ solution). An in-depth derivation is given in Appendix \ref{sec:DimRed}.
\item We then present a three-parameter family of type IIB backgrounds, derived by performing an ATD of the two-parameter IIA background. This solution contains within it a one-parameter family of $\mathcal{N}=1$ solutions, as well as a one-parameter family of $S^2$ preserved $\mathcal{N}=0$ solutions. Investigations at the boundary were conducted in the same manner as the IIA case, finding the presence of orbifold singularities. In analogy with the IIA solutions, NS5 and D7 branes are present in all cases. However, D5 branes only appear in backgrounds which descend from the $S^2$ preserved IIA solutions. In studying the ATD for multiple cases, with full calculations given in Appendix \ref{sec:ATD}, the TST solution of  \cite{Nunez:2019gbg} is re-derived - now implying this is an $\mathcal{N}=0$ solution in general, enhancing to $\mathcal{N}=1$ for $\gamma=-1$. 
\item This then motivated the uplifting of a TST solution, deriving a deformation of the M-Theory Gaiotto-Maldacena solution (believed to be supersymmetry breaking, and given in Appendix \ref{sec:TSTGM}). It proved useful to perform an ATD and TST for a general form, deriving the ATD solutions for many backgrounds at once. These forms are presented in Appendix \ref{sec:ATD}. The breaking of quantization seems to occur without the presence of an orbifold singularity in some solutions, a proposal for the possible interpretation of this is given in Appendix \ref{sec:RotatingBranes}- involving rotated D-branes.
\item Some additional IIB backgrounds are presented, including a zero-parameter $\mathcal{N}=1$ solution. Both families of $\mathcal{N}=1$ IIB solutions presented in this work have the interesting property of a zero five-form flux. See \cite{Couzens:2016iot} for some context.
\item The Holographic Central charge was derived for the general metric, showing as in \cite{us}, the deformations are marginal in the dual CFT. See Appendix \ref{sec:HCC}. The G-Structure forms for all $\mathcal{N}=1$ solutions are derived, and presented in Appendix \ref{sec:N=1Gstructures}.
\end{itemize}
 As outlined throughout this paper, there is still analysis which could and should be conducted in future work. In addition to the list presented in \cite{us}, 
\begin{itemize}
\item Investigate more thoroughly the backgrounds presented in this paper, including the charge quantization, behaviour at the boundaries and stability analysis, where required.
\item Use the TST formula presented in Appendix \ref{sec:ATD} to check whether performing a TST transformation on these multi-parameter backgrounds leads to an additional (non-trivial) parameter.
\item It would be interesting to take a closer look at the deformed Gaiotto-Maldacena solution in M-Theory.
\item It would be nice to investigate specific Rank Function examples for these backgrounds, including the Triangular, Trapezium and Sfetsos-Thompson cases.
\item It would be interesting to use the multi-parameter solutions presented here within the context of Black Holes. Notably, the presence of Spindles in these solutions.
\item It would be nice to investigate whether something can be done along the lines of \cite{Bah:2017wxp}\cite{Couzens:2022aki}, in the context of 5D Minimal Gauged Supergravity. 
\end{itemize}
Hopefully, further work related to these topics will follow in the near future.
\section*{Acknowledgements}
I thank Carlos Nunez for his continued help and guidance. I thank Niall T. Macpherson for many very informative discussions, especially regarding G-Structures. I thank Christopher Couzens for an insightful discussion.

\appendix

\section{G-Structure Calculations}\label{sec:GStructureCalcs}
In order to validate the supersymmetry preservation of the Type IIA and Type IIB theories presented throughout the paper, our goal is to study the G-Structure analysis in each case. We must first calculate the Gaiotto-Maldacena G-Structures for each $\mathcal{N}=1$ reduction frame in turn, allowing one to derive the corresponding IIA and IIB G-Structures as required.
\subsection{Gaiotto-Maldacena}\label{sec:GMGstructures}

One can use the B$\ddot{\text{a}}$cklund transformation to calculate the GM G-Structure forms from the LLM G-Structure analysis presented in Appendix D of \cite{Macpherson:2016xwk}. This calculation derives the G-Structures corresponding to the $\chi$ reduction frame, given in \eqref{eqn:GstructureForms} and \eqref{eqn:GM-G}. We now wish to rotate these forms in a general fashion to a reduction frame which accounts for the $SL(3,\mathds{R})$ transformations performed in this paper. One can then easily restrict to the $\mathcal{N}=1$ reduction frames of interest.
 
 Following the $SL(3,\mathds{R})$ transformation given in \eqref{eqn:S2breakingdefns}, one needs to re-write the metric in the following form (for a dimensional reduction along $\phi_3$)
   \begin{align}
        ds_{11}^2&=f_1\bigg[4ds^2(\text{AdS}_5)+f_2d\theta^2+f_4(d\sigma^2+d\eta^2)\bigg]+f_1^2e^{-\frac{4}{3}\Phi_{IIA}}ds^2_2+ e^{\frac{4}{3}\Phi_{IIA}}(d\phi_3+ C_{1,\phi_1}d\phi_1+C_{1,\phi_2}d\phi_2)^2\nn\\[2mm]
    ds^2_2 &=  h_{\phi_1}(\eta,\sigma,\theta)d\phi_1^2 + h_{\phi_2}(\eta,\sigma,\theta)d\phi_2^2 + h_{\phi_1\phi_2}(\eta,\sigma,\theta)d\phi_1 d\phi_2 \quad\quad\quad\quad\quad\quad\quad\quad\quad\quad\quad\quad\quad \nn\\[2mm]
    &= h_{\phi_1}(\eta,\sigma,\theta)\bigg(d\phi_1 + \frac{1}{2}\frac{h_{\phi_1\phi_2}(\eta,\sigma,\theta)}{h_{\phi_1}(\eta,\sigma,\theta)}d\phi_2\bigg)^2 + \bigg( h_{\phi_2}(\eta,\sigma,\theta)- \frac{1}{4} \frac{h_{\phi_1\phi_2}(\eta,\sigma,\theta)^2}{h_{\phi_1}(\eta,\sigma,\theta)}\bigg)d\phi_2^2,
    \end{align}
   with $(\phi_1,\phi_2,\phi_3)$ representing any arrangement of $(\beta,\chi,\phi)$ as required. The general G-Structure forms for the GM background then read
  \begin{align}\label{eqn:Gstructuregenforms}
K&= \frac{\kappa\, e^{-2\rho}}{f_1}d(\cos\theta e^{2\rho}\dot{V}),\nn\\[2mm]
E_1 &= f_1 e^{-\frac{2}{3}\Phi}\sqrt{h_1(\eta,\sigma,\theta)}  \bigg[b_1(\eta,\sigma,\theta)d\sigma +b_2(\eta,\sigma,\theta)d\eta + b_3(\eta,\sigma,\theta)d\rho +b_4(\eta,\sigma,\theta)d\theta \nn\\[2mm]
&~~~~~~~~~~~ -i  \bigg(d\phi_1 + \frac{1}{2}\frac{h_{\phi_1\phi_2}(\eta,\sigma,\theta)}{h_{\phi_1}(\eta,\sigma,\theta)}d\phi_2\bigg)\bigg],\nn\\[2mm]
E_2&= e^{i\bar{\phi}} \bigg[b_5(\eta,\sigma,\theta)d\sigma +b_6(\eta,\sigma,\theta)d\eta + b_7(\eta,\sigma,\theta)d\rho +b_8(\eta,\sigma,\theta)d\theta\nn\\[2mm]
&~~~~~~~~~~ +i\,  f_1 e^{-\frac{2}{3}\Phi}\sqrt{ h_{\phi_2}(\eta,\sigma,\theta)- \frac{1}{4} \frac{h_{\phi_1\phi_2}(\eta,\sigma,\theta)^2}{h_{\phi_1}(\eta,\sigma,\theta)}}  d\phi_2\bigg],\nn\\[2mm]
E_3 &=-\sqrt{f_1}e^{i\bar{\chi}} \,\Bigg[b_9(\eta,\sigma,\theta)d\sigma +b_{10}(\eta,\sigma,\theta)d\eta + b_{11}(\eta,\sigma,\theta)d\rho + b_{12}(\eta,\sigma,\theta)d\theta \nn\\[2mm]
&~~~~~~~~~~~~~~ + i \frac{e^{\frac{2}{3}\Phi}}{\sqrt{f_1}}\big(d\phi_3 + C_{1,\phi_1}d\phi_1+C_{1,\phi_2}d\phi_2\big)\Bigg],\nn\\[8mm]
 &\bar{\phi}=  s\,\chi +v\,\beta +u\,\phi,~~~~~~~~~~~~~~~~~~~~~~~~~\bar{\chi}= p\, \chi +q\, \beta + m \,\phi,\nn
\end{align}
 where $(b_1,..,~b_{12}) $ are functions which must be derived in each case. \\\\
 Notice here that $K$ remains intact under the rotation of frames because it is independent of $(\beta,\chi,\phi)$. Hence, in order to still satisfy all four G-Structure equations in \eqref{eqn:Geqns}, $J$ must also remain intact under a frame rotation. After enforcing this requirement, expressions for $(b_1,..,~b_{12}) $ are easily derived by performing the $SL(3,\mathds{R})$ transformation \eqref{eqn:S2breakingdefns} on the original forms for the $\chi$ reduction frame, given in \eqref{eqn:GstructureForms} and \eqref{eqn:GM-G}. This process must be repeated for each arrangement of $(\phi_1,\phi_2,\phi_3)$ in \eqref{eqn:Gstructuregenforms}, giving a separate set of functions for each. These functions, in their most general form,  depend on the nine $SL(3,\mathds{R})$ parameters and the derivatives of $V(\eta,\sigma)$. These results are too cumbersome to justify including here, but one can easily derive them using Mathematica and the method just outlined. \\
 
To derive these functions for the $\beta$ reduction frame, we could choose $(\phi_1,\phi_2,\phi_3)=(\chi,\phi,\beta)$. One must then specify the relevant values of the nine $SL(3,\mathds{R})$ parameters. For the $\mathcal{N}=2$ $\beta$ reduction frame presented in \eqref{eqn:GstructureForms} (and given in \cite{us}), we fix $(p,b,u)=1$ with all other parameters set to zero. Analogously, for the $\chi$ reduction forms, we pick $(\phi_1,\phi_2,\phi_3)=(\beta,\phi,\chi)$, re-deriving \eqref{eqn:GM-G} with $(p,b,u)=1$ and all other parameters set to zero\footnote{Alternatively, one can derive the $\chi$ reduction forms (up to $E_2\rightarrow -E_2$) from the $\beta$ reduction case just discussed by utilising the $SL(3,\mathds{R})$ transformation, with $a=1,q=1,u=-1$ and all others set to zero (noting that from the determinant in \eqref{eqn:S2breakingdefns}, we now require $qau=-1$). Hence,  in this case, we have made the transformations $\beta\rightarrow\chi,~\chi\rightarrow\beta,~\phi\rightarrow-\phi$.}. In the $\phi$ reduction case, we can simply cycle the roles of $E_i$ such that $(E_1,E_2,E_3)\rightarrow (E_2,E_3,E_1)$, meaning we get the following two alternative forms (from \eqref{eqn:GM-G} and \eqref{eqn:GstructureForms}, respectively)
 \begin{subequations}\label{eqn:subeqns}
   \begin{align}
 & ~~~~~~~~~~~~~~~~~~~~~~~~~~~~~~~~~~~~~~~~~~~~~~~~~~~~~~~~~~~~~~~~  \textbf{$\phi$ reduction frame}\nonumber\\[2mm]
&K= \frac{\kappa\, e^{-2\rho}}{f_1}d(\cos\theta e^{2\rho}\dot{V}),~~~~~~~~~~~~~~~~~~
E_3=e^{i\phi}\bigg[ \frac{\kappa  }{f_1} e^{-2\rho}d(\sin \theta e^{2\rho}\dot{V})+i\,\sqrt{f_1f_2}\,\sin \theta d\phi\bigg],\\
&\text{with}~~~~~~~~~~~~~~~~~~~~~~~~~~~~~~~~~~~~~~~~~~~~~~~~~~~~~~~~~~~~~~~~~~~~~~~~~~~~~~~~~~~~~~~~~~~~~~~~~~~~~~~~~~~~~~~~~~~~~~~~~~~~~~~~~~~~~~~~~~~~~~~~~~~~~~~~~~~~~~~~~~~~~~~
\nonumber\\
&E_1 =-e^{i\chi} \sqrt{f_1 f_5\Big(f_6^2+\frac{f_3}{f_5}\Big)} \,\Bigg[ d\rho - \frac{V''}{(\dot{V}')^2-\ddot{V}V''} d(\dot{V})+ i\,\bigg(d\chi +  \frac{f_6}{f_6^2 +\frac{f_3}{f_5}}  d\beta\bigg)\Bigg],\label{eqn:subeq1}\\[2mm]
&E_2   = \sqrt{\frac{f_1f_3}{f_6^2 +\frac{f_3}{f_5}}}  \bigg( d(V') -i\,d\beta\bigg),\nonumber \\
&\text{or}~~~~~~~~~~~~~~~~~~~~~~~~~~~~~~~~~~~~~~~~~~~~~~~~~~~~~~~~~~~~~~~~~~~~~~~~~~~~~~~~~~~~~~~~~~~~~~~~~~~~~~~~~~~~~~~~~~~~~~~~~~~~~~~~~~~~~~~~~~~~~~~~~~~~~~~~~~~~~~~~~~~~~~~
\nonumber\\
&E_1 =-e^{i\chi}\sqrt{f_1f_5} \,\Bigg[-\frac{1}{4}f_3 \frac{\dot{V}'}{\sigma} d\sigma -V''d\eta +f_6 d\rho  + i \Big(d\beta +f_6d\chi\Big)\Bigg], \label{eqn:subeq2} \\[2mm]
&E_2  =  -\sqrt{f_1f_3}\bigg(\frac{1}{\sigma} d\sigma+ d\rho   +i d\chi\bigg),\nonumber
 \end{align}
\end{subequations}
where $(E_1\bar{E_1}+E_2\bar{E_2}$), $(E_1\wedge \bar{E_1}+E_2\wedge \bar{E_2})$ and $E_1\wedge E_2$ for \eqref{eqn:subeq1} and \eqref{eqn:subeq2} are equivalent (as required). Alternatively, \eqref{eqn:subeq1} and \eqref{eqn:subeq2} are derived directly from \eqref{eqn:Gstructuregenforms} with $(\phi_1,\phi_2,\phi_3)=(\chi,\beta,\phi)$ and $(\phi_1,\phi_2,\phi_3)=(\beta,\chi,\phi)$, respectively (with $(p,b,u)=1$ and all other parameters set to zero).
 
 The forms presented in \eqref{eqn:GstructureForms}, \eqref{eqn:GM-G} and \eqref{eqn:subeqns} correspond to naive dimensional reductions along $\beta,\chi$ and $\phi$ respectively, with no $SL(3,\mathds{R})$ transformations taking place. Of course, as we've discussed throughout this paper, in order to derive $\mathcal{N}=1$ Type IIA and Type IIB backgrounds, one must include an appropriate transformation prior to reduction. The framework just outlined is now general enough to easily derive the G-Structure forms for all such solutions, allowing for the verification of the preservation of SUSY in each case. One could use the G-Structure reductions given in \eqref{eqn:11Dto10Dforms} to find analogous general IIA forms in terms of these 12 functions. Supersymmetry will only be preserved in a special set of cases, so in general, such general forms would be largely redundant.

 To then derive the IIA forms, we simply use \eqref{eqn:DimRed} and \eqref{eqn:11Dto10Dforms}.
% \subsection{Type IIA forms}
\subsection{G-Structure rules for Abelian T-Duality}\label{sec:IIBGstructures}
 
 Here we present the Abelian T-Duality of the G-Structure conditions. Note that in this calculation the convention $d_{H_3}=d+H_3\wedge$ is required - in order to use the minus sign convention (used throughout the rest of the paper) we would need to appropriately flip the sign of the $B$ field in the T-Dual rules given in \eqref{eqn:Tdualeq} (such that $E_\mathcal{B}^y=e^{-C^\mathcal{A}}(dy+B_1^\mathcal{A})$). 
 
Motivated by the Type IIA G-Structure condition given in \eqref{eqn:newform}, quoted here for convenience

\begin{equation*}
  \text{Vol}_4\wedge d_{H^\mathcal{A}_3}(e^{4A-\Phi_\mathcal{A}}\text{Im}\Psi^\mathcal{A}_-)=    \frac{1}{8}( F_6+F_8+F_{10}),
\end{equation*}
one can use the T-Dual rules of \eqref{eqn:Tdualeq} to make the decomposition 
\begin{equation*}
 \Psi_\pm^\mathcal{A}=\Psi_{\pm_{\perp}}^\mathcal{A} + \Psi_{\pm_{||}}^\mathcal{A}\wedge E^y_\mathcal{A},~~~~~~~~~~~~~~~~~~~~~~~~~~~
  \Psi_\mp^\mathcal{B}=e^{C^\mathcal{A}}\Psi_{\pm_{||}}^\mathcal{A} + \Psi_{\pm_{\perp}}^\mathcal{A}\wedge (dy-B_1^\mathcal{A}),
\end{equation*}
 with  $E^y_\mathcal{A}=  e^{C^\mathcal{A}}(dy+A_1^\mathcal{A})$ and $E^y_\mathcal{B}=  e^{C^\mathcal{B}}(dy+A_1^\mathcal{B}) =  e^{-C^\mathcal{A}}(dy-B_1^\mathcal{A}) $. The above form of $\Psi_\mp^\mathcal{B}$ is an initial ansatz - the final form will be derived at the end of the following subsection.
\subsubsection*{G-Structure conditions}

We begin by transforming the G-Structure condition under Abelian T-Duality from IIA to IIB. For convenience, we will summarise the left-hand side of the following IIA G-Structure conditions simply as $d_{H^\mathcal{A}_3}(e^{\alpha A-\Phi_\mathcal{A}}\Psi_\pm^\mathcal{A}) $,
\begin{equation*}
\begin{aligned}
d_{H^\mathcal{A}_3}(e^{3A-\Phi_\mathcal{A}}\Psi^\mathcal{A}_+)&=0,\\
d_{H^\mathcal{A}_3}(e^{2A-\Phi_\mathcal{A}}\text{Re}\Psi^\mathcal{A}_-)&=0,\\
d_{H^\mathcal{A}_3}(e^{4A-\Phi_\mathcal{A}}\text{Im}\Psi^\mathcal{A}_-)&=\frac{e^{4A}}{8}*_6\lambda(g).
\end{aligned}
\end{equation*}
 with the choice of $\alpha\in (2,3,4)$ depending on the specific condition.
Now, the condition which will transform in the same manner as the Ramond fields is
\beq
  \text{Vol}_4\wedge d_{H^\mathcal{A}_3}(e^{\alpha A-\Phi_\mathcal{A}}\Psi_\pm^\mathcal{A}) =   \text{Vol}_4\wedge\Big[d(e^{\alpha A-\Phi_\mathcal{A}}\Psi_\pm^\mathcal{A})  + dB^\mathcal{A} \wedge e^{\alpha A-\Phi_\mathcal{A}}\Psi_\pm^\mathcal{A}\Big].
\eeq
From the T-Dual rules \eqref{eqn:Tdualeq}, we have
\beq
\begin{aligned}
dB^\mathcal{A}&=dB_2^\mathcal{A}+dB_1^\mathcal{A}\wedge dy\\
&=dB_2^\mathcal{A} +e^{-C^\mathcal{A}}dB_1^\mathcal{A}\wedge e^{C^\mathcal{A}}(dy+A_1^\mathcal{A})-dB_1^\mathcal{A}\wedge A_1^\mathcal{A}\\
&=(dB_2^\mathcal{A} -dB_1^\mathcal{A}\wedge A_1^\mathcal{A})+ e^{-C^\mathcal{A}}dB_1^\mathcal{A}\wedge E^y_\mathcal{A},
\end{aligned}
\eeq
and using the decomposition for $\Psi_\pm^\mathcal{A}$, we get
\beq
\begin{aligned}
 dB^\mathcal{A} \wedge e^{\alpha A-\Phi_\mathcal{A}}\Psi_\pm^\mathcal{A}&=[(dB_2^\mathcal{A} -dB_1^\mathcal{A}\wedge A_1^\mathcal{A})+ e^{-C^\mathcal{A}}dB_1^\mathcal{A}\wedge E^y_\mathcal{A}]\wedge [e^{\alpha A-\Phi_\mathcal{A}}\Psi_{\pm_{\perp}}^\mathcal{A} +e^{\alpha A-\Phi_\mathcal{A}} \Psi_{\pm_{||}}^\mathcal{A}\wedge E^y_\mathcal{A}]\\
 &=(dB_2^\mathcal{A} -dB_1^\mathcal{A}\wedge A_1^\mathcal{A})\wedge e^{\alpha A-\Phi_\mathcal{A}}\Psi_{\pm_{\perp}}^\mathcal{A} \\
 &~~~~+\Big[(dB_2^\mathcal{A} -dB_1^\mathcal{A}\wedge A_1^\mathcal{A}) \wedge e^{\alpha A-\Phi_\mathcal{A}} \Psi_{\pm_{||}}^\mathcal{A} -   e^{\alpha A-\Phi_\mathcal{A}}\Psi_{\pm_{\perp}}^\mathcal{A} \wedge e^{-C^\mathcal{A}}dB_1^\mathcal{A} \Big]\wedge E^y_\mathcal{A}.
\end{aligned}
\eeq
%where the $\pm$ in the final line corresponds to the cases of $\Psi_{\pm_{\perp}}^\mathcal{A} $ respectively, because $\Psi_{+_{\perp}}^\mathcal{A} $ is even dimensional and $\Psi_{-_{\perp}}^\mathcal{A} $ is odd dimensional. 
In addition,
\beq
\begin{aligned}
 d(e^{\alpha A-\Phi_\mathcal{A}}\Psi_\pm^\mathcal{A})  &=  d(e^{\alpha A-\Phi_\mathcal{A}}\Psi_{\pm_{\perp}}^\mathcal{A} + e^{\alpha A-\Phi_\mathcal{A}}\Psi_{\pm_{||}}^\mathcal{A}\wedge E^y_\mathcal{A}) \\
 &=  d(e^{\alpha A-\Phi_\mathcal{A}}\Psi_{\pm_{\perp}}^\mathcal{A})+d( e^{\alpha A-\Phi_\mathcal{A}}\Psi_{\pm_{||}}^\mathcal{A})\wedge E^y_\mathcal{A}+ e^{\alpha A-\Phi_\mathcal{A}}\Psi_{\pm_{||}}^\mathcal{A} \wedge \Big(d(e^{C^\mathcal{A}})\wedge dy +d(e^{C^\mathcal{A}}A_1^\mathcal{A})\Big)\\
 % &=  d(e^{\alpha A-\Phi_A}\Psi_{\pm_{\perp}}^A)+d( e^{\alpha A-\Phi_A}\Psi_{\pm_{||}}^A)\wedge E^y_A\\
  %&~~~~~+ e^{\alpha A-\Phi_A}\Psi_{\pm_{||}}^A \wedge \Big(e^{-C^A}d(e^{C^A})\wedge E^y_A +d(e^{C^A}A_1^A)-d(e^{C^A})\wedge A_1^A\Big)\\
%    &=  d(e^{\alpha A-\Phi_A}\Psi_{\pm_{\perp}}^A)+d( e^{\alpha A-\Phi_A}\Psi_{\pm_{||}}^A)\wedge E^y_A\\
  %&~~~~~+ e^{\alpha A-\Phi_A}\Psi_{\pm_{||}}^A \wedge \Big(e^{-C^A}d(e^{C^A})\wedge E^y_A +e^{C^A}d(A_1^A)\Big)\\
     &=  d(e^{\alpha A-\Phi_\mathcal{A}}\Psi_{\pm_{\perp}}^\mathcal{A})+e^{\alpha A-\Phi_\mathcal{A}}\Psi_{\pm_{||}}^\mathcal{A} \wedge e^{C^\mathcal{A}}d(A_1^\mathcal{A})\\
  &~~~~~+\Big[d( e^{\alpha A-\Phi_\mathcal{A}}\Psi_{\pm_{||}}^\mathcal{A}) + e^{\alpha A-\Phi_\mathcal{A}}\Psi_{\pm_{||}}^\mathcal{A} \wedge e^{-C^\mathcal{A}}d(e^{C^\mathcal{A}})\Big]\wedge E^y_\mathcal{A} .
 \end{aligned}
\eeq
Hence, we have
\beq\label{eqn:Gammas}
\begin{gathered}
  \text{Vol}_4\wedge d_{H^\mathcal{A}_3}(e^{\alpha A-\Phi_\mathcal{A}}\Psi_\pm^\mathcal{A}) =  \text{Vol}_4\wedge(\Gamma_{\perp}^\mathcal{A}+\Gamma_{||}^\mathcal{A}\wedge E^y_\mathcal{A}),\\\\
  \Gamma_{\perp}^\mathcal{A} =  d(e^{\alpha A-\Phi_\mathcal{A}}\Psi_{\pm_{\perp}}^\mathcal{A})+e^{\alpha A-\Phi_\mathcal{A}}\Psi_{\pm_{||}}^\mathcal{A} \wedge e^{C^\mathcal{A}}d(A_1^\mathcal{A})+ (dB_2^\mathcal{A} -dB_1^\mathcal{A}\wedge A_1^\mathcal{A})\wedge e^{\alpha A-\Phi_\mathcal{A}}\Psi_{\pm_{\perp}}^\mathcal{A} ,\\
  \Gamma_{||}^\mathcal{A} = d( e^{\alpha A-\Phi_\mathcal{A}}\Psi_{\pm_{||}}^\mathcal{A}) + e^{\alpha A-\Phi_\mathcal{A}}\Psi_{\pm_{||}}^\mathcal{A} \wedge e^{-C^\mathcal{A}}d(e^{C^\mathcal{A}})~~~~~~~~~~~~~~~~~~~~~~~~~~~~~~~~~~~~~~~~~~~~~~~~~~~~~~~~~~~ \\
  +(dB_2^\mathcal{A} -dB_1^\mathcal{A}\wedge A_1^\mathcal{A}) \wedge e^{\alpha A-\Phi_\mathcal{A}} \Psi_{\pm_{||}}^\mathcal{A} -e^{\alpha A-\Phi_\mathcal{A}}\Psi_{\pm_{\perp}}^\mathcal{A}\wedge e^{-C^\mathcal{A}}dB_1^\mathcal{A}  .
 \end{gathered}
\eeq
Now, applying the T-Dual rules, one gets for the IIB equations
\beq
\text{Vol}_4\wedge\Big(e^{C^\mathcal{A}}\Gamma_{||}^\mathcal{A} + \Gamma_{\perp}^\mathcal{A}\wedge (dy-B_1^\mathcal{A})\Big).
\eeq
Before proceeding, from the transformation rules
\beq
B^\mathcal{B}=B_2^\mathcal{B}+B_1^\mathcal{B}\wedge dy =B_2^\mathcal{A} -A_1^\mathcal{A}\wedge (dy-B_1^\mathcal{A})  ,
\eeq
the following result will be required
\beq
dB_2^\mathcal{A} -dB_1^\mathcal{A}\wedge A_1^\mathcal{A} = dB^\mathcal{B} +dA_1^\mathcal{A} \wedge (dy-B_1^\mathcal{A}),
\eeq
which we substitute directly into \eqref{eqn:Gammas}, giving

\beq 
\begin{gathered}
  \Gamma_{\perp}^\mathcal{A} =  d(e^{\alpha A-\Phi_\mathcal{A}}\Psi_{\pm_{\perp}}^\mathcal{A})+  dB^\mathcal{B} \wedge e^{\alpha A-\Phi_\mathcal{A}}\Psi_{\pm_{\perp}}^\mathcal{A}+e^{\alpha A-\Phi_\mathcal{A}}\Psi_{\pm_{||}}^\mathcal{A} \wedge e^{C^\mathcal{A}}d(A_1^\mathcal{A})-dA_1^\mathcal{A}\wedge e^{\alpha A-\Phi_\mathcal{A}}\Psi_{\pm_{\perp}}^\mathcal{A} \wedge (dy-B_1^\mathcal{A})  ,\\
  \Gamma_{||}^\mathcal{A} = d( e^{\alpha A-\Phi_\mathcal{A}}\Psi_{\pm_{||}}^\mathcal{A}) + e^{\alpha A-\Phi_\mathcal{A}}\Psi_{\pm_{||}}^\mathcal{A} \wedge e^{-C^\mathcal{A}}d(e^{C^\mathcal{A}})~~~~~~~~~~~~~~~~~~~~~~~~~~~~~~~~~~~~~~~~~~~~~~~~~~~~~~~~~~~ \\
  +\Big( dB^\mathcal{B} +dA_1^\mathcal{A} \wedge (dy-B_1^\mathcal{A})\Big) \wedge e^{\alpha A-\Phi_\mathcal{A}} \Psi_{\pm_{||}}^\mathcal{A} - e^{\alpha A-\Phi_\mathcal{A}}\Psi_{\pm_{\perp}}^\mathcal{A}\wedge e^{-C^\mathcal{A}}dB_1^\mathcal{A} ,
 \end{gathered}
\eeq
hence
\beq
\begin{aligned}
e^{C^\mathcal{A}}\Gamma_{||}^\mathcal{A} + \Gamma_{\perp}^\mathcal{A}\wedge (dy-B_1^\mathcal{A})&=e^{C^\mathcal{A}}d( e^{\alpha A-\Phi_\mathcal{A}}\Psi_{\pm_{||}}^\mathcal{A}) + e^{\alpha A-\Phi_\mathcal{A}}\Psi_{\pm_{||}}^\mathcal{A} \wedge d(e^{C^\mathcal{A}})+e^{C^\mathcal{A}} dB^\mathcal{B} \wedge e^{\alpha A-\Phi_\mathcal{A}} \Psi_{\pm_{||}}^\mathcal{A}\\
&~~~~~~~ -e^{C^\mathcal{A}} dA_1^\mathcal{A}\wedge e^{\alpha A-\Phi_\mathcal{A}} \Psi_{\pm_{||}}^\mathcal{A} \wedge (dy-B_1^\mathcal{A})  - e^{\alpha A-\Phi_\mathcal{A}}\Psi_{\pm_{\perp}}^\mathcal{A}\wedge dB_1^\mathcal{A}\\
+&\bigg[ d(e^{\alpha A-\Phi_\mathcal{A}}\Psi_{\pm_{\perp}}^\mathcal{A})+  dB^\mathcal{B} \wedge e^{\alpha A-\Phi_\mathcal{A}}\Psi_{\pm_{\perp}}^\mathcal{A}+e^{\alpha A-\Phi_\mathcal{A}}\Psi_{\pm_{||}}^\mathcal{A} \wedge e^{C^\mathcal{A}}d(A_1^\mathcal{A})\bigg]\wedge (dy-B_1^\mathcal{A})\\
&=d( e^{\alpha A-\Phi_\mathcal{A}}e^{C^\mathcal{A}}\Psi_{\pm_{||}}^\mathcal{A}) + dB^\mathcal{B} \wedge e^{\alpha A-\Phi_\mathcal{A}}e^{C^\mathcal{A}} \Psi_{\pm_{||}}^\mathcal{A}\\
&~~~~~~+  d\Big(e^{\alpha A-\Phi_\mathcal{A}}\Psi_{\pm_{\perp}}^\mathcal{A}\wedge (dy-B_1^\mathcal{A})\Big)+  dB^\mathcal{B} \wedge e^{\alpha A-\Phi_\mathcal{A}}\Psi_{\pm_{\perp}}^\mathcal{A}\wedge (dy-B_1^\mathcal{A})\\
&= d_{H^\mathcal{B}_3}\Big[ e^{\alpha A-\Phi_\mathcal{A}} \Big(e^{C^\mathcal{A}}\Psi_{\pm_{||}}^\mathcal{A} +\Psi_{\pm_{\perp}}^\mathcal{A}\wedge (dy-B_1^\mathcal{A}) \Big)\Big]\\
&= d_{H^\mathcal{B}_3}( e^{\alpha A-\Phi_\mathcal{A}} \Psi_\mp^\mathcal{B} ).
\end{aligned}
\eeq

The IIB G-Structure equations now read
\begin{equation*}
\begin{aligned}
d_{H^\mathcal{B}_3}(e^{3A-\Phi_\mathcal{A}}\Psi^\mathcal{B}_-)&=0,\\
d_{H^\mathcal{B}_3}(e^{2A-\Phi_\mathcal{A}}\text{Re}\Psi^\mathcal{B}_+)&=0,\\
d_{H^\mathcal{B}_3}(e^{4A-\Phi_\mathcal{A}}\text{Im}\Psi^\mathcal{B}_+)&=\frac{e^{4A}}{8}*_6\lambda(g). 
\end{aligned}
\end{equation*}
Of course, by adjusting the transformation as follows
\begin{equation*}
e^{-\Phi_\mathcal{B}}  \Psi_\mp^\mathcal{B}=e^{-\Phi_\mathcal{A}}\Big[e^{C^\mathcal{A}}\Psi_{\pm_{||}}^\mathcal{A} + \Psi_{\pm_{\perp}}^\mathcal{A}\wedge (dy-B_1^\mathcal{A})\Big],
\end{equation*}
one can re-write the above G-Structure conditions in terms of $\Phi_\mathcal{B}$, 
\begin{equation*}
\begin{aligned}
d_{H^\mathcal{B}_3}(e^{3A-\Phi_\mathcal{B}}\Psi^\mathcal{B}_-)&=0,\\
d_{H^\mathcal{B}_3}(e^{2A-\Phi_\mathcal{B}}\text{Re}\Psi^\mathcal{B}_+)&=0,\\
d_{H^\mathcal{B}_3}(e^{4A-\Phi_\mathcal{B}}\text{Im}\Psi^\mathcal{B}_+)&=\frac{e^{4A}}{8}*_6\lambda(g). 
\end{aligned}
\end{equation*}
%this is purely for aesthetic purposes!
\subsubsection*{G-Structure forms}
      Now we wish to calculate the IIB pure spinors which we need for the G-Structure conditions just derived. The $SU(2)$ pure spinors for IIA are given in \eqref{eqn:Psi}, and re-written here for clarity
      
      \begin{equation*}
\Psi^\mathcal{A}_+=\frac{1}{8} e^{\frac{1}{2}z^\mathcal{A}\wedge \overline{z}^\mathcal{A}}\wedge \omega^\mathcal{A},~~~~~\Psi^\mathcal{A}_-=\frac{i}{8}  e^{-i j^\mathcal{A}}\wedge z^\mathcal{A}.
\end{equation*}
 In what follows, it will prove useful to make the following decompositions
\beq\label{eqn:decomps}
\begin{gathered}
\omega^\mathcal{A}=\omega^\mathcal{A}_{\perp}+\omega^\mathcal{A}_{||}\wedge E^y_\mathcal{A},~~~~~~~~~~~~~~~~~
j^\mathcal{A}=j^\mathcal{A}_{\perp}+j^\mathcal{A}_{||}\wedge E^y_\mathcal{A},~~~~~~~~~~~~~
z^\mathcal{A}=z^\mathcal{A}_{\perp}+z^\mathcal{A}_{||}\wedge E^y_\mathcal{A},\\
z^\mathcal{A}_{\perp}=u^\mathcal{A}_{\perp}+i\,v^\mathcal{A}_{\perp},~~~~~~~~~~~~~~~~~~z^\mathcal{A}_{||}=u^\mathcal{A}_{||}+i\,v^\mathcal{A}_{||}.
\end{gathered}
\eeq  
 We then note
\beq
\begin{aligned}
e^{\frac{1}{2}z^\mathcal{A}\wedge \overline{z}^\mathcal{A}} &= 1+\frac{1}{2}z^\mathcal{A}\wedge \overline{z}^\mathcal{A}\\
&= \Big(1+\frac{1}{2} z^\mathcal{A}_{\perp}\wedge \overline{z}^\mathcal{A}_{\perp}\Big) +\frac{1}{2}( z^\mathcal{A}_{\perp}\wedge \overline{z}^\mathcal{A}_{||} -z^\mathcal{A}_{||}\wedge \overline{z}^\mathcal{A}_{\perp})\wedge E^y_\mathcal{A},\\
\end{aligned}
\eeq
as $( z^\mathcal{A}\wedge \overline{z}^\mathcal{A})\wedge ( z^\mathcal{A}\wedge \overline{z}^\mathcal{A})  =0$, and 

\beq
\hspace{-0.5cm}
\begin{aligned}
e^{-ij^\mathcal{A}} &= 1+(-ij^\mathcal{A})+\frac{1}{2}(-ij^\mathcal{A})\wedge (-ij^\mathcal{A}) +\frac{1}{3!}(-ij^\mathcal{A})\wedge (-ij^\mathcal{A})\wedge (-ij^\mathcal{A})+....\\
&=1-ij^\mathcal{A} -\frac{1}{2}j^\mathcal{A}\wedge j^\mathcal{A} +\frac{i}{3!}j^\mathcal{A}\wedge j^\mathcal{A} \wedge j^\mathcal{A}+....\\
&=1-i(j^\mathcal{A}_{\perp}+j^\mathcal{A}_{||}\wedge E^y_\mathcal{A}) -\frac{1}{2}(j^\mathcal{A}_{\perp} \wedge j^\mathcal{A}_{\perp} +2j^\mathcal{A}_{\perp}\wedge j^\mathcal{A}_{||}\wedge E^y_\mathcal{A}) +\frac{i}{3!}(j^\mathcal{A}_{\perp}\wedge j^\mathcal{A}_{\perp}\wedge j^\mathcal{A}_{\perp} + 3j^\mathcal{A}_{\perp}\wedge j^\mathcal{A}_{\perp}\wedge j^\mathcal{A}_{||}\wedge E^y_\mathcal{A})+ ..\\
&=\Big(1- i j^\mathcal{A}_{\perp} - \frac{1}{2}j^\mathcal{A}_{\perp} \wedge j^\mathcal{A}_{\perp}+\frac{i}{3!}j^\mathcal{A}_{\perp}\wedge j^\mathcal{A}_{\perp}\wedge j^\mathcal{A}_{\perp}+... \Big) +\Big(-i j^\mathcal{A}_{||} -j^\mathcal{A}_{\perp}\wedge j^\mathcal{A}_{||} +\frac{i}{2}j^\mathcal{A}_{\perp}\wedge j^\mathcal{A}_{\perp}\wedge j^\mathcal{A}_{||}+... \Big)\wedge E^y_\mathcal{A}\\
&=e^{-ij^\mathcal{A}_{\perp}}\wedge (1-i j^\mathcal{A}_{||}\wedge E^y_\mathcal{A}).
\end{aligned}
\eeq

Hence, we get
\beq
\hspace{0.5cm}
\begin{gathered}
 \Psi_\pm^\mathcal{A}=\Psi_{\pm_{\perp}}^\mathcal{A} + \Psi_{\pm_{||}}^\mathcal{A}\wedge E^y_\mathcal{A}\\
 \Psi_{+_{\perp}}^\mathcal{A}=\frac{1}{8}\Big(1+\frac{1}{2}z^\mathcal{A}_{\perp}\wedge \overline{z}^\mathcal{A}_{\perp}\Big)\wedge \omega^\mathcal{A}_{\perp},~~~~~~~\Psi_{+_{||}}^\mathcal{A}=\frac{1}{8}\bigg[\Big(1+\frac{1}{2}z^\mathcal{A}_{\perp}\wedge \overline{z}^\mathcal{A}_{\perp}\Big)\wedge \omega^\mathcal{A}_{||}+\frac{1}{2}( z^\mathcal{A}_{\perp}\wedge \overline{z}^\mathcal{A}_{||} -z^\mathcal{A}_{||}\wedge \overline{z}^\mathcal{A}_{\perp})\wedge \omega^\mathcal{A}_{\perp}\bigg]\\
 \Psi_{-_{\perp}}^\mathcal{A}=\frac{i}{8}e^{-ij^\mathcal{A}_{\perp}}\wedge z^\mathcal{A}_{\perp},~~~~~~~~~~~~~~~~~~~\Psi_{-_{||}}^\mathcal{A}=\frac{i}{8}e^{-ij^\mathcal{A}_{\perp}}\wedge (z^\mathcal{A}_{||}+ij^\mathcal{A}_{||}\wedge z^\mathcal{A}_{\perp}).
\end{gathered}
\eeq
Now, recalling
\begin{equation*}
\Psi_\mp^\mathcal{B}=e^{\Phi_\mathcal{B}-\Phi_\mathcal{A}}\Big[e^{C^\mathcal{A}}\Psi_{\pm_{||}}^\mathcal{A} + \Psi_{\pm_{\perp}}^\mathcal{A}\wedge (dy-B_1^\mathcal{A})\Big],
\end{equation*}
%\begin{equation*}
  %\Psi_\mp^\mathcal{B}=e^{C^\mathcal{A}}\Psi_{\pm_{||}}^\mathcal{A} + \Psi_{\pm_{\perp}}^\mathcal{A}\wedge (dy-B_1^\mathcal{A}).
%\end{equation*}
we finally arrive at
\beq
\begin{aligned}\label{eqn:IIBforms}
  \Psi_-^\mathcal{B}%&=\frac{1}{8}e^{\Phi_\mathcal{B}-\Phi_\mathcal{A}}\bigg[\Big(1+\frac{1}{2}z^\mathcal{A}_{\perp}\wedge \overline{z}^\mathcal{A}_{\perp}\Big)\wedge\Big( e^{C^\mathcal{A}}\omega^\mathcal{A}_{||}+   \omega^\mathcal{A}_{\perp}\wedge (dy-B_1^\mathcal{A})\Big)+e^{C^\mathcal{A}}\frac{1}{2}( z^\mathcal{A}_{\perp}\wedge \overline{z}^\mathcal{A}_{||} -z^\mathcal{A}_{||}\wedge \overline{z}^\mathcal{A}_{\perp})\wedge \omega^\mathcal{A}_{\perp}\bigg] \\
  &=\frac{1}{8}e^{\Phi_\mathcal{B}-\Phi_\mathcal{A}}\bigg[e^{\frac{1}{2}z^\mathcal{A}_{\perp}\wedge \overline{z}^\mathcal{A}_{\perp}}\wedge\Big( e^{C^\mathcal{A}}\omega^\mathcal{A}_{||}+   \omega^\mathcal{A}_{\perp}\wedge (dy-B_1^\mathcal{A})\Big)+e^{C^\mathcal{A}}\frac{1}{2}( z^\mathcal{A}_{\perp}\wedge \overline{z}^\mathcal{A}_{||} -z^\mathcal{A}_{||}\wedge \overline{z}^\mathcal{A}_{\perp})\wedge \omega^\mathcal{A}_{\perp}\bigg] \\
  %&=\frac{1}{8} \Big(1+\frac{1}{2}z^\mathcal{A}_{\perp}\wedge \overline{z}^\mathcal{A}_{\perp}\Big)\wedge\Big( e^{C^\mathcal{A}}\omega^\mathcal{A}_{||}+   \omega^\mathcal{A}_{\perp}\wedge (dy-B_1^\mathcal{A})+e^{C^\mathcal{A}}\frac{1}{2}( z^\mathcal{A}_{\perp}\wedge \overline{z}^\mathcal{A}_{||} -z^\mathcal{A}_{||}\wedge \overline{z}^\mathcal{A}_{\perp})\wedge \omega^\mathcal{A}_{\perp}\Big)  \\
 %   &=\frac{1}{8}\bigg[e^{\frac{1}{2}z^\mathcal{A}_{\perp}\wedge \overline{z}^\mathcal{A}_{\perp}}\wedge\Big( e^{C^\mathcal{A}}\omega^\mathcal{A}_{||}+   \omega^\mathcal{A}_{\perp}\wedge (dy-B_1^\mathcal{A})+e^{C^\mathcal{A}}\frac{1}{2}( z^\mathcal{A}_{\perp}\wedge \overline{z}^\mathcal{A}_{||} -z^\mathcal{A}_{||}\wedge \overline{z}^\mathcal{A}_{\perp})\wedge \omega^\mathcal{A}_{\perp}\Big)\bigg]\\
 %       &=\frac{1}{8} e^{\Phi_\mathcal{B}-\Phi_\mathcal{A}} e^{\frac{1}{2}z^\mathcal{A}_{\perp}\wedge \overline{z}^\mathcal{A}_{\perp}}\wedge\Big(  \tilde{ \omega}^\mathcal{B}+e^{C^\mathcal{A}}\frac{1}{2}( z^\mathcal{A}_{\perp}\wedge \overline{z}^\mathcal{A}_{||} -z^\mathcal{A}_{||}\wedge \overline{z}^\mathcal{A}_{\perp})\wedge \omega^\mathcal{A}_{\perp}\Big)  
   \Psi_+^\mathcal{B}&=\frac{i }{8} e^{\Phi_\mathcal{B}-\Phi_\mathcal{A}}e^{-ij^\mathcal{A}_{\perp}}\wedge \bigg[ (e^{C^\mathcal{A}}z^\mathcal{A}_{||}+ z^\mathcal{A}_{\perp}\wedge (dy-B_1^\mathcal{A}))+i  e^{C^\mathcal{A}} j^\mathcal{A}_{||}\wedge z^\mathcal{A}_{\perp}\bigg].
\end{aligned}
\eeq
%where $ \tilde{ \omega}^\mathcal{B}$ is no longer a 2-form but instead an odd polyform (hence the inclusion of a tilde to make this clear).

     \subsection{$\mathcal{N}=1$ G-Structures}\label{sec:N=1Gstructures}
     We will now present the G-Structure results for the IIA and IIB $\mathcal{N}=1$ solutions given in the main body of the paper, beginning with the 11D forms in each case.
     \subsubsection{Unique Solutions}\label{sec:UniqIIAGstructures}

          \subsubsection*{11D Forms}

 \begin{align}
 K&= -\frac{\kappa e^{-2\rho}}{f_1}d(e^{2\rho}\dot{V}\cos\theta)\nn\\[2mm]
 E_1&=\frac{2\kappa^2 f_1^{-\frac{5}{2}}f_5^{-1}}{ \sqrt{\Sigma}\sqrt{f_3+f_2\sin^2\theta}}\bigg[\dot{V}f_2\sin^2\theta d\eta +2\sigma^2\Big(1+\frac{1}{4}f_2\sin^2\theta\Big)d(V')\nn\\[2mm]
 &~~~~-\frac{\dot{V}}{4}f_5f_6 \Big(e^{-6\rho}d(e^{6\rho}\dot{V}\sin^2\theta)+\sin^2\theta d(\dot{V})\Big) 
 -\frac{i}{2\kappa^2}f_1^3f_5(f_3+f_2\sin^2\theta)\Big(d\beta +\frac{f_2f_6\sin^2\theta}{f_3+f_2\sin^2\theta}d\phi\Big)\bigg],\nn\\[2mm]
 E_2&= e^{i\phi}\Bigg[  \frac{\kappa}{f_1\sqrt{\frac{1}{4}\big(1+\frac{4}{ \sigma^2f_4}\big)f_2 \sin^2\theta +1}}\Big(e^{-3\rho}d(e^{3\rho}\dot{V}\sin\theta) +\frac{\dot{V}}{\sigma}\sin\theta \,d\sigma\Big)   + i\,\sqrt{\frac{f_1f_2f_3}{f_3+f_2\sin^2\theta}}\sin\theta d\phi \Bigg],\nn\\[2mm]
 E_3&=   -   \bigg[   \frac{\sqrt{8f_1} (f_5\Sigma)^{-\frac{1}{4}}}{\big(1+\frac{f_2}{4} (\sin^2\theta-4)\big)^{\frac{1}{4}}}\Big[d\rho + \frac{1}{8}f_2 \Big(d(\sin^2\theta) -\frac{2}{\dot{V}}e^{-2\rho}d\big(e^{2\rho}\dot{V}(\sin^2\theta+2)\big)\Big)\Big]+i\sqrt{f_1f_5\Sigma}(d\chi+C_1)\bigg],\nn\\[8mm]
 &~~~~~~~~~~~~~~~~~~~~~~~~~~~~~~~~~~~\Sigma=f_6^2 +\frac{f_3}{f_5}+\frac{f_2}{f_5}\sin^2\theta,~~~~~~~~~~~~~~~~  C_1=\frac{ 1}{ \Sigma}\bigg( f_6 d\beta  - \frac{f_2}{f_5}  \sin^2\theta d\phi\bigg)
 \end{align}

          \subsubsection*{IIA Forms}
 The G-Structure forms for the IIA $\mathcal{N}=1$ solution given in \eqref{eqn:N=1IIA2} then read
    
    \begin{align}\label{eqn:UnqIIAGs}
 u^\mathcal{A}&=\frac{2\sqrt{2} f_1^{\frac{3}{4}}}{\Big(1+\frac{f_2}{4} (\sin^2\theta-4)\Big)^{\frac{1}{4}}}\bigg[d\rho + \frac{1}{8}f_2 \bigg(d(\sin^2\theta) -\frac{2}{\dot{V}}e^{-2\rho}d\Big(e^{2\rho}\dot{V}(\sin^2\theta+2)\Big)\bigg)\bigg],\nn\\[2mm]
 v^\mathcal{A}&=-\kappa\sqrt{2}f_1^{-\frac{3}{4}}\Big(1+\frac{f_2}{4} (\sin^2\theta-4)\Big)^{\frac{1}{4}} e^{-2\rho}d(e^{2\rho}\dot{V}\cos\theta),\nn\\[2mm]
 j^\mathcal{A}&= \frac{\kappa f_2^{\frac{1}{2}}}{2\Big(1+\frac{f_2}{4} (\sin^2\theta-4)\Big)^{\frac{1}{2}}}\Bigg[\bigg( (1-f_2) \Big(d(2\dot{V}\sin^2\theta)+e^{-6\rho}  \sin^2\theta d(2e^{6\rho} \dot{V})\Big) -2f_2 \sin^2\theta d(\dot{V})\bigg)\wedge d\phi\nn\\[2mm]
 &~~~~~~~~~~~~ + \bigg[f_5f_6\Big(e^{-3\rho}d(\sin^2\theta e^{3\rho}\dot{V}) -\frac{1}{2}d(\sin^2\theta \dot{V}) +\sin^2\theta d(\dot{V})\Big)- \frac{\Lambda f_3}{\dot{V}} d(V')-2\sin^2\theta d\eta \bigg]\wedge d\beta \Bigg],\nn\\[2mm]
 \omega^\mathcal{A}&=- 2\kappa   \frac{\sqrt{f_2}}{\dot{V}} \bigg( \sigma \dot{V} e^{-3\rho}  d(V')\wedge d(e^{3\rho}\sin\theta e^{i\phi})+   \tilde{\Delta} \Big(1-\frac{3 }{2}f_2\Big) e^{i\phi} \sin\theta d\sigma\wedge d\eta  \nn\\[2mm]
 &~~~~~~~~~~~~~~~+i e^{-3\rho} \, d(\sigma \dot{V}e^{3\rho}e^{i\phi}\sin\theta )\wedge d\beta \bigg).
\end{align}

                    \subsubsection*{IIB Forms}
 Performing an ATD of \eqref{eqn:N=1IIA2} along $\beta$ leads to the $\mathcal{N}=1$ IIB solution given in \eqref{eqn:uniqueN1IIB}. To then verify the preservation of supersymmetry, we note from \eqref{eqn:IIBforms}
                        \beq
     \begin{aligned}
     \Psi_-^\mathcal{B}&=\frac{1}{8}e^{\Phi_\mathcal{B}-\Phi_\mathcal{A}}\bigg(
     \Big(1+\frac{1}{2}z^\mathcal{A}_\perp\wedge \bar{z}^\mathcal{A}_\perp\Big)\wedge \Big(e^{C^\mathcal{A}}\omega^\mathcal{A}_{||}+   \omega^\mathcal{A}_{\perp}\wedge (dy-B_1^\mathcal{A})\Big) +\frac{1}{2}e^{C^\mathcal{A}}(z^\mathcal{A}_\perp\wedge \bar{z}^\mathcal{A}_{||}-z^\mathcal{A}_{||}\wedge \bar{z}^\mathcal{A}_\perp)\bigg),\\[2mm]
      \Psi_+^\mathcal{B}&=\frac{i}{8}\Big(1-i j_\perp^\mathcal{A}-\frac{1}{2}j_\perp^\mathcal{A}\wedge j_\perp^\mathcal{A}\Big)\wedge \Big(e^{C^\mathcal{A}}z_{||}^\mathcal{A} +z_\perp^\mathcal{A}\wedge (dy-B_1^\mathcal{A})+i e^{C^\mathcal{A}}j_{||}^\mathcal{A}\wedge z_\perp^\mathcal{A} \Big),
     \end{aligned}
     \eeq
     with
     \beq
     \begin{aligned}
     \text{Re}\Psi_+^\mathcal{B}&=\frac{1}{8} \bigg[j^\mathcal{A}_\perp\wedge \Big(e^{C^\mathcal{A}}u^\mathcal{A}_{||}+u^\mathcal{A}_\perp \wedge  (dy-B_1^\mathcal{A}) -e^{C^\mathcal{A}}j^\mathcal{A}_{||}\wedge v^\mathcal{A}_\perp\Big) -\Big(e^{C^\mathcal{A}}v^\mathcal{A}_{||}+e^{C^\mathcal{A}}j^\mathcal{A}_{||}\wedge u^\mathcal{A}_\perp +v^\mathcal{A}_\perp \wedge (dy-B_1^\mathcal{A}) \Big)\\[2mm]
  &~~~~~~~~~~~~~~~~~+\frac{1}{2}j^\mathcal{A}_\perp\wedge j^\mathcal{A}_\perp\wedge \Big(e^{C^\mathcal{A}} v^\mathcal{A}_{||}+v^\mathcal{A}_\perp\wedge (dy-B_1^\mathcal{A}) + e^{C^\mathcal{A}}j^\mathcal{A}_{||}\wedge u^\mathcal{A}_\perp \Big) \bigg],\\[2mm]
  \text{Im}\Psi_+^\mathcal{B}&=\frac{1}{8} \bigg[j^\mathcal{A}_\perp\wedge \Big(e^{C^\mathcal{A}}v^\mathcal{A}_{||}+v^\mathcal{A}_\perp \wedge  (dy-B_1^\mathcal{A}) +e^{C^\mathcal{A}}j^\mathcal{A}_{||}\wedge u^\mathcal{A}_\perp\Big) +\Big(e^{C^\mathcal{A}}u^\mathcal{A}_{||}-e^{C^\mathcal{A}}j^\mathcal{A}_{||}\wedge v^\mathcal{A}_\perp +u^\mathcal{A}_\perp \wedge (dy-B_1^\mathcal{A}) \Big)\\[2mm]
  &~~~~~~~~~~~~~~~~~-\frac{1}{2}j^\mathcal{A}_\perp\wedge j^\mathcal{A}_\perp\wedge \Big(e^{C^\mathcal{A}} u^\mathcal{A}_{||}+u^\mathcal{A}_\perp\wedge (dy-B_1^\mathcal{A}) - e^{C^\mathcal{A}}j^\mathcal{A}_{||}\wedge v^\mathcal{A}_\perp \Big) \bigg].
     \end{aligned}
     \eeq

For our present discussion, we have $dy\equiv d\beta$, with $(u^\mathcal{A}, v^\mathcal{A}, j^\mathcal{A},\omega^\mathcal{A})$ given in \eqref{eqn:UnqIIAGs} and the decompositions given in \eqref{eqn:decomps}. One can immediately see by observation that $u^\mathcal{A}_{||}=v^\mathcal{A}_{||}=0$ and $u^\mathcal{A}_\perp=u^\mathcal{A},~v^\mathcal{A}_\perp=v^\mathcal{A}$.\footnote{Hence, now we note
     \begin{equation*}
     \Psi_-^\mathcal{B}=\frac{1}{8}e^{\Phi_\mathcal{B}-\Phi_\mathcal{A}} 
     \Big(1+\frac{1}{2}z^\mathcal{A}_\perp\wedge \bar{z}^\mathcal{A}_\perp\Big)\wedge \tilde{\omega}^\mathcal{B} = \frac{1}{8}e^{\Phi_\mathcal{B}-\Phi_\mathcal{A}} 
     \Big(1- i\, u^\mathcal{A}_\perp \wedge v^\mathcal{A}_\perp \Big)\wedge \tilde{\omega}^\mathcal{B}  , 
     \end{equation*}
     with
          \begin{equation*}
    \tilde{\omega}^\mathcal{B}  = e^{C^\mathcal{A}}\omega^\mathcal{A}_{||}+   \omega^\mathcal{A}_{\perp}\wedge (dy-B_1^\mathcal{A}).
          \end{equation*}
      }

Recall the following relations (see \eqref{eqn:A1Aval} with $\phi_1=\beta,\phi_2=\phi$)
\beq
 E^\beta_\mathcal{A}=  e^{C^\mathcal{A}}(d\beta +A_1^\mathcal{A}),~~~~~~~~~~~A_1^\mathcal{A}=\frac{1}{2}\frac{h_{\beta\phi}}{h_{\beta}}d\phi = \frac{f_2f_6\sin^2\theta}{f_3+f_2\sin^2\theta}d\phi,
 \eeq
using \eqref{eqn:chi-Gen} as appropriate. Then to derive the remaining elements we re-write the forms given in \eqref{eqn:UnqIIAGs}, as follows
\beq
\begin{gathered}
j^\mathcal{A}= j_1^\mathcal{A}\wedge d\beta + j_2^\mathcal{A}\wedge d\phi = e^{-C^\mathcal{A}}j_1^\mathcal{A}\wedge E^\beta_\mathcal{A} +\Big(j_2^\mathcal{A} -\frac{1}{2}\frac{h_{\beta\phi}}{h_{\beta}} j_1^\mathcal{A} \Big)\wedge d\phi,\\
\omega^\mathcal{A} =  \omega_1^\mathcal{A}\wedge d\beta+\omega_2^\mathcal{A} = e^{-C^\mathcal{A}}\omega_1^\mathcal{A}\wedge E^\beta_\mathcal{A} +\Big(\omega_2^\mathcal{A}-\frac{1}{2}\frac{h_{\beta\phi}}{h_{\beta}} \omega_1^\mathcal{A} \wedge d\phi\Big),
\end{gathered}
\eeq
leading to
\beq
\hspace{-0.5cm}
\begin{aligned}
& j_{||}^\mathcal{A}= \frac{ e^{-C^\mathcal{A}} \kappa f_2^{\frac{1}{2}}}{2\Big(1+\frac{f_2}{4} (\sin^2\theta-4)\Big)^{\frac{1}{2}}}\bigg[f_5f_6\Big(e^{-3\rho}d(\sin^2\theta e^{3\rho}\dot{V}) -\frac{1}{2}d(\sin^2\theta \dot{V}) +\sin^2\theta d(\dot{V})\Big)- \frac{\Lambda f_3}{\dot{V}} d(V')-2\sin^2\theta d\eta \bigg],\\[2mm]
& j_\perp^\mathcal{A}=  \frac{  \kappa f_2^{\frac{1}{2}}}{2\Big(1+\frac{f_2}{4} (\sin^2\theta-4)\Big)^{\frac{1}{2}}}\Bigg[\bigg( (1-f_2) \Big(d(2\dot{V}\sin^2\theta)+e^{-6\rho}  \sin^2\theta d(2e^{6\rho} \dot{V})\Big) -2f_2 \sin^2\theta d(\dot{V})\bigg) \\[2mm]
&~~~~~- \frac{f_2f_6\sin^2\theta}{f_3+f_2\sin^2\theta} \bigg[f_5f_6\Big(e^{-3\rho}d(\sin^2\theta e^{3\rho}\dot{V}) -\frac{1}{2}d(\sin^2\theta \dot{V}) +\sin^2\theta d(\dot{V})\Big)- \frac{\Lambda f_3}{\dot{V}} d(V')-2\sin^2\theta d\eta \bigg] \Bigg]\wedge d\phi,\nn\\[2mm]
&\omega_{||}^\mathcal{A}= -2i\,e^{-C^\mathcal{A}}\kappa   \frac{\sqrt{f_2}}{\dot{V}}  e^{-3\rho} \, d(\sigma \dot{V}e^{3\rho}e^{i\phi}\sin\theta ) ,\\[2mm]
& \omega_\perp^\mathcal{A} = - 2\kappa   \frac{\sqrt{f_2}}{\dot{V}} \bigg( \sigma \dot{V} e^{-3\rho}  d(V')\wedge d(e^{3\rho}\sin\theta e^{i\phi})+   \tilde{\Delta} \Big(1-\frac{3 }{2}f_2\Big) e^{i\phi} \sin\theta d\sigma\wedge d\eta \\[2mm]
 &~~~~~~~~~~~~~~~~~~~~~~~~~~~~~~~~~~~~~~~~~~~~~~~~~~~~~~~~~~~~~~~~~~~~~ -i \frac{f_2f_6\sin^2\theta}{f_3+f_2\sin^2\theta}\, e^{-3\rho} \, d(\sigma \dot{V}e^{3\rho}e^{i\phi}\sin\theta )\wedge d\phi \bigg),
\end{aligned}
\eeq
and we note $(dy-B_1^\mathcal{A})=(d\beta -f_8\sin\theta d\theta)$. Notice that the $C^\mathcal{A}$ dependence drops out of the expressions.
     \subsubsection{One-parameter Type IIB}\label{sec:oneparamfam}
     We now present the G-Structures corresponding to the one-parameter family of $\mathcal{N}=1$ Type IIB backgrounds given in \eqref{eqn:IIBN=1-1}. We first present the 11D and IIA results (where for clarity, we have left $\gamma$ general - deriving the forms presented in \cite{us} when $\gamma=0$). In the case of the IIB results, we must fix $\gamma=-1$ as outlined throughout the paper. It is worth noting that in the case of the 11D and IIA forms themselves, the extra $\gamma$ plays only a trivial role, and one can use the forms presented in \cite{us}. This $\gamma$ is only necessary when deriving the IIB G-Structures, so must be included in each step.
     \subsubsection*{11D Forms}
   %  \textcolor{red}{Investigate whether imply sending $\phi\rightarrow \phi+\gamma \chi$ in the $\gamma=0$ results gives a correct set also....}

 \begin{align}
 &K= -\frac{\kappa e^{-2\rho}}{f_1}d(e^{2\rho}\dot{V}\cos\theta),\nn\\[2mm]
& E_1=-\sqrt{\frac{f_1f_3}{\Xi\,\Sigma_1}} \bigg[\frac{1}{\sigma}d\sigma +d\rho+i  \Big( \Sigma_1 \,d\chi +\Sigma_2\,\sin^2\theta d\phi \Big)  +\xi \,d(V')-\frac{\gamma\xi}{2\sigma^2}\dot{V}\sin^2\theta\, d\eta\nn\\[2mm]
 &~~~~~ +\frac{e^{-2\rho}}{2\dot{V}}\bigg( \Sigma_2\, d(e^{2\rho}\dot{V}\sin^2\theta) +\bigg(\Sigma_2 +\xi \frac{f_2}{f_3} \Big(\gamma f_6 -4\xi (1+\gamma)\frac{f_2}{f_5}\Big)\bigg)\sin^2\theta d(e^{2\rho}\dot{V}) \nn\\[2mm]
 &~~~~~+ \frac{4\xi^2(1+\gamma)f_2}{f_3f_5} \sin^2\theta \, \dot{V}d(e^{2\rho})\bigg)\bigg],\nn\\[2mm]
& E_2=\frac{e^{i(\phi+\gamma\chi)}}{\sqrt{\Sigma_1}} \Bigg[\frac{\kappa}{f_1} \bigg[e^{-2\rho} d(e^{2\rho}\dot{V}\sin\theta) - \dot{V}\sin\theta \bigg(\xi (\gamma+1) d(V') + \gamma \Big(\frac{1}{\sigma}d\sigma+d\rho\Big)\bigg)\bigg]+i \sqrt{f_1f_2}\sin\theta d\phi \Bigg] ,\nn\\[2mm]
& E_3= -e^{i\chi} \,\Xi^{\frac{1}{2}}\sqrt{f_1f_5}\bigg[ \frac{1}{\Xi}\left(-\frac{f_3 \dot{V}'}{4 \sigma}d\sigma -V''d\eta+f_6d\rho + \frac{\xi}{f_5}\Big(\frac{\kappa^2e^{-4\rho}}{2f_1^3}d(e^{4\rho}\dot{V}^2(\cos^2\theta-3))+4 d\rho\Big)\right)\nn\\[2mm]
 &~~~~~~~~~+i \Big(d\beta + C_1\Big)\bigg], \nn\\[8mm]
 &\Sigma_1=1+\frac{f_2}{f_3}\bigg(\big(\gamma+\xi(1+\gamma)f_6\big)^2+\xi^2(1+\gamma)^2\frac{f_3}{f_5}\bigg)\sin^2\theta\equiv \frac{h_\chi}{f_3f_5},\nn\\[2mm]
  &\Sigma_2=\frac{f_2}{f_3}\bigg((1+\xi f_6)\Big(\gamma+\xi (1+\gamma)f_6\Big)+\xi^2(1+\gamma)\frac{f_3}{f_5}\bigg)\equiv \frac{h_{\chi\phi}}{2f_3f_5 \sin^2\theta },\nn\\[2mm]
   & \Xi=(1+\xi f_6)^2 +\xi^2 \frac{f_3}{f_5}+\xi^2 \frac{f_2}{f_5}\sin^2\theta,~~~~~~~~~~~~~C_1= \frac{1}{\Xi}\bigg[\bigg(f_6+ \xi \Big(f_6^2+ \frac{f_3}{f_5}-\gamma  \frac{f_2}{f_5}\sin^2\theta\Big)\bigg)d\chi -\xi \frac{f_2}{f_5}\sin^2\theta d\phi\bigg].
 \end{align}	
     Notice that in the case of $E_3$, $\gamma$ only appears in the $d\chi$ term of $C_1$.
          \subsubsection*{IIA Forms}
The easiest way to derive the IIA forms here is to simply replace $\phi\rightarrow\phi+\gamma\, \chi$ in the forms given in \cite{us}, as follows
          \begin{align}\label{eqn:uandv}
&v=  \kappa e^{-2\rho}f_5^{\frac{1}{4}}f_1^{-\frac{3}{4}} \Xi^{\frac{1}{4}}d\left(e^{2\rho} \dot{V}\cos\theta\right),\nn\\[2mm]
&u=  (f_1f_5)^{\frac{3}{4}}\Xi^{-\frac{1}{4}}\left(\frac{f_3 \dot{V}'}{4 \sigma}d\sigma +V''d\eta-f_6d\rho - \frac{\xi}{f_5}\Big(\frac{\kappa^2e^{-4\rho}}{2f_1^3}d(e^{4\rho}\dot{V}^2(\cos^2\theta-3))+4 d\rho\Big)\right),\nn\\[2mm]
&\omega=  -2\kappa^2  f_1^{-\frac{3}{2}}e^{-3\rho} d\Big(e^{2\rho}\dot{V} e^{-\xi V'}e^{i(\phi+\gamma\chi)}\sin\theta\, d(e^{i\chi}e^{\rho}e^{\xi V'}\sigma)\Big) ,\nn\\[2mm]
%&=  -2\kappa^2  f_1^{-\frac{3}{2}}e^{-3\rho}\bigg[d\Big(e^{2\rho}\dot{V} e^{-\xi V'}e^{i\phi}\sin\theta\, d(e^{i\chi}e^{\rho}e^{\xi V'}\sigma)\Big) -\gamma \,e^{2\rho}\dot{V} e^{-\xi V'}e^{i\phi}\sin\theta \,d\Big(\sigma e^\rho e^{\xi V'}d(e^{i\chi})\Big)\bigg],\nn\\[2mm]
&j=\frac{1}{\sqrt{\Xi}}\Bigg[ \frac{f_1^{\frac{3}{2}}f_5^{\frac{1}{2}} f_3}{\sigma} e^{-\rho}e^{-\xi V'}d\left(e^{\rho}\sigma e^{\xi V'}\right)\wedge d\chi+ \kappa f_2 f_3^{\frac{1}{2}}X_1 \wedge  \Big((\gamma+1) d\chi+ d\phi\Big) +X_2\wedge (d\phi+\gamma\, d\chi)\Bigg]\nn,\\\\[2mm]
%&=\frac{1}{\sqrt{\Xi}}\Bigg[ \frac{f_1^{\frac{3}{2}}f_5^{\frac{1}{2}} f_3}{\sigma} e^{-\rho}e^{-\xi V'}d\left(e^{\rho}\sigma e^{\xi V'}\right)\wedge d\chi+ \kappa f_2 f_3^{\frac{1}{2}}  (X_1 \wedge d\phi+ X_2\wedge  d\chi) \nn\\[2mm]
%&+\kappa\Bigg(\frac{\sigma f_2 e^{-4\rho}}{f_3^{\frac{1}{2}} \dot{V}^2}(1+\xi f_6)d(e^{4\rho}\sin^2\theta \dot{V}^2)-\frac{\xi}{2} f_2f_3^{\frac{1}{2}}\sin^2\theta\left(\dot{V}'\left(d\sigma-\frac{2 \dot{V}}{\sigma V''}d\rho\right)+\frac{d\eta}{\sigma}\left(2 \dot{V}- \ddot{V}\right)\right)\Bigg)\wedge d\phi\Bigg]\nn,\\\\[2mm]
&X_1=\xi \frac{\dot{V}' e^{-4\rho}}{2 \dot{V} V''\sigma}d(e^{4\rho}\sin^2\theta\dot{V}^2)+\xi^2\frac{(-V'' \ddot{V}+(\dot{V}')^2 )e^{-6\rho}}{2\sigma \dot{V} V''}d(e^{6\rho}\sin^2\theta\dot{V}^2)+\xi^2 \dot{V}\sin^2\theta\Big(V'' d\sigma-\frac{\dot{V}'}{\sigma}d\eta\Big)\nn,\\[2mm]
&X_2=\kappa\Bigg(\frac{\sigma f_2 e^{-4\rho}}{f_3^{\frac{1}{2}} \dot{V}^2}(1+\xi f_6)d(e^{4\rho}\sin^2\theta \dot{V}^2)-\frac{\xi}{2} f_2f_3^{\frac{1}{2}}\sin^2\theta\left(\dot{V}'\left(d\sigma-\frac{2 \dot{V}}{\sigma V''}d\rho\right)+\frac{d\eta}{\sigma}\left(2 \dot{V}- \ddot{V}\right)\right)\Bigg)\nn.\\[2mm]
%X_2&= \frac{ e^{-4\rho}\big(2\gamma +\xi (5\gamma+2)\dot{V}'\big)}{4 \dot{V} V''\sigma}d(e^{4\rho}\sin^2\theta\dot{V}^2)+\xi^2(\gamma+1)\frac{(-V'' \ddot{V}+(\dot{V}')^2 )e^{-6\rho}}{2\sigma \dot{V} V''}d(e^{6\rho}\sin^2\theta\dot{V}^2)\nn \\[2mm]
%&+\xi^2(\gamma+1) \dot{V}\sin^2\theta\Big(V'' d\sigma-\frac{\dot{V}'}{\sigma}d\eta\Big)  +\gamma \bigg(\frac{\sigma\, e^{-4\rho}}{4\dot{V}^2}d(e^{4\rho}\sin^2\theta\dot{V}^2) -\frac{\xi}{2} \frac{1}{\sigma V''}\Big(\tilde{\Delta} \sin^2\theta d\eta +\frac{1}{2}\dot{V}\dot{V}' d(\sin^2\theta)\Big)\bigg).\nn
\end{align}
          Noting that $v$ and $u$ are independent of $\phi$, so independent of $\gamma$. To derive the IIB G-Structures, we must now fix $\gamma=-1$ in $\omega$ and $j$, giving
                    \begin{align}\label{eqn:jandomega}
\omega&=  -2\kappa^2  f_1^{-\frac{3}{2}}e^{-3\rho} d\Big(e^{2\rho}\dot{V} e^{-\xi V'}e^{i(\phi-\chi)}\sin\theta\, d(e^{i\chi}e^{\rho}e^{\xi V'}\sigma)\Big)\nn \\[2mm]
&=  -2\kappa^2  f_1^{-\frac{3}{2}}e^{-3\rho} \bigg[d\Big(e^{2\rho}\dot{V}e^{-\xi V'} e^{i\phi} \sin\theta d(e^\rho e^{\xi V'}\sigma)\Big)+i \,d(e^{3\rho}\dot{V}e^{i\phi}\sigma \sin\theta )\wedge d\chi \bigg],\nn\\[2mm]
j&=\frac{1}{\sqrt{\Xi}}\Bigg[\bigg( \frac{f_1^{\frac{3}{2}}f_5^{\frac{1}{2}} f_3}{\sigma} e^{-\rho}e^{-\xi V'}d\left(e^{\rho}\sigma e^{\xi V'}\right)-X_2\bigg)\wedge d\chi+ \Big(\kappa f_2 f_3^{\frac{1}{2}}X_1 +X_2 \Big)\wedge d\phi \Bigg] .
\end{align}
               \subsubsection*{IIB Forms}
We now perform an ATD along $\chi$, meaning $dy\equiv d\chi$, with $u^\mathcal{A},v^\mathcal{A}$ given in \eqref{eqn:uandv} and $\omega^\mathcal{A},j^\mathcal{A}$ given in \eqref{eqn:jandomega}. Once again, we have $u^\mathcal{A}_{||}=v^\mathcal{A}_{||}=0$ and $u^\mathcal{A}_\perp=u^\mathcal{A},~v^\mathcal{A}_\perp=v^\mathcal{A}$. Recall again the following relations (see \eqref{eqn:A1Aval} with $\phi_1=\chi,\phi_2=\phi$)
\beq
 E^\chi_\mathcal{A}=  e^{C^\mathcal{A}}(d\chi +A_1^\mathcal{A}),~~~~~~~~~~~A_1^\mathcal{A}=\frac{1}{2}\frac{h_{\chi\phi}}{h_{\chi}}d\phi =-\frac{f_2(1+\xi f_6)}{f_3+f_2\sin^2\theta}\sin^2\theta d\phi,
 \eeq
now using \eqref{eqn:beta-Gen} to calculate $A_1^\mathcal{A}$ \footnote{With $(p,b,u)=1,(a,c,m)=0,q=-v=\xi,s=\gamma=-1$.}. Then to derive the remaining elements we re-write the forms given in \eqref{eqn:UnqIIAGs}, as follows
\beq
\begin{gathered}
j^\mathcal{A}= j_1^\mathcal{A}\wedge d\chi + j_2^\mathcal{A}\wedge d\phi = e^{-C^\mathcal{A}}j_1^\mathcal{A}\wedge E^\chi_\mathcal{A} +\Big(j_2^\mathcal{A} -\frac{1}{2}\frac{h_{\chi\phi}}{h_{\chi}} j_1^\mathcal{A} \Big)\wedge d\phi,\\
\omega^\mathcal{A} =  \omega_1^\mathcal{A}\wedge d\chi+\omega_2^\mathcal{A} = e^{-C^\mathcal{A}}\omega_1^\mathcal{A}\wedge E^\chi_\mathcal{A} +\Big(\omega_2^\mathcal{A}-\frac{1}{2}\frac{h_{\chi\phi}}{h_{\chi}} \omega_1^\mathcal{A} \wedge d\phi\Big),
\end{gathered}
\eeq
leading to
  \beq\label{eqn:IIBfamGs}
\begin{aligned}
j_{||}^\mathcal{A}&=   \frac{ e^{-C^\mathcal{A}}}{\sqrt{\Xi}} \bigg( \frac{f_1^{\frac{3}{2}}f_5^{\frac{1}{2}} f_3}{\sigma} e^{-\rho}e^{-\xi V'}d\left(e^{\rho}\sigma e^{\xi V'}\right)-X_2\bigg) ,\\[2mm]
j_\perp^\mathcal{A} &=\frac{1}{\sqrt{\Xi}}\Bigg[\Big(\kappa f_2 f_3^{\frac{1}{2}}X_1 +X_2 \Big) +\frac{f_2(1+\xi f_6)}{f_3+f_2\sin^2\theta} \bigg( \frac{f_1^{\frac{3}{2}}f_5^{\frac{1}{2}} f_3}{\sigma} e^{-\rho}e^{-\xi V'}d\left(e^{\rho}\sigma e^{\xi V'}\right)-X_2\bigg)\sin^2\theta    \Bigg] \wedge d\phi \\[2mm]
\omega_{||}^\mathcal{A}&=  -2i \,\kappa^2   e^{-C^\mathcal{A}}  f_1^{-\frac{3}{2}}e^{-3\rho}   d(e^{3\rho}\dot{V}e^{i\phi}\sigma \sin\theta )    ,\\[2mm]
 \omega_\perp^\mathcal{A} &= -2\kappa^2  f_1^{-\frac{3}{2}}e^{-3\rho} \bigg[d\Big(e^{2\rho}\dot{V}e^{-\xi V'} e^{i\phi} \sin\theta d(e^\rho e^{\xi V'}\sigma)\Big)+ i \frac{f_2(1+\xi f_6)\sin^2\theta}{f_3+f_2\sin^2\theta} \,d(e^{3\rho}\dot{V}e^{i\phi}\sigma \sin\theta ) \wedge d\phi \bigg].
\end{aligned}
\eeq

\section{Dimensional Reductions}\label{sec:DimRed}

In this Appendix, we present in some detail the various dimensional reductions which were performed following the $SL(3,\mathds{R})$ transformation of the three $U(1)$ directions of Gaiotto-Maldacena $(\beta,\chi,\phi)$, given in \eqref{eqn:S2breakingdefns}. The $U(1)_r$ component of the $SU(2)_R\times U(1)_r$ R-Symmetry is now transformed as follows
\beq\label{eqn:GMU1}
U(1)_r=\chi+\phi \rightarrow (p+s)\chi +(q+v)\beta +(m+u)\phi,
\eeq
and plays a central role in determining whether supersymmetry is preserved under each reduction. In addition, the $C_3$ given in \eqref{eqn:GM} now transforms to
 \beq
 \begin{aligned}
 C_3 &=\sin\theta \,\bigg[(up-sm)f_7+(ua-sc)f_8 \bigg]d\chi  \wedge d\theta \wedge d\phi\\
 &~~~~~~~~~~~~+\sin\theta \,\bigg[\Big((v p -sq )f_7+(v a -sb)f_8 \Big)d\chi  +\Big((v m - uq)f_7 +(v c-ub)f_8 \Big)d\phi \bigg] \wedge d\theta. \wedge d\beta.
 \end{aligned}
 \eeq
 
The following subsections will correspond to a dimensional reduction along each of the three $U(1)$ directions in turn, beginning with an initial form which includes all nine transformation parameters, namely \eqref{eqn:beta-Gen},\eqref{eqn:chi-Gen},\eqref{eqn:phi-Gen}. These results are of course too general because one can eliminate many of the parameters without loss of generality, leaving the three free parameters in each case (corresponding to the three $U(1)$ directions being mixed). However, these general forms prove very useful, allowing one to simply plug in the desired values of the parameters for each unique example, with all of the calculation already in place. \\
 The remaining three free parameters will be labelled $(\xi,\zeta,\gamma)$, in which $\zeta$ will keep track of the SUSY under dimensional reduction to Type IIA, and $\gamma$ will keep track of the SUSY under an Abelian T-duality to Type IIB. %We will refer to these as `SUSY parameters'.As we will soon see, 
 They provide us with the option to preserve $\mathcal{N}=1$ SUSY under reduction to Type IIA and under a subsequent abelian T-Duality to Type IIB. The remaining parameter, $\xi$, will be left over as a free parameter in the resulting backgrounds.

We first present the $\beta$ reduction case, repeating the derivation of the results given in \cite{us}. We will then turn to the $\chi$ and $\phi$ reductions in turn, where the analysis will be largely analogous to the $\beta$ reduction. %Throughout this discussion, we will set up the necessary conditions which lead to $\mathcal{N}=1$ preserving abelian T-Duality to Type IIB, foreshadowing the arguments presented in Appendix ..... 
In each case, we use the reduction formula given in \eqref{eqn:DimRed}, and we will fix $(p,b,u)=1$ to trivially absorb them into the definitions of $(\chi,\beta,\phi)$, respectively (see \eqref{eqn:S2breakingdefns}). The $U(1)_r$ R-Symmetry component given in \eqref{eqn:GMU1} then reduces to
\beq\label{eqn:newU(1)}
U(1)_r = (1+s)\chi +(q+v)\beta +(m+1)\phi.
\eeq
Each reduction must preserve this component in order to preserve supersymmetry, giving $\mathcal{N}=1$ Type IIA solutions (with a $U(1)$ R-Symmetry) when the $SU(2)_R$ component is broken.

\subsection*{$\beta$ reduction}

The general form following a dimensional reduction along $\beta$ (before eliminating the nine transformation parameters to three free parameters) reads 
  \begin{align}\label{eqn:beta-Gen}
 &       ds_{10,st}^2=e^{\frac{2}{3}\Phi}f_1\bigg[4ds^2(\text{AdS}_5)+f_2d\theta^2+f_4(d\sigma^2+d\eta^2)\bigg]+f_1^2 e^{-\frac{2}{3}\Phi} ds^2_2,\nn\\[2mm]
  &e^{\frac{4}{3}\Phi}=  f_1 \Big[q^2(f_3+f_5f_6^2)+b^2 f_5 +2bqf_5f_6 + v^2 \sin^2\theta f_2\Big] ,~~~~~~~~~~~~~~~~~~  B_2 =\Big(B_{2,\chi}d\chi  +B_{2,\phi}d\phi \Big) \wedge d\theta,\nn\\[2mm]
&C_1= \Big(C_{1,\chi} d\chi  + C_{1,\phi} d\phi\Big)  ,~~~~~~~~~~~~~~~~~
       C_3=C_{3,\chi\phi} d\chi \wedge d\theta \wedge d\phi,\nn\\[2mm]
    &   B_{2,\chi} =   \sin\theta \Big((vp -sq )f_7+(va -sb)f_8 \Big),~~~~~~~~B_{2,\phi} = \sin\theta \Big((vm - uq)f_7 +(vc-ub)f_8 \Big),\nn\\[2mm]
     &  C_{1,\chi} =   f_1 e^{-\frac{4}{3}\Phi}\Big(bf_5 (a+pf_6) +p\,q f_3 + qf_5f_6(a+pf_6) +s\,v \sin^2\theta f_2\Big), \nn\\[2mm]
    &   C_{1,\phi} =  f_1 e^{-\frac{4}{3}\Phi} \Big(bf_5 (c+mf_6) +m\,q f_3 + qf_5f_6(c+mf_6) +u\,v \sin^2\theta f_2\Big),\nn\\[2mm]
    &   C_{3,\chi\phi}=\sin\theta \,\bigg[u (p f_7+a f_8) - s(mf_7 +cf_8)  \bigg], \nn\\[8mm]
   &    ds^2_2 = h_\chi(\eta,\sigma,\theta)d\chi^2 + h_\phi(\eta,\sigma,\theta)d\phi^2 + h_{\chi\phi}(\eta,\sigma,\theta)d\chi d\phi\nn\\[2mm]
   &    h_\chi(\eta,\sigma,\theta) = f_3f_5 (bp-aq)^2 +\sin^2\theta f_2\Big[b^2s^2 f_5 +(pv-qs)^2f_3 - 2bsf_5\big((pv-qs)f_6 +av\big) \nn\\[2mm]
   &   ~~~~~~~~~~~~~~~~~~~~~~~~~~~~~~ +f_5\big((pv-qs)f_6 +av\big)^2\Big],\nn\\[2mm]
    &          h_\phi(\eta,\sigma,\theta) = f_3f_5 (bm-cq)^2 +\sin^2\theta f_2\Big[c^2v^2 f_5 +(qu-mv)^2f_3 - 2cv f_5\big((qu-mv)f_6 +bu\big)\nn\\[2mm]
    &          ~~~~~~~~~~~~~~~~~~~~~~~~~~~~~~~~~ +f_5\big((qu-mv)f_6 +bu\big)^2\Big],\nn\\[2mm]
    &                        h_{\chi\phi}(\eta,\sigma,\theta) =2\bigg[ f_3f_5 (bp-aq)(bm-cq)  +\sin^2\theta f_2 h_4(\eta,\sigma)\bigg],\nn\\[2mm]
   &                       h_4(\eta,\sigma) =  cv f_5\Big((pv-qs)f_6+av\Big) +b^2suf_5 -(qu-mv) \Big[(pv-qs)f_3+f_5f_6 \Big((pv-qs)f_6+av\Big)\Big]\nn\\[2mm]
   &~~~~~~~~~~~~~~~~~~ - bf_5\bigg(sv(c+mf_6)+u\Big(av+(pv-2qs)f_6\Big)\bigg).
    \end{align}

In this case, one is motivated to keep $(q,v)$ free, allowing one to preserve the $U(1)_r$ component given in \eqref{eqn:GMU1}. This is the first case which we will now consider.
\begin{itemize}
\item \textbf{Keeping $(q,v)$ free}\\
From the determinant in \eqref{eqn:S2breakingdefns}, this requires $(a,c)=0$. The determinant then becomes 
\beq\label{eqn:pums}
pu-ms=1,
\eeq
which in turn becomes $ms=0$ after fixing $(p,b,u)=1$. This gives the third free parameter (which we label as $\gamma$). One then derives the two-parameter family of solutions given in \eqref{eqn:generalresult1} (with $q\equiv \xi,v\equiv \zeta$). We have eliminated this third free parameter from the Type IIA backgrounds by setting $\gamma=0$ without loss of generality. More specifically, when $(s\equiv\gamma,m=0)$ one derives \eqref{eqn:generalresult1} with $\phi=\phi+\gamma \chi$. Alternatively, when $(s=0,m\equiv \gamma)$, we derive \eqref{eqn:generalresult1} but with $\chi\rightarrow \chi+\gamma\phi$. So $\gamma$ is carried through the calculation trivially, and can be fixed to zero without loss of generality\footnote{Alternatively, we can satisfy \eqref{eqn:pums} with $pu=0$ and $ms=-1$. One still derives the solution given in \eqref{eqn:generalresult1} but now we have trivially re-defined the $U(1)$ directions amongst themselves. This is clear from the $SL(3,\mathds{R})$ transformation. Specifically, when $(p=0,u\equiv\gamma,s=-1,m=1)$ we require $(\chi\rightarrow\phi,~\phi\rightarrow -\chi+\gamma \phi)$ to map to \eqref{eqn:generalresult1}, and when $(p\equiv\gamma, u=0,s=-1,m=1)$ we require the mapping $(\chi\rightarrow\phi +\gamma \chi,~\phi\rightarrow \chi)$.}. We can then preserve the $U(1)_r$ component under a $\beta$ reduction by fixing $\zeta=-\xi$. When $\zeta\neq0$ however, the $SU(2)_R$ R-Symmetry component is broken. So the 2-parameter family of solutions contains within it the $\mathcal{N}=2$ solution of \cite{Nunez:2019gbg} (with $\zeta=-\xi=0$), an $\mathcal{N}=1$ solution (with $\zeta=-\xi\neq 0$), and $\mathcal{N}=0$ solutions otherwise. This solution is presented in \cite{us}.

\item{\textbf{Alternative parameters}}\\
We now relax the condition that both $q$ and $v$ are free, and demonstrate that all roads lead back to the two-parameter family given in \eqref{eqn:generalresult1} (presented in \cite{us}).
\begin{itemize}

\item \textbf{Keeping $(q,c)$ free}\\
With $(p,b,u)=1$, to keep $q$ as a free parameter, one can alternatively fix $(a,s)=0$. The determinant now becomes $vc=0$. To avoid the previous case, we fix $v=0$ with $(q\equiv\xi,m\equiv\gamma)$. The solution derived can be mapped to the $\zeta=0$ solution of \eqref{eqn:generalresult1}, via
\beq
C_1\rightarrow C_1+ c\, d\phi,~~~~\chi\rightarrow \chi+(\gamma-c\,\xi)\phi.
\eeq

\item \textbf{Keeping $(v,s)$ free}\\
The other possibility to keep $v$ a free parameter is to fix $(c,m)=0$, leaving the determinant $qa=0$. We of course should fix $q=0$ with $(v\equiv \zeta,~s\equiv \gamma)$. This now derives the $\xi=0$ solution of \eqref{eqn:generalresult1}, following
\beq
C_1\rightarrow C_1+ a\, d\chi,~~~~~~C_3\rightarrow C_3 + a\,d\chi \wedge B_2,~~~~~~\phi\rightarrow \phi+(\gamma-a\,\zeta)\phi.
\eeq

\item \textbf{Keeping $(a,c)$ free}\\
The final possibility is to enforce that neither $q$ or $v$ are free parameters by fixing both to zero. This then makes $(a,c)$ free, with the determinant becoming $ms=0$. This solution re-derives the $\mathcal{N}=2$ background (\eqref{eqn:generalresult1} with $(\xi,\zeta)=0$), with the gauge transformations
\beq
C_1\rightarrow C_1+(a d\chi+c d\phi),~~~~~~~~~~~~C_3\rightarrow C_3+(a d\chi+c d\phi)\wedge B_2,
\eeq
and $\phi\rightarrow \phi+s\chi$ or $\chi\rightarrow\chi+m\phi$ (depending on which choice of parameter is made).
\end{itemize}
\end{itemize}

This completes the discussion on the $\beta$ reduction, we now turn to the $\chi$ reduction.

\subsection*{$\chi$ reduction}\label{sec:chireduction}

Now we consider a dimensional reduction along $\chi$. The general background in this case is given by
  \begin{align}\label{eqn:chi-Gen}
  &      ds_{10,st}^2=e^{\frac{2}{3}\Phi}f_1\bigg[4ds^2(\text{AdS}_5)+f_2d\theta^2+f_4(d\sigma^2+d\eta^2)\bigg]+f_1^2 e^{-\frac{2}{3}\Phi} ds^2_2,\nn\\[2mm]
    &    e^{\frac{4}{3}\Phi}= f_1 \Big[p^2(f_3+f_5f_6^2)+a^2 f_5 +2apf_5f_6 + s^2 \sin^2\theta f_2\Big] ,~~~~~~~~~~~
       B_2=(B_{2,\beta}d\beta +B_{2,\phi}d\phi)\wedge d\theta, \nn\\[2mm]
    &   C_1=C_{1,\beta}d\beta+ C_{1,\phi}d\phi,~~~~~~~~~~C_3 = C_{3,\beta\phi} d\beta \wedge d\theta \wedge d\phi, \nn\\[2mm]
   &    B_{2,\beta} = \sin\theta\Big((sq-vp)f_7+(s b-va)f_8\Big),~~~~~~~B_{2,\phi}= \sin\theta\Big( (s m-up )f_7 +(s c-ua) f_8\Big), \nn\\[2mm]
      % B_2 = \bigg[\sin\theta\Big((sq-tp)f_7+(s b-ta)f_8\Big)d\beta +\sin\theta\Big( (s m-up )f_7 +(s c-ua) f_8\Big)d\phi\bigg] \wedge \, d\theta,\\
   &    C_{1,\beta}=  f_1 e^{-\frac{4}{3}\Phi}\Big(bf_5 (a+pf_6) +p\,q f_3 + qf_5f_6(a+pf_6) +s\,v \sin^2\theta f_2\Big),  \nn\\[2mm]
   &    C_{1,\phi} =    f_1 e^{-\frac{4}{3}\Phi}\Big(af_5 (c+mf_6) +m\,p f_3 + pf_5f_6(c+mf_6) +u\,s \sin^2\theta f_2\Big)   ,\nn\\[2mm]
     &   C_{3,\beta\phi} =\sin\theta\Big[u (qf_7+bf_8) - v(mf_7 +cf_8)  \Big],  \nn\\[8mm]
   %    C_3=\sin\theta \bigg[ (qf_7+bf_8)d\beta +(mf_7 +cf_8)d\phi \bigg] \wedge \, d\theta \wedge (t\,d\beta +u\,d\phi),\\\\
   &    ds^2_2 = h_\beta(\eta,\sigma,\theta)d\beta^2 + h_\phi(\eta,\sigma,\theta)d\phi^2 + h_{\beta\phi}(\eta,\sigma,\theta)d\beta d\phi\nn\\[2mm]
   &    h_\beta(\eta,\sigma,\theta) = f_3f_5 (bp-aq)^2 +\sin^2\theta f_2\Big[b^2s^2 f_5 +(pv-qs)^2f_3 - 2bsf_5\big((pv-qs)f_6 +av\big) \nn\\[2mm]
   &    ~~~~~~~~~~~~~~~~~~~~~~~~~~~~~~+f_5\big((pv-qs)f_6 +av\big)^2\Big],\nn\\[2mm]
  &            h_\phi(\eta,\sigma,\theta) = f_3f_5 (pc - am)^2 +\sin^2\theta f_2\Big[c^2s^2 f_5 +(pu-ms)^2f_3 - 2cs f_5\big((pu-ms)f_6 +au\big) \nn\\[2mm]
    &          ~~~~~~~~~~~~~~~~~~~~~~~~~~~~~~~+f_5\big((pu-ms)f_6 +au\big)^2\Big],\nn\\[2mm]
     &                       h_{\beta\phi}(\eta,\sigma,\theta) =2\bigg[ f_3f_5 (bp-aq)(pc-am)  +\sin^2\theta f_2 h_4(\eta,\sigma)\bigg],\nn\\[2mm]
    &                      h_4(\eta,\sigma) =-  sb f_5\Big((pu-ms)f_6-cs\Big) +a^2vuf_5 +(pv-qs) \Big[(pu-ms)f_3+f_5f_6 \Big((pu-ms)f_6-cs\Big)\Big]\nn\\[2mm]
    &~~~~~~~~~~~~~~~~~~ - af_5\bigg(sv(c+mf_6)+u\Big(bs+(qs-2pv)f_6\Big)\bigg).
    \end{align}
    
After fixing $(p,b,u)=1$, the $U(1)_r$ component reduces to \eqref{eqn:newU(1)}. Now that we are reducing along $\chi$, we must fix $s=-1$ in order to preserve this component under reduction. Notice that this SUSY preserving condition takes a slightly different form to the $\beta$ reduction case, where the condition was $v=-q$. The same will be true for the $\phi$ reduction (where the condition will be $m=-1$). This is a consequence of the $U(1)_r$ component being $\chi+\phi$ prior to the $SL(3,\mathds{R})$ transformation, and why the only $\mathcal{N}=2$ preserving reduction is along $\beta$. That is, in fixing $s=-1$ (or $m=-1$ in the $\phi$ reduction case), the $S^2$ (and hence the $SU(2)_R$ R-Symmetry component) is broken due to the mixing of $\phi$. See \eqref{eqn:S2breakingdefns}. Hence, the maximum supersymmetry which can be achieved under a $\chi$ or $\phi$ reduction is $\mathcal{N}=1$, because the very condition required to preserve the $U(1)_r$ component is what breaks the $SU(2)_R$ component.\\
We will now choose $s$ to be one of our three free parameters, allowing us to turn on/off the $\mathcal{N}=1$ at will.

\begin{itemize}
\item \textbf{Keeping $s$ free}\\
With $s$ a free parameter, from the determinant given in \eqref{eqn:S2breakingdefns}, we must fix $m=0$, with either $q=0$ (where $a$ is free) or $c=0$ (where $v$ is free). We will now investigate both options in turn.

\begin{itemize}
\item \textbf{$(m,q)=0$ with $(s,a)$ free parameters}\\
%\textcolor{red}{This case is equivalent to the $\beta$ reduction case in...... from the coordinate transformations.....} 
In this case, the determinant in \eqref{eqn:S2breakingdefns} leads to the condition $vc=0$. Hence, the third free parameter will be $v$ (with $c=0$) or $c$ (with $v=0$)\footnote{In keeping $c$ free, one gets \eqref{eqn:chireduction1} with $\beta=\beta+c\,\phi$. Alternatively, when keeping $v$ free, one gets \eqref{eqn:chireduction1} with $\phi=\phi+v\,\beta$. Hence, one can set both parameters to zero without loss of generality. To investigate whether this parameter becomes important when considering SUSY preserving abelian T-Duality, we must revisit \eqref{eqn:newU(1)}, hence we momentarily keep $v$ free.}. 
This leads to the following 11D transformation (re-labelling $s\equiv \zeta, a\equiv \xi,v\equiv \gamma$ for notational consistency)
\beq\label{eqn:chi1}
\begin{gathered}
d\beta = d\beta + \xi \, d\chi,~~~~~~~~~~~~~~~~~d\chi = d\chi  ,~~~~~~~~~~~~~~~~~~d\phi = d\phi+ \zeta \, d\chi + \gamma d\beta ,\\
U(1)_r = (1+\zeta)\chi +\gamma \beta +\phi,
\end{gathered}
\eeq

which, following a $\chi$ reduction, gives the following two-parameter background 
     \begin{align}\label{eqn:chireduction1}
  &      ds_{10,st}^2=e^{\frac{2}{3}\Phi}f_1\bigg[4ds^2(\text{AdS}_5)+f_2d\theta^2+f_4(d\sigma^2+d\eta^2)\bigg]+ f_1^2 e^{-\frac{2}{3}\Phi}  ds^2_2,\nn\\[2mm]
 &          ds^2_2 = \big(f_3f_5 +\zeta^2  \sin^2\theta f_2 f_5\big) d\beta^2 +\sin^2\theta f_2\big(f_3+f_5(f_6 +\xi )^2\big)d\phi^2 -2\zeta  f_2  f_5(f_6 +\xi) \sin^2\theta  d\beta d\phi,\nn\\[2mm]
 &       e^{\frac{4}{3}\Phi}= f_1 \Big[f_3+f_5(f_6+\xi)^2 + \zeta^2 \sin^2\theta f_2\Big] ,\nn\\[2mm]
    &   B_2 = \sin\theta \bigg[  \zeta f_8 d\beta -\Big( f_7+ \xi f_8\Big)d\phi\bigg] \wedge \, d\theta,~~~~~~~~
      C_1=  f_1 e^{-\frac{4}{3}\Phi} \bigg[f_5 (f_6+\xi) d\beta  +\zeta \sin^2\theta f_2 d\phi\bigg]  ,\nn\\[2mm]
    &   C_3=  f_8  \sin\theta \,d\beta  \wedge \, d\theta \wedge d\phi,
    \end{align}
    
    where $\phi=\phi+\gamma\beta$, allowing one to set $\gamma$ to zero without loss of generality. In addition, from the $U(1)_r$ component given in \eqref{eqn:chi1}, fixing $\gamma=0$ will allow for a SUSY preserving T-Duality along $\beta$ - we will return to this discussion later.

The Type IIA solutions here are summarised in Table \ref{table:2}. 

        \begin{table}[h!]
     \begin{center}
\begin{tabular}{c | c c c c c }
$\zeta$ &$\mathcal{N}$&$U(1)_r$&$SU(2)_R$  \\
\hline
$-1$&$1$& \checkmark&$\times$  \\
$0$&$0$&$\times$ &$\checkmark$ \\
$   \mathds{Z}/\{0,-1\}$&$0$&$\times$ &$\times$  
\end{tabular}
\end{center}
\caption{$\chi$ Reduction}
\label{table:2}
\end{table}

To map to \eqref{eqn:generalresult1}, one requires the following transformations
    \begin{equation}\label{eqn:Transformation}
    \begin{gathered}
    g_{MN} \rightarrow k_1 g_{MN},~~~~~~~~~~~~~H_3 \rightarrow k_1 H_3,~~~~~~~~~F_n \rightarrow k_1^{\frac{n}{2}} F_n,~~~~~~~~~e^{-\Phi} \rightarrow k_1^{\frac{1}{2}} e^{-\Phi},\\\\
    F_n \rightarrow k_2 F_n,~~~~~~~~~e^{-\Phi} \rightarrow k_2 e^{-\Phi},
    \end{gathered}
    \end{equation}
    with $k_1=\xi^{-1},~k_2=\xi^{2}$, giving
    \beq\label{eqn:mapping1}
        g_{MN} \rightarrow \frac{1}{\xi} g_{MN},~~~~~~~~~~~~~B_2 \rightarrow \frac{1}{\xi} B_2,~~~~~~~~~C_1\rightarrow \xi\, C_1,~~~~~~~~~~~C_3\rightarrow C_3,~~~~~~~~~e^{\frac{4}{3}\Phi} \rightarrow \frac{1}{\xi^2} e^{\frac{4}{3}\Phi},
    \eeq
followed by (in order)
\beq\label{eqn:mapping1-1}
\begin{gathered}
C_1\rightarrow C_1 -\,d\beta,~~~~~~~~~~~~~~~C_3\rightarrow C_3- \, d\beta \wedge B_2,\\
\zeta \rightarrow \xi\, \zeta,~~~~~~~~~~~~~~\phi\rightarrow \phi-\zeta\,\xi\,\chi,~~~~~~~~~~~~~~\beta \rightarrow  - \xi \,\chi,\\
\xi\rightarrow\frac{1}{\xi},
\end{gathered}
\eeq
Notice that we require $\xi \rightarrow 1/\xi$, despite $\xi\in \mathds{Z}$. This is somewhat non-trivial, fixing $\xi=0$ gives a unique solution - given in \eqref{eqn:IIAnew1}.\\

We now turn to the first $\mathcal{N}=1$ background explicitly presented in this section, derived by fixing $\zeta=-1$ in \eqref{eqn:chireduction1}. This preserves a $U(1)_r$ R-Symmetry under a $\chi$ reduction, breaking the $SU(2)_R$ component (as detailed in Table \ref{table:2}). The background reads
      \begin{align}\label{eqn:chiN=1}
&        ds_{10,st}^2=e^{\frac{2}{3}\Phi}f_1\bigg[4ds^2(\text{AdS}_5)+f_2d\theta^2+f_4(d\sigma^2+d\eta^2)\bigg]+ f_1^2 e^{-\frac{2}{3}\Phi}  ds^2_2,\nn\\[2mm]
  &        ds^2_2 = \big[f_3f_5 +  \sin^2\theta f_2 f_5\big] d\beta^2 +\sin^2\theta f_2\big[f_3+f_5(f_6 +\xi )^2\big]d\phi^2 +2  \sin^2\theta f_2  f_5(f_6 +\xi) d\beta d\phi,\nn\\[2mm]
&        e^{\frac{4}{3}\Phi}= f_1 \Big[f_3+f_5(f_6+\xi)^2 +  \sin^2\theta f_2\Big] ,~~~~~~~~~~~    B_2 =- \sin\theta \bigg[   f_8 d\beta +( f_7+ \xi f_8)d\phi\bigg] \wedge \, d\theta,\nn\\[2mm]
 &      C_1= f_1 e^{-\frac{4}{3}\Phi} \bigg[f_5 (f_6+\xi) d\beta  - \sin^2\theta f_2 d\phi\bigg]  ,~~~~~~~~~   C_3=  f_8  \sin\theta \,d\beta  \wedge \, d\theta \wedge d\phi.
    \end{align}
Thus, we must fix $\zeta=-1$ to dimensionally reduce to an $\mathcal{N}=1$ Type IIA theory, and set $\gamma=0$ to perform a SUSY preserving abelian T-duality along $\beta$. Notice that, for the parameters chosen here, no SUSY preserving ATD along $\phi$ can be performed. The IIA backgrounds given in \eqref{eqn:chireduction1} are summarised in Table \ref{table:2}, deriving the $\mathcal{N}=1$ solution given in \eqref{eqn:chiN=1} when $\zeta=-1$. One can map this solution to the $\mathcal{N}=1$ background derived from \eqref{eqn:generalresult1} (with $\zeta=-\xi$), given in \cite{us}\footnote{Specifically, one requires the transformation outlined in \eqref{eqn:mapping1}, followed by
\begin{equation*}
\begin{gathered}
C_1\rightarrow C_1 -\,d\beta,~~~~~~~~~~~~~~~C_3\rightarrow C_3- \, d\beta \wedge B_2,\\
\phi\rightarrow \phi +\,\chi,~~~~~~~~~~~~~~\beta \rightarrow  - \xi \,\chi,~~~~~~~~~~~~~~~~\xi\rightarrow\frac{1}{\xi}.
\end{gathered}
\end{equation*}}. We therefore leave out the explicit G-Structure forms for this solution.

%equation \eqref{eqn:chireduction1}.
%     \begin{equation}\label{eqn:chireduction1}
 % \hspace{-1.5cm}
  %  \begin{gathered}
  %      ds_{10,st}^2=e^{\frac{2}{3}\Phi}f_1\bigg[4ds^2(\text{AdS}_5)+f_2d\theta^2+f_4(d\sigma^2+d\eta^2)\bigg]+ f_1^2\frac{e^{-\frac{2}{3}\Phi}}{n^2} ds^2(\Xi^2),\\
   %        ds^2(\Xi^2) = \big[f_3f_5 +\zeta^2  \sin^2\theta f_2 f_5\big] d\beta^2 +\sin^2\theta f_2\big[f_3+f_5(f_6 +\xi )^2\big]d\phi^2 -2\zeta \big[ \sin^2\theta f_2  f_5(f_6 +\xi) \big]d\beta d\phi,\\
   %     e^{\frac{4}{3}\Phi}=\frac{1}{n^2}f_1 \Big[f_3+f_5(f_6+\xi)^2 + \zeta^2 \sin^2\theta f_2\Big] ,\\
   %    B_2 =\frac{1}{n}\sin\theta \bigg[  \zeta f_8 d\beta -\Big( f_7+ \xi f_8\Big)d\phi\bigg] \wedge \, d\theta,~~~~~~~~
   %    C_1= \frac{1}{n}f_1 e^{-\frac{4}{3}\Phi} \bigg[f_5 (f_6+\xi) d\beta  +\zeta \sin^2\theta f_2 d\phi\bigg]  ,\\
    %   C_3=  f_8  \sin\theta \,d\beta  \wedge \, d\theta \wedge d\phi,
  %  \end{gathered}
   % \end{equation} 

%        \begin{table}[h!]
 %    \begin{center}
%\begin{tabular}{c | c c c c  }
%$\zeta$ &$\mathcal{N}$&$SU(2)$  \\
%\hline
%$-1$&$1$ &$\times$  \\
%$0$&$0$ &$\checkmark$ \\
%$   \mathds{Z}/\{0,-1\}$&$0$ &$\times$  
%\end{tabular}
%\end{center}
%\caption{$\chi$ Reduction: \textcolor{red}{add in the U(1) column}}
%\label{table:2}
%\end{table}

\item \textbf{$(m,c)=0$ with $(s,v)$ free parameters}\\
The determinant given in \eqref{eqn:S2breakingdefns} now reduces to $qa=0$. To avoid repeating the previous case, we must fix $a=0$ with $q$ free. After re-labelling $(s\equiv \zeta,q\equiv \xi, v\equiv \gamma)$, we derive a solution which maps to the $\xi=0$ solution of \eqref{eqn:chireduction1}, with
\beq
C_1\rightarrow C_1 +\xi d\beta,~~~~~~~~~ C_3\rightarrow C_3+\xi d\beta \wedge B_2,~~~~~~~~~~~\phi\rightarrow \phi-\zeta\xi \beta.
\eeq

\end{itemize}

  We now investigate the case where $s$ is not a free parameter (which will of course break supersymmetry in all cases).

\item \textbf{Fixing $s=0$ (with $m$ free)}\\
The other possibility is to ensure $s$ is not a free parameter by fixing it to zero. In these cases, $m$ is now the free parameter.
\begin{itemize}
\item  \textbf{$(s,a)=0$ with $(m,q)$ free parameters}\\
Here the determinant reduces to $vc=0$. We now look at each case in turn. These cases can be derived from the $(\xi,\zeta)=0$ solution of \eqref{eqn:chireduction1}, by the following transformations.
\begin{itemize}
\item \textbf{$c=0$ (with $v$ free)}: Re-defining $\phi\rightarrow \phi+ v\beta$ followed by
\beq
C_1\rightarrow C_1+q\, d\beta+m\, d\phi,~~~~~~~~~~~~~C_3\rightarrow C_3+(q \,d\beta+m\, d\phi)\wedge B_2.
\eeq
\item \textbf{$v=0$ (with $c$ free)}: Re-defining $\beta\rightarrow \beta+ c\phi$ followed by 
\beq
C_1\rightarrow C_1+q\, d\beta+m\, d\phi,~~~~~~~~~~~~~C_3\rightarrow C_3+(q \,d\beta+m\, d\phi)\wedge B_2.
\eeq
\end{itemize}
\item  \textbf{$(s,v,q)=0$ with $(m,c,a)$ free parameters}\\
The remaining case can be derived from the $\zeta=0$ solution of \eqref{eqn:chireduction1} by re-defining $\xi\equiv a$ then $\beta\rightarrow \beta+(c-a\,m)\phi$, followed by
\beq
C_1 \rightarrow C_1+m\,d\phi,~~~~~~~~~~~~~C_3 \rightarrow C_3 +m\,d\phi \wedge B_2.
\eeq
\end{itemize} 

\end{itemize}
\subsection*{$\phi$ reduction}
We now finally consider a dimensional reduction along $\phi$, with the following general background

   \begin{align}\label{eqn:phi-Gen}
  &      ds_{10,st}^2=e^{\frac{2}{3}\Phi}f_1\bigg[4ds^2(\text{AdS}_5)+f_2d\theta^2+f_4(d\sigma^2+d\eta^2)\bigg]+f_1^2 e^{-\frac{2}{3}\Phi} ds^2_2,\nn\\[2mm]
 &       e^{\frac{4}{3}\Phi}= f_1 \Big[m^2(f_3+f_5f_6^2)+c^2 f_5 +2cmf_5f_6 + u^2 \sin^2\theta f_2\Big] ,~~~~~~~~~B_2=(B_{2,\chi}d\chi+B_{2,\beta}d\beta)\wedge d\theta,\nn\\[2mm]
&        C_1=C_{1,\chi}d\chi+C_{1,\beta}d\beta,~~~~~~~~~~C_3 = C_{3,\chi\beta}d\chi\wedge d\theta \wedge d\beta,\nn\\[2mm]
   &    B_{2,\chi} = \sin\theta\Big((up -sm) f_7+(ua-sc) f_8\Big),~~~~~~~~~~~~~~ B_{2,\beta} =\sin\theta \Big((uq-vm)f_7+(ub-vc)f_8\Big)  ,\nn\\[2mm]
 &      C_{1,\chi}=    f_1 e^{-\frac{4}{3}\Phi}\Big(af_5 (c+mf_6) +p\,m f_3 + pf_5f_6(c+mf_6) +s\,u \sin^2\theta f_2\Big), \nn\\[2mm]
   &    C_{1,\beta}=  f_1 e^{-\frac{4}{3}\Phi}\Big(bf_5 (c+mf_6) +m\,q f_3 + qf_5f_6(c+mf_6) +u\,v \sin^2\theta f_2\Big)   ,\nn\\[2mm]
  &      C_{3,\chi\beta}=\sin\theta \bigg[v(p f_7+a f_8)-s (qf_7+bf_8) \bigg] ,\nn\\[8mm]
       %C_3=\sin\theta \bigg[(p f_7+a f_8)d\chi + (qf_7+bf_8)d\beta  \bigg] \wedge \, d\theta \wedge (s\, d\chi +t\,d\beta ),\\\\
 &      ds^2_2 = h_\chi(\eta,\sigma,\theta)d\chi^2 + h_\beta(\eta,\sigma,\theta)d\beta^2 + h_{\chi\beta}(\eta,\sigma,\theta)d\chi d\beta, \nn\\[2mm]
 &      h_\chi(\eta,\sigma,\theta) =    f_3f_5 (am-pc)^2 +\sin^2\theta f_2\Big[c^2s^2 f_5 +(pu-ms)^2f_3 - 2cs f_5\big((pu-ms)f_6 +au\big) \nn\\[2mm]
  &     ~~~~~~~~~~~~~~~~~~~~~~~~~~~+f_5\big((pu-ms)f_6 +au\big)^2\Big],\nn\\[2mm]
   &           h_\beta(\eta,\sigma,\theta) = f_3f_5 (bm-cq)^2 +\sin^2\theta f_2\Big[c^2v^2 f_5 +(qu-mv)^2f_3 - 2cv f_5\big((qu-mv)f_6 +bu\big) \nn\\[2mm]
 &             ~~~~~~~~~~~~~~~~~~~~~~~~~~~~+f_5\big((qu-mv)f_6 +bu\big)^2\Big], \nn\\[2mm]
     &                       h_{\chi\beta}(\eta,\sigma,\theta) =2\bigg[ f_3f_5 (am-pc)(bm-cq)  +\sin^2\theta f_2 h_4(\eta,\sigma)\bigg],\nn\\[2mm]
     &                     h_4(\eta,\sigma) =  ub f_5\Big((pu-ms)f_6+au\Big) +c^2sv f_5 +(qu-mv) \Big[(pu-ms)f_3+f_5f_6 \Big((pu-ms)f_6+au\Big)\Big]\nn\\[2mm]
     &~~~~~~~~~~~~~~~~~~ - cf_5\bigg(uv(a+pf_6)+s\Big(ub+(qu-2mv)f_6\Big)\bigg).
    \end{align}
     Here we follow the same steps as in the $\chi$ reduction case. Fixing $(p,b,u)=1$ leads to \eqref{eqn:newU(1)} for the $U(1)_r$ component, and to preserve SUSY under reduction, one must now set $m=-1$. We begin by keeping $m$ a free parameter.
  
  \begin{itemize}
\item \textbf{Keeping $m$ free}\\
From the determinant given in \eqref{eqn:S2breakingdefns}, we now must fix $s=0$, with either $a=0$ or $v=0$. We now investigate both options in turn.

\begin{itemize}
\item \textbf{$(s,a)=0$ with $(m,q)$ free parameters}\\
The determinant in \eqref{eqn:S2breakingdefns} reduces to $vc=0$, meaning the three free parameters are now $(m,q,v\text{ or }c)$. 
In the following IIA backgrounds, one can set $q=0$ without loss of generality, where $\chi=\chi+q\, \beta$ for either $v$ or $c$ free. For now however, we leave it free.
\begin{itemize}
\item \textbf{$v=0$ (with $c$ free)}\\
Here we apply the following 11D transformations (with $m\equiv \zeta, c\equiv \xi, q\equiv \gamma$)
    \beq
\begin{gathered}
d\beta = d\beta +\xi\, d\phi,~~~~~~~~~~~~~~~~~d\chi =  d\chi +\gamma\, d\beta + \zeta \,d\phi,~~~~~~~~~~~~~~~~~~d\phi =d\phi ,\\
U(1)_r = \chi +\gamma \beta + (\zeta+1)\phi.
\end{gathered}
\eeq
Hence, one would need to fix $\gamma=0$ to T-Dualise in a SUSY preserving manner. Indeed, as already discussed, we can fix $\gamma=0$ without loss of generality in the IIA background,
      \begin{align}\label{eqn:phieqmatch}
    &    ds_{10,st}^2=e^{\frac{2}{3}\Phi}f_1\bigg[4ds^2(\text{AdS}_5)+f_2d\theta^2+f_4(d\sigma^2+d\eta^2)\bigg]+f_1^2 e^{-\frac{2}{3}\Phi}  ds^2_2,\nn\\[2mm]
    &        ds^2_2 = \Big( \xi^2 f_3f_5  +\sin^2\theta f_2( f_3   +f_5 f_6^2)\Big) d\chi^2 +\Big(\zeta^2 f_3f_5  +\sin^2\theta f_2 f_5\Big) d\beta^2 \nn\\[2mm]
    &~~~~~~~~~-2f_5\Big(\xi\zeta f_3   -\sin^2\theta f_2  f_6  \Big) d\chi d\beta,\nn\\[2mm]
        &e^{\frac{4}{3}\Phi}=  f_1 \Big[f_5(\xi  +\zeta f_6)^2 + \zeta^2 f_3+\sin^2\theta f_2\Big] ,~~~~~~~~~~~~   B_2= \sin\theta\Big(   f_7 d\chi+  f_8 d\beta\Big)\wedge d\theta,\nn\\[2mm]
    &   C_1=  f_1 e^{-\frac{4}{3}\Phi} \Big[  \Big( \zeta f_3 + f_5f_6(\xi+\zeta f_6) \Big)  d\chi+ f_5 (\xi+\zeta f_6) d\beta\Big],    ~~~~~~~~~~~   C_3 =0
    \end{align}
    
with $\chi=\chi+\gamma \beta$, allowing $\gamma=0$ without loss of generality. Once again, this is the condition which is required to preserve the $U(1)_r$ R-Symmetry component under an ATD. \\

To map to \eqref{eqn:generalresult1}, one requires the transformations of \eqref{eqn:Transformation}
    with $k_1=\xi^{-1},~k_2=\xi^{2}$, giving
    \beq\label{eqn:mappingphi1}
        g_{MN} \rightarrow \frac{1}{\xi} g_{MN},~~~~~~~~~~~~~B_2 \rightarrow \frac{1}{\xi} B_2,~~~~~~~~~C_1\rightarrow \xi\, C_1,~~~~~~~~~~~C_3\rightarrow C_3,~~~~~~~~~e^{\frac{4}{3}\Phi} \rightarrow \frac{1}{\xi^2} e^{\frac{4}{3}\Phi},
    \eeq
followed by (in order)
\beq\label{eqn:mappingphi2}
\begin{gathered}
C_1\rightarrow C_1 +\xi \,d\phi,~~~~~~~~~~~~~~~C_3\rightarrow C_3+\xi \, d\phi \wedge B_2,\\
\zeta \rightarrow \frac{\tilde{\zeta}}{\zeta},~~~~~~~~~~~~~\xi\rightarrow \frac{1}{\zeta},\\
\tilde{\zeta}\rightarrow \xi,~~~~~~~\chi\rightarrow \chi-\frac{\xi}{\zeta}\phi,~~~~~~~~~~~~~~\beta \rightarrow -\frac{1}{\zeta}\phi.
\end{gathered}
\eeq
%\textcolor{red}{Here I think we just have a new U(1) which is $-\frac{1}{\zeta}\phi$ - is this now a $GL(2,\mathds{R})$ transformation?}\\
To map to \eqref{eqn:chireduction1}, one requires  
\beq
\begin{gathered}
g_{MN}\rightarrow g_{MN},~~~~~~~B_2\rightarrow B_2,~~~~~~~C_1\rightarrow \zeta^2 C_1,~~~~~~~~e^{\frac{4}{3}\Phi} \rightarrow \frac{1}{\zeta^2} e^{\frac{4}{3}\Phi},\\
C_1\rightarrow C_1-\zeta d\chi,~~~~~~~~~~C_3\rightarrow C_3-\zeta d\chi \wedge B_2,\\
\zeta\rightarrow \frac{1}{\zeta},~~~~~~\xi\rightarrow \frac{\xi}{\zeta},~~~~~\chi\rightarrow -\phi,~~~~~\beta\rightarrow \zeta \beta-\xi \phi,
\end{gathered}
\eeq
which does not appear to fit into the conditions of \eqref{eqn:Transformation}. The mapping to \eqref{eqn:generalresult1} requires both $1/\xi$ and $1/\zeta$. Hence, fixing $\zeta=0$ and $\xi=0$ derives the new and unique solutions given in \eqref{eqn:newIIA2} and \eqref{eqn:newIIA3}, respectively.\\

%  \begin{equation}\label{eqn:phieqmatch}
 % \hspace{-1.5cm}
  %  \begin{gathered}
   %     ds_{10,st}^2=e^{\frac{2}{3}\Phi}f_1\bigg[4ds^2(\text{AdS}_5)+f_2d\theta^2+f_4(d\sigma^2+d\eta^2)\bigg]+f_1^2\frac{e^{-\frac{2}{3}\Phi} }{n^2}ds^2(\Xi^2),\\
   %         ds^2(\Xi^2) = \Big[ \xi^2 f_3f_5  +\sin^2\theta f_2( f_3   +f_5 f_6^2)\Big] d\chi^2 +\Big[\zeta^2 f_3f_5  +\sin^2\theta f_2 f_5\Big] d\beta^2 -2f_5(\xi\zeta f_3   -\sin^2\theta f_2  f_6  ) d\chi d\beta
   %     \\e^{\frac{4}{3}\Phi}=\frac{1}{n^2} f_1 \Big[f_5(\xi  +\zeta f_6)^2 + \zeta^2 f_3+\sin^2\theta f_2\Big] ,\\
   %    B_2=\frac{1}{n}\sin\theta\Big(   f_7 d\chi+  f_8 d\beta\Big)\wedge d\theta,~~~~~~~~~~~   C_3 =0\\
   %     C_1=\frac{1}{n} f_1 e^{-\frac{4}{3}\Phi} \Big[  \Big( \zeta f_3 + f_5f_6(\xi+\zeta f_6) \Big)  d\chi+ f_5 (\xi+\zeta f_6) d\beta\Big],
  %  \end{gathered}
  %  \end{equation} 

By fixing $\zeta=-1$, one derives an $\mathcal{N}=1$ background which re-derives \eqref{eqn:chiN=1} via an appropriate set of gauge and coordinate transformations, which read 
\beq\label{eqn:N1trans}
\begin{gathered}
C_1\rightarrow -(C_1+d\chi),~~~~~~~~B_2\rightarrow -B_2,~~~~~~~~~~~~~~~C_3\rightarrow C_3 -d\chi\wedge B_2,\\
\beta\rightarrow \beta-\xi \phi,~~~~~~~~~\chi\rightarrow \phi,~~~~~~~~~~\xi\rightarrow -\xi.
\end{gathered}
\eeq

%      \begin{equation}\label{eqn:phiN=1-2}
 %     \hspace{-3cm}
 %   \begin{gathered}
 %       ds_{10,st}^2=e^{\frac{2}{3}\Phi}f_1\bigg[4ds^2(\text{AdS}_5)+f_2d\theta^2+f_4(d\sigma^2+d\eta^2)\bigg]+f_1^2 e^{-\frac{2}{3}\Phi}  ds^2_2,\\
 %           ds^2_2 = \Big[ \xi^2 f_3f_5  +\sin^2\theta f_2( f_3   +f_5 f_6^2)\Big] d\chi^2 +\Big[ f_3f_5  +\sin^2\theta f_2 f_5\Big] d\beta^2 +2f_5(\xi f_3   +\sin^2\theta f_2  f_6  ) d\chi d\beta
    %    \\e^{\frac{4}{3}\Phi}=  f_1 \Big[f_5(\xi  - f_6)^2 +   f_3+\sin^2\theta f_2\Big] ,\\
  %     B_2= \sin\theta\Big(   f_7 d\chi+  f_8 d\beta\Big)\wedge d\theta,~~~~~~~~~~~   C_3 =0\\
  %      C_1=  f_1 e^{-\frac{4}{3}\Phi} \Big[  \Big( - f_3 + f_5f_6(\xi- f_6) \Big)  d\chi+ f_5 (\xi- f_6) d\beta\Big],
  %  \end{gathered}
  %  \end{equation} 

\item  \textbf{$c=0$ (with $v$ free)}\\
In this case, we now make the following re-definitions $(m\equiv \zeta, v\equiv \xi,q\equiv \gamma)$, which re-derives the $\xi=0$ solution of \eqref{eqn:phieqmatch} with 
\beq
C_1\rightarrow C_1+\xi d\beta,~~~~~~~~~~~~~~~~C_3\rightarrow C_3+\xi d\beta \wedge B_2,~~~~~~~~~~~\chi\rightarrow \chi-\zeta\xi \beta.
\eeq

\end{itemize}
\item \textbf{$(s,v)=0$ with $(m,c)$ free parameters}\\
The determinant in \eqref{eqn:S2breakingdefns} now becomes $qa=0$. Of course, fixing $a=0$ corresponds to the case just studied, and taking $q=0$ with $a$ free re-derives \eqref{eqn:phieqmatch} as well (with $\beta=\beta+a\chi$, i.e. one can set $a=0$ without loss of generality).
  \end{itemize}
  We now investigate the case where $m$ is not a free parameter (which will of course break supersymmetry in all cases).
  
\item \textbf{Fixing $m=0$ (with $s$ free)}\\
The other possibility is to ensure $m$ is not a free parameter by fixing it to zero. In these cases, $s$ is now the free parameter.

\begin{itemize}
\item  \textbf{$(m,c)=0$ with $(s,v)$ free parameters}\\
Here the determinant reduces to $qa=0$. These cases can be derived from the $(\xi,\zeta)=0$ solution of \eqref{eqn:phieqmatch} by the following transformations
\begin{itemize}
\item \textbf{$a=0$ (with $q$ free)}: Re-defining $\chi\rightarrow \chi+ q\beta$ followed by 
\beq
C_1\rightarrow C_1 +v\,d\beta+s\,d\chi,~~~~~~~~~~~C_3\rightarrow C_3+(v\,d\beta+s\,d\chi)\wedge B_2.
\eeq
\item  \textbf{$q=0$ (with $a$ free)}: Re-defining $\beta\rightarrow \beta+a\chi$ followed by
\beq
C_1\rightarrow C_1 +v\,d\beta+s\,d\chi,~~~~~~~~~~~C_3\rightarrow C_3+(v\,d\beta+s\,d\chi)\wedge B_2.
\eeq
\end{itemize}
\item  \textbf{$(m,q,v)=0$ with $(s,a,c)$ free parameters}\\
The remaining case can be derived from the $\zeta=0$ solution of \eqref{eqn:phieqmatch} after re-defining $\xi\equiv c$ and $\beta \rightarrow \beta+(a-s\,c)\chi$, followed by
\beq
C_1\rightarrow C_1+s\,d\chi,~~~~~~~~~~~~C_3\rightarrow C_3 +s\,d\chi\wedge B_2.
\eeq
\end{itemize} 

\end{itemize}
 \newpage
 \section{Abelian T-Duality}\label{sec:ATD}
 We will now outline the Abelian T-Duality (ATD) calculations performed in this work. Throughout this section, we will use T-Dual rules presented in \cite{Kelekci:2014ima} and given in  \eqref{eqn:Tdualeq} (re-written here for convenience). Begin by making the following decomposition in Type IIA
   \begin{equation}\label{eqn:TD1}
  ds_{10}^2=ds_{9,\mathcal{A}}^2+e^{2C}(dy+A_1)^2,~~~~~~~B=B_2+B_1 \wedge dy,~~~~~~~~~F=F_{\perp}+F_{||}\wedge E^y,
  \end{equation}
  with $E^y =e^{C}(dy+A_1)$. Then the Type IIB T-dual theory is defined as follows
  \begin{equation}\label{eqn:TD2}
  \begin{gathered}
  ds_{9,\mathcal{B}}^2=ds_{9,\mathcal{A}}^2,~~~~~\Phi^\mathcal{B}=\Phi^\mathcal{A}-C^\mathcal{A},~~~~~~~C^\mathcal{B}=-C^\mathcal{A},\\
  B_2^\mathcal{B}=B_2^\mathcal{A}+A_1^\mathcal{A} \wedge B_1^\mathcal{A},~~~~~~~~A_1^\mathcal{B}=-B_1^\mathcal{A},~~~~~~~~B_1^\mathcal{B}=-A_1^\mathcal{A},\\
  F_{\perp}^\mathcal{B}=e^{C^\mathcal{A}}F_{||}^\mathcal{A},~~~~~~~F_{||}^\mathcal{B}=e^{C^\mathcal{A}}F_{\perp}^\mathcal{A}.
  \end{gathered}
  \end{equation}
   We will begin by looking at the most general results for backgrounds with the make-up of the Gaiotto-Maldacena solutions.
  
 \subsection*{General forms}
 
We will first use the following general form for the IIA backgrounds presented in this work

      \begin{equation}\label{eqn:GenGenIIA}
    \begin{aligned}
 & ~~~~~~~~~~~~~~~~~~~~~~~~~~~~~~~~~~~~~~~~~~~~~~~~~~~~~~~  \textbf{\underline{IIA}}\\[2mm]
   &     ds_{10,st}^2= ds_7^2 + \,ds^2_3,~~~~~~~~~~~~~~~~    \Phi_A,\\[2mm]
  &         ds^2_3 = h_{\phi_1}(\eta,\sigma,\theta)d\phi_1^2 + h_{\phi_2}(\eta,\sigma,\theta)d\phi_2^2 + h_{\theta}(\eta,\sigma,\theta)d\theta^2 + h_{\phi_1\phi_2}(\eta,\sigma,\theta)d\phi_1 d\phi_2 \\[2mm]
  &~~~~~~~~~~~+ h_{\phi_1\theta}(\eta,\sigma,\theta)d\phi_1 d\theta+ h_{\phi_2\theta}(\eta,\sigma,\theta)d\phi_2 d\theta,\\[2mm]
   &   B_2 = B_{2,\phi_1\theta}d\phi_1 \wedge d\theta  +B_{2,\phi_2\theta}d\phi_2 \wedge d\theta +B_{\phi_1,\phi_2}d\phi_1 \wedge d\phi_2  ,\\[2mm]
    &    C_1= C_{1,\phi_1} d\phi_1  + C_{1,\phi_2} d\phi_2+ C_{1,\theta} d\theta    ,~~~~~~~~~~~~~~~~~
       C_3=C_{3,\phi_1\theta\phi_2} d\phi_1 \wedge d\theta \wedge d\phi_2.
    \end{aligned}
    \end{equation} 
    In fact, this is a little too general for the GM backgrounds themselves (where for example $h_{\phi_1\theta}=h_{\phi_2\theta}=0$)\footnote{This more general form is needed for a second TST along $\phi_2$ (following one conducted along $\phi_1$) - which turns out to be a trivial re-definition of the TST parameter. }.\\
 The first step to T-Dualise will be to re-write the metric in the following form (for an ATD along $\phi_1$)
    \beq
\begin{gathered}
     ds^2_3 = h_{\phi_1}\Big(d\phi_1+\frac{1}{2h_{\phi_1}}(h_{\phi_1\phi_2}d\phi_2+h_{\phi_1\theta}d\theta)\Big)^2+\frac{4h_\theta h_{\phi_1}-h_{\phi_1\theta}^2}{4h_{\phi_1}}d\theta^2~~~~~~~~~~~\\
  +\frac{4h_{\phi_1} h_{\phi_2}-h_{\phi_1\phi_2}^2}{4h_{\phi_1}}d\phi_2^2 +\frac{2h_{\phi_1}h_{\phi_2\theta}-h_{\phi_1\theta}h_{\phi_1\phi_2}}{2h_{\phi_1}}d\theta d\phi_2,
     \end{gathered}
    \eeq
 noting
  \begin{equation}
\begin{aligned}
F_2&=dC_1 =dC_{1,\phi_1}\wedge  d\phi_1  + dC_{1,\phi_2} \wedge d\phi_2+ dC_{1,\theta} \wedge d\theta,\\
F_4&=dC_3-H_3\wedge C_1\\
&= \Big( dC_{3,\phi_1\theta\phi_2} + C_{1,\phi_1}dB_{2,\phi_2\theta}  + C_{1,\theta}dB_{2,\phi_1\phi_2}- C_{1,\phi_2} dB_{2,\phi_1\theta}\Big)\wedge  d\phi_1 \wedge d\theta \wedge d\phi_2 .
\end{aligned}
  \end{equation}
  
  We then use the rules in \eqref{eqn:TD1} and \eqref{eqn:TD2} to derive the following IIB general solution
        \begin{align}\label{eqn:betaredchitdualgeneral}
 &  ~~~~~~~~~~~~~~~~~~~~~~~~~~~~~~~~~~~~~~~~~~~~~~~~~~~~~~~~~~~~~   \textbf{\underline{IIB (ATD along $\phi_1$)}}\nn\\
   &    ds_{10,B}^2=ds_9^2   +     \,  \frac{1}{ h_{\phi_1}}\Big(d\phi_1+B_{2,\phi_1\theta}d\theta + B_{2,\phi_1\phi_2}d\phi_2\Big)^2,\nn\\
  &     ds_9^2=ds_7^2+\bigg(\frac{4h_\theta h_{\phi_1}-h_{\phi_1\theta}^2}{4h_{\phi_1}}\bigg)d\theta^2+\bigg(\frac{4h_{\phi_1} h_{\phi_2}-h_{\phi_1\phi_2}^2}{4h_{\phi_1}}\bigg)d\phi_2^2 +\bigg(\frac{2h_{\phi_1}h_{\phi_2\theta}-h_{\phi_1\theta}h_{\phi_1\phi_2}}{2h_{\phi_1}}\bigg)d\theta d\phi_2 ,\nn\\
%e^{2\Phi^B}=\frac{(e^{\frac{4}{3}\Phi_A})^2}{f_1^2h_1 },
& e^{\frac{4}{3}\Phi_B}=\Big(\frac{1}{h_{\phi_1}}\Big)^{\frac{2}{3}}e^{\frac{4}{3}\Phi_A},
~~~~~~~~~~~~~~~~~~~~~~~C_0=C_{1,\phi_1},\\
& B_2=\bigg(B_{2,\phi_2\theta}+ \frac{1}{2h_{\phi_1}}  (h_{\phi_1\theta}  B_{2,\phi_1\phi_2}-h_{\phi_1\phi_2}  B_{2,\phi_1\theta}) \bigg) d\phi_2 \wedge d\theta -\frac{1}{2h_{\phi_1}} (h_{\phi_1\phi_2}d\phi_2+h_{\phi_1\theta}d\theta)\wedge d\phi_1 ,\nn\\
 &  C_2=\bigg[ C_{3,\phi_1\theta\phi_2}+B_{2,\phi_1\phi_2} \Big(C_{1,\theta} - \frac{1}{2}\frac{h_{\phi_1\theta}}{h_{\phi_1}} C_{1,\phi_1} \Big) - B_{2,\phi_1\theta} \Big(C_{1,\phi_2} - \frac{1}{2}\frac{h_{\phi_1\phi_2}}{h_{\phi_1}} C_{1,\phi_1} \Big) \bigg] d\theta \wedge d\phi_2\nn \\
&~~~~~~~~~~  + \bigg(C_{1,\phi_2} -\frac{1}{2}\frac{h_{\phi_1\phi_2}}{h_{\phi_1}}C_{1,\phi_1}\bigg) d\phi_2\wedge d\phi_1  + \bigg(C_{1,\theta} -\frac{1}{2}\frac{h_{\phi_1\theta}}{h_{\phi_1}}C_{1,\phi_1}\bigg) d\theta\wedge d\phi_1.\nn
  \end{align}  
  The idea here is that one need only plug in the functions corresponding to the specific IIA example in question, and get out the IIB ATD without needing to perform the calculation each time. In addition, we now write the TST transformation by making the coordinate transformation $\phi_2 \rightarrow \phi_2+\gamma_1 \phi_1$ in \eqref{eqn:betaredchitdualgeneral}, before performing the ATD along $\phi_1$ for a second time (to return to a IIA theory)
       \begin{align}\label{eqn:TSTgenlgeneral}
  &    ~~~~~~~~~~~~~~~~~~~~~~~~~~~~~~~~~~~~~~~~~~~~~~~~~~~~~~~~~~~~~~~~~~~~~~~~~~~~~~~~  \textbf{\underline{TST }}\nn\\[2mm]
 &      ds_{10,A}^2=ds_9^2+\frac{h_{\phi_1}(1-\gamma_1 \alpha_1)}{1+\gamma_1 B_{2,\phi_1\phi_2}}\bigg[d\phi_1 +\frac{h_{\phi_1\phi_2}}{2h_{\phi_1}}d\phi_2+\frac{h_{\phi_1\theta}}{2h_{\phi_1}}d\theta+\gamma_1 \Big[B_{2,\phi_2\theta}+\frac{1}{2h_{\phi_1}}(h_{\phi_1\theta}B_{2\phi_1\phi_2}-h_{\phi_1\phi_2}B_{2,\phi_1\theta})\Big]d\theta\bigg]^2  \nn \\[2mm]
  &     ds_9^2=ds_7^2 + \frac{1}{h_{\phi_1}}\bigg(B_{2,\phi_1\theta}^2 +h_\theta h_{\phi_1} -\frac{1}{4}h_{\phi_1\theta}^2-\alpha_3\alpha_2^2\bigg)d\theta^2 + \bigg(\frac{h_{\phi_1}h_{\phi_2}-\frac{1}{4}h_{\phi_1\phi_2}^2}{h_{\phi_1}\alpha_3}\bigg) d\phi_2^2~~~~~~~~~~~~~~~~~~~~~~~~~~~~~~~~ \nn\\[2mm]
 &    ~~~~~~~  + \frac{1}{h_{\phi_1}\alpha_3}\bigg[(1+\gamma_1 B_{2,\phi_1\phi_2}) \bigg(h_{\phi_1}h_{\phi_2\theta}-\frac{1}{2}h_{\phi_1\theta}h_{\phi_1\phi_2}\bigg) - 2\gamma_1 B_{2,\phi_1\theta}  \bigg(h_{\phi_1}h_{\phi_2}-\frac{1}{4} h_{\phi_1\phi_2}^2\bigg) \bigg] d\theta d\phi_2 ,\nn\\[2mm]
 &      e^{\frac{4}{3}\Phi_A^{TST}}=\bigg(\frac{ 1-\gamma_1 \alpha_1}{1+\gamma_1 B_{2,\phi_1\phi_2}}\bigg)^{\frac{2}{3}} e^{\frac{4}{3}\Phi_A},\nn\\[2mm]
   &    B_2= \bigg[B_{2,\phi_2\theta} +\frac{(1-\gamma_1 \alpha_1)}{2h_{\phi_1}} \Big(h_{\phi_1\theta} B_{2,\phi_1\phi_2} -h_{\phi_1\phi_2} B_{2,\phi_1\theta}\Big) +\alpha_2 \frac{h_{\phi_1\phi_2}}{2h_{\phi_1}}-\alpha_1 \bigg(\frac{h_{\phi_1\theta}}{2h_{\phi_1}}+\gamma_1 B_{2,\phi_2\theta}\bigg)\bigg]d\phi_2\wedge d\theta \nn\\[2mm]
  &  ~~~~~~~~~~   -(\alpha_1 d\phi_2 + \alpha_2d\theta)\wedge d\phi_1,\nn\\[2mm]
  &     C_1 = C_{1,\phi_2}d\phi_2 +C_{1,\phi_1}d\phi_1 +C_{1,\theta}d\theta+\gamma_1 \Big(C_{3,\phi_1\theta\phi_2} + C_{1,\theta}B_{2,\phi_1\phi_2}+ C_{1,\phi_1}B_{2,\phi_2\theta} -C_{1,\phi_2}B_{2,\phi_1\theta}\Big)d\theta,\nn\\[2mm]
  &     C_3 = \bigg[(1-\gamma_1 \alpha_1) \bigg(C_{3,\phi_1\theta\phi_2} +C_{1,\theta} B_{2,\phi_1\phi_2} -C_{1,\phi_2}B_{2,\phi_1\theta} +\frac{C_{1,\phi_1}}{2h_{\phi_1}}\Big( h_{\phi_1\phi_2}B_{2,\phi_1\theta} -h_{\phi_1\theta}B_{2,\phi_1\phi_2}\Big)\bigg) \nn\\[2mm]
    &~~~~~~~~~~~~~   + \alpha_1\bigg(C_{1,\phi_1} \frac{h_{\phi_1\theta}}{2h_{\phi_1}} - C_{1,\theta}\bigg) -\alpha_2 \bigg(C_{1,\phi_1} \frac{h_{\phi_1\phi_2}}{2h_{\phi_1}} - C_{1,\phi_2}\bigg)\bigg] d\theta \wedge d\phi_2 \wedge d\phi_1,\nn\\[8mm]
  &     \alpha_1= \frac{1}{\alpha_3}\Big[B_{2,\phi_1\phi_2}(1+\gamma_1 B_{2,\phi_1\phi_2}) +\gamma_1 \Big(h_{\phi_1}h_{\phi_2} -\frac{1}{4}h_{\phi_1\phi_2}^2\Big)\Big],\nn\\[2mm]
 &      \alpha_2=\frac{1}{\alpha_3}\Big[B_{2,\phi_1\theta}(1+\gamma_1 B_{2,\phi_1\phi_2})  +\frac{\gamma_1}{2} \Big(h_{\phi_1}h_{\phi_2\theta} -\frac{1}{2}h_{\phi_1\theta}h_{\phi_1\phi_2}\Big) \Big],\nn\\[2mm]
    &   \alpha_3=(1+\gamma_1 B_{2,\phi_1\phi_2})^2 +\gamma_1^2 \Big(h_{\phi_1}h_{\phi_2} -\frac{1}{4}h_{\phi_1\phi_2}^2\Big).
 %      G_1=(1+\gamma B_{2\phi_1\phi_2})^2+\gamma^2(h_{\phi_1}h_{\phi_2}-\frac{1}{4}h_{\phi_1\phi_2}^2)
  \end{align}

  \subsection*{Less general forms}
  For the calculations performed in this section, the previous results are a little too general and so are unnecessarily cumbersome. The Type IIA backgrounds presented throughout this paper all fit into the following form, where $(\phi_1,\phi_2)$ are the two $U(1)$ directions

      \begin{equation}\label{eqn:GenIIA}
    \begin{aligned}
 &~~~~~~~~~~~~~~~~~~~~~~~~~~~~~~~~~~~~~~~~~~~~~~~~~~~~~~~   \textbf{\underline{IIA}}\\[2mm]
&        ds_{10,st}^2= ds_8^2+\Gamma \,ds^2_2,~~~~~~~~~~~~~~~~~~~~\Phi_A,\\[2mm]
&           ds^2_2 = h_{\phi_1}(\eta,\sigma,\theta)d\phi_1^2 + h_{\phi_2}(\eta,\sigma,\theta)d\phi_2^2 + h_{\phi_1\phi_2}(\eta,\sigma,\theta)d\phi_1 d\phi_2,\\[2mm]
 &     B_2 =  \Big(B_{2,\phi_1}d\phi_1  +B_{2,\phi_2}d\phi_2 \Big) \wedge d\theta,~~~~~~~~~~~~~~~~~
        C_1= C_{1,\phi_1} d\phi_1  + C_{1,\phi_2} d\phi_2  ,\\[2mm]
  &     C_3=C_{3,\phi_1\phi_2} d\phi_1 \wedge d\theta \wedge d\phi_2,
    \end{aligned}
    \end{equation} 
      where for the backgrounds in question,
 \begin{equation} 
  \begin{aligned}
 &ds_8^2=  ds_7^2 +f_\theta d\theta^2,~~~~~~~~~~~~~~~~~  ds_7^2=e^{\frac{2}{3}\Phi_A}f_1\bigg[4ds^2(\text{AdS}_5)+f_4(d\sigma^2+d\eta^2)\bigg],   \\[2mm]
 &  \Gamma=f_1^2e^{-\frac{2}{3}\Phi_A},~~~~~~~~~~~~f_\theta=e^{\frac{2}{3}\Phi_A}f_1 f_2 ,
  \end{aligned}
  \end{equation}
  noting that in this case,
  \begin{equation}
\begin{aligned}
F_2&=dC_1 =dC_{1,\phi_1}\wedge  d\phi_1  + dC_{1,\phi_2} \wedge d\phi_2,\\[2mm]
F_4&=dC_3-H_3\wedge C_1%\\
%&=  dC_{3,\chi\phi}\wedge  d\chi \wedge d\theta \wedge d\phi -  \Big(dB_{2,\chi}\wedge d\chi  +dB_{2,\phi}\wedge d\phi \Big) \wedge d\theta \wedge \Big(C_{1,\chi} d\chi  + C_{1,\phi} d\phi\Big)\\
= \Big( dC_{3,\phi_1\phi_2} + C_{1,\phi_1}dB_{2,\phi_2} - C_{1,\phi_2} dB_{2,\phi_1}\Big)\wedge  d\phi_1 \wedge d\theta \wedge d\phi_2.
\end{aligned}
  \end{equation}
  We play the same game as before, re-writing the metric as follows
    \beq\label{eqn:A1Aval}
  \begin{aligned}
    ds^2_2 &= h_{\phi_1} d\phi_1^2 + h_{\phi_2} d\phi_2^2 + h_{\phi_1\phi_2} d\phi_1 d\phi_2 = h_{\phi_1}\bigg(d\phi_1 +\frac{1}{2}\frac{h_{\phi_1\phi_2}}{h_{\phi_1}}d\phi_2\bigg)^2+\bigg(h_{\phi_2}-\frac{1}{4}\frac{h_{\phi_1\phi_2}^2}{h_{\phi_1}}\bigg)d\phi_2^2,
    \end{aligned}
  \eeq
before using the ATD rules given in \eqref{eqn:TD1} and \eqref{eqn:TD2}. Of course, one could instead use the more general forms in the previous subsection. Now, performing a T-Duality along $\phi_1$ on \eqref{eqn:GenIIA}, one gets
    \begin{align}\label{eqn:betaredchitdual}
 &   ~~~~~~~~~~~~~~~~~~~~~~~~~~~~~~~~~~~~~~~~~~~~~~~~~~~~~~~~~~~~~~~~~~~~  \textbf{\underline{IIB (ATD along $\phi_1$)}}\nn\\[2mm]
 &      ds_{10,B}^2=ds_8^2+\delta_1\, d\phi_2^2+\delta_2\,\bigg(d\phi_1 +B_{2,\phi_1} d\theta\bigg)^2,~~~~~~~~~~~~~~~  \delta_1=\Gamma  \bigg(h_{\phi_2}-\frac{1}{4}\frac{h_{\phi_1\phi_2}^2}{h_{\phi_1}}\bigg),%~~~~~~~~~~\beta=\frac{e^{\frac{2}{3}\Phi_A}}{f_1^2h_1 },
       ~~~~~~~~~~~~~~~~\delta_2=\frac{1}{h_{\phi_1} \Gamma },\nn\\[2mm]
%e^{2\Phi^B}=\frac{(e^{\frac{4}{3}\Phi_A})^2}{f_1^2h_1 },
&e^{\frac{4}{3}\Phi_B}=\delta_2^{\frac{2}{3}}e^{\frac{4}{3}\Phi_A},
~~~~~~~~~~B_2=\bigg(B_{2,\phi_2}- \frac{1}{2}\frac{h_{\phi_1\phi_2}}{h_{\phi_1}}  B_{2,\phi_1} \bigg) d\phi_2 \wedge d\theta -\frac{1}{2}\frac{h_{\phi_1\phi_2}}{h_{\phi_1}}d\phi_2\wedge d\phi_1 ,~~~~~~~~~~~~~~  C_0=C_{1,\phi_1},\nn\\[2mm]
&  C_2=\bigg( C_{3,\phi_1\phi_2}- B_{2,\phi_1} \Big(C_{1,\phi_2} - \frac{1}{2}\frac{h_{\phi_1\phi_2}}{h_{\phi_1}} C_{1,\phi_1} \Big) \bigg) d\theta \wedge d\phi_2 + \bigg(C_{1,\phi_2} -\frac{1}{2}\frac{h_{\phi_1\phi_2}}{h_{\phi_1}}C_{1,\phi_1}\bigg) d\phi_2\wedge d\phi_1.
    \end{align}
Once again, we now write the TST solution by making the coordinate transformation $\phi_2 \rightarrow \phi_2+\gamma_1 \phi_1$ in \eqref{eqn:betaredchitdual}, before performing the ATD along $\phi_1$ for a second time (to return to a IIA theory)
     \begin{align}\label{eqn:TSTgenl}
 &    ~~~~~~~~~~~~~~~~~~~~~~~~~~~~~~~~~~~~~~~~~~~~~~~~~~~~~~~~~~~~~~~~~~~~~~~~~~~~~~   \textbf{\underline{TST }}\nn\\[2mm]
&       ds_{10,A}^2=ds_8^2 +\frac{1}{\delta_2 + \gamma_1^2 \delta_1}\bigg[\delta_1 \delta_2(d\phi_2 -\gamma_1 B_{2,\phi_1}d\theta)^2+ \bigg(d\phi_1 +\frac{1}{2}\frac{h_{\phi_1\phi_2}}{h_{\phi_1}}d\phi_2 +\gamma_1 \Big(B_{2,\phi_2}-\frac{1}{2}\frac{h_{\phi_1\phi_2}}{h_{\phi_1}}B_{2,\phi_1}\Big)d\theta\bigg)^2\,\bigg],\nn\\[2mm]
 &      e^{\frac{4}{3}\Phi_A^{TST}}= \bigg(\frac{\delta_2}{\delta_2 +\gamma_1^2 \delta_1}\bigg)^{\frac{2}{3}}e^{\frac{4}{3}\Phi_A},~~~~~~~~~~~~B_2= \frac{\delta_2}{\delta_2+\gamma_1^2\delta_1}(B_{2,\phi_2}d\phi_2 +B_{2,\phi_1}d\phi_1)\wedge d\theta -\frac{\gamma_1 \delta_1}{\delta_2+\gamma_1^2\delta_1}d\phi_2 \wedge d\phi_1,\nn\\[2mm]
&       C_1 = C_{1,\phi_2}d\phi_2 +C_{1,\phi_1}d\phi_1 +\gamma_1 (C_{3,\phi_1\phi_2} + C_{1,\phi_1}B_{2,\phi_2} -C_{1,\phi_2}B_{2,\phi_1})d\theta,\nn\\[2mm]
 &      C_3=\frac{\delta_2}{\delta_2+\gamma_1^2 \delta_1}C_{3,\phi_1\phi_2}d\theta \wedge d\phi_2 \wedge d\phi_1,~~~~~~~~~~~
       \delta_1=\Gamma  \bigg(h_{\phi_2}-\frac{1}{4}\frac{h_{\phi_1\phi_2}^2}{h_{\phi_1}}\bigg),%~~~~~~~~~~\beta=\frac{e^{\frac{2}{3}\Phi_A}}{f_1^2h_1 },
       ~~~~~~~~~~\delta_2=\frac{1}{h_{\phi_1} \Gamma }.
  \end{align}
   In doing such a calculation, one picks up the parameter $\gamma_1$. One can indeed see by observation that setting $\gamma_1=0$ in \eqref{eqn:TSTgenl} gives \eqref{eqn:GenIIA}.

 In what follows, we will utilise \eqref{eqn:betaredchitdual} to derive Type IIB theories, keeping track of the supersymmetry along the way. We will see that one can indeed get $\mathcal{N}=1$ IIB theories using this method, but one must be careful in order to preserve the $U(1)_r$ R-Symmetry component. We do not calculate Type IIA TST solutions of our backgrounds, this is left for future investigations. We will however utilise \eqref{eqn:TSTgenl} to find a TST deformation of the GM background following an uplift. See Appendix \ref{sec:TSTGM}.
 
We split this section into three, with the seed IIA theories corresponding to the $\beta,~\chi$ and $\phi$ dimensional reduction, respectively. This will make things cleaner, as in each of the three cases, our definitions of $(\phi_1,\phi_2)$ will differ.
  
  \subsection*{T-Dualising the $\beta$ Reduction}
 Recall that the most general 11D G-structure forms must preserve the $U(1)_r$ R-Symmetry component given in \eqref{eqn:U(1)} under dimensional reduction, in order to preserve supersymmetry. In the case of the $\beta$ reduction, this corresponds to the condition $\zeta=-\xi$ (with $c=a=0$ and $q\equiv \xi,~v\equiv \zeta$). This derives $\mathcal{N}=1$ solutions for $\xi\neq 0$, promoting to $\mathcal{N}=2$ for $\xi=0$ (where the $SU(2)_R$ component is recovered). The $U(1)_r$ component now becomes in general $(p+s)\chi+(m+u)\phi$, which in turn must be preserved under T-Duality to preserve the supersymmetry in Type IIB. Thus, for a T-duality along $\chi$, we must fix $p+s=0$; for a T-duality along $\phi$, we must fix $m+u=0$. Recalling that the determinant given in \eqref{eqn:S2breakingdefns} becomes $pu-ms=1$, so we must either have $(pu=1,~ms=0)$ or $(pu=0,~ms=-1)$.\\
 We look first at the case where $pu=1$ and $ms=0$ (where $m$ or $s$ is a free parameter and can be set to zero in the IIA case without loss of generality - see Appendix \ref{sec:DimRed}). Following the conventions we have been using throughout this paper, we will use $\gamma$ for this free parameter. Hence, we now have the following possibilities (using the $SL(3,\mathds{R})$ transformation given in \eqref{eqn:S2breakingdefns})
 \begin{enumerate}
\item $(p=1,~s=\gamma,~u=1,~m=0)$ with a T-duality along $\chi$
  \beq\label{eqn:1}
d\chi = d\chi+ \xi\, d\beta ,~~~~~~~~~~~~~~~~~~d\phi =d\phi+ \gamma\, d\chi +\zeta \,d\beta ,\\
  \eeq
  with $\gamma=-1,~\zeta=-\xi$ for $\mathcal{N}=1$ (noting $p=-1,~\gamma=1$ corresponds to the same background after $\chi\rightarrow-\chi$ and $\phi\rightarrow -\phi$).
  \item $(u=1,~m=\gamma,~p=1,~s=0)$ with a T-duality along $\phi$
    \beq\label{eqn:2}
d\chi = d\chi+ \gamma d\phi +\xi\, d\beta ,~~~~~~~~~~~~~~~~~~d\phi=d\phi+\zeta \,d\beta ,\\
  \eeq
   with $\gamma=-1,~\zeta=-\xi$ for $\mathcal{N}=1$ (noting as above, $u=-1,~\gamma=1$ corresponds to the same background after $\chi\rightarrow-\chi$ and $\phi\rightarrow -\phi$).
\end{enumerate}
 We now instead fix $pu=0$ and $ms=-1$ (where in this case $\gamma=1$ corresponds to supersymmetry preservation), giving
 \begin{enumerate}[resume]
\item  $(s=1,m=-1,p=0,u=\gamma)$ with a T-duality along $\phi$
\beq \label{eqn:3}
d\chi =   -d\phi +\xi\, d\beta ,~~~~~~~~~~~~~~~~~~d\phi =  d\chi + \gamma d\phi+\zeta d\beta .\\
\eeq
  \item $(m=1,s=-1,u=0,p=\gamma)$ with a T-duality along $\chi$
  \beq\label{eqn:4}
  d\chi = d\phi +\gamma d\chi +\xi\, d\beta ,~~~~~~~~~~~~~~~~~~d\phi =- d\chi +\zeta\,d\beta.
  \eeq
\end{enumerate}

We now see that \eqref{eqn:3} maps to \eqref{eqn:1} by $\gamma\rightarrow-\gamma,~\phi \rightarrow -\chi,~\chi\rightarrow \phi$ and \eqref{eqn:4} maps to \eqref{eqn:2} by $\gamma\rightarrow-\gamma,~\chi \rightarrow -\phi,~\phi\rightarrow \chi$. The consequence of this is that the coordinate transformations of \eqref{eqn:2} followed by a T-duality along $\phi$ is equivalent to the coordinate transformations of \eqref{eqn:4} followed by a T-duality along $\chi$. Thus, calculating the most general form for a T-duality along $\chi$ will automatically contain within it the T-duality along $\phi$, and vice versa. Analogous arguments should of course hold for the $\chi$ and $\phi$ reductions, with ATDs along $(\beta,\phi)$ and $(\beta,\chi)$, respectively.

We will now perform an ATD on the two-parameter family of solutions given in \eqref{eqn:generalresult1} (and derived in \cite{us}), where we have already fixed $(p,u)=1$. For this discussion it will be instructive to switch on the SUSY parameter $\gamma$ (which plays a trivial role in the Type IIA background and set to zero) as it could become vital in the following analysis. Of course, there are many $\mathcal{N}=0$ solutions contained within the mathematics, but we will focus here on deriving an $\mathcal{N}=1$ background. The type IIA background in question now have the following $U(1)_r$ component
\beq
U(1)_r = (1+s)\chi+(1+m)\phi,~~~~~~~~~~~~~~~ms=0.
\eeq
Hence, to preserve the R-Symmetry under ATD, one must either fix $(s\equiv \gamma=-1,m=0)$ and T-dualise along $\chi$, or fix $(m\equiv \gamma=-1,s=0)$ and T-Dualise along $\phi$. We will see in fact that both approaches lead to the same $\mathcal{N}=1$ IIB background, following appropriate transformations.

\begin{itemize}
\item \underline{\textbf{$s\equiv \gamma$ }}\\
We begin with fixing $s\equiv \gamma$ (with $m=0$). We will start with the ATD which will give rise to an $\mathcal{N}=1$ background, which in this case is along $\chi$. We then T-dualise along the other $U(1)$ direction, which is of course along $\phi$ in this case. The parameters are now $(p,b,u)=1,~(a,c,m)=0,~q\equiv \xi,v\equiv \zeta,s\equiv \gamma$.
\begin{itemize}
\item \textbf{Performing an ATD along $\chi$}\\
Using \eqref{eqn:betaredchitdual}, we now derive the three-parameter family of solutions given in \eqref{eqn:ATD1}. As we've explained, we derive the $\mathcal{N}=1$ background by fixing $\zeta=-\xi$ and setting $\gamma=-1$ (noting that $\gamma\xi-\zeta=0$). One then derives the solution given in \eqref{eqn:IIBN=1-1}. 

\item \textbf{Performing an ATD along $\phi$}\\
Performing the ATD along $\phi$ leads to a two parameter $\mathcal{N}=0$ family of solutions.
 This background will be re-derived as the $\gamma=0$ solution of \eqref{eqn:ATD2} (up to appropriate gauge transformation). %We leave the above background in for instructional purposes.

\end{itemize}

\item \underline{\textbf{$m\equiv \gamma$ }}\\
We now investigate the solutions with $m\equiv \gamma$ (and $s=0$). Once again, we begin with the case which will derive a SUSY background, which is now an ATD along $\phi$. We then once again T-Dualise along the other $U(1)$ in question, now $\chi$. The parameters now read \\$(p,b,u)=1,~(a,c,s)=0,~q\equiv\xi,v\equiv\zeta,m\equiv\gamma$.
\begin{itemize}
\item \textbf{Performing an ATD along $\phi$}\\
One now gets (noting $C_{3,\phi\chi}=-C_{3,\chi\phi}$)\footnote{By the discussion at the start of this subsection, one can derive this result by fixing $p=\gamma,u=0,m=1,s=-1,b=1,c=a=0,q=\xi,v=\zeta$, performing an ATD along $\chi$ and then redefining $\gamma=-\gamma$ with $\phi\rightarrow \chi, \chi\rightarrow -\phi$. }
    \begin{align}\label{eqn:ATD2}
      & ds_{10,B}^2=e^{\frac{2}{3}\Phi_\mathcal{A}}f_1\Bigg[4ds^2(\text{AdS}_5)+f_2d\theta^2+f_4(d\sigma^2+d\eta^2) \nn\\[2mm]
       &~~~~~~~~~~~~~~~~~~~+ \frac{ 1}{\hat{\Xi}}\bigg(f_2 f_3 f_5\sin^2\theta\, d\chi^2+\frac{1}{f_1^3}\,\Big(d\phi - \big(f_8+(\xi-\zeta \gamma)f_7\big)\sin\theta d\theta\Big)^2\bigg) \Bigg],\nn\\[2mm]
   %     \alpha=\Gamma  \bigg(h_2-\frac{1}{4}\frac{h_3^2}{h_1}\bigg),%~~~~~~~~~~\beta=\frac{e^{\frac{2}{3}\Phi_A}}{f_1^2h_1 },
   %    ~~~~~~~~~~~~~~~~\beta=\frac{1}{h_1 \Gamma },\\
%e^{2\Phi^B}=\frac{(e^{\frac{4}{3}\Phi_A})^2}{f_1^2h_1 },
&\hat{\Xi}= \gamma^2 f_3 f_5 +f_2\Big((\xi-\gamma\zeta)^2 f_3+f_5\big(1+(\xi-\gamma\zeta)f_6\big)^2\Big)\sin^2\theta ,\nn\\[2mm]
&e^{2\Phi_\mathcal{B}}=\frac{1}{\hat{\Xi}}\Big[f_5(1+\xi f_6)^2 +\xi^2f_3+ \zeta^2 f_2 \sin^2\theta\Big]^2,~~~~
 e^{\frac{4}{3}\Phi_\mathcal{A}}= f_1 \Big[f_5(1+\xi f_6)^2 +\xi^2f_3+ \zeta^2 f_2 \sin^2\theta\Big] ,\nn\\[2mm]
&C_0= f_1 e^{-\frac{4}{3}\Phi_\mathcal{A}} \Big(\gamma f_5 f_6 (1+\xi f_6) +\gamma \xi f_3 +\zeta f_2 \sin^2\theta\Big),\nn\\[2mm]
& B_2=\bigg(\zeta f_7+ \frac{1}{\hat{\Xi}}\big(f_8+(\xi-\gamma \zeta)f_7\big)\Big(\gamma f_3 f_5~~~~~~~~~~~~~~~~~~~~~~~~~~~~~~~~~~~~~~~~~~~~~~~~~ \nn\\[2mm]
&~~~~~~-\zeta f_2 \big(f_5f_6(1+(\xi-\gamma \zeta)f_6 ) +(\xi-\gamma \zeta)f_3\big)\sin^2\theta\Big) \bigg) \sin\theta d\chi \wedge d\theta\nn\\[2mm]
&~~~~~~- \frac{1}{\hat{\Xi}}\Big(\gamma f_3 f_5 -\zeta f_2 \big(f_5f_6(1+(\xi-\gamma \zeta)f_6 ) +(\xi-\gamma \zeta)f_3\big)\sin^2\theta\Big)  d\chi\wedge d\phi ,\nn\\[2mm]
 & C_2=%\frac{\gamma^2f_3f_5f_7+f_2\big[f_3 (\xi-\gamma\zeta)\big(f_8+2(\xi-\gamma\zeta)f_7\big) +f_5\big(1+(\xi-\gamma\zeta)f_6\big)\big(f_6f_8+f_7(1+2(\xi-\gamma\zeta)f_6)\big)\big]\sin^2\theta}{\gamma^2 f_3 f_5 +f_2\big[(\xi-\gamma\zeta)^2 f_3+f_5\big(1+(\xi-\gamma\zeta)f_6\big)^2\big]\sin^2\theta} \,\sin\theta\,d\theta \wedge d\chi \\
  - \frac{1}{\hat{\Xi}}\Big(\gamma^2 f_3 f_5 f_7+f_2\sin^2\theta \big[f_5(f_7-f_6f_8)\big(1+(\xi-\gamma\zeta)f_6\big)-(\xi-\gamma\zeta)f_3f_8\big] \Big)  \sin\theta d\theta \wedge d\chi\nn\\[2mm]
&  ~~~~~~~~ + \frac{1}{\hat{\Xi}} f_2  \big(f_5f_6(1+(\xi-\gamma \zeta)f_6 ) +(\xi-\gamma \zeta)f_3\big)\sin^2\theta   d\chi \wedge d\phi.
    \end{align}
  
 One can map this solution to \eqref{eqn:ATD1} via the transformations given in \eqref{eqn:Transformation} \\(with $k_1=1,k_2=\frac{1}{\gamma}$)\footnote{As before, when $\gamma=-1$ and $\zeta=-\xi$ (where $\xi -\gamma\zeta=0$), one derives an $\mathcal{N}=1$ background. This solution re-derives \eqref{eqn:IIBN=1-1} via the following transformations
  \begin{equation*}
  \begin{gathered}
  C_0\rightarrow -C_0,~~~~~~~~~~C_2\rightarrow -C_2,~~~~~~~~~~~~~~~~B_2\rightarrow B_2+d\phi\wedge d\chi,\\
  \phi\rightarrow -\chi,~~~~~~~~~~\chi\rightarrow \phi.
  \end{gathered}
  \end{equation*}
}
 \beq\label{eqn:IIBtrans1}
 \begin{gathered}
 e^{2\Phi_B}\rightarrow \gamma^2 e^{2\Phi_B},~~~~~~~~~C_0\rightarrow \frac{1}{\gamma} C_0,~~~~~~~C_2\rightarrow \frac{1}{\gamma}C_2,~~~~~~~~B_2\rightarrow B_2-\frac{1}{\gamma}d\phi\wedge d\chi,\\
 \phi\rightarrow \gamma \chi,~~~~~~\chi\rightarrow -\gamma \phi,~~~~~~~~~~~\gamma\rightarrow \frac{1}{\gamma}.
 \end{gathered}
 \eeq
One can now say that the $\gamma=0$ case is a new and unique family of $\mathcal{N}=0$ solutions, %, given in \eqref{eqn:newIIBcase1}.
%\subsubsection*{ATD along $\phi$ of the $\beta$ reduction}
%Performing an ATD of the $\beta$ reduction along $\phi$, one derives eqn. \eqref{eqn:ATD2}. To map this solution to eqn. \eqref{eqn:ATD1}, one requires $\gamma\rightarrow 1/\gamma$. Hence, by fixing $\gamma=0$, one derives a unique 2-parameter family of $\mathcal{N}=0$ solutions \footnote{Because the $\mathcal{N}=1$ condition of eqn. \eqref{eqn:ATD2} is $\gamma=-1$ and $\zeta=-\xi$}. It reads
\beq\label{eqn:newIIBcase1}
\hspace{-0.5cm}
    \begin{aligned}
&       ds_{10,B}^2=e^{\frac{2}{3}\Phi_\mathcal{A}}f_1\Bigg[4ds^2(\text{AdS}_5)+f_2d\theta^2+f_4(d\sigma^2+d\eta^2)~~~~~~~~~~~~~~~~~~~~~~~~~~~~~~~~~~~~~~~~~~~~~~~~~~~~~~~~~~~~~\\[2mm]
&       ~~~~~~~~~~~~~~~~~~~~~~~~~~~~~+ \frac{ 1}{\hat{\Xi}}\bigg(f_2 f_3 f_5\sin^2\theta\, d\chi^2+\frac{1}{f_1^3}\,\Big(d\phi - \big(f_8+\xi f_7\big)\sin\theta d\theta\Big)^2\bigg) \Bigg],\\[2mm]
&\hat{\Xi}=  f_2\Big(\xi^2 f_3+f_5\big(1+\xi f_6\big)^2\Big)\sin^2\theta ,~~~~~~~~~~~~~~~~C_0=\zeta f_1  f_2 e^{-\frac{4}{3}\Phi_\mathcal{A}}  \sin^2\theta,\\[2mm]
&e^{2\Phi_\mathcal{B}}=\frac{1}{\hat{\Xi}}\Big[f_5(1+\xi f_6)^2 +\xi^2f_3+ \zeta^2 f_2 \sin^2\theta\Big]^2,  ~~~~~  e^{\frac{4}{3}\Phi_\mathcal{A}}= f_1 \Big[f_5(1+\xi f_6)^2 +\xi^2f_3+ \zeta^2 f_2 \sin^2\theta\Big] ,\\[2mm]
&B_2=\zeta \bigg( f_7-  \frac{f_2}{\hat{\Xi}}\big(f_8+\xi f_7\big)  \big(f_5f_6(1+\xi f_6 ) +\xi f_3\big)\sin^2\theta \bigg) \sin\theta d\chi \wedge d\theta\\[2mm]
&~~~~~~~~~~~~~~~ +\zeta \frac{f_2}{\hat{\Xi}} \big(f_5f_6(1+\xi f_6 ) +\xi f_3\big)\sin^2\theta d\chi\wedge d\phi ,\\[2mm]
&  C_2=  \frac{-f_2}{\hat{\Xi}}\bigg[ \Big(f_5(f_7-f_6f_8)\big(1+\xi f_6\big)-\xi f_3f_8\Big)  \sin^3\theta d\theta \wedge d\chi  - \big(f_5f_6(1+\xi f_6 ) +\xi f_3\big)\sin^2\theta   d\chi \wedge d\phi\bigg]
  \end{aligned} 
 \eeq
  Typically, performing an analogous T-Duality along the $U(1)$ of the $S^2$ generates singularities in the dual description. Because of this, we refrain from including this solution in the main body of the paper, but include it here for completeness.

\item \textbf{Performing an ATD along $\chi$}\\
Now, by T-dualising along $\chi$, one derives a background which corresponds to the $\gamma=0$ solution of \eqref{eqn:ATD1} (up to a gauge transformation in $B_2$).

%    \begin{equation}%equation C.22 in mathematica
 %       \hspace{-3cm}
 % \begin{gathered}
  %     ds_{10,st}^2=e^{\frac{2}{3}\Phi_A}f_1\bigg[4ds^2(\text{AdS}_5)+f_2d\theta^2+f_4(d\sigma^2+d\eta^2)+ \frac{1}{\Xi_4}\Big(f_2 f_3f_5  \sin^2\theta d\phi^2   +\frac{1}{f_1^3 }(d\chi +\zeta f_7\sin\theta  d\theta)^2 \Big)\bigg] \\
%e^{2\Phi^B}=\frac{1}{\Xi_4} \big(   f_5(1+\xi  f_6)^2+\xi^2 f_3 + \zeta^2f_2 \sin^2\theta\big)^2,~~~~~~~~~~~~~~       e^{\frac{4}{3}\Phi_A}= f_1 \Big[  f_5(1+\xi  f_6)^2+\xi^2 f_3 + \zeta^2f_2 \sin^2\theta \Big],\\
%%B_2=\frac{1}{f_5(f_6+\xi)^2+f_3} \bigg[\Big(f_5(f_6+\xi)(\gamma f_6+\gamma\xi-\zeta)+\gamma f_3\Big)d\phi-\zeta \Big(f_3f_8 +f_5(f_6f_8-f_7)(f_6+\xi)\Big)\sin\theta d\theta \bigg] \wedge d\beta ,\\
%B_2=-\frac{ 1}{\Xi_4} \bigg[\zeta f_2 \Big(f_5f_6(1+\xi f_6)+\xi f_3\Big)\sin^2\theta d\chi  ~~~~~~~~~~~~~~~~~~~~~~~~~~~~~~~~~~~~~~~~~~~~~~~~~~~~~~~~~~~~~~~~~~~~~~~~~~~~~~~~~~~\\
%~~~~~~~~~~~~~~~~~~~~~~~-\Big(f_3f_5(f_8+\xi f_7)+\zeta^2 f_2 \Big(f_3f_8+f_5f_6(f_6f_8-f_7)\Big)\sin^2\theta \Big)\sin\theta d\theta \bigg] \wedge d\phi + \gamma d\chi \wedge d\phi ,\\
 % C_2=-\frac{1}{\Xi_4}\sin\theta\bigg[ \zeta f_2(f_3+f_5f_6^2)\sin\theta d\chi - f_3f_5f_7 d\theta\bigg] \wedge d\phi,\\
  %   \Xi_4= f_3f_5+\zeta^2 f_2(f_3+f_5f_6^2)\sin^2\theta ,~~~~~~~~~~~~~~~~~~~  C_0= f_1   e^{-\frac{4}{3}\Phi_A}\Big(f_5f_6(1+\xi f_6)+\xi f_3\Big) .
  %\end{gathered}
 % \end{equation}
\end{itemize}
\end{itemize}

\subsection{Deriving the TST of NRSZ}
Let us now fix $\zeta=-\xi$ in \eqref{eqn:ATD1}, recalling that this is the first of the two necessary conditions to preserve $\mathcal{N}=1$ in the IIB theory (along with $\gamma=-1$). This of course derives a two parameter family of $\mathcal{N}=0$ backgrounds, promoting to the $\mathcal{N}=1$ solution in \eqref{eqn:IIBN=1-1} for $\gamma=-1$. However, let us for the moment keep $\gamma$ free and instead fix $\xi=0$, one then gets
\beq
\hspace{-0.5cm}
     \begin{aligned} 
&       ds_{10,st}^2= f_1^{\frac{3}{2}}f_5^{\frac{1}{2}}\bigg[4ds^2(\text{AdS}_5)+f_2d\theta^2+f_4(d\sigma^2+d\eta^2) + \frac{f_2f_3}{f_3 +\gamma^2 f_2\sin^2\theta}\sin^2\theta d\phi^2+\frac{(d\chi  -\gamma f_8  \sin\theta d\theta)^2}{f_1^3( f_3f_5  +\gamma^2\sin^2\theta f_2 f_5) }\bigg], \\[2mm]
&e^{2\Phi^B}=\frac{f_5}{ f_3 +\gamma^2 f_2\sin^2\theta },~~~~~~~~~~B_2=  \frac{  f_3  f_8 }{ f_3 +\gamma^2 f_2\sin^2\theta }     \sin\theta d\theta \wedge   d\phi  +\frac{\gamma f_2   }{f_3 +\gamma^2\sin^2\theta f_2} \sin^2\theta  d\chi  \wedge   d\phi  ,  \\[2mm]
&C_0=f_6 ,~~~~~~~~~~~~~~ 
  C_2=\frac{f_3f_7+\gamma^2f_2(f_7-f_6f_8)\sin^2\theta}{f_3 +\gamma^2 f_2\sin^2\theta} d\theta \wedge d\phi +\frac{\gamma f_2f_6}{f_3 +\gamma^2 f_2\sin^2\theta} \sin^2\theta   d\chi \wedge  d\phi .
  \end{aligned}
  \eeq
  As we will now demonstrate,  we have re-derived the TST background presented in \cite{Nunez:2019gbg}. Rewriting the GM warp factors in terms of the definitions used in \cite{Nunez:2019gbg} (which we label with a bar for clarity) 
  \begin{equation}
  \begin{gathered}
  f_3 =\frac{\bar{f}_4}{\bar{f}_1},~~~~~~~~~  f_2 =\frac{\bar{f}_3}{\bar{f}_1},~~~~~~~~~~~ f_5 =\frac{\bar{f}_8}{\bar{f}_1},~~~~~~~~~~~ f_4 =\frac{\bar{f}_2}{\bar{f}_1},~~~~~~~~~~~ f_8=\bar{f}_5,\\
   f_6=\bar{f}_6,~~~~~~~~~~~ f_7=\bar{f}_7,~~~~~~~~~~~  f_1^{\frac{3}{2}}  f_5^{\frac{1}{2}}=\kappa^2\Lambda^{\frac{1}{2}} = \kappa^2 \bar{f}_1,
   \end{gathered}
  \end{equation} 
   one arrives at 
            \begin{align}
 &       ds_{10,B}^2=\kappa^2\bigg[4\bar{f}_1ds^2(\text{AdS}_5)+ \bar{f}_3d\theta^2+ \bar{f}_2(d\sigma^2+d\eta^2)~~~~~~~~~~~~~~~~~~~~~~~~~~~~~~~~~~~~~~~~~ \nn\\[2mm]
 &  ~~~~~~~~~~~~~~~~~~~~~~~     +\frac{1}{\bar{f}_4+\gamma^2  \bar{f}_3 \sin^2\theta}\Big(\bar{f}_3 \bar{f}_4\sin^2\theta d\phi^2+\frac{1}{\kappa^4}(d\chi-\gamma\,\bar{f}_5 \sin\theta  d\theta )^2\Big)\bigg], \nn\\[2mm]
 &e^{2 \Phi^B}=\frac{\bar{f}_8}{\bar{f}_4+\gamma^2 \bar{f}_3\sin^2\theta},~~~~~~~~~~~~~
               B^B  =\frac{\gamma \bar{f}_3 \sin^2\theta}{\bar{f}_4+\gamma^2\bar{f}_3\sin^2\theta}( d\chi -\gamma \bar{f}_5\sin\theta d\theta)\wedge d\phi +\bar{f}_5\sin\theta   d\theta\wedge  d\phi  , \nn\\[2mm]
&C_0=\bar{f}_6,~~~~~~~~~~~
C_2=\bar{f}_7 \sin\theta d\theta \wedge d\phi + \frac{\gamma \bar{f}_6\bar{f}_3 \sin^2\theta}{\bar{f}_4+\gamma^2\bar{f}_3\sin^2\theta}( d\chi -\gamma \bar{f}_5\sin\theta d\theta)\wedge d\phi
    \end{align} 
with $(\theta =\bar{\chi},~\chi =\bar{\beta},~\phi =\bar{\xi})$. Here, we have jumped in at the `S' stage of their TST transformation, making a coordinate transformation in IIA before T-Dualising to IIB.
%\textcolor{red}{double check all this and explain why - ie it is mid way through the TST calculation.}
We can now say that the TST solution of \cite{Nunez:2019gbg} is in fact an $\mathcal{N}=1$ background when $\gamma=-1$, and SUSY broken otherwise. It is also a specific example of the three parameter family of solutions given in \eqref{eqn:ATD1} (with $\zeta=-\xi=0$). The G-Structure conditions for the $\gamma=-1$ solution are thus given in \eqref{eqn:IIBfamGs} (with $\xi=0$).

  \subsection*{T-Dualising the $\chi$ Reduction}
  
  In the $\chi$ reduction case, we will follow the same procedure as the previous subsection. Once again, in order to preserve supersymmetry under T-Duality, we must leave the $U(1)_r$ component intact. As we will see, only an ATD along $\beta$ (with $\zeta=-1$ and $\gamma=0$) will lead to a SUSY preserved solution. This is a little different to the $\beta$ reduction case, where one could perform an ATD along both $\chi$ and $\phi$ to derive an $\mathcal{N}=1$ background. This is a remnant of the GM $U(1)_r$ component being $\chi+\phi$ prior to the $SL(3,\mathds{R})$ transformation. We now effectively pick up where Appendix \ref{sec:DimRed} left off, and investigate the ATD for the $\chi$ reduction with $a\equiv \xi$ and $q \equiv \xi$, in turn.

%\subsubsection*{Performing an ATD long $\beta$}
\begin{itemize}
\item \underline{$a\equiv \xi$}\\
In this case, we have (from \eqref{eqn:U(1)})
  \begin{equation}
U(1)_r = (1+\zeta)\chi +\gamma \beta +\phi,
\end{equation}
 with $(p,b,u)=1,~(q,c,m)=0,~a\equiv \xi,s\equiv \zeta,v\equiv \gamma$.
 
 Using \eqref{eqn:betaredchitdual}, we now investigate an ATD along $\beta$ and $\phi$ in turn.
\begin{itemize}
\item \textbf{Performing an ATD long $\beta$}\\
Now performing an ATD along $\beta$, we get the following three-parameter family of solutions
     \begin{align}\label{eqn:ATD3}
   &    ds_{10,st}^2=e^{\frac{2}{3}\Phi_\mathcal{A}}f_1\bigg[4ds^2(\text{AdS}_5)+f_2d\theta^2+f_4(d\sigma^2+d\eta^2)+ \frac{1}{\hat{\Xi}}\bigg(f_2 f_3f_5 \sin^2\theta d\phi^2~~~ \nn\\[2mm]
 & ~~~~~~~~~~~~~~~~~~~~~~~~~~~~~~      +\frac{1}{f_1^3 }\Big(d\beta +\big((\zeta-\gamma\xi)f_8-\gamma f_7\big)\sin\theta  d\theta\Big)^2\bigg)\bigg] , \nn\\[2mm]
&e^{2\Phi_\mathcal{B}}=\frac{ 1}{\hat{\Xi}}\big(  f_5( f_6+\xi)^2+f_3 + \zeta^2f_2 \sin^2\theta\big)^2,~~~~~~~~~~~~~~    e^{\frac{4}{3}\Phi_\mathcal{A}}= f_1 \Big[  f_5( f_6+\xi)^2+f_3 + \zeta^2f_2 \sin^2\theta \Big],  \nn\\[2mm]
&B_2= \sin\theta \frac{1}{\hat{\Xi}} \bigg[f_2\Big(\gamma f_3 +f_5(f_6+\xi)(\gamma f_6+\gamma\xi-\zeta)\Big)\sin\theta d\beta      \nn\\[2mm]
&  ~~~+ \Big[f_3f_5(f_7+\xi f_8)+\zeta f_2\Big(\gamma f_3f_8+f_5(f_6f_8-f_7)(\gamma f_6+\gamma\xi-\zeta)\Big)\sin^2\theta\Big] d\theta\bigg] \wedge d\phi , \nn\\[2mm]
&  C_2=  \sin\theta \frac{1}{\hat{\Xi}}\bigg[f_2f_5(\gamma f_6+\gamma\xi-\zeta)\sin\theta d\beta    \nn\\[2mm]
& ~~~~~~~~~~ +\Big[f_3f_5f_8 +\gamma f_2 \Big(\gamma f_3f_8+f_5(f_6f_8-f_7)(\gamma f_6+\gamma\xi-\zeta)\Big)\sin^2\theta\Big]d\theta\bigg] \wedge d\phi , \nn\\[2mm]
 &     C_0=f_1    e^{-\frac{4}{3}\Phi_\mathcal{A}}\Big( f_5 (f_6+\xi) +\gamma \zeta f_2\sin^2\theta\Big) ,~~~~~   \hat{\Xi}=f_3f_5+f_2\big(\gamma^2f_3 +f_5(\gamma f_6+\gamma\xi-\zeta)^2\big)\sin^2\theta.
  \end{align}

Once again we find that one can map this solution to \eqref{eqn:ATD1} via the transformations given in \eqref{eqn:Transformation} (with $k_1=\frac{1}{\xi},k_2=\xi^{\frac{5}{2}}$)
 \beq\label{eqn:IIBtrans2}
 \begin{gathered}
g_{MN}\rightarrow \frac{1}{\xi} g_{MN},~~~~~~~~~ e^{2\Phi_B}\rightarrow \frac{1}{\xi^4} e^{2\Phi_B},~~~~~~~~~C_0\rightarrow \xi^2 C_0,~~~~~~~C_2\rightarrow \xi C_2,~~~~~~~B_2\rightarrow \frac{1}{\xi}  B_2,\\
B_2\rightarrow- B_2,~~~~~~~~~~~~~C_0\rightarrow -C_0+\xi,~~~~~~~~~~~C_2\rightarrow C_2 - \xi B_2,\\
 \phi\rightarrow -\phi ,~~~~~~\beta \rightarrow - \chi,~~~~~~~~~~~\zeta \rightarrow \zeta \xi,~~~~\gamma\rightarrow \zeta -\frac{\gamma}{\xi},~~~~\xi\rightarrow \frac{1}{\xi}.
 \end{gathered}
 \eeq
 It does not appear possible to map this to \eqref{eqn:ATD2}. We therefore argue that the $\xi=0$ solution is new and unique - given in \eqref{eqn:newIIBcase2}.

After fixing $(\zeta=-1,\gamma=0)$, one derives the following $\mathcal{N}=1$ solution %given in equation \eqref{eqn:IIBN=1-2}.

 %  \subsubsection{$\mathcal{N}=1$ Type IIB - Background 2}
  
     \begin{align}\label{eqn:IIBN=1-2}
&       ds_{10,st}^2=e^{\frac{2}{3}\Phi_A}f_1\bigg[4ds^2(\text{AdS}_5)+f_2d\theta^2+f_4(d\sigma^2+d\eta^2)+ \frac{f_2 f_3}{f_3+f_2\sin^2\theta} \sin^2\theta d\phi^2 \nn\\[2mm]
 &   ~~~~~~~~~~~~~~~~~~~~~~~~   +\frac{(d\beta -f_8\sin\theta  d\theta)^2}{f_1^3(f_3f_5  +\sin^2\theta f_2   f_5) }\bigg],~~~~~~~~~~~~~~
       e^{\frac{4}{3}\Phi_A}= f_1 \Big[  f_5( f_6+\xi)^2+f_3 + f_2 \sin^2\theta \Big], \nn\\[2mm]
&e^{2\Phi^B}=\frac{ \big(  f_5( f_6+\xi)^2+f_3 +  \sin^2\theta f_2\big)^2}{ f_3f_5  +\sin^2\theta f_2   f_5},~~~~~~~~~~~~~~  C_0=\frac{  f_5 (f_6+\xi) }{  f_5( f_6+\xi)^2+f_3 +  \sin^2\theta f_2}, \nn\\[2mm]
&B_2=\frac{\sin\theta}{f_3+f_2\sin^2\theta} \bigg(f_2(f_6+\xi)\sin\theta d\beta + \Big(f_3(f_7+\xi f_8)+f_2(f_7-f_6f_8)\sin^2\theta\Big) d\theta\bigg) \wedge d\phi , \nn\\[2mm]
&  C_2=\frac{\sin\theta}{f_3+f_2\sin^2\theta}\Big(f_2\sin\theta d\beta +f_3f_8 d\theta\Big) \wedge d\phi .
  \end{align}
  
  Using the transformations given in \eqref{eqn:Transformation}, with $k_1=\xi^{-1},~k_2=\xi^{\frac{5}{2}}$ 
        \begin{equation}\label{eqn:N=1transforms}
    g_{MN} \rightarrow \frac{1}{\xi}\, g_{MN},~~~~~~~~~~~~~B_2 \rightarrow  \frac{1}{\xi}\,  B_2~~~~~~~~~C_0 \rightarrow \xi^2\, C_0,~~~~~~~~~~~~C_2 \rightarrow \xi \,C_2~~~~~~~~~e^{2\Phi} \rightarrow \frac{1}{\xi^4}e^{2\Phi},
    \end{equation}
  followed by
  \beq\label{eqn:N=1transforms2}
  C_0\rightarrow -C_0+\xi,~~~~~~~~~~~~~C_2\rightarrow -C_2 + \xi B_2,~~~~~~~~~\xi\rightarrow \frac{1}{\xi},~~~~~~~~~~~~\beta\rightarrow -\chi,
  \eeq
one maps to \eqref{eqn:IIBN=1-1}. Again, we argue that the $\xi=0$ case here is new and unique - given in \eqref{eqn:uniqueN1IIB}.

\item \textbf{Performing an ATD long $\phi$}\\
Performing an ATD along $\phi$ leads to the following $\mathcal{N}=0$ family of solutions
%\subsubsection*{ATD along $\phi$ of the $\chi$ reduction}
%Performing an ATD along $\phi$ of the $\chi$ reduction leads to the $\mathcal{N}=0$ family of solutions
     \begin{align}\label{eqn:chiredphiatd}
&       ds_{10,st}^2=e^{\frac{2}{3}\Phi_A}f_1\bigg[4ds^2(\text{AdS}_5)+f_2d\theta^2+f_4(d\sigma^2+d\eta^2)~~~~~~~~~~~~~~~~~~~~~~~~~~~~~~~~~~~~~~~~~~~~~~~~~~~~~~~~~~~~~~~~~~~~~~\nn\\[2mm]
&       +\frac{1}{\hat{\Xi}}\bigg( f_2f_3f_5 \sin^2\theta  d\beta^2
       +\frac{1}{f_1^3 } \Big(d\phi -(f_7+\xi f_8)\sin\theta  d\theta\Big)^2 \bigg)\bigg] , \\[2mm]
&  e^{2\Phi^B}=\frac{1 }{\hat{\Xi}}\big(  f_5( f_6+\xi)^2+f_3 + \zeta^2f_2 \sin^2\theta\big)^2,~~~~~~~~~~~~~~     e^{\frac{4}{3}\Phi_A}= f_1 \Big[  f_5( f_6+\xi)^2+f_3 + \zeta^2f_2 \sin^2\theta \Big],\nn\\[2mm]
%B_2=\frac{1}{f_5(f_6+\xi)^2+f_3} \bigg[\Big(f_5(f_6+\xi)(\gamma f_6+\gamma\xi-\zeta)+\gamma f_3\Big)d\phi-\zeta \Big(f_3f_8 +f_5(f_6f_8-f_7)(f_6+\xi)\Big)\sin\theta d\theta \bigg] \wedge d\beta ,\\
&B_2=-\frac{ \zeta }{\hat{\Xi}} f_2\sin^2\theta \bigg[ f_5(f_6+\xi)d\phi+\Big(f_3f_8 +f_5(f_6f_8-f_7)(f_6+\xi)\Big)\sin\theta d\theta \bigg] \wedge d\beta + \gamma d\phi \wedge d\beta ,\nn\\[2mm]
 % C_2=\frac{-1}{f_5(f_6+\xi)^2+f_3}\bigg[ f_5(f_6+\xi)d\phi + \Big(f_3f_8+f_5(f_6f_8-f_7)(f_6+\xi)\Big)\sin\theta d\theta \bigg] \wedge d\beta 
& C_2= -\frac{\gamma}{\zeta} d\phi\wedge d\beta,~~~~~~~~~~~  C_0=   \zeta f_1 f_2  e^{-\frac{4}{3}\Phi_A} \sin^2\theta -\frac{1}{\zeta} ,~~~~~~~~~\hat{\Xi} = f_2\big(f_5(f_6+\xi)^2+f_3\big) \sin^2\theta,\nn
  \end{align}
  where we can see that $\gamma$ only emerges as a large gauge transformation of $B_2$. We have introduced a gauge transformation into $C_0$ to simplify $C_2$. Once again, we refrain from including this in the main body because typically, analogous T-Duals along a $U(1)$ of an $S^2$ lead to singularities in the dual description.
% given in \eqref{eqn:chiredphiatd}.
%\textcolor{red}{could do with verifying this on mathematica and MAP TO PREVIOUS CASES}

 \end{itemize}

\item \underline{$q\equiv \xi$}

We now have (from \eqref{eqn:U(1)})
  \begin{equation}
U(1)_r= (1+\zeta)\chi +(\xi+\gamma)\beta +\phi,
\end{equation}
with $(p,b,u)=1,~(a,c,m)=0,~q\equiv \xi,s\equiv \zeta,v\equiv \gamma$.

Using \eqref{eqn:betaredchitdual}, we once again investigate an ATD along $\beta$ and $\phi$ in turn.

\begin{itemize}
\item \textbf{Performing an ATD long $\beta$}\\
Performing an ATD along $\beta$ calculates a background which is derived from \eqref{eqn:ATD3} by fixing $\xi\rightarrow 0,\gamma\rightarrow \gamma-\zeta \xi$, before applying the following gauge transformations
\beq
C_0\rightarrow C_0+\xi,~~~~~~~~~C_2\rightarrow C_2+\xi B_2.
\eeq

One then re-derives the existing $\mathcal{N}=1$ solution.

\item \textbf{Performing an ATD long $\phi$}

Here one derives a solution which matches \eqref{eqn:chiredphiatd}  with $\xi=0$, followed by the following gauge transformations
\beq
C_2\rightarrow C_2-\xi d\phi\wedge d\beta,~~~~~~~~~B_2\rightarrow B_2 -\zeta\xi d\phi\wedge d\beta.
\eeq

\end{itemize}

\end{itemize}

\newpage
  \subsection*{T-Dualising the $\phi$ Reduction}
This subsection will largely mirror the previous one, following the same logic. Once again, one must perform the ATD along $\beta$ in order to preserve SUSY.
\begin{itemize}

\item  \underline{$c\equiv \xi$}\\
We now have 
\beq
U(1)_r = \chi + \gamma\beta +(\zeta+1)\phi,
\eeq
 with $(p,b,u)=1,~(s,a,v)=0,~c\equiv \xi,m\equiv \zeta,q\equiv \gamma$.
 
  \begin{itemize}
 \item \textbf{Performing an ATD along $\beta$}\\
 One now derives the following three parameter family of solutions
 \footnote{Fixing $(\gamma=0,\zeta=-1)$ derives an $\mathcal{N}=1$ solution  which re-derives \eqref{eqn:IIBN=1-2} via the following transformations
  \begin{equation*}
  \begin{gathered}
  C_2\rightarrow C_2-d\beta \wedge d\chi,~~~~~~~~~~~~B_2\rightarrow-B_2+\xi d\beta \wedge d\chi,~~~~~~C_0\rightarrow-C_0,\\
  \chi\rightarrow \phi,~~~~~~\beta\rightarrow -\beta,~~~~~~\xi\rightarrow-\xi.
  \end{gathered}
  \end{equation*}\label{footnote_ref1}}
       \begin{align}\label{eqn:ATD6}
 &      ds_{10,st}^2=e^{\frac{2}{3}\Phi_\mathcal{A}}f_1\bigg[4ds^2(\text{AdS}_5)+f_2d\theta^2+f_4(d\sigma^2+d\eta^2)~~~~~~~~~~~~~~~~~~~~~~~~~~~~~~~~~~~~~~~~~~~~~~~~~~~~~~~~~   \nn\\[2mm]
  &   ~~~~~~~~~~~~~~~~~~~~~~~~~~~~~~~~~~~~  +\frac{1}{\hat{\Xi}}\bigg(  f_2 f_3 f_5 \sin^2\theta d\chi^2  +\frac{1}{f_1^3  }\Big(d\beta +(f_8+\gamma f_7)\sin\theta  d\theta\Big)^2\bigg)\bigg] ,   \nn\\[2mm]
  &    \hat{\Xi}= (\zeta-\gamma\xi)^2f_3f_5  +f_2  \Big( \gamma^2 f_3+f_5(1+\gamma f_6)^2\Big)\sin^2\theta ,   \nn\\[2mm]
&e^{2\Phi_\mathcal{B}}=\frac{1}{\hat{\Xi}}\big(   f_5( \zeta f_6+\xi)^2+\zeta^2 f_3 + f_2 \sin^2\theta \big)^2 ,~~~~~~~~~~        e^{\frac{4}{3}\Phi_\mathcal{A}}= f_1 \Big[  f_5( \zeta f_6+\xi)^2+\zeta^2 f_3 + f_2 \sin^2\theta \Big],   \nn\\[2mm]
&B_2=\frac{1}{\hat{\Xi}} \bigg[\Big(\xi(\gamma\xi-\zeta)f_3f_5 +f_2(\gamma f_3+f_5f_6(1+\gamma f_6))\sin^2\theta\Big) d\beta~~~~~~~~~~~~~~~~~~~~~~~~~~~~~~~~~~~~~~~~~~~~~~~~~~~~~~~~~~    \nn\\[2mm]
&~+ \sin\theta\Big((\gamma\xi-\zeta)f_3f_5(\zeta f_7+\xi f_8) -f_2 \Big(-\gamma f_3f_8 +f_5(f_7-f_6f_8)(1+\gamma f_6)\Big)\sin^2\theta\Big) d\theta\bigg] \wedge d\chi ,   \nn\\[2mm]
 & C_0= f_1   e^{-\frac{4}{3}\Phi_\mathcal{A}} \Big( \gamma\zeta f_3+f_5(1+\gamma f_6)(\zeta f_6+\xi) \Big) ,   \nn\\[2mm]
 &  C_2=\frac{(\gamma\xi-\zeta)}{\hat{\Xi}}f_3f_5 \Big( d\beta +(f_8+\gamma f_7) \sin\theta d\theta\Big) \wedge d\chi .
  \end{align}
  
  Once again we find that one can map this solution to \eqref{eqn:ATD1} via the transformations given in \eqref{eqn:Transformation} (with $k_1=\frac{1}{\xi},k_2=\frac{\xi^{\frac{5}{2}}}{\zeta-\gamma \xi}$)
 \beq\label{eqn:IIBtrans3}
 \begin{gathered}
g_{MN}\rightarrow \frac{1}{\xi} g_{MN},~~~~~~~~~ e^{2\Phi_B}\rightarrow \frac{(\zeta-\gamma \xi)^2}{\xi^4} e^{2\Phi_B},~~~~~~~~~C_0\rightarrow \frac{\xi^2}{\zeta-\gamma\xi} C_0,\\
C_2\rightarrow \frac{\xi}{\zeta-\gamma\xi} C_2,~~~~~~~B_2\rightarrow \frac{1}{\xi}  B_2,\\
 C_0\rightarrow -C_0+\frac{\xi}{\zeta-\gamma\xi},~~~~~~~~~~~C_2\rightarrow -C_2 + \frac{\xi}{\zeta-\gamma\xi} B_2,\\
 \chi\rightarrow -(\zeta-\gamma\xi)\phi ,~~~~~~\beta \rightarrow -(\zeta-\gamma\xi) \chi,~~~~~~~~~~~\zeta \rightarrow \frac{\bar{\xi}}{\bar{\zeta}},~~~~\xi\rightarrow \frac{1}{\bar{\zeta}},~~~~\gamma\rightarrow- \frac{1}{\bar{\gamma}} (\bar{\zeta} - \bar{\gamma}\bar{\xi}).
 \end{gathered}
 \eeq
 
 Here we see that $1/\xi,1/\zeta,1/\gamma,1/(\zeta-\gamma\xi)$ are required. We therefore argue that fixing each of these to zero in turn derives new and unique solutions, given in \eqref{eqn:newIIBcase3}, \eqref{eqn:newIIBcase4}, \eqref{eqn:newIIBcase5} and \eqref{eqn:newIIBcase6}, respectively.  It does not appear possible to map this solution to \eqref{eqn:ATD2} or \eqref{eqn:ATD3}.
  
%  then fixing $(\gamma=0,\zeta=-1)$ derives an $\mathcal{N}=1$ solution
%      \begin{equation}
 %      \hspace{-2.5cm}
 % \begin{gathered}
  %     ds_{10,st}^2=e^{\frac{2}{3}\Phi_A}f_1\bigg[4ds^2(\text{AdS}_5)+f_2d\theta^2+f_4(d\sigma^2+d\eta^2)+ \frac{f_2 f_3}{f_3+f_2\sin^2\theta} \sin^2\theta d\chi^2+\frac{(d\beta +f_8\sin\theta  d\theta)^2}{f_1^3(f_3f_5  +\sin^2\theta f_2   f_5) }\bigg] \\
   %    e^{\frac{4}{3}\Phi_A}= f_1 \Big[  f_5( f_6-\xi)^2+f_3 + f_2 \sin^2\theta \Big],\\
%e^{2\Phi^B}=\frac{ \big(  f_5( f_6-\xi)^2+f_3 + f_2 \sin^2\theta \big)^2}{ f_3f_5  +\sin^2\theta f_2   f_5},~~~~~~~~~~~~~~  C_0=\frac{  f_5 (-f_6+\xi) }{  f_5( f_6-\xi)^2+f_3 +  f_2 \sin^2\theta},\\
%B_2=\frac{1}{f_3+f_2\sin^2\theta} \bigg((\xi f_3+f_2f_6\sin^2\theta) d\beta + \sin\theta\Big(-f_3f_7+\xi f_3f_8-f_2(f_7-f_6f_8)\sin^2\theta\Big) d\theta\bigg) \wedge d\chi ,\\
%  C_2=\frac{f_3}{f_3+f_2\sin^2\theta}\Big( d\beta +f_8 \sin\theta d\theta\Big) \wedge d\chi .
 % \end{gathered}
 % \end{equation}
%  which maps to  equation \eqref{eqn:IIBN=1-2} under the following transformations
  
%  \beq
%  \begin{gathered}
%  C_2\rightarrow C_2-d\beta \wedge d\chi,~~~~~~~~~~~~B_2\rightarrow-B_2+\xi d\beta \wedge d\chi,~~~~~~C_0\rightarrow-C_0,\\
 % \chi\rightarrow \phi,~~~~~~\beta\rightarrow -\beta,~~~~~~\xi\rightarrow-\xi.
 % \end{gathered}
% \eeq

 \item \textbf{Performing an ATD long $\chi$}\\
 Here we have the $\mathcal{N}=0$ solution given in \eqref{eqn:ATD2param6}.
 
 \end{itemize}

\item \underline{$v\equiv \xi$}\\
We now have (from \eqref{eqn:U(1)})
\beq
U(1)_r = \chi +(\gamma+\xi)\beta +(\zeta+1)\phi,
\eeq
 with $(p,b,u)=1,~(s,a,c)=0,~v\equiv \xi,m\equiv \zeta,q\equiv \gamma$.
 \begin{itemize}
 \item \textbf{Performing an ATD along $\beta$}\\
Here one computes a background which is derived via \eqref{eqn:ATD6} by fixing $\xi\rightarrow 0,\gamma\rightarrow \gamma-\zeta \xi$, before applying the following gauge transformations
 \beq
 C_0\rightarrow C_0+\xi,~~~~~~~~~~~~C_2\rightarrow C_2 +\xi B_2.
 \eeq

 Then, fixing $(\gamma=-\xi,~\zeta=-1)$ we derive another $\mathcal{N}=1$ solution
%      \begin{equation}
 %       \hspace{-2.5cm}
 % \begin{gathered}
 %      ds_{10,st}^2=e^{\frac{2}{3}\Phi_A}f_1\bigg[4ds^2(\text{AdS}_5)+f_2d\theta^2+f_4(d\sigma^2+d\eta^2)+ \frac{f_2 f_3}{ f_3+f_2\sin^2\theta} \sin^2\theta d\chi^2+\frac{(d\beta +f_8\sin\theta  d\theta)^2}{f_1^3( f_3f_5  +\sin^2\theta f_2   f_5) }\bigg] \\
  %     e^{\frac{4}{3}\Phi_A}= f_1 \Big[  f_5f_6^2+f_3 + f_2 \sin^2\theta \Big],\\
%e^{2\Phi^B}=\frac{\big(  f_5f_6^2+f_3 + f_2 \sin^2\theta \big)^2}{ f_3f_5  +\sin^2\theta f_2   f_5},~~~~~~~~~~~~~~  C_0=\frac{ f_5f_6(-1+\xi f_6) +\xi f_3 +\xi f_2 \sin^2\theta}{  f_5f_6^2+f_3 + f_2 \sin^2\theta},\\
%B_2=\frac{\sin\theta}{f_3+f_2\sin^2\theta} \bigg(f_2f_6\sin\theta d\beta + \Big(f_3f_7+f_2(f_7-f_6f_8)\sin^2\theta\Big) d\theta\bigg) \wedge d\chi ,\\
%  C_2=\frac{1}{f_3+f_2\sin^2\theta}\Big[(f_3+\xi f_2 f_6\sin^2\theta) d\beta +\sin\theta \Big(f_3(f_8-\xi f_7) -\xi f_2(f_7-f_6f_8)\sin^2\theta\Big)d\theta\Big] \wedge d\chi .
 % \end{gathered}
 % \end{equation}
  which once again maps to the solutions already given.\footnote{Via 
  %maps to equation \eqref{eqn:IIBN=1-3} under the following transformations
\beq
 \begin{gathered}
  C_2\rightarrow C_2-d\beta\wedge d\chi,~~~~~~~B_2\rightarrow-B_2,~~~~~~~~C_0\rightarrow -C_0,\\
  \phi\rightarrow \chi,~~~~~~~\chi\rightarrow \phi,~~~~~~\beta\rightarrow-\beta,~~~~~~~~\xi\rightarrow-\xi.
  \end{gathered}
  \eeq
 }
 \item \textbf{Performing an ATD along $\chi$}\\
 Performing an ATD along $\chi$ derives a solution where $\gamma$ and $\xi$ are carried through the calculation as gauge transformations of $B_2$ and $C_2$. The background can be derived from \eqref{eqn:ATD2param6} by fixing $\xi=0$ %derives equation \eqref{eqn:phiredchiatd1} - after 
before applying the following gauge transformations
  \beq
  B_2\rightarrow B_2-\zeta \xi d\chi\wedge d\beta,~~~~~~~~~~C_2\rightarrow C_2+\xi d\beta \wedge d\chi.
  \eeq

%     \begin{equation}\label{eqn:phiredchiatd1}
 %       \hspace{-2.5cm}
 % \begin{gathered}
 %      ds_{10,st}^2=e^{\frac{2}{3}\Phi_\mathcal{A}}f_1\bigg[4ds^2(\text{AdS}_5)+f_2d\theta^2+f_4(d\sigma^2+d\eta^2)+\frac{1}{\Xi_9}\bigg( f_2f_3 f_5\sin^2\theta  d\beta^2+\frac{1}{f_1^3 }(d\chi +f_7\sin\theta  d\theta)^2\bigg) \bigg] \\
%e^{2\Phi_\mathcal{B}}=\frac{1}{\Xi_9}\big(\zeta^2  f_5f_6^2+\zeta^2f_3 + f_2 \sin^2\theta \big)^2,~~~~~~~~~~~~~~        e^{\frac{4}{3}\Phi_\mathcal{A}}= f_1 \Big[\zeta^2  f_5f_6^2+\zeta^2f_3 + f_2 \sin^2\theta \Big],\\
% C_0= \zeta f_1  e^{-\frac{4}{3}\Phi_\mathcal{A}} ( f_3+f_5f_6^2)  ,~~~~~~~~~~~~~~~~~\Xi_9 =f_2( f_3+f_5f_6^2)\sin^2\theta \\
%B_2=\frac{1}{\Xi_9} f_2\sin^2\theta \bigg(f_5f_6 d\chi - \Big(f_3f_8+f_5f_6(f_6f_8-f_7)\Big) \sin\theta d\theta\bigg) \wedge d\beta +(\gamma-\zeta \xi)d\chi\wedge d\beta,\\
%  C_2=\xi d\beta \wedge d\chi .
 % \end{gathered}
%  \end{equation}
% \textcolor{red}{does this match another background?}
 \end{itemize}

\end{itemize}

\section{TST deformations of Gaiotto-Maldacena}\label{sec:TSTGM}

\begin{figure}[h!]
\begin{center} 
 \begin{tikzpicture}

\draw (0,2.4) node[above] {\textbf{GM}};

\draw (0.2,1) node[right] {$\phi_3$};

\draw (0,-0.5) node[below] {\textbf{IIA}};

\draw (2.2,-0.5) node[above] {\textbf{T}};
\draw (2.2,-1) node[below] {ATD $\phi_1$};

\draw (4.5,-0.5) node[below] {\textbf{IIB}};

\draw (6.7,-0.5) node[above] {\textbf{S}};
\draw (6.7,-1) node[below] { $\phi_2\rightarrow \phi_2+\gamma\phi_1$};

\draw (9,-0.5) node[below] {\textbf{IIB}};

\draw (11.2,-0.5) node[above] {\textbf{T}};
\draw (11.2,-1) node[below] {ATD $\phi_1$};

\draw (13.5,-0.5) node[below] {\textbf{IIA}};

\draw (13.5,2.4) node[above] {\textbf{TST deformed GM} };

%\draw (6.7,-2.5) node[below] {\textbf{$\phi_1$ TST}};

\begin{scope}[ every node/.style={sloped,allow upside down}]

 \draw ((0,2)-- node {\midarrow} (0,0);
  \draw ((0.75,-0.75)-- node {\midarrow} (3.75,-0.75);
     \draw ((5.25,-0.75)-- node {\midarrow} (8.25,-0.75);
          \draw ((9.75,-0.75)-- node {\midarrow} (12.75,-0.75);
          
           \draw ((13.5,0)-- node {\midarrow} (13.5,2);
           \draw (13.3,1) node[left] {$\phi_3$};

\end{scope}

% \draw [decorate,decoration = {brace,amplitude=8pt}] (13.5,-2) --  (0,-2) ;

% \draw (6.7,-4.5) node[below] {$~$};
 
\end{tikzpicture}
\end{center}
\caption{Dimensionally reducing along $\phi_3$, performing a TST along $\phi_1$, before uplifting along $\phi_3$.}
\label{fig:GMTSTtransformations }
\end{figure}
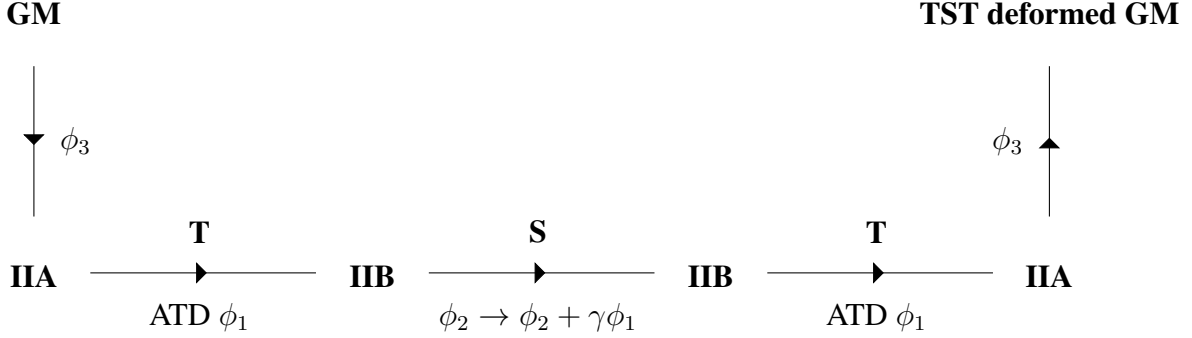

   Using the TST formulas given in \eqref{eqn:TSTgenlgeneral} and \eqref{eqn:TSTgenl}, one can simply uplift along $\phi_3$ to get deformations of the GM background up in 11 dimensions, see Figure \ref{fig:GMTSTtransformations }. 
Using the reduction formula given in \eqref{eqn:DimRed} (with $\psi\equiv \phi_3$), we uplift \eqref{eqn:TSTgenl} as follows
       \begin{align}
 &      ds^2= \bigg(\frac{\delta_2 +\gamma_1^2 \delta_1}{\delta_2}\bigg)^{\frac{1}{3}}e^{-\frac{2}{3}\Phi_A}\Bigg[ds_8^2 +\frac{1}{\delta_2 + \gamma_1^2 \delta_1}\bigg[\delta_1 \delta_2(d\phi_2 -\gamma_1 B_{2,\phi_1}d\theta)^2~~~~~~~~~~~~~~~~~~~~~~~~~~~~~~~~~~~~~~~~~~~~~~~~~~~~~~\nn\\[2mm]
  & ~~~~~~~~~~~~~    + \bigg(d\phi_1 +\frac{1}{2}\frac{h_{\phi_1\phi_2}}{h_{\phi_1}}d\phi_2 +\gamma_1 \Big(B_{2,\phi_2}-\frac{1}{2}\frac{h_{\phi_1\phi_2}}{h_{\phi_1}}B_{2,\phi_1}\Big)d\theta\bigg)^2\,\bigg]\Bigg]\nn\\[2mm]
 &   ~~~~~~~~~~~~~   + \bigg(\frac{\delta_2}{\delta_2 +\gamma_1^2 \delta_1}\bigg)^{\frac{2}{3}}e^{\frac{4}{3}\Phi_A} \bigg(d\phi_3 + C_{1,\phi_2}d\phi_2 +C_{1,\phi_1}d\phi_1 +\gamma_1 (C_{3,\phi_1\phi_2} + C_{1,\phi_1}B_{2,\phi_2} -C_{1,\phi_2}B_{2,\phi_1})d\theta\bigg)^2,\nn\\[2mm]
&C_3=\frac{\delta_2}{\delta_2+\gamma_1^2 \delta_1}C_{3,\phi_1\phi_2}d\theta \wedge d\phi_2 \wedge d\phi_1 + \frac{\delta_2}{\delta_2+\gamma_1^2\delta_1}(B_{2,\phi_2}d\phi_2 +B_{2,\phi_1}d\phi_1)\wedge d\theta \wedge d\phi_3\nn\\[2mm]
&~~~~~~~~-\frac{\gamma_1 \delta_1}{\delta_2+\gamma_1^2\delta_1}d\phi_2 \wedge d\phi_1\wedge d\phi_3,\nn
 ~~~~~~~~~~~~~~~~~~~~~~~     \delta_1=\Gamma  \bigg(h_{\phi_2}-\frac{1}{4}\frac{h_{\phi_1\phi_2}^2}{h_{\phi_1}}\bigg),%~~~~~~~~~~\beta=\frac{e^{\frac{2}{3}\Phi_A}}{f_1^2h_1 },
       ~~~~~~~~~~\delta_2=\frac{1}{h_{\phi_1} \Gamma }, 
  \end{align}
which is sufficient for our purposes, of course the more general form given in \eqref{eqn:TSTgenlgeneral} follows in the same manner.

For the GM background, we have six iterations corresponding to $(\phi_1,\phi_2,\phi_3)=(\chi,\phi,\beta)$ etc. This is the choice which we will use in the following analysis. In fact, this will be sufficient, as it seems each of the six alternatives will lead to the same background (up to relabelling $\gamma\rightarrow-\gamma$). 

Explicitly, we have
\begin{align}
&ds_8^2= f_1 e^{\frac{2}{3}\Phi_A}  \Big[4ds^2(AdS_5)+f_2d\theta^2 +f_4(d\sigma^2+d\eta^2)\Big],\nn\\[2mm]
&e^{\frac{4}{3}\Phi_A} = f_1f_5,~~~~~~~~~~~~~~~~~~~~\Gamma = f_1 e^{\frac{2}{3}\Phi_A},~~~~~~~~~~~C_{1,\phi_1}=C_{1,\chi}=f_6,~~~~~~~~~~C_{1,\phi_2}=C_{1,\phi}=0,\nn\\[2mm]
&B_{2,\phi_1} = B_{2,\chi}=0  ,~~~~~~~~~~~~~~~
 B_{2,\phi_2} = B_{2,\phi}=-f_8\sin\theta,~~~~~~~~~~~~~C_{3,\phi_1\phi_2}=C_{3,\chi\phi} = f_7\sin\theta, \nn\\[2mm]
&h_{\phi_1} = h_\chi=f_3,~~~~~~~~~~~~~~~~~ h_{\phi_2} =h_\phi=f_2\sin^2\theta,~~~~~~~~ h_{\phi_1\phi_2} =h_{\chi\phi}=0,
\end{align}
 giving
    \begin{align}
    &   ds^2= f_1 (1+\gamma^2 f_1^3f_2f_3f_5\sin^2\theta)^{\frac{1}{3}}  \Bigg[  4ds^2(AdS_5)+f_2d\theta^2 +f_4(d\sigma^2+d\eta^2)  \quad\quad\quad\quad\quad\quad\quad\quad\quad\quad\quad\quad\quad\quad\quad \nn\\[2mm]
 &      ~+ \frac{1}{1+\gamma^2 f_1^3f_2f_3f_5\sin^2\theta} \bigg[ f_2\sin^2\theta d\phi^2 +f_3 (d\chi-\gamma f_8 \sin\theta d\theta )^2  +f_5\Big( d\beta +f_6 d\chi +\gamma (f_7-f_6f_8)\sin\theta d\theta \Big)^2 \bigg]  \Bigg] ,\nn\\[2mm]
&C_3=\frac{1}{1+\gamma^2 f_1^3f_2f_3f_5\sin^2\theta}\bigg[ f_7\sin\theta d\theta \wedge d\phi \wedge d\chi + f_8\sin\theta d\theta \wedge d\phi \wedge  d\beta- \gamma   f_1^3f_2f_3f_5\sin^2\theta d\phi \wedge d\chi\wedge d\beta\bigg].
  \end{align}
  Following the $U(1)_r$ component through such a calculation shows that, at best, it is necessarily broken by the TST transformation prior to the uplift. Hence, $\gamma\neq 0$ breaks the $\mathcal{N}=2$ solution to $\mathcal{N}=0$. %We now briefly investigate the boundary.

     \section{Brane rotation interpretation}\label{sec:RotatingBranes}
     
A possible interpretation of broken quantization in the absence of an orbifold singularity is now presented. For the sake of this appendix, we assume the first gauge transformation of the D5 branes is chosen, leading to the charge given in \eqref{eqn:D5charge} and \eqref{eqn:D7charge}. First define the `true charges' for $\xi=0$ in each interval 
\beq
\begin{aligned}
Q^T_{D5,k} &=   N_k,~~~~~~~~~~~~~~~~~Q^T_{D7,k+1} &=  2N_{k+1}-N_k -N_{k+2}.
\end{aligned}
\eeq
We can then interpret the charges \eqref{eqn:D5charge} as a projected value of the true charges following a rotation of the branes in each interval - with the magnitude of rotation depending on $\xi$. For the D5s, we then get\footnote{
Using the charges given in \eqref{eqn:N=0IIBcase1}, the magnitude of rotation in each interval is given by the following ratio
\beq
\frac{C_0}{C_{0_{\xi=0}}}\bigg|_{\sigma\rightarrow 0} = \frac{f_6(1+\xi f_6)+\xi \frac{f_3}{f_5}}{f_6\Big((1+\xi f_6)^2 +\xi^2 \frac{f_3}{f_5}\Big)}\bigg|_{\sigma\rightarrow 0} =  \frac{1}{1+\xi \mathcal{R}'(\eta)}.
\eeq
\indent Note also, $\frac{e^{ \frac{\Phi}{2}_{\xi=0}}}{e^{ \frac{\Phi}{2}}}\bigg|_{\sigma\rightarrow 0}$ leads to the same result.} 
\begin{equation}\label{eqn:D5charge2}
Q_{D5,k} = Q_{D5,k}^T \cos\varphi^k,~~~~~~~~~~~~~~~~\cos\varphi^k=\frac{1}{l_k},~~~~~~~~~l_k=1+\xi (N_{k+1}-N_{k}).
\end{equation}
In the case of the D7s, they are rotated by the same amount as the D5 in each interval (namely, $\varphi^k$ in some direction) plus an additional rotation in a plane perpendicular ($\varphi^{k+1}$), i.e.
\begin{equation}
\begin{gathered}
Q_{D7,k+1}=Q^T_{D7,k+1} \cos\varphi^k\cos\varphi^{k+1} .
\end{gathered}
\end{equation}

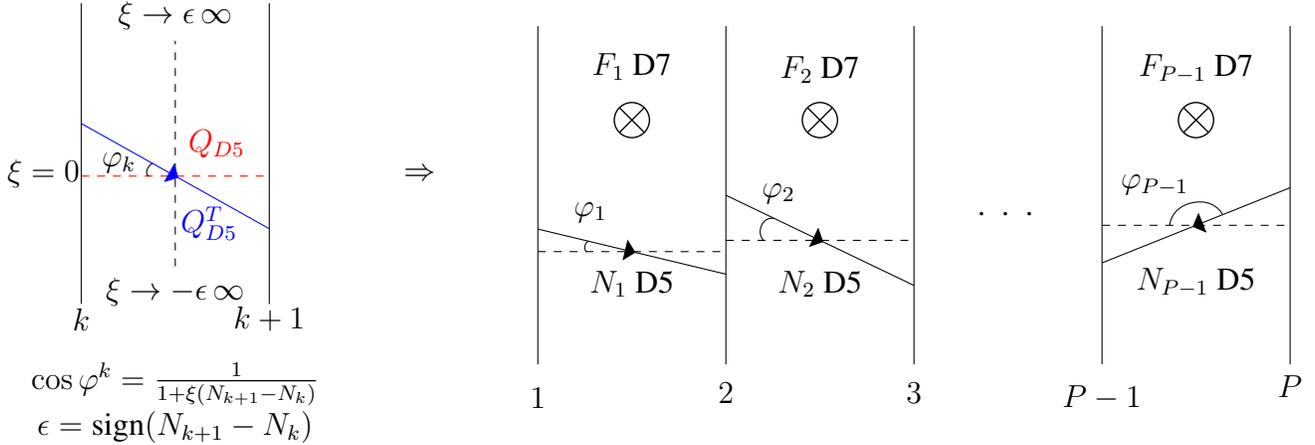
\begin{figure}[h!]
\centering
\begin{center}
\hspace*{-4cm}  
 \begin{minipage}{.4\textwidth}
 \begin{tikzpicture}
 %------ D6 branes --------

%\node[label=above:$F_1$ D6] at (1.25,1.5){\LARGE $\otimes$};

%-----------NS5 Branes-----------

\draw (0,-2) -- (0,2);
%\draw (0.25,-2) -- (-0.25,2.5);
\draw (0,-2.5) node[above] {$k$};
%\draw (0,-2.16) node[below] {$1$};
%
\draw(2.5,-2) -- (2.5,2);
%\draw (2.75,-2) -- (2.25,2.5);
\draw (2.5,-2.5) node[above] {$k+1$};
%\draw (2.5,-2.09) node[below] {$2$};

\draw[dashed](1.25,-1.5) -- (1.25,1.5);
\draw (1.25,-1.5) node[below] {$\xi \rightarrow -\epsilon\, \infty$};
\draw (1.25,1.5) node[above] {$\xi \rightarrow \epsilon\, \infty$};

%------------ D5 Branes ------------------

\draw[red,dashed] (0,-0.3) -- (2.5,-0.3);
%\draw[blue] (0,0.4) -- (2.5,-1);
\begin{scope}[ every node/.style={sloped,allow upside down}]
\draw[blue] (0,0.4) -- node{\midarrow}  (2.5,-1); 
\end{scope}

 \draw  plot [smooth, tension=1.5] coordinates {(0.9,-0.3) (0.875,-0.22) (0.95,-0.13)};

\draw (1.8,-0.2) node[above] {\textcolor{red}{$Q_{D5}$}};
\draw (1.7,-1.3) node[above] {\textcolor{blue}{$Q^T_{D5}$}};

\draw (0.5,-0.40) node[above] {$\varphi_k$};

%\draw (1.25,-3.5) node[above] {$\cos(\theta_k)=\frac{1}{1+\xi \mathcal{R}'(\eta)}$ }; %\bigg|_k^{k+1}
\draw (1.25,-3.5) node[above] {$\cos \varphi^k=\frac{1}{1+\xi (N_{k+1}-N_k)}$ }; %\bigg|_k^{k+1}
\draw (1.25,-4) node[above] {$\epsilon =\text{sign}(N_{k+1}-N_k)$ };

\draw (-0.5,-0.6) node[above] {$\xi=0$};

\draw (4.5,-0.5) node[above] {$\Rightarrow$};
 
    \end{tikzpicture}
  \end{minipage}%
  \begin{minipage}{.4\textwidth}
\begin{tikzpicture}

%------ D6 branes --------

\node[label=above:$F_1$ D7] at (1.25,1.25){\LARGE $\otimes$};
%\node[label=above:] at (2.5,1.5){\LARGE $\otimes$};
%
\node[label=above:] at (3.75,1.25){\LARGE $\otimes$};
 
\draw (3.75,1.6) node[above] {$F_2$ D7};
\node[label=above:] at (8.75,1.25){\LARGE $\otimes$};
 
\draw (8.75,1.6) node[above] {$F_{P-1}$ D7};

%-----------NS5 Branes-----------

\draw (0,-2) -- (0,2.5);
 
\draw (0,-2.16) node[below] {$1$};
\draw(2.5,-2) -- (2.5,2.5);
 
\draw (2.5,-2.09) node[below] {$2$};
\draw (5,-2) -- (5,2.5);
 
\draw (5,-2.1) node[below] {$3$};
\draw (7.5,-2) -- (7.5,2.5);
 
\draw (7.5,-2.13) node[below] {$P-1$};
\draw (10,-2) -- (10,2.5);
 
\draw (10,-2) node[below] {$P$};

%------------ D5 Branes ------------------

\draw[dashed] (0,-0.5) -- (2.5,-0.5);
%\draw (0,-0.2) -- (2.5,-0.8);
%
 
\draw[dashed] (2.5,-0.35) -- (5,-0.35);
%\draw (2.5,0.25) -- (5,-0.95);
%
\draw (5.7,0) node[right] {.~~.~~.};
 
\draw[dashed] (7.5,-0.15) -- (10,-0.15);
%\draw (7.5,-0.65) -- (10,0.35);

\begin{scope}[ every node/.style={sloped,allow upside down}]
\draw (0,-0.2) --node{\midarrow} (2.5,-0.8);
\draw (2.5,0.25) --node{\midarrow} (5,-0.95);
\draw (10,0.35) -- node{\midarrow}(7.5,-0.65);
\end{scope}

 \draw  plot [smooth, tension=1.5] coordinates {(0.65,-0.5) (0.625,-0.43) (0.675,-0.36)};
  \draw  plot [smooth, tension=1.5] coordinates {(3,-0.35) (2.975,-0.15) (3.136,-0.06)};
    \draw  plot [smooth, tension=1.5] coordinates {(8.4,-0.15) (8.7,0.15) (9.1,0)};

\draw (1.25,-1.25) node[above] {$N_1$ D5};
\draw (3.75,-1.25) node[above] {$N_2$ D5};
\draw (8.75,-1.25) node[above] {$N_{P-1}$ D5};

\draw (0.7,-0.25) node[above] {$\varphi_1$};
\draw (3.2,0) node[above] {$\varphi_2$};
\draw (8.2,0.1) node[above] {$\varphi_{P-1}$};

\end{tikzpicture}
  \end{minipage}
\end{center}
\caption{ The D5 branes still have a charge $Q_{D5}^T$ but are rotated as a function of $\xi$. Hence, when calculating the charge along the
usual cycle, this value is distorted due to the rotation - meaning, the charge calculated, $Q_{D5}$, is just the quantity of charge projected onto
this cycle.}
\label{fig:D5rotations}
\end{figure}

It is worth noting that the NS5 branes are untouched. Notice from \eqref{eqn:D5charge2} that when $\cos\varphi^k=-1$, the D5 charge flips in sign, equivalent to a $\pi$ rotation - this is consistent with the flipping of orientation we're used to. The same is true in the D7 case when $ \cos\varphi^k\cos\varphi^{k+1} =-1$. In order to avoid divergences, the choice of Rank Function naturally constrains the possible choices of $\xi$. 
In each interval, each brane is rotated by a unique amount, dictated by $N_{k+1}-N_k$ and $N_{k+2}-N_{k+1}$. A Hannany-Witten diagram showing a schematic of D5 brane rotations is given in Figure \ref{fig:D5rotations}. The magnitude of rotation in each interval is highly constrained by the fact $(\xi,~N_k) \in \mathds{Z}$, and cannot be arbitrarily small. In the $\xi\rightarrow \infty$ limit, the D5 branes are rotated by $\frac{\pi}{2}$.

\newpage
  \section{Holographic Central Charge for Type IIB}\label{sec:HCC}
The general Holographic Central Charge, $c_{hol}$, is given by (see \cite{Henningson:1998gx,Macpherson:2014eza,Bea:2015fja,Klebanov:2007ws} for further details)
\begin{equation}
\begin{gathered}
ds^2=\alpha(\rho,\vv{\theta})\Big(dx_{1,d}^2+\beta(\rho)d\rho^2\Big)+g_{ij}(\rho,\vv{\theta})d\theta^id\theta^j,\\
c_{hol}=\frac{d^d}{G_N}\beta^{d/2}\frac{H^{\frac{2d+1}{2}}}{(H')^d},~~~~~~~~~~~~~~~~~~~~~~H=V_{int}^2,~~~~~~~~~~~~~~~~~~~~~~~~V_{int}=\int d\vv{\theta}\sqrt{\text{det}[g_{ij}]e^{-4\Phi}\alpha^d},
\end{gathered}
\end{equation}
with $G_N=8\pi^6 \alpha'^4g_s^2=8\pi^6$, and $d=3$ for the present discussion.

We now calculate this quantity for the form of the IIB metrics given throughout this paper, which read in general 
  \begin{equation}
  \begin{gathered}
       ds_{10,B}^2= 4\rho^2 e^{\frac{2}{3}\Phi_\mathcal{A}}f_1 \Big(dx_{1,3}^2+\frac{1}{\rho^4}d\rho^2\Big)+e^{\frac{2}{3}\Phi_\mathcal{A}}f_1\bigg[f_2d\theta^2+f_4(d\sigma^2+d\eta^2)~~~~~~~~~~~~~~~~~~~~~~~~~~~~~~~~~~~~~~~~~~~~~~~~~~~~~~~~~~~~~~~~~\\
     ~~~~~~~~~~~~~  +\frac{1}{\Xi }\bigg(f_2f_3f_5\sin^2\theta\, d\phi_1^2+\frac{1}{f_1^3}\,\big[d\phi_2+ g(\eta,\sigma) \sin\theta  d\theta\big]^2\bigg)\bigg],~~~~~~~~~~~~~
  e^{2\Phi_\mathcal{B}}=\frac{1}{\Xi f_1^2}e^{\frac{8}{3}\Phi_\mathcal{A}} ,
     \end{gathered}
     \end{equation}
%     \begin{equation}
 %    \hspace{-1cm}
 % \begin{gathered}
  %     ds_{10,B}^2= 4\rho^2 e^{\frac{2}{3}\Phi_\mathcal{A}}f_1 \Big(dx_{1,3}^2+\frac{1}{\rho^4}d\rho^2\Big)+e^{\frac{2}{3}\Phi_\mathcal{A}}f_1\bigg[f_2d\theta^2+f_4(d\sigma^2+d\eta^2)~~~~~~~~~~~~~~~~~~~~~~~~~~~~~~~~~~~~~~~~~~~~~~~~~~~~~~~~~~~~~~~~~\\
   %  ~~~~~~~~~~~~~~~~~~~~~~~~~~~~  +\frac{1}{\Xi_1 }\bigg(f_2f_3f_5\sin^2\theta\, d\phi^2+\frac{1}{f_1^3}\,\big[d\chi -\big((\gamma \xi -\zeta )f_7 +\gamma f_8 \big)\sin\theta  d\theta\big]^2\bigg)\bigg],\\
  %   \Xi_1 = f_3f_5+f_2\big[f_3(\gamma\xi-\zeta)^2 +f_5\big(\gamma +(\gamma\xi-\zeta)f_6\big)^2\big]\sin^2\theta,\\
 % e^{2\Phi_\mathcal{B}}=\frac{1}{\Xi_1 }\Big[f_5(1+\xi f_6)^2 +\xi^2f_3+ \zeta^2 f_2 \sin^2\theta\Big]^2 ,~~~~~~~~~~~~~~~~~~~~~~~~~~~~ e^{\frac{4}{3}\Phi_\mathcal{A}}= f_1 \Big[f_5(1+\xi f_6)^2 +\xi^2f_3+ \zeta^2 f_2 \sin^2\theta\Big],
  %   \end{gathered}
   %  \end{equation}
     where for the three parameter family of solutions given in \eqref{eqn:ATD1} we have $\phi_1=\phi,~\phi_2=\chi,$\\ $g(\eta,\sigma)= -\big((\gamma \xi -\zeta )f_7 +\gamma f_8 \big)$. We now calculate the following quantities
\beq
\begin{gathered}
\text{det}[g_{ij}] = \frac{e^{\frac{10}{3}\Phi_\mathcal{A}}}{\Xi^2}f_1^2f_4^2f_2^2f_3f_5\sin^2\theta,~~~~~~~~~~\alpha =  4\rho^2 e^{\frac{2}{3}\Phi_\mathcal{A}}f_1,\\
\Rightarrow \sqrt{\text{det}[g_{ij}]e^{-4\Phi_\mathcal{B}}\alpha^3} =8\rho^3 f_1^{\frac{9}{2}}f_3^{\frac{1}{2}}f_5^{\frac{1}{2}}f_4f_2\text{Vol}(S^1)\text{Vol}(S^2),
\end{gathered}
\eeq
with all dilaton and $\Xi$ dependence dropping out neatly. This matches exactly the IIA Holographic Central charges discussed in \cite{us}, and is independent of all transformation parameters.  We hence simply quote the result derived in \cite{us},
\beq
c_{hol}=\frac{\kappa^3}{\pi^4}\sum_{k=1}^\infty P \mathcal{R}_k^2.
\eeq
This is unsurprising given the arguments presented in Section 4.3 of \cite{Macpherson:2014eza}. 
  %   \newpage
\section{Values of $f_i$ at the boundaries}\label{sec:fs}
Here we simply quote the results given in \cite{us} of $f_i$ at each boundary in turn.

\subsubsection*{At $\sigma\rightarrow \infty$}
To leading order
  \beq\label{eqn:Vlimit}
  V=-\mathcal{R}_1 e^{-\frac{\pi}{P}\sigma}\sqrt{\frac{P}{2\sigma}}\sin\bigg(\frac{2\pi}{P}\eta\bigg)+...,
  \eeq
      with 
\begin{align}\label{eqn:finf}
&f_1=\left(\frac{\pi^3 {\cal R}^2_1 \kappa^2 \sigma^2}{4 P^2}e^{-\frac{2\pi}{P}\sigma}\right)^{\frac{1}{3}},~~~f_2=\frac{2P}{\pi\sigma}\sin^2\left(\frac{\pi\eta}{P}\right),~~~f_3=4,~~~~f_4=\frac{2\pi}{P\sigma},~~~f_5=\frac{4P^2}{\pi^3 {\cal R}_1^2}e^{\frac{2\pi}{P}\sigma}\\[2mm]
&f_6=\sqrt{\frac{2}{P\sigma}}\pi {\cal R}_1 e^{-\frac{\pi}{P}\sigma}\cos\left(\frac{\pi\eta}{P}\right),~~~f_7=-2\kappa{\cal R}_1 \sqrt{\frac{2 P}{\sigma}}e^{-\frac{\pi}{P}\sigma}\sin^3\left(\frac{\pi \eta}{P}\right),~~~f_8=\kappa\left(-2 \eta+\frac{P}{\pi}\sin\left(\frac{2\pi\eta}{P}\right)\right)\nn.
\end{align}

\subsubsection*{At $\eta=0$ with $\sigma\neq 0$}
        \begin{align}\label{eqn:feq}
 &   \dot{V}'=f,~~~\dot{V}=f\eta,~~~V''=-\frac{\eta}{\sigma^2}\dot{f},~~~\ddot{V}=\eta \dot{f},~~~f(\sigma)=\frac{\pi^2}{P^2}\sum_{n=1}^\infty \mathcal{R}_n\,\sigma\,n^2K_1\Big(\frac{n\pi}{P}\sigma\Big),\nn\\[2mm]
&f_1=\left(\frac{\kappa^2 \sigma^2 f^3}{-2\dot{f}}\right)^{\frac{1}{3}},~~~f_2=\frac{-2\eta^2\dot{f}}{\sigma^2 f},~~~f_3=\frac{-4 \dot{f}}{2f- \dot{f}},~~~f_4=-\frac{2 \dot{f}}{f\sigma^2},~~~f_5=\frac{2(2f -\dot{f})}{f^3},\\[2mm]
&f_6=\frac{2 f^2}{2f-\dot{f}},~~~f_7=\frac{4\kappa\eta \dot{f}}{\sigma^2},~~~f_8=2\kappa\left(-\eta+\frac{\eta}{f}\right),~~~~~~~~~~  \text{where}~~ |\dot{f}|=-\dot{f}. \nn
\end{align}
At $\eta=P$ the behaviour is qualitatively equivalent.

\subsubsection*{At $\sigma=0,~\eta\in (k,k+1)$}
     Along the $\sigma=0$ boundary, $\ddot{V}=0$ to leading order, hence (using the boundary condition given in \eqref{eqn:BCs} in the final step), we first note
       \begin{equation}
  \frac{f_2}{f_5}\Big|_{\sigma\rightarrow0}=\frac{\dot{V}^2V''}{2\dot{V}-\ddot{V}} \Big|_{\sigma\rightarrow0} = \frac{1}{2}\dot{V}V''=\frac{1}{2}\mathcal{R}(\eta)V'',
  \end{equation}
and using \eqref{eq:alternative}, in this limit
     \begin{align}
V''&= P_k \label{eqPdef}\\[2mm]
P_k&= \sum_{j=k+1}^P\frac{b_j}{j-\eta}+\frac{1}{2P}\sum_{j=1}^Pb_j\left(\psi\left(\frac{\eta+j}{2P}\right)-\psi\left(\frac{\eta-j}{2P}\right)+\frac{\pi}{2}\left(\cot\left(\frac{\pi(\eta+j)}{2P}\right)-\cot\left(\frac{\pi(\eta-j)}{2P}\right)\right)\right),\nn
\end{align}
with $\psi$ the digamma function. This doesn't vanish or blow up between these bounds. One finds
\begin{align}\label{eqn:fsat0}
f_1&=\left(\frac{\kappa^{2}{\cal R}(2 {\cal R}P_k +({\cal R}')^2)}{2 P_k}\right)^{\frac{1}{3}},~~~f_2=\frac{2 {\cal R} P_k}{2 {\cal R} P_k+({\cal R}')^2},~~~f_3= \frac {2 \sigma^2 P_k}{\cal R},~~~f_4=\frac{ 2 P_k}{{\cal R}},\\[2mm]
f_5&=\frac{4}{2{\cal R}P_k+({\cal R}')^2},~~~f_6={\cal R}',~~~f_7=-\frac{4\kappa {\cal R}^2 P_k}{2{\cal R} P_k+({\cal R}')^2},~~~f_8=2\kappa\left(-\eta+\frac{{\cal R}{\cal R}'}{2{\cal R}P_k+({\cal R}')^2}\right),\nn
\end{align}
recalling ${\cal R}=N_k+(N_{k+1}-N_{k})(\eta-k)$.

           \subsubsection*{At $\sigma=0,~\eta=0$} To approach this boundary, adopt the coordinate change $(\eta=r \cos\alpha,~\sigma=r \sin\alpha)$, expanding about $r=0$. To leading order
  \begin{align}\label{eqQdef}
&\dot{V}=N_1 r \cos\alpha,~~~~~~\dot{V}'= N_1,~~~V''=\frac{1}{4P^2}r\cos\alpha\sum_{j=1}^Pb_k\left(2 \psi^1\left(\frac{j}{2P}\right)-\pi^2\csc^2\left(\frac{j\pi}{2P}\right)\right),\nn,\\[2mm]
&f_1=\left(\frac{\kappa^2 N_1^3 }{2 Q}\right)^{\frac{1}{3}},~~~f_2=\frac{2 r^2 Q \cos^2\alpha }{N_1},~~~f_3= \frac{2 r^2 Q \sin^2\alpha }{N_1},~~~f_4=\frac{2 Q}{N_1},\\[2mm]
&f_5= \frac{4}{N_1^2},~~~f_6= N_1,~~~f_7=-4\kappa Q r^3 \cos^3\alpha,~~~f_8=0,\nn
\end{align}
   where $Q$ is extracted via $V''=r Q\,\cos\alpha$. Notice that $f_5,f_6$ remain finite whereas $f_3$ and $f_2$ vanish.
     
          \subsubsection*{At $\sigma=0,~\eta=k$}
           Now one should make the following coordinate change $(\eta=k-r \cos\alpha,~\sigma=r \sin\alpha)$ (where $0<k<P$ and $k\in \mathds{Z}$). To leading order  
\begin{align}\label{eqn:fsfork}
&f_1=(\kappa N_k)^{\frac{2}{3}},~~~f_2=1,~~~f_3=\frac{r^2\sin^2\alpha}{N_k}\frac{b_k}{r},~~~f_4=\frac{1}{N_k}\frac{b_k}{r},~~~f_5=\frac{4}{N_k}\frac{r}{b_k},~~~f_7=-2\kappa N_k,~~~f_8=-2\kappa k.\nn\\[2mm]
&f_6=\frac{b_k}{2}(1+\cos\alpha)+N_{k+1}-N_k=\cos^2\Big(\frac{\alpha}{2}\Big) (N_k-N_{k-1})+\sin^2\Big(\frac{\alpha}{2}\Big) (N_{k+1}-N_{k}) \equiv g(\alpha).
\end{align} 
noting the use of $b_k=2N_k-N_{k+1}-N_{k-1}$ to re-write $f_6$.

\end{document}